\documentclass[longbibliography,aps,rmp,reprint,amsmath,amssymb,floatfix]{revtex4-1}

\usepackage{graphicx}
\usepackage{bm}
\usepackage[latin1]{inputenc}
\usepackage{amstext}
\usepackage{amsxtra}
\newcommand{\bs}{\boldsymbol}
\newcommand{\be}{\begin{equation}}
\newcommand{\ee}{\end{equation}}
\newcommand{\bpm}{\begin{pmatrix}}
\newcommand{\epm}{\end{pmatrix}}
\newcommand{\bea}{\begin{eqnarray}}
\newcommand{\eea}{\end{eqnarray}}

\newcommand{\D}{{\rm d}}
\newcommand{\I}{{\rm i}}
\newcommand{\E}{{\rm e}}
\newcommand{\ie}{{\it i.e.\ }}
\newcommand{\eg}{{\it e.g.\ }}
\newcommand{\vv}[1]{{\bm #1}}
\newcommand{\deq}{=}

\newcommand{\bra}[1]{\left< #1\right|}
\newcommand{\ket}[1]{\left|#1\right>}

\usepackage[normalem]{ulem}
\usepackage{tikz}
\usepackage{tikzscale}
\usetikzlibrary{arrows.meta}
\usepackage{color}

\usepackage{hyperref}
\hypersetup{
    bookmarks=false,         
    unicode=false,          
    pdftoolbar=false,        
    pdfmenubar=true,        
    pdffitwindow=false,     
    pdfstartview={FitH},    
    pdftitle={Topological Bands for Ultracold Atoms},    
    pdfauthor={N. R. Cooper, J. Dalibard, I. B. Spielman },     
    pdfsubject={},   
    pdfcreator={},   
    pdfproducer={}, 
    pdfkeywords={topology} {optical lattices} {cold atoms}, 
    pdfnewwindow=true,      
    colorlinks=true,       
    linkcolor=black,          
    citecolor=blue,        
    filecolor=magenta,      
    urlcolor=blue           
}

\begin{document}

\title{Topological Bands for Ultracold Atoms}
\date{\today}
\author{N. R. Cooper}
\affiliation{T.C.M. Group, Cavendish Laboratory, J.~J.~Thomson Avenue, Cambridge CB3~0HE, United Kingdom}
\author{J. Dalibard}
\affiliation{Laboratoire Kastler Brossel, Coll{\`e}ge de France, CNRS, ENS-Universit{\'e} PSL, Sorbonne Universit{\'e}, 11 place Marcelin Berthelot, 75005, Paris, France}
\author{I. B. Spielman}
\affiliation{Joint Quantum Institute, National Institute of Standards and Technology and University of Maryland, Gaithersburg, Maryland, 20899, USA}

\begin{abstract}

There have been significant recent advances in realizing band structures with geometrical and topological features in experiments on cold atomic gases. We provide an overview of these developments, beginning with a summary of the key concepts of geometry and topology for Bloch bands. We describe the different methods that have been used to generate these novel band structures for cold atoms, as well as the physical observables that have allowed their characterization. We focus on the physical principles that underlie the different experimental approaches, providing a conceptual framework within which to view these developments. However, we also describe how specific experimental implementations can influence physical properties. Moving beyond single-particle effects, we describe the forms of inter-particle interactions that emerge when atoms are subjected to these energy bands, and some of the many-body phases that may be sought in future experiments.

\end{abstract}

\maketitle

\tableofcontents

\section{Introduction}

Topology is a mathematical concept that refers to certain
  properties that are preserved under continuous deformations. One
  familiar example is the number of twists put into a belt before its
  buckle is fastened. Usually we aim to fasten a belt without any
  twists. But if we were to introduce a single twist we would produce
  a M{\" o}bius strip. No continuous deformation of the closed belt
    would get rid of this uncomfortable twist. The number of twists is
    said to be a ``topological invariant'' of the closed belt.

The importance of topological invariants in stabilizing spatial deformations and defects is also well known in  physics, in diverse areas ranging from cosmology to condensed matter.
For a superfluid confined to a ring, 
the number of times that the superfluid phase $\phi$ changes by $2\pi$ around the ring
\begin{equation}
N = \frac{1}{2\pi} \oint \vv{\nabla}\phi \cdot \D\vv{l}
\label{eq:sfwinding}
\end{equation}
is a topological invariant.  This winding number cannot change under smooth deformations of the superfluid: a change would require the superfluid density to vanish somewhere -- such that $\phi$ is ill-defined -- requiring the motion of a quantized vortex line across the ring. Here, the topological stability arises from the interplay of the underlying space (the ring) and the form of the local order parameter (the phase of the superfluid wavefunction).
  
In recent years it has come to be understood that topology enters
physics in another, very fundamental way, through the nature of the
quantum states of particles moving through crystalline lattices.  The
energy eigenstates of electrons moving through periodic potentials are
well-known to form energy bands, a result that follows from the
existence of a conserved crystal momentum via Bloch's theorem.
Remarkably, under certain circumstances, each
Bloch energy band can be assigned a robust integer-valued topological invariant related to
how the quantum wavefunction of the electron ``twists'' as a function
of crystal momentum.
This integer is invariant under continuous
changes of material properties: ``continuous'' meaning that the energy
gaps to other bands should not close. The first example of such a
topological invariant for Bloch energy bands arose from a highly
original analysis of the integer quantum Hall effect in a
two-dimensional (2D) lattice~\cite{thoulesschern}. 
Recent
theoretical breakthroughs have shown that, once additional
symmetries are included, topological invariants are much more widespread. These
ideas underpin a recent revolution in our understanding of
band insulators and superconductors~\cite{hasankane,qizhang}. 
{The topological nature of these materials endows them with physical characteristics that are insensitive to microscopic details, a notable example being the exact quantization of the Hall resistance in 2D irrespective of the presence or form of a random disorder potential.}

A great deal of current research focuses on understanding the
physical consequences of these new materials, and experimental studies
of topological insulators and superconductors in solid state systems
continue apace. Furthermore, there is significant activity in
exploring the nature of the strongly correlated phases of matter that
arise in these materials, notably to construct strong-correlation
variants of these topological states of weakly interacting electrons. Theory suggests many interesting
possibilities, which are still seeking experimental realization and verification. 

Such questions are ideally placed to be addressed using realizations with cold atomic gases.
Cold atomic gases allow strongly
interacting phases of matter to be explored in controlled experimental
settings.  However, a prerequisite for quantum simulations of such
issues is the ability to generate topological energy bands for cold
atoms. This poses a significant challenge, even at this single particle level. Realizing topological
energy bands typically requires either the introduction of effective
orbital magnetic fields acting on neutral atoms and/or the
introduction of a spin-orbit coupling between the internal spin states
of an atom and its center-of-mass motion.  This is an area of research
that has attracted significant attention over the last years, both
theoretical and experimental. Much progress has been made in developing techniques to generate
artificial magnetic fields and spin-orbit coupling for neutral
atoms~\cite{Dalibard:2016,Aidelsburger:2017,zhaireview}. The use of these techniques in the setting of optical lattices has led to the realization and characterization of topological Bloch bands for cold atoms.

In this review we describe how topological energy bands can be
generated and probed in cold atom systems.
We focus on
existing experimental studies, for which the essential behavior can
be understood in terms of non-interacting particles.
We start by 
explaining the concepts underpinning topological energy bands in Sec.~\ref{sec:topology}. We describe 
the key
physical effects that are required to generate these bands and how
these can be engineered in cold atom systems in Sec.~\ref{sec:implementations}.
Our  emphasis is on recent experimental developments. In Sec.~\ref{sec:experimental} 
we describe the principal
observables that have been used to characterize the geometrical and topological characters of the resulting energy
bands. In Sec.~\ref{sec:interactions} we move beyond single-particle physics,
to discuss some of the theoretical
understanding of the consequences of interactions in these novel
optical lattices, and to describe some of the interacting many-body
phases that can be sought in future experiments on these systems.
We conclude in Sec.~\ref{sec:outlook} with 
comments on the outlook for future work and point out connections 
 to broader research areas.
 
Throughout this review we explain just the essential physics underlying recent developments, so the content is necessarily incomplete. The review should be used as a starting point from which to explore the literature, rather than as a comprehensive survey of the field. We note that we focus on topological bands for atoms in periodic lattices. Related phenomena can appear for photons and hybrid light-matter objects (cavity polaritons) in novel optical materials, and we refer the interested readers to the review by \textcite{photonicreview}. We note also that all our discussions are based on the notion of band structure, meaning that we restrict to periodic materials and that we do not address the case of atoms or photons moving in quasiperiodic systems~\cite{Kunz1986,Kraus2012,tanese2014fractal,Dareau2017}.

\section{Topology of Bloch Bands}

\label{sec:topology}

In this section we provide an introduction to how  topology enters into band theory for particles moving in periodic potentials. We focus on quantities and models that are relevant to later sections. More comprehensive accounts are available in review articles on topological insulators~\cite{hasankane,qizhang,RevModPhys.88.035005}.

\subsection{Band theory}

Here we describe the essential properties of a quantum particle moving through a periodic lattice potential.  Our aim is to understand both the resulting eigenenergies (giving the conventional band structure) along with the underlying structure of the eigenstates (from which the band topology is derived).

Any lattice potential is invariant under displacements by a set of lattice vectors $\vv{R}$, which can be written  $\vv{R} = \sum_i m_i \vv{a}_i$ in terms of integer $\{m_i\}$ multiples of basis vectors $\{\vv{a}_i\}$ (with $i=1,\ldots d$ in  dimension $d$).  From Bloch's theorem, the energy eigenstates take the form
\begin{equation}
\psi^{(n)}_\vv{q}(\vv{r})  =   e^{\I\vv{q}\cdot \vv{r}} u^{(n)}_\vv{q}(\vv{r}) \,,
\end{equation}
with $u^{(n)}_{\vv{q}}(\vv{r}) = u^{(n)}_{\vv{q}}(\vv{r}+\vv{R})$ having the periodicity of the lattice. These eigenstates are characterized by the band index, $n$, and the crystal momentum, $\vv{q}$. Although we shall retain $\hbar$ in equations, we shall refer to momentum and wavevector interchangeably.
Starting from the Schr{\" o}dinger equation for the energy eigenstate $\psi^{(n)}_\vv{q}(\vv{r})$
\begin{eqnarray}
\left[-\frac{1}{2m}\vv{\nabla}^2 + V(\vv{r})\right]\psi^{(n)}_{\vv{q}}(\vv{r}) & = & E^{(n)}_\vv{q}\psi^{(n)}_{\vv{q}}(\vv{r})\,,
\end{eqnarray}
one readily finds that the Bloch states $u^n_\vv{q}(\vv{r})$ are eigenstates of the $\vv{q}$-dependent Hamiltonian 
\begin{equation}
\hat{H}_{\vv q} = \left[-\frac{1}{2m}\left(\vv{\nabla}+\I\vv{q}\right)^2 + V(\vv{r})\right] \,.
\end{equation}
 If there are internal degrees of freedom, \eg spin, then one should replace $u^n_\vv{q}(\vv{r})$ by $u^n_{\vv{q}}(\alpha,\vv{r})$ with $\alpha$ labelling these additional degrees of freedom.

States that differ in crystal momentum $\vv{q}$ by a reciprocal lattice vector, $\vv{G}= \sum_i m_i \vv{G}_i$, are physically equivalent. (The reciprocal lattice is constructed from basis vectors  $\{\vv{G}_i\}$  defined by the condition  $\vv{G}_i \cdot \vv{a}_j = 2\pi \delta_{ij}$.) Thus, $\vv{q}$ can be chosen to be restricted to the first Brillouin zone (BZ): the locus of points $\vv{q}$ that are closer to the origin than to any reciprocal lattice vector $\vv{G}$.  For example, for a 2D lattice with basis vectors $\vv{a}_1=(a_1,0)$ and $\vv{a}_2=(0,a_2)$ the reciprocal lattice has basis vectors $\vv{G}_1 = (2\pi/a_1,0)$ and  $\vv{G}_2 = (0,2\pi/a_2)$, and the BZ is the region $-\pi < q_xa_1\leq \pi$ and $-\pi < q_ya_2\leq \pi$. 

Many of the cases we shall describe will be tight-binding models, for which there are $M$ single-particle orbitals within each unit cell of the lattice. These could be multiple lattice sites and/or other internal degrees of freedom. The system Hamiltonian $\hat{H}$ acts on the set of $M\times N_c$ single-particle states $|\alpha,\vv{R}\rangle = |\alpha\rangle\otimes|\vv{R}\rangle$ with $\alpha=1,\ldots M$ labeling the states within a unit cell and $\vv{R}$ the position of each of a total of $N_c$ unit cells. 
For periodic boundary conditions, the energy eigenstates are plane waves
\begin{equation}
|\psi_\vv{q}^{(n)}\rangle =  \frac{1}{\sqrt{N_c}}
\sum_{\vv{R}}
\E^{\I \vv{q}\cdot \vv{R}} 
|u^{(n)}_{\vv{q}}\rangle\otimes |\vv{R}
\rangle\,,
\end{equation}
where $n=1,\ldots M$ is the band index and $|u^{(n)}_{\vv{q}}\rangle$ are the associated Bloch wavefunctions in the $M$-dimensional internal space. The band energies and Bloch wavefunctions follow from the spectrum of a wave-vector-dependent Hamiltonian, $\hat{H}_\vv{q}$ that acts in this internal space. For example, $\hat{H}_\vv{q}$  can be represented by the $M\times M$ matrix 
\begin{equation}
H^{\alpha\beta}_{\vv{q}} = \frac{1}{N_c} \sum_{\vv{R},\vv{R}'}
\E^{\I \vv{q}\cdot (\vv{R}-\vv{R}')} \langle \alpha,\vv{R}'|\hat{H} |\beta, \vv{R}\rangle \,.
\end{equation}
Note that the Bloch Hamiltonian, $\hat{H}_\vv{q}$, and thus its matrix representation, is not uniquely defined: it can be replaced by 
$\hat{H}_{\vv{q}}' = \hat{\cal U}_\vv{q}\hat{H}_\vv{q}\hat{\cal U}^\dag_\vv{q}$, where $\hat{\cal U}_\vv{q}$ is a wavevector-dependent unitary transformation.  Because the states at wavevectors  $\vv{q}$ and $\vv{q}+\vv{G}$ are physically equivalent, one can always choose the Hamiltonian to have the same periodicity as the BZ, $\hat{H}_\vv{q} = \hat{H}_{\vv{q}+\vv{G}}$. We shall make this choice throughout this section. However, as will be explained later, some care is required in doing so when considering geometrical properties of the bands. Other choices are discussed in Appendix~\ref{subsubsec:gaugeinvariancezak}.

\subsection{Geometrical phase}

At its most basic level, the band structure of a periodic potential is specified by the energies $E^{(n)}_\vv{q}$.  However, the bands are also characterized by geometrical and topological features. These relate to how the wavefunctions $|u^{(n)}_\vv{q}\rangle$ vary with wavevector $\vv{q}$ across the BZ.  No variations can occur for simple tight-binding models with one orbital per unit cell $M=1$ (\ie one-band models). However, topological features arise already for  two orbitals per unit cell, $M=2$. We shall describe topological classifications of Bloch bands in 1D and 2D, and illustrate these using two-band tight-binding models.

Our presentation will rely heavily on the concept of the geometrical phase~\cite{Berry}, which we briefly review.
Consider a Hamiltonian $\hat{\mathcal{H}}(\vv{X})$ which depends on a set of parameters $\vv{X}$,
and with nondegenerate spectrum
\begin{equation}
\hat{\mathcal{H}}(\vv{X}) |\Psi^{(n)}(\vv{X})\rangle  = E^{(n)}(\vv{X}) |\Psi^{(n)}(\vv{X})\rangle \,.
\end{equation}
The system is prepared in an eigenstate $|\Psi^{(n)}(\vv{X})\rangle$ at an initial time $t_{\rm i}$, and the parameters $\vv{X}_t$ are changed slowly in time $t$ such that the state   evolves adiabatically,  following an instantaneous  eigenstate of $\hat{\mathcal{H}}(\vv{X}_t)$. The parameters are taken around a cycle such that 
$\vv{X}_{t_{\rm f}} = \vv{X}_{t_{\rm i}}$. Since the Hamiltonian returns to the initial form at $t_{\rm f}$,  so too must any eigenstate up to an overall phase factor (in this case of a nondegenerate spectrum). This phase has both dynamical and geometrical contributions
\begin{eqnarray}
|\Psi^{(n)}_{\rm f}\rangle & = & \E^{\I [\gamma^{(n)}_{\rm dyn}+\gamma^{(n)}_{\rm geo}]} |\Psi^{(n)}_{\rm i}\rangle
\\
\gamma^{(n)}_{\rm dyn} & = & -\frac{1}{\hbar}\int_{t_{\rm i}}^{t_{\rm f}} E^{(n)}(\vv{X}_{t'}) \D t'\\
\gamma^{(n)}_{\rm geo}  & = & \oint   \I \langle \Psi^{(n)} | \partial_\vv{X} \Psi^{(n)}\rangle \cdot \D\vv{X} \,.
\label{eq:geometricphase}
\end{eqnarray}
The geometrical phase is the integral of the {\it Berry connection}
\begin{equation}
\vv{A}^{(n)}(\vv{X}) \equiv \I \langle \Psi^{(n)} | \partial_\vv{X} \Psi^{(n)}\rangle 
\end{equation}
around the closed loop in parameter space $\vv{X}$. 
The Berry connection plays a role similar to that of the vector potential for a magnetic field. It is gauge-dependent,  varying under local gauge transformations, $|\Psi^{(n)}\rangle \to \E^{\I \Phi(\vv{X})}|\Psi^{(n)}\rangle$.
However, if $\vv{X}$ has more than one component, 
one can define a gauge-invariant {\it Berry curvature}
\begin{equation}
\Omega_{ij}^{(n)} \equiv  \partial_{X_i} A^{(n)}_j  -  \partial_{X_j} A^{(n)}_i \,.
\end{equation}
If the closed loop $\vv{X}$ can be viewed as the boundary of a 2D surface $(X_1,X_2)$ on which the Berry curvature is everywhere well-defined, then the geometrical phase (\ref{eq:geometricphase})
is the flux of the Berry curvature 
\begin{equation}
\Omega^{(n)} \equiv \epsilon_{ij} \partial_{X_i} A^{(n)}_j = \epsilon_{ij} \I \partial_{X_i} \langle \Psi^{(n)} | \partial_{X_j} \Psi^{(n)}\rangle 
\label{eq:berrycurvaturedefine}
\end{equation}
through this 2D surface. Here $\epsilon_{ij}$ is the antisymmetric tensor of two indices, $\epsilon_{xy}=-\epsilon_{yx}=1$, and summation over repeated indices is assumed.

We shall apply these concepts to physical situations in which the role of external parameters $\vv{X}$ is played either by the crystal momentum $\vv{q}$ or by the real-space position $\vv{r}$. In both cases, the Berry connection and Berry curvature will define local {\it geometric} properties of the quantum states. The integrals of these geometric quantities over a closed manifold -- the BZ for $\vv{q}$, or the unit cell of the lattice for $\vv{r}$ -- will give rise to {\it topological} properties. As discussed in later sections, trajectories of $\vv{q}$ can be imposed in physical systems by the application of external forces, inducing adiabatic dynamics of the Bloch states. 

\subsection{Topological invariants}
\label{subsec:topologicalinvariant}

\subsubsection{The Zak phase}
\label{sec:zak_phase}

Owing to the periodicity of the BZ  under the addition of   any reciprocal lattice vector $\vv{G}$, a trajectory in wavevector from $\vv{q}_{\rm i}$ to $\vv{q}_{\rm f}=\vv{q}_{\rm i}+\vv{G}$ is a closed loop. Since the Hamiltonian is periodic, $\hat{H}_\vv{q}= \hat{H}_{\vv{q}+\vv{G}}$, the ideas of Berry apply directly. 
The integral of the Berry connection along such closed loops 
\begin{equation}
\label{eq:zak}
\phi^{(n)}_{\rm Zak} =  \int_{\vv{q}_{\rm i}}^{\vv{q}_i+\vv{G}}   \I \langle u^{(n)} | \partial_\vv{q} u^{(n)}\rangle \cdot \D\vv{q}
\end{equation}
was proposed by  Zak as a way to characterize the energy bands~\cite{Zak:1989}.  
 
A first glimpse of how the mathematics of topological invariants can arise
in band theory is provided by computing the Zak phase for simple
two-band tight-binding models in one dimension (1D). We shall
illustrate this for the Su-Schrieffer-Heeger (SSH) model.

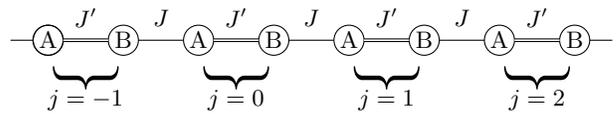
\begin{figure}[t]
\begin{center}
\begin{tikzpicture}
\draw (0,0) circle (0.2) ;
\draw (0,0) circle (0.2) ;
\draw (1,0) circle (0.2) ;
\draw (2,0) circle (0.2) ;
\draw (3,0) circle (0.2) ;
\draw (4,0) circle (0.2) ;
\draw (5,0) circle (0.2) ;
\draw (6,0) circle (0.2) ;
\draw (7,0) circle (0.2) ;
\draw (0,0) node{A};
\draw (1,0) node{B};
\draw (2,0) node{A};
\draw (3,0) node{B};
\draw (4,0) node{A};
\draw (5,0) node{B};
\draw (6,0) node{A};
\draw (7,0) node{B};
\draw (-0.2,0) -- (-0.5,0) ;
\draw [double] (0.2,0) -- (0.8,0) ;
\draw              (1.2,0) -- (1.8,0) ;
\draw [double] (2.2,0) -- (2.8,0) ;
\draw               (3.2,0) -- (3.8,0) ;
\draw [double] (4.2,0) -- (4.8,0) ;
\draw               (5.2,0) -- (5.8,0) ;
\draw [double] (6.2,0) -- (6.8,0) ;
\draw               (7.2,0) -- (7.5,0) ;
\draw (0.5,0.3) node{$J'$};
\draw (1.5,0.3) node{$J$};
\draw (2.5,0.3) node{$J'$};
\draw (3.5,0.3) node{$J$};
\draw (4.5,0.3) node{$J'$};
\draw (5.5,0.3) node{$J$};
\draw (6.5,0.3) node{$J'$};
\draw (0.5,-0.5) node[rotate=90] {\Huge \{};
\draw (2.5,-0.5) node[rotate=90] {\Huge \{};
\draw (4.5,-0.5) node[rotate=90] {\Huge \{};
\draw (6.5,-0.5) node[rotate=90] {\Huge \{};
\draw (0.5,-0.8) node{$j=-1$};
\draw (2.5,-0.8) node{$j=0$};
\draw (4.5,-0.8) node{$j=1$};
\draw (6.5,-0.8) node{$j=2$};
\end{tikzpicture}
\end{center}
\caption{Tight-binding (SSH) model of the polyacetylene molecule. For the SSH model (\ref{eq:hssh}), the onsite energies for A and B are supposed to be equal. This constraint will be relaxed for the Rice--Mele model (\ref{eq:hrm1}).}
\label{fig:sshmodel1}
\end{figure}

The SSH model is a tight-binding model in which there are
two sites in the unit cell, which we label A and B. The sites are connected by alternating tunnel couplings, $J$ and $J'$, see Fig.~\ref{fig:sshmodel1}. 
The single-particle Hamiltonian reads
\begin{equation}
\hat H=-\sum_j \left(  J'\, \hat a_j^\dagger \hat b_j +J\, \hat a_j^\dagger \hat b_{j-1}+ \mbox{H.c.}\right)\,,
\label{eq:SSH_Hamiltonian}
\end{equation}
where $\hat a_j^\dagger$ and $\hat b_j^\dagger$ create a particle on the A and B sites of the $j$-th unit cell. 
For a bulk system with periodic boundary conditions, we look for the energy eigenstates using the Bloch wave form
\begin{equation}
|\psi_q\rangle =\sum_j \E^{\I j a q} \left( u_q^{\rm A} |{\rm A}_j\rangle + u_q^{\rm B} |{\rm B}_j\rangle \right)\,,
\label{eq:blochwaveform}
\end{equation}
with $a$ the lattice constant. The problem reduces 
to finding the eigenvectors, $(u^{\rm A}_q,u^{\rm B}_q)^{\rm T}$, of the Hamiltonian in reciprocal space, which has matrix representation
\begin{equation}
\hat{H} _q \deq - \begin{pmatrix} 0 &J'+J\E^{-\I qa}\\ J'+J\E^{\I qa} & 0 \end{pmatrix}
\label{eq:hssh}
\end{equation}
within the BZ $-\pi/a < q \leq \pi/a$. 

The Hamiltonian (\ref{eq:hssh})
 can be written in the general form 
\begin{equation}
\label{eq:sshham1}
  \hat{H}_q = - \vv{h}(q)\cdot \hat{\vv{\sigma}},
\end{equation}
where $\hat{\sigma}_{\alpha}$ are the Pauli operators. Throughout this review, we shall use the conventional matrix representation of the Pauli operators in which $\hat{\sigma}_z$ is diagonal. Then, the SSH model has
\begin{equation}
h_x(q)+\I h_y(q)=J'+J\E^{\I qa}, \qquad h_z=0\,,
\label{eq:sshham21}
\end{equation}
with $h_{x,y}(q)$ real periodic functions of $q$. The energy spectrum is composed of the two bands 
\begin{equation}
E_q^{(\pm)} = \pm |\vv{h}(q)| =\pm \left[ J^2+J'^2+2JJ'\cos(qa)\right]^{1/2},
\label{eq:energies_SSH}
\end{equation}
which are separated by a gap provided $|\vv{h}|$ does not vanish at any $q$. Assuming $J,J'>0$, the gap $2|J-J'|$ closes only when $J=J'$, for the quasi-momentum $q=\pi/a$. 

Provided there is a non-zero gap, \ie $|\vv{h}|\neq 0$ for all $q$, one can write $h_x+\I h_y = |\vv{h}| \E^{\I\phi_q} $ with a well-defined $\phi_q$, and the Bloch states are 
\begin{equation}
\begin{pmatrix}
{u}^{\rm A}_q\\
{u}^{\rm B}_q
\end{pmatrix}
=\frac{1}{\sqrt{2}}\left(\begin{array}{c} 1\\ \mp \E^{\I\phi_q}\end{array}\right).
\label{eq:1dwavefunction1}
\end{equation}
In this pseudospin representation, the fact that $h_z=0$ in (\ref{eq:sshham1}-\ref{eq:sshham21}) for the SSH model entails that these eigenstates lie on the equator of the Bloch sphere. 
The resulting Zak phases of the bands (\ref{eq:zak}) are
\begin{eqnarray} 
\phi^{(\pm)}_{\rm Zak} &  = &  - \frac{1}{2}\int_{\rm BZ} \frac{\partial \phi_q}{\partial q} \; \D q\,.
 \label{eq:zakphasewinding1}
 \end{eqnarray}
Thus, 
\begin{eqnarray}
 N & \equiv  & -\frac{1}{\pi} 
\phi^{(\pm)}_{\rm Zak} =
 \frac{1}{2\pi} \int_{\rm BZ} \frac{\partial\phi_q}{\partial q} \D q
 \label{eq:windingnumber1}
 \end{eqnarray}
 is the number of times  $\phi_q$ changes by $2\pi$ as $q$ runs over the BZ. Since the Hamiltonian is periodic, $\phi_q = \phi_{q+G}$ modulo $2\pi$,  $N$ is an integer winding number, analogous to that for  the phase of a superfluid around a ring (\ref{eq:sfwinding}). It measures the solid angle drawn by the pseudo-spins $u_q^{(\pm)}$ along the equator of the Bloch sphere when $q$ spans the BZ.  

\begin{figure}
\begin{center}
\includegraphics{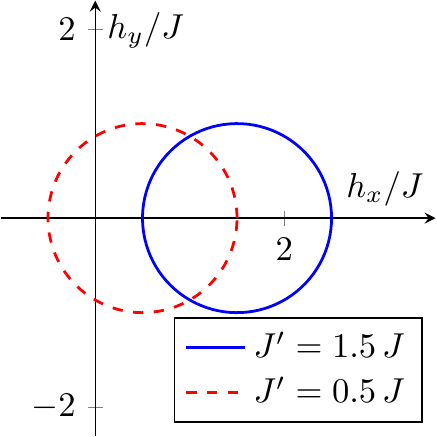}
\end{center}
\caption{Illustration of the winding number for the SSH model. The  curves plot the locus of $(h_x,h_y)$ as $q$ runs over the BZ.  For $J'/J>1$ the  vector $(h_x,h_y)$ does not encircle the origin, but for $J'/J<1$ it 
encircles the origin once, indicating that these two cases have winding numbers (\ref{eq:windingnumber1}) of $N=0$ and $N=1$ respectively.}
\label{fig:sshwinding1}
\end{figure}

The winding number (\ref{eq:windingnumber1}) is a topological
invariant of 1D band insulators arising from Hamiltonians of the form
(\ref{eq:sshham1}): $N$ cannot be changed unless $|\vv{h}(q)|$
vanishes at some $q$, \ie unless the band-gap closes.  The case of
the SSH model is illustrated in Figure~\ref{fig:sshwinding1}, which
shows the locus of $(h_x,h_y)$ as $q$ runs over the BZ. For $J'/J<1$
the curve encircles the origin once, $N=1$; while for $J'/J>1$ the
curve does not encircle the origin, $N=0$. These two curves cannot be
smoothly interconverted without crossing $h_x=h_y=0$, \ie without the
bandgap closing. One could evade this conclusion by including terms proportional to 
$\hat{\sigma}_z$ in the Hamiltonian.  Then, two gapped states with different
$N$ could be continuously deformed into each other. (We shall explore this below for the Rice--Mele model.)
However, as explained in Sec.~\ref{subsec:topoinsulator},
including $\hat{\sigma}_z$ terms would break an underlying ``chiral'' symmetry of the SSH model.  The winding number in 1D is an example of a topological
invariant whose existence relies on  an underlying symmetry.

A perceptive reader will notice that the topological invariant we
have constructed (\ref{eq:windingnumber1}) appears to be rather
unphysical. The two parameter regimes of the SSH model which have
different winding numbers, at $J'/J>1$ and $J'/J<1$, could be trivially
related by reversing the labelling of the A and B sublattices in Fig.~\ref{fig:sshmodel1}, at least deep in the bulk of the system. 
In what sense are
these two parameter regimes topologically distinct?  A key point to note is
that the winding number (\ref{eq:windingnumber1}) has a basis-dependent offset. 
In place of (\ref{eq:blochwaveform}) we could just as well have sought an energy eigenstate of the form 
\begin{eqnarray}
|\psi_q\rangle & = & \sum_j \E^{\I j a q} \left( \tilde{u}_q^{\rm A} |{\rm A}_{j+N_{\rm A}}\rangle + \tilde{u}_q^{\rm B} |{\rm B}_{j+N_{\rm B}}\rangle \right)
\label{eq:Zak_gauge_change1}
\\
 & = & \sum_j \E^{\I j a q} \left( \tilde{u}_q^{\rm A} \E^{-\I N_{\rm A} aq}|{\rm A}_j\rangle + \tilde{u}_q^{\rm B} \E^{-\I N_B aq} |{\rm B}_j\rangle \right)\,,
\label{eq:Zak_gauge_change}
\end{eqnarray}
associating site A in the cell $j+N_{\rm A}$ with site B in cell $j+N_{\rm B}$. Comparing (\ref{eq:Zak_gauge_change}) with (\ref{eq:blochwaveform}) shows that this amounts to replacing $(u^{\rm A}_q,u^{\rm B}_q)^{\rm T}$ by $(\tilde{u}^{\rm A}_q \E^{-\I N_{\rm A} aq},\tilde{u}^{\rm B}_q \E^{-\I N_{\rm B} aq})^{\rm T}$, \ie to a $q$-dependent unitary transformation, $\hat{\cal U}_q$.
In this new basis, $(\tilde{u}^{\rm A}_q,\tilde{u}^{\rm B}_q )^{\rm T}$, the eigenstates (\ref{eq:1dwavefunction1}) are replaced by
\begin{equation}
\begin{pmatrix}
\tilde{u}^{\rm A}_q\\
\tilde{u}^{\rm B}_q
\end{pmatrix}
= \frac{1}{\sqrt{2}}\left(\begin{array}{c} \E^{\I q a N_{\rm A}}\\ \mp \E^{\I\phi_q}\E^{\I q a N_{\rm B}}\end{array}\right)\,.
\label{eq:1dwavefunction-gauge}
\end{equation}
For $N_{\rm A}=N_{\rm B}$ this transformation may be viewed as a reciprocal-space gauge transformation. A direct calculation of the  Zak phase  for (\ref{eq:1dwavefunction-gauge}) shows that the winding number, $N \equiv -\phi_{\rm Zak}/\pi$, is increased by $(N_{\rm A}+N_{\rm B})$ as compared to that for (\ref{eq:1dwavefunction1}).
This example shows that the absolute value of the winding number cannot be physically meaningful. 
Instead, physical
consequences can only involve {\it differences}  of winding numbers, 
which are well-defined when computed within the same reciprocal-space basis choice. In Sec.~\ref{subsec:edge} we shall describe how the winding number
difference at a boundary between two regions influences the
spectrum of states on the edge.  However, note that we have already
found one physical consequence of the winding number difference: the
parameter space of microscopic couplings becomes disconnected, in the
sense that there is no way to continuously change the physical parameters to evolve between regions of 
different winding numbers without crossing a critical point at which the band-gap closes.  
Such gap closings between topologically distinct regions have direct physical consequences in measurements of the bulk excitation spectrum. In particular, for a set of non-interacting fermions filling one such band, the gap closing implies a  thermodynamic phase transition between two insulating regimes, separated by a semi-metallic state.
We have illustrated this gap closing for the SSH model.
  However, this holds also for
any 1D model in which $\vv{h} = (h_x,h_y,0)$, \ie with chiral symmetry.
The requirement of the closing of a band-gap in order to
change the winding number is a defining feature of a topological
invariant of the energy band.

\subsubsection{The Chern number}
\label{subsubsec:The Chern number}

A topological classification of (non-degenerate) energy bands in 2D exists without requiring any underlying symmetry, and without ambiguities related to the choice of basis in reciprocal space. The topological invariant is the Chern number\begin{equation}
 {\cal C} = \frac{1}{2\pi} \int_{\rm BZ}
 \I   \epsilon_{ij}\partial_{q_i} \langle u_\vv{q} | \partial_{q_j} |u_\vv{q}\rangle 
  \; \D^2q \,.
 \label{eq:chern}
 \end{equation}
 We suppress the band index, $n$, but note that a Chern number exists for each band.
This topological invariant, and terminology,  arises from the mathematics of fibre bundles~\cite{stonegoldbart}. However, it can be readily interpreted more physically in terms of 
the Berry curvature (\ref{eq:berrycurvaturedefine})  of the Bloch states
\begin{equation}
\label{eq:berrycurvature}
\Omega(\vv{q})  \equiv  \vv{\nabla}_\vv{q}\times \vv{A}(\vv{q}) \cdot \vv{e}_z = \I  \epsilon_{ij}\partial_{q_i} \langle u_\vv{q} | \partial_{q_j} |u_\vv{q}\rangle \,.
\end{equation}    
The Chern number is related to the flux of the  Berry curvature $\Omega(\vv{q})$ through the BZ. Just as the Dirac quantization condition requires the magnetic flux through a closed surface to be quantized (in units of $h/e$), the flux of the Berry curvature through the BZ (a closed surface with the topology of a torus) is quantized (in units of $2\pi$). 
This follows by using Stokes's theorem to relate the integral of $\Omega(\vv{q})$ over the BZ to the line integral of  $\vv{A}(\vv{q})$ around its boundary.
Since the BZ is a closed surface this line integral must  be an integer multiple of $2\pi$. This integer, ${\cal C}$, is the number of flux quanta of Berry curvature through the BZ. 
Since the Berry curvature is a gauge-invariant quantity, so too is the Chern number.

The Chern number vanishes for systems with time-reversal symmetry (TRS), for which  $u_\vv{q}(\vv{r}) \propto u^*_{-\vv{q}}(\vv{r})$ and hence  $\Omega(\vv{q}) = - \Omega({-\vv{q}})$. The realization of Chern bands  therefore requires a means to  break TRS.
In Section~\ref{sec:implementations} we shall describe ways in which this can be achieved for cold-atom systems. 
One  class of implementation involves tight-binding models, in which a (spinless) particle hops on a lattice with complex tunneling matrix elements, \eg to represent the Peierls phase factors for a charged particle in a magnetic field. TRS is broken if, for some closed loop on the lattice, 
 the phases acquired on encircling the loop in the clockwise ($+\phi_{\rm AB}$) and  anticlockwise ( $-\phi_{\rm AB}$) directions differ modulo $2\pi$,  \ie provided $\phi_{\rm AB}$ is not an integer multiple of $\pi$. 
 Another class of implementation exploits  internal atomic states, labeled by index $\alpha$. Quite generally, the (non-dissipative) action of laser light on an atom is to couple  state $\alpha$ to  state $\alpha'$ with a  well-defined momentum transfer $\vv{\kappa}$.  An optical lattice is defined by
 a set of such couplings, $V^{\alpha'\alpha}_\vv{\kappa}$, which (to preserve periodicity) must build up a regular lattice in momentum space. 
 The  couplings  $V^{\alpha'\alpha}_\vv{\kappa}$ therefore define a tight-binding model in momentum space
with amplitudes and phases determined by the laser fields. For shallow lattices, the net phase acquired on encircling a closed loop on this momentum-space lattice determines the integrated Berry curvature through that loop, allowing lattices with broken TRS $\Omega(\vv{q}) \neq - \Omega({-\vv{q}})$ to be directly constructed~\cite{coopermoessner}.

To illustrate how a non-zero Chern number arises in tight-binding models, we discuss here the key features of the {\it Haldane model}~\cite{haldanehoneycomb}. This is  a two-band model, so the Bloch Hamiltonian is a $2\times 2$ Hermitian matrix
\begin{equation}
\hat{H}_\vv{q} = -{\vv{h}}(\vv{q}) \cdot \hat{\vv{\sigma}}
\label{eq:haldaneham}
\end{equation}
with ${\vv{h}}(\vv{q})$ a three-component vector coupled to  the Pauli matrices, $\hat{\sigma}_{x,y,z}$.
Just as in the SSH model, the energy bands are $E_\vv{q}^{(\pm)} = \pm |{\vv{h}}(\vv{q})|$ and there is a gap provided $|{\vv{h}}_{\vv{q}}|$ does not vanish at any $\vv{q}$. However, now the Bloch state of the lower band depends on the three-component unit vector ${{\vv{e}}}(\vv{q})  \equiv {\vv{h}}(\vv{q})/|{\vv{h}}(\vv{q})|$, \eg 
\begin{equation}
|u_\vv{q}\rangle = \left(\begin{array}{l} \cos(\theta_\vv{q}/2)\\ \sin(\theta_\vv{q}/2) \,\E^{\I\phi_q}\end{array}\right)
\label{eq:2dwavefunction}
\end{equation}
for ${{\vv{e}}}(\vv{q})= (\sin\theta_\vv{q}\cos\phi_\vv{q},\sin\theta_\vv{q}\sin\phi_\vv{q},\cos\theta_\vv{q})$.
The Chern number can be written in terms of this unit vector as
\begin{equation}
{\cal C} = N_{\rm 2D} \equiv -\frac{1}{8\pi}\epsilon_{ij} \int_{\rm BZ}  {{\vv{e}}} \cdot \frac{\partial {{\vv{e}}}}{\partial{q_i}}\times \frac{\partial {{\vv{e}}}}{\partial{q_j}}\; \D^2q
\end{equation}
 $N_{\rm 2D}$ counts the number of times that the unit vector ${{\vv{e}}(\vv{q})}$ wraps over the unit sphere as $\vv{q}$ spans the BZ; it is the 2D analogue of the 1D winding number (\ref{eq:windingnumber1}). 

\begin{figure}
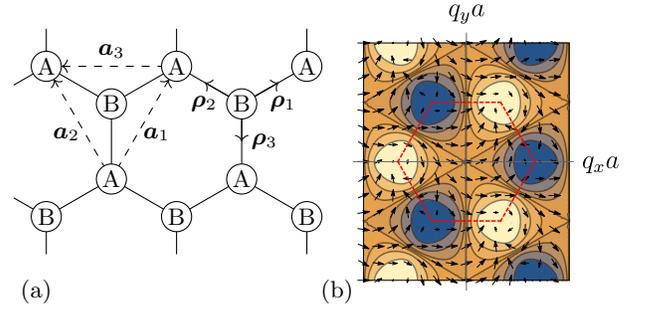

\include{figures/fig_haldane_model_lattice-revised}
\caption{(a) Real space honeycomb lattice for the Haldane model. (b) Illustration of the unit vector ${\vv{e}}(\vv{q})$ in the topological phase of the Haldane model. The dashed line shows the conventional BZ, whose corners are the points $\vv{Q}_\pm$. The arrows show the projection of  $\vv{e}(\vv{q})$ on the plane, $(e_x,e_y)$, while the colors and contours indicate $e_z$. The unit vector wraps once over the sphere within the BZ, indicating the topological character
with ${\cal C}=1$.}
\label{fig:haldanewrap}
\end{figure}

The Haldane model is described in detail in Section~\ref{sec:implementations}.
It is defined on a honeycomb lattice, for which the unit cell contains two sites, which we label A and B, as in Fig.~\ref{fig:haldanewrap}(a).
Nearest-neighbor tunneling is off-diagonal in the sublattice index, and leads to 
\begin{equation}
h_x(\vv{q})+\I h_y(\vv{q}) = J \left(\E^{\I \vv{q}\cdot \vv{a}_1} + \E^{\I \vv{q}\cdot \vv{a}_2}+1\right)
\label{eq:haldanennperiodic}
\end{equation}
in (\ref{eq:haldaneham}), where $\vv{a}_{1,2}$ are the lattice vectors marked on Fig.~\ref{fig:haldanewrap}a. The Hamiltonian has been chosen to be periodic under the addition of reciprocal lattice vectors. Note, however, that other choices can be made, related to the Hamiltonian via unitary transformations,
$\hat{H}_\vv{q} \to \hat{\cal{U}}_\vv{q}\hat{H}_\vv{q}\hat{\cal{U}}_\vv{q}^\dag$.
Indeed, in Section~\ref{sec:implementations} we shall replace
\begin{equation}
h_x(\vv{q})+\I h_y(\vv{q})\to J \left(\E^{\I \vv{q}\cdot \vv{\rho}_1} + \E^{\I \vv{q}\cdot \vv{\rho}_2} + \E^{\I \vv{q}\cdot \vv{\rho}_3}\right)
\label{eq:haldanenn}
\end{equation}
which arises for $\hat{\cal U}_q =  \exp(-\I\, \hat{\sigma}_z\,\vv{q}\cdot\vv{\rho}_3/2)$, with $\vv{\rho}_3$ 
the nearest-neighbor lattice vector in Fig.~\ref{fig:haldanewrap}a. This transformation does not change the energy spectrum nor the topology of the band. However, it can be more helpful to work with (\ref{eq:haldanenn}) when considering physical observables~\cite{Bena2009}. As discussed in Appendix~\ref{app:embedding}, unitary transformations of the form $\hat{\cal U}_q$ affect the definition of the force and current operators as well as the local Berry curvature.

The resulting band structure is well known from studies of graphene: there are two points at the corners of the BZ, $\vv{q}=\vv{Q}_\pm$ with $\vv{Q}_\pm = \pm \frac{4\pi}{3\sqrt{3}a}(1,0)$,  at which the bands touch, \ie $h_x+\I h_y =0$.  Close to either band-touching point, with ${\tilde{\vv{q}}} \equiv \vv{q}-\vv{Q}_{\pm}$,
the Hamiltonian has the 2D Dirac form
 \begin{equation}
 \label{eq:lowenergyhaldane}
\hat{H}^\pm_{\vv{q}} \approx \hbar v \left[\pm{\tilde q}_x \hat{\sigma}_x + {\tilde q}_y \hat{\sigma}_y\right] + h_z^\pm\hat{\sigma}_z
\end{equation}
with velocity $v=(3/2)Ja/\hbar$.  The terms $h_z^\pm$ arise from effects other than nearest-neighbor tunneling, and open gaps at the Dirac points between the two bands.
One such effect is an energy splitting, $\Delta$, between A and B sublattices. In this case the coefficients $h_z^+$ and $h_z^-$ are both equal to $\Delta$ and the gap openings at the two Dirac points are equivalent. This causes the resulting Bloch bands to have vanishing Chern number, so the two resulting bands are non-topological. This is consistent with  the fact that  this model has TRS. Introducing next-nearest-neighbor hopping with  Aharonov-Bohm phase, $\phi_{\rm AB}\neq 0 \mod \pi$, breaks TRS and provides a term with $h_z^+ = -h_z^-$ for which the  Chern number of the bands are $1$ and $-1$. 
Figure~\ref{fig:haldanewrap}
shows  how ${{\vv{e}}(\vv{q})}$ varies in reciprocal space in this topological phase.

The preceding discussions are somewhat abstract, focusing on mathematical aspects of the energy bands in 2D. However, the Chern number has direct physical consequences. As first shown by~\textcite{thoulesschern}, a band insulator exhibits the integer quantum Hall effect if the total Chern number of the filled bands is nonzero, with a Hall conductivity quantized at  ${\cal C}$ times the fundamental unit of conductance $e^2/h$.   
This result can be extended to systems subjected to disorder and inter-particle interactions, by defining a Chern number for the many-body groundstate in a geometry with periodic boundary conditions~\cite{Niu1985}. For non-interacting systems without translational invariance, a local Chern marker can be defined~\cite{Bianco2011}.
The quantization of the Hall conductance is intimately related to the existence of edge states, which we will now discuss.

\subsection{Edge states}

\label{subsec:edge}

Topological band insulators have the generic feature that although they are bulk insulators -- owing to the energy gap between the filled and empty bands --   they host gapless states on their surfaces. 

\begin{figure}

\begin{tikzpicture}
\draw[black, very thick, fill=blue!5!white] (36mm,24mm) rectangle (0,0);

\draw[blue, thick] (18mm,12mm) arc (0:360:3mm);
\draw[->,blue, thick] (18mm,12mm) arc (0:-45:3mm);

\draw[blue, thick] (36mm,0) arc (270:90:3mm);
\draw[->,blue, thick] (36mm,0) arc (270:150:3mm);
\draw[blue, thick] (36mm,6mm) arc (270:90:3mm);
\draw[->,blue, thick] (36mm,6mm) arc (270:150:3mm);
\draw[blue, thick] (36mm,12mm) arc (270:90:3mm);
\draw[->,blue, thick] (36mm,12mm) arc (270:150:3mm);
\draw[blue, thick] (36mm,18mm) arc (270:90:3mm);
\draw[->,blue, thick] (36mm,18mm) arc (270:150:3mm);

\draw[blue, thick] (0,6mm) arc (90:-90:3mm);
\draw[->,blue, thick] (0,6mm) arc (90:-30:3mm);
\draw[blue, thick] (0,12mm) arc (90:-90:3mm);
\draw[->,blue, thick] (0,12mm) arc (90:-30:3mm);
\draw[blue, thick] (0,18mm) arc (90:-90:3mm);
\draw[->,blue, thick] (0,18mm) arc (90:-30:3mm);
\draw[blue, thick] (0,24mm) arc (90:-90:3mm);
\draw[->,blue, thick] (0,24mm) arc (90:-30:3mm);

\draw[blue, thick] (0,0) arc (180:0:3mm);
\draw[->,blue, thick] (0,0) arc (180:60:3mm);
\draw[blue, thick] (6mm,0) arc (180:0:3mm);
\draw[->,blue, thick] (6mm,0) arc (180:60:3mm);
\draw[blue, thick] (12mm,0) arc (180:0:3mm);
\draw[->,blue, thick] (12mm,0) arc (180:60:3mm);
\draw[blue, thick] (18mm,0) arc (180:0:3mm);
\draw[->,blue, thick] (18mm,0) arc (180:60:3mm);
\draw[blue, thick] (24mm,0) arc (180:0:3mm);
\draw[->,blue, thick] (24mm,0) arc (180:60:3mm);
\draw[blue, thick] (30mm,0) arc (180:0:3mm);
\draw[->,blue, thick] (30mm,0) arc (180:60:3mm);
\draw[blue, thick] (6mm,24mm) arc (360:180:3mm);
\draw[->,blue, thick] (6mm,24mm) arc (360:240:3mm);
\draw[blue, thick] (12mm,24mm) arc (360:180:3mm);
\draw[->,blue, thick] (12mm,24mm) arc (360:240:3mm);
\draw[blue, thick] (18mm,24mm) arc (360:180:3mm);
\draw[->,blue, thick] (18mm,24mm) arc (360:240:3mm);
\draw[blue, thick] (24mm,24mm) arc (360:180:3mm);
\draw[->,blue, thick] (24mm,24mm) arc (360:240:3mm);
\draw[blue, thick] (30mm,24mm) arc (360:180:3mm);
\draw[->,blue, thick] (30mm,24mm) arc (360:240:3mm);
\draw[blue, thick] (36mm,24mm) arc (360:180:3mm);
\draw[->,blue, thick] (36mm,24mm) arc (360:240:3mm);
\end{tikzpicture}

\caption{Illustration of skipping orbits for a charged particle in a uniform magnetic field. In the bulk, the semiclassical dynamics of a wavepacket leads to a circular cyclotron orbit. The reflection of such an orbit from the (hardwall) edge leads to a skipping motion around the sample edge.}
\label{fig:skipping}
\end{figure}
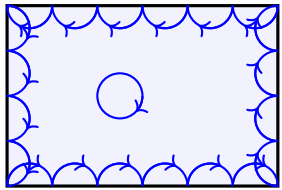

The existence of gapless edge modes for 2D systems with nonzero Chern number is well-known from studies of the integer quantum Hall effect~\cite{halperingauge}. Each filled Landau level gives rise to a chiral edge mode.
This can be understood semiclassically in terms of the skipping orbit of the cyclotron motion around the edge of the sample, Fig.~\ref{fig:skipping}.  These semiclassical skipping orbits consist of two features. 
The rapid skipping motion at the cyclotron frequency is a feature which in a quantum description arises when the wavefunction has nonzero amplitude in more than one Landau level, such that its time-dependence involves the Landau level energy spacing (cyclotron energy); it is therefore related to the inter-Landau level excitation known as a ``magnetoplasmon". The drift of the guiding center of the orbit around the perimeter of the sample is a feature that exists for quantum states within a single Landau level and that represents the chiral edge mode of that Landau level.
The existence of these edge modes is required for consistency of the quantized bulk Hall conductance~\cite{laughlingauge}. 

More generally,  gapless edge states occur at the boundary between two insulating regions with different values of a topological invariant. 
A simple semiclassical view of these gapless regions is provided by considering the boundary to arise from a smooth spatial variation in the parameters of the Hamiltonian, between two phases with different topological indices. Since the two insulators far-to-the-left and far-to-the-right of the boundary are topologically distinct, at some point in space the gap between the filled and empty bands must close. This gap closing motivates the existence of gapless edge states. 
While this semiclassical argument applies only for smooth variations in space, the result is a robust feature for any form of boundary, referred to as the bulk-boundary correspondence~\cite{hasankane}. The only restriction is that, in cases where the  topological invariant relies on an underlying symmetry, this symmetry must be preserved also in the boundary region. This is illustrated below for the edge state of the SSH model.

We shall demonstrate the emergence of edge states in the SSH  and Haldane models within continuum approximations for which the edge state wavefunctions take simple analytic forms.

\subsubsection{SSH model}

\label{sec:sshedge}

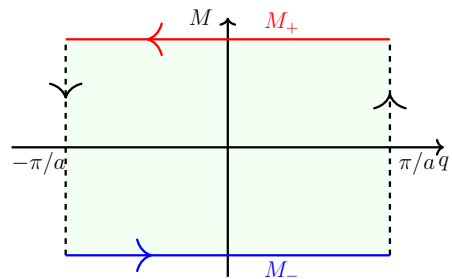
\begin{figure}
\resizebox{0.7\columnwidth}{!}{\begin{tikzpicture}
\draw[white, fill=green!5!white] (3,2) rectangle (-3,-2);

\draw[black,very thick,dashed](3,-2)--(3,2);
\node at (3.5,-0.3) {\large $\pi/a$};

\draw[black,very thick,dashed](-3,-2)--(-3,2);
\node at (-3.5,-0.3) {\large $-\pi/a$};

\draw[-{>[scale=2]},black,very thick](-3,1)--(-3,0.9);
\draw[-{>[scale=2]},black,very thick,](3,0.9)--(3,1);
\draw[-{>[scale=2]},blue,very thick](-1.5,-2)--(-1.4,-2);
\draw[-{>[scale=2]},red,very thick,](-1.4,2)--(-1.5,2);

\draw[->,black,very thick](-4,0)--(4,0);
\node at (4,-0.3) {\large $q$};

\draw[->,black,very thick](0,-2.4)--(0,2.4);
\node at (-0.5,2.4) {\large $M$};

\draw[red,very thick](-3,2)--(3,2);
\node at (1,2.3) {\textcolor{red}{\large $M_+$}};

\draw[blue,very thick](-3,-2)--(3,-2);
\node at (1,-2.3) {\textcolor{blue}{\large $M_-$}};

\end{tikzpicture}}
\caption{An edge of the SSH model is represented by $h_x+ \I h_y= |h_{q,M}|\E^{\I\phi({q,M})}$, which depends both on the wavevector $q$ and on some control parameter $M$ that varies smoothly between $M_-$ at $x \ll 0$ and $M_+$  at $x\gg 0 $. The number of vortices of $h_x+\I h_y$ within this region 
can be computed by integrating  $\vv{\nabla}\phi({q,M})$ around the boundary. The integrals along $q=\pm \pi/a$ cancel (since the BZ is periodic), leaving only the integrals along $M=M_\pm$, which give the difference of winding numbers $N_--N_+$.}
\label{fig:sshedge1}
\end{figure}

Consider a 1D band insulator formed by filling the lower energy band of a Hamiltonian of the form (\ref{eq:sshham1}), which is purely non-diagonal in the sublattice basis
and parameterized by the two-component vector $(h_x,h_y)$. 
This restriction on the form of the Hamiltonian arises from the chiral symmetry of the model, as discussed further in Sec.~\ref{subsec:topoinsulator}.
Let the properties of the system depend on some quantity $M$ such that $h_x+\I h_y \equiv |h({q,M})|\E^{\I \phi({q,M})}$   is a function of both wavevector $q$ and $M$.
We introduce a boundary between two gapped phases at position $x=0$ by allowing
 $M(x)$ to vary slowly in space, as compared to the lattice constant $a$, and setting   $M(x\ll 0) = M_-$ and   $M(x\gg 0) = M_+$.
The two phases are characterized by winding numbers $N_\pm$ (\ref{eq:windingnumber1}) computed from 
 $\phi({q,M_\pm})$. It is straightforward to show that  $h_x+\I h_y$ must have $N_--N_+$ vortices within the boundary region. 
  (See Fig.~\ref{fig:sshedge1}.) So if $N_+\neq N_-$ then $|h({q,M})|$ must vanish at certain points $(q,M)$: these are the gap-closing points, discussed above semiclassically, which lead to gapless edge states.

For the SSH model,  $h_x +\I h_y =  J'+J\E^{\I qa}$, the gap-closing point ($J'/J=1$, $qa=\pi$) hosts a single vortex. Defining $M = 1-J'/J$ and  $\tilde{q} = q- \pi/a$, we expand the Hamiltonian around $M=\tilde{q}=0$ to first order in $\tilde{q}$ (suitable for the continuum limit, $|\tilde{q}|\ll1/a$) 
  to give 
  \begin{eqnarray}
\hat{H}^{\rm SSH}/J  &  \approx &  \tilde{q}a\hat{\sigma}_y + M(x) \hat{\sigma}_x \\
 & 
= & 
 -\I a \hat{\sigma}_y \partial_x + M(x) \hat{\sigma}_x\,.
 \label{eq:jr1}
 \end{eqnarray}
In the last line we have replaced  $\tilde{q}$ by $-\I\partial_x$ to give a 1D Dirac Hamiltonian in which $M(x)$ can be viewed as a  spatially varying ``mass". This is a model studied by \textcite{Jackiw:1976}, who showed that, provided $M(x)$ changes sign, there is a solution that is localized at the boundary with exactly zero energy. For $M_+>0$ and $M_-<0$, this is
\begin{equation}
| \Psi\rangle \deq \begin{pmatrix}
\E^{-\int_0^x M(x') \D x'/a} \E^{\I\pi x/a} \\
0 
\end{pmatrix}\,.
\label{eq:jrzero1}
\end{equation}
That the state is an eigenstate of $\hat{\sigma}_z$ with 
 $|\sigma_z=+1\rangle$ indicates that it has non-zero amplitude only on the A sublattice. For $M_+<0$ and $M_->0$, the non-zero amplitude would be on the B sublattice.
This is the gapless edge mode of the SSH model which arises because $M_\pm$ describe different topological phases.
That the state has energy $E=0$ is a consequence of a ``chiral" symmetry that protects the relevant 1D topological invariant, as discussed below in Sec.~\ref{subsec:topoinsulator}. If the Hamiltonian were to depart from this chiral form, for example by including terms $h_z$ that couple to $\hat{\sigma}_z$, the energy of this subgap state would depart from $E=0$.  This could occur either by a change of the bulk Hamiltonian, or due to breaking of the symmetry near the edge: for example arising from an onsite potential that shifts the energy of the A site relative to the B site near the edge.

Although we have derived this edge state for a continuum model, the properties are robust to lattice effects provided
the chiral symmetry is preserved. 
A derivation of the edge state of the SSH model for a sharp boundary is provided in 
Appendix~\ref{subsec:SSH}.

\subsubsection{Haldane model}

One can use the above solution for the  edge state of the SSH model to construct a 1D band of edge states on a surface of the (topological) band insulator formed from the Haldane model. Consider the low-energy theory for the Haldane model (\ref{eq:lowenergyhaldane}) in a topological phase with $h^\pm_z = \mp |H|$. We impose a boundary to a non-topological phase at $x> x_{R}$ by  adding a spatially dependent energy offset $\Delta(x)\hat{\sigma}_z$, such that $h^\pm_z(x) = \Delta(x) \mp |H|$. 
The low-energy Hamiltonians close to $\vv{Q}_\pm$ become
 \begin{eqnarray}
 \hat{H}^\pm_{\vv{q}}  & \approx &  \mp \I \hbar v  \hat{\sigma}_x  \partial_x + h_z^\pm(x)\hat{\sigma}_z + \hbar v {\tilde q}_y  \hat{\sigma}_y
 \label{eq:haldaneedge}
 \end{eqnarray}
where we have replaced  ${\tilde q}_x \to -\I \partial_x$. Translational invariance is maintained along the $y$-direction, such that $q_y$ (and therefore ${\tilde {q}}_y = {q}_y-{Q}^{\pm}_y$) is  conserved. 

\begin{figure}

\begin{tikzpicture}

\fill[fill=blue!5!white] (1,1.2) rectangle (-1,-1.2);

\draw[->,red,thick] (-1,1.2) -- (-1,-0.4);
\draw[red,thick] (-1,1.2) -- (-1,-1.2);

\draw[->,green,thick] (1,-1.2) -- (1,0.4);
\draw[green,thick] (1,-1.2) -- (1,1.2);

\draw[->,black,thick] (-1.3,0) -- (1.3,0);
\draw[->,black,thick] (0,-1.5) -- (0,1.5);

\node at (1.4,-0.2) {$x$};
\node at (-0.2,1.5) {$y$};

\node at (1,-1.6) {$x_R$};
\node at (-1,-1.6) {$x_L$};

\node at (-1.3,-2) {(a)};

\begin{scope}[shift={(4,0)}]

\draw[black, fill=black!5!white] (-1,2) .. controls (0,0) and (0,0) .. (1,2);
\draw[black, fill=black!5!white] (-1,-2) .. controls (0,0) and (0,0) .. (1,-2);

\draw[red,thick] (-0.6,1.2) -- (0.6,-1.2);
\draw[green,thick] (0.6,1.2) -- (-0.6,-1.2);

\draw[->,black,thick] (-1.5,0) -- (1.5,0);

\node at (1.3,0.2) {${\tilde q}_y$};

\node at (-1.3,-2) {(b)};

\end{scope}

\end{tikzpicture}

\caption{Illustration of the low-energy spectrum of the Haldane model on a finite-width strip. (a) The strip is bounded in the $x$-direction, but uniform along $y$ such that the wavevector ${\tilde q}_y = q_y-Q_y^+$ is conserved.  (b) The spectrum has a continuum of states in the bulk, shown shaded. The bulk bands are topological, with unit Chern number, so a single edge state connects between these bulk bands: red (green) band corresponds to the edge mode on the left (right) boundary. We show only the part of the spectrum close to the $\vv{Q}^+$ point, at which the boundaries force a gap-closing and at which the edge states appear. 
}
\label{fig:haldaneedgemodes}
\end{figure}
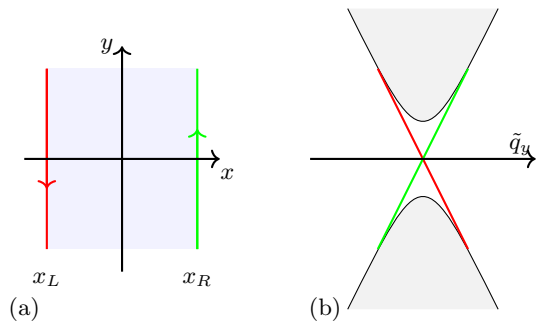

For  $\Delta(x)$ an increasing function of $x$ at $x>0$ there is a boundary between topological ($x<x_{R}$) and non-topological ($x>x_{R}$) regions where  $\Delta(x_{R})=|H|$. At this point,  $h_z^+(x_{R})=0$ so the gap at  $\vv{Q}^+$ vanishes. 
The low-energy Hamiltonian close to $\vv{Q}^+$ (\ref{eq:haldaneedge}) takes the form of the 
Jackiw-Rebbi Hamiltonian (\ref{eq:jr1}), just under a permutation of the Pauli matrices [$(\hat{\sigma}_x,\hat{\sigma}_y,\hat{\sigma}_z)
\to
(\hat{\sigma}_y,\hat{\sigma}_z,\hat{\sigma}_x)$], plus a term $\hbar v {\tilde q}_y\hat{\sigma}_y$ for which the zero-mode (\ref{eq:jrzero1}) is also an eigenstate (noting the permutation of the Pauli matrices), with eigenvalue $E^{R}_{q_y} = \hbar v{\tilde q}_x$. This is the chiral edge mode, which 
 propagates along the right hand boundary $x_{R}$ at velocity $v\vv{e}_y$. For a finite width strip, with another boundary to a non-topological phase at $x<x_{L}$, there is an edge mode with the opposite velocity, $E^{L}_{q_y} = -\hbar v{\tilde q}_y$. The dispersion is illustrated in Figure~\ref{fig:haldaneedgemodes}.

\subsection{Topological insulators}

\label{subsec:topoinsulator}

\subsubsection{Discrete Symmetries}

\label{subsubsec:discretesymmetries}

The Chern number for 2D systems is an example of a topological
invariant for energy bands in a setting where no symmetries constrain
the form of the Hamiltonian.
The field of topological insulators arose from the realization that,
when one constrains the Hamiltonian to have additional symmetries, new
topological invariants emerge in various spatial dimensions.
The general theory of
symmetry-protected topological invariants for gapped fermionic systems
concerns global symmetries -- time-reversal, particle-hole symmetry,
chiral symmetry -- which make no reference to spatial structure or
dimension\cite{Altland1997}. These define ten distinct symmetry classes which, combined
with the spatial dimensionality, determine the possible topological
invariants~\cite{RevModPhys.88.035005}.
An important example is time-reversal symmetry which leads to the
existence of a $\mathbb{Z}_2$ topological invariant in 2D and 3D for
electronic systems with spin-orbit coupling.  This led to the
discovery of materials which behave as topological insulators in these
dimensions~\cite{hasankane,qizhang}.
Furthermore, this formalism also allows a unified description of 
topological superfluids, formed from the pairing of fermions. 
Within the BCS mean-field theory, the spectrum of
quasi-particle excitations of such superfluids is obtained from the
Bogoliubov-de Gennes (BdG) theory, which takes the form of a
non-interacting fermionic system albeit with anomalous terms that mix
particle and hole excitations.  The intrinsic particle-hole symmetry of the BdG
theory stabilizes gapped topological
superfluids. We shall discuss such topological superfluids in 
Section~\ref{sec:topologicalsuperfluids}.

Here we explore the physics of symmetry-protected topological band insulators in the context of the  SSH model described above.
This provides an example of a topological
band insulator that arises in 1D when there is an underlying ``chiral"
symmetry.
This chiral symmetry arises from the existence of a
unitary operator $\hat{U}$ that anticommutes with the Hamiltonian $\hat{H}$\begin{equation}
\hat{H}\hat{U} = -\hat{U} \hat{H}\,.
\label{eq:chiral1}
\end{equation}
Then, for an energy eigenstate $|\Psi\rangle$ of eigenvalue $E$, 
the state $\hat{U}|\Psi\rangle$ is readily shown to be an energy eigenstate with  energy
$-E$. 
 Thus, the chiral symmetry enforces a symmetry on the spectrum
about $E=0$.
  This rather formal construction arises in tight-binding
models, such as the SSH model, in which the Hamiltonian only involves terms that hop
between two different sublattices (which we label A and B).  Defining $\hat{P}_{\rm A/B}$ as
projectors onto the A/B sublattices, then
$\hat{U} \equiv (\hat{P}_{\rm A} - \hat{P}_{\rm B})$ satisfies (\ref{eq:chiral1}) if $\hat{P}_{\rm A} \hat{H}\hat{P}_{\rm A} = \hat{P}_{\rm B}
\hat{H}\hat{P}_{\rm B} = 0$,
\ie  provided the Hamiltonian $\hat{H}$ only couples A and B sublattices.
The chiral symmetry constrains the Hamiltonian for the
SSH model (\ref{eq:sshham1}) to be off-diagonal in the sublattice
basis. As described in Sec.~\ref{sec:zak_phase} this form ensures that the
winding number is a topological invariant. Furthermore, the symmetry of the energy spectrum
ensures that the edge mode, found in Sec.~\ref{sec:sshedge}, is at exactly zero energy, $E=0$, and is also an eigenstate of $\hat{U} = \hat{P}_{\rm A} - \hat{P}_{\rm B}$. This implies that there is at least one zero energy in-gap mode, but other non-zero energy in-gap modes are allowed by chiral symmetry provided they come in $\pm E$ pairs. For boundaries that break the chiral symmetry, the in-gap states will have no particular relation to zero energy. {A disordered potential that acts differently on the two sublattice sites will break chiral symmetry and also shift the energies of in-gap states away from zero energy. However, if the disorder is such that (\ref{eq:chiral1}) is preserved, the topology of this model is robust and the in-gap states remain at zero energy.}

\subsubsection{The Rice--Mele model}

\label{subsubsec:rm}

\begin{figure}
\begin{minipage}[c]{.6\columnwidth}
\includegraphics[width=\columnwidth]{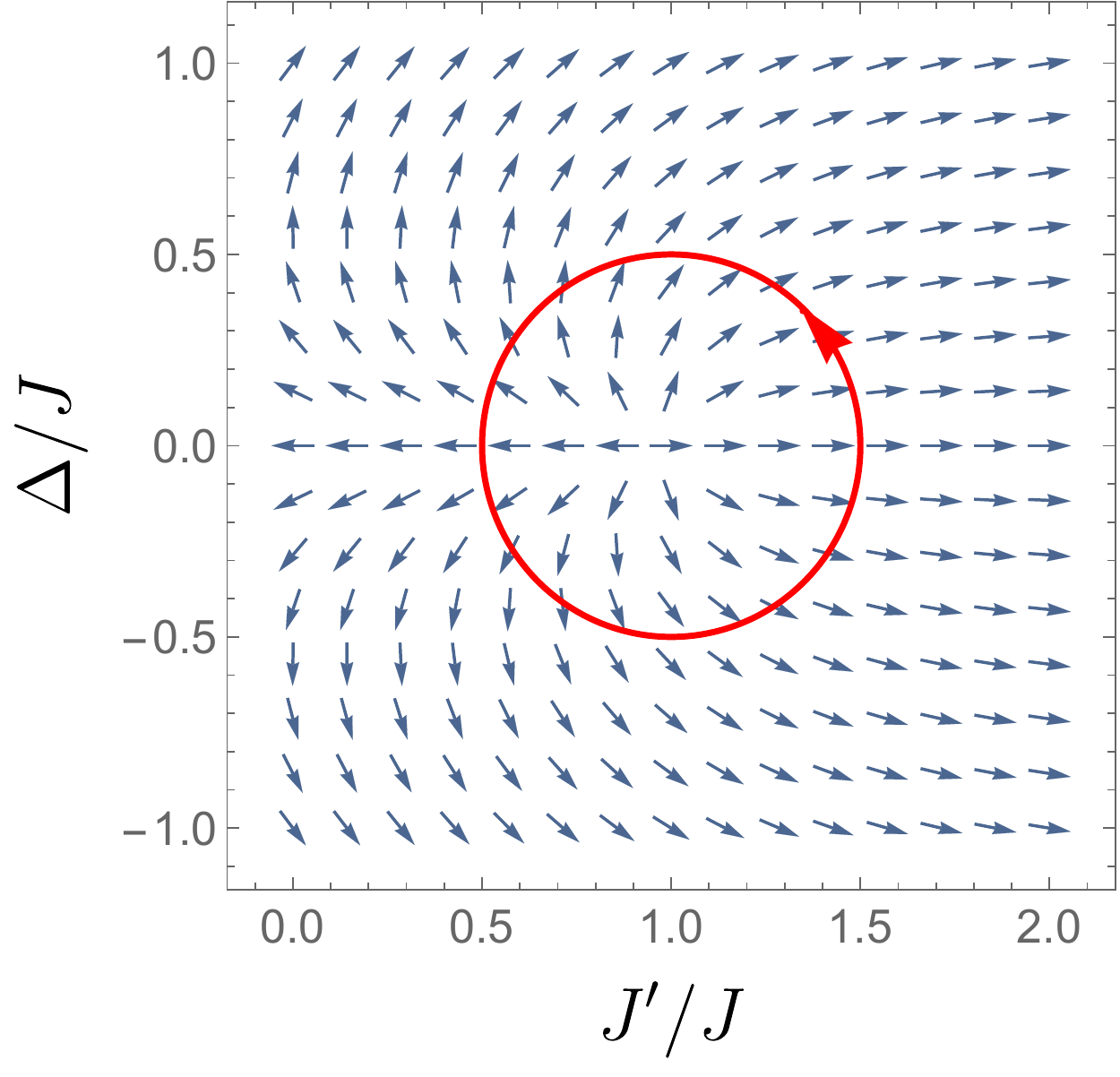}
  \end{minipage} \hfill
\begin{minipage}[c]{.35\columnwidth}
\null 
\includegraphics[width=\columnwidth]{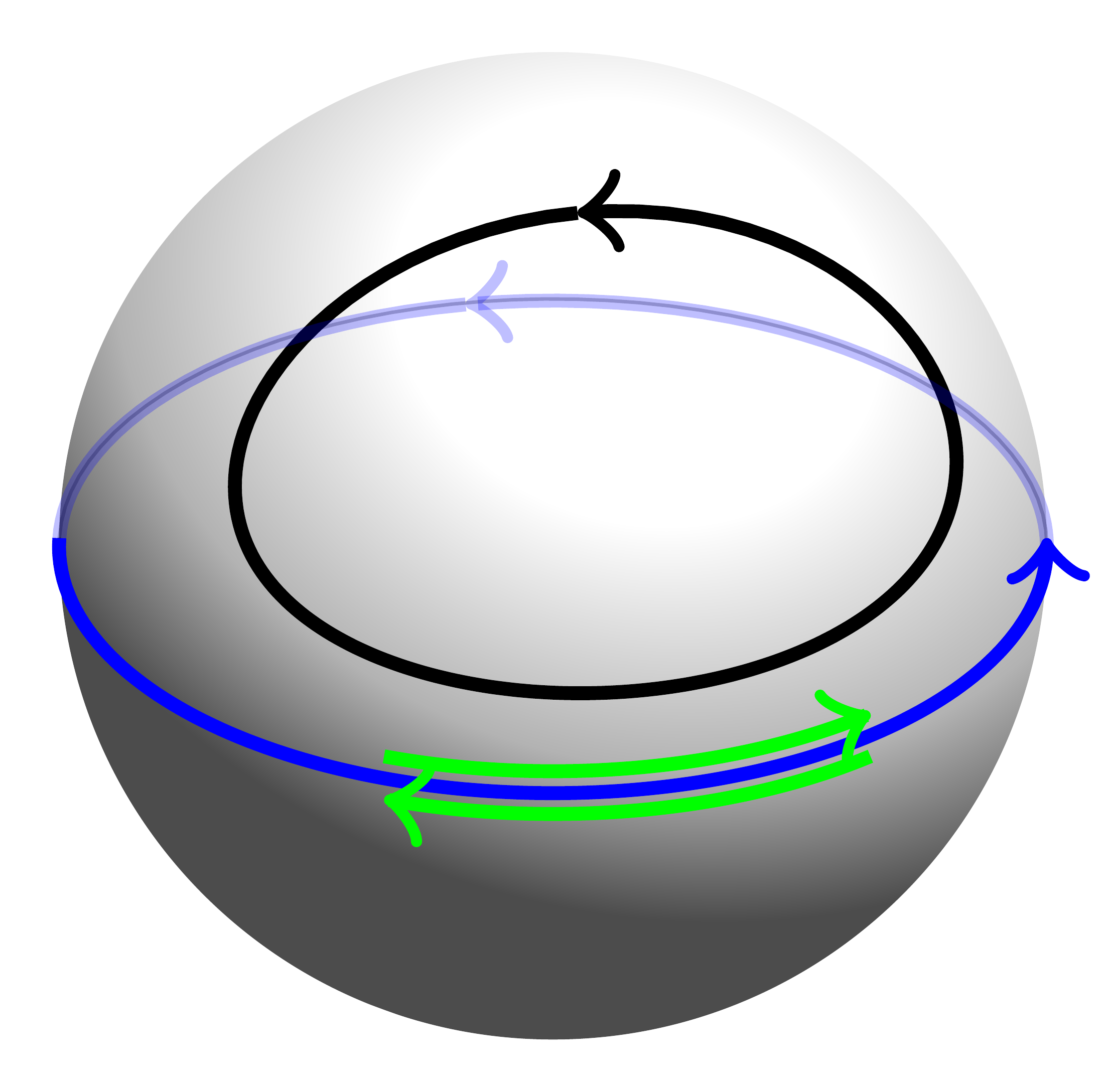}
  \end{minipage}
\caption{Left: Zak phase of the lower band of the Rice--Mele model, represented by the unit vector $(\cos\phi_{\rm Zak},\sin\phi_{\rm Zak})$. The phase changes by $2\pi$ around the point $\Delta=0$ and $J'/J=1$ at which the  gap closes. The loop denotes a possible pumping cycle. Right: Representation of the eigenstates (\ref{eq:eigenstates_uq}) on the Bloch sphere. Blue: Closed trajectory on the equator for the SSH model in the topological {case ($\Delta=0$, $J'/J<1$)}, obtained when $q$ scans the Brillouin zone.  The Zak phase, given by half the subtended solid angle, is equal to $\pi$. Green: Trajectory in the non-topological case of the SSH model ($\Delta=0$, $J'/J>1$) with a zero subtended solid angle. Black: Closed trajectory obtained for the Rice-Mele model with a non-zero energy offset $\Delta$. A continuous variation of $\Delta$ and $J'/J$ in the Rice--Mele model allows one to connect the topological and non-topological SSH trajectories, without going through the singular point $J'/J=1$, $\Delta=0$. 
}
\label{fig:zakphaserm1}
\end{figure}

Since the topological invariant of the SSH model is ensured by a (chiral) symmetry, it is possible to smoothly evolve between states with different topological invariants 
by breaking this symmetry. 
To illustrate this, we study the Rice--Mele (RM) model which generalizes the SSH model by relaxing the assumption that the A and B sites have the same energy, and assigning energy offsets $\pm\Delta$. The Hamiltonian (\ref{eq:hssh}) is replaced by
\begin{equation}
\hat{H}^{\rm RM} _q \deq \left( \begin{array}{cc} \Delta & -J'-J\E^{-\I qa}\\ -J'-J\E^{\I qa} &-\Delta \end{array}\right)
\label{eq:hrm1}
\end{equation}
within the BZ $-\pi/a < q \leq \pi/a$. 
The 
two bands have energies
\begin{equation}
E_q^{(\pm)} = \pm \left[\Delta^2 +|J'+J\E^{-\I qa}|^2\right]^{1/2},
\label{eq:energies_Rice_Mele}
\end{equation}
so are separated by a gap provided $\Delta\neq 0$. (For $\Delta =0$ the gap closes at the topological transition of the SSH model, $J'/J=1$.)
The eigenstate of the lower band can be written 
\begin{equation}
|u^{(-)}_q\rangle \deq \left(\begin{array}{c} \sin(\gamma_q/2)\\ -\E^{\I\phi_q}\cos(\gamma_q/2)\end{array}\right)
\label{eq:eigenstates_uq}
\end{equation}
with $\gamma_q$ and $\phi_q$  defined by  $-(J'+J\E^{-\I qa})/\Delta \equiv \tan\gamma_q \,\E^{-\I\phi_q}$. 
The  resulting Zak phase (\ref{eq:zak}) is shown
in Figure~\ref{fig:zakphaserm1} (left) as a function of $J'/J$ and $\Delta/J$. Along the line $\Delta=0$ the Zak phase steps by $\pi$ across $J'/J=1$, consistent with the change in winding number by one at the topological transition of the SSH model, and Eq.~(\ref{eq:windingnumber1}). However, note that  
the gapped band insulators at $\Delta =0$, $J'/J>1$ and  $\Delta =0$, $J'/J<1$  can now be continuously connected by tracing out a path  that 
has $\Delta \neq 0$ when $J'/J=1$ (Fig.~\ref{fig:zakphaserm1}, right).  

\subsection{Adiabatic pumping}

\label{subsec:pumping}

Consider again the Rice--Mele model, whose Zak phase is illustrated in Fig.~\ref{fig:zakphaserm1}.
There is a vortex in $\phi_{\rm Zak}$ around the gap-closing point ($\Delta=0$ and $J'/J=1$) at which the Zak phase is undefined. The winding of $\phi_{\rm Zak}$ by $2\pi$ around a closed loop encircling the gapless point, such as the loop shown in Fig.~\ref{fig:zakphaserm1}, is a topological invariant of the model:  this winding is preserved under smooth variations of the Rice--Mele Hamiltonian that do not cause the  gap to close on this loop. 

The existence of this invariant is at the basis of the  concept of a quantized pump~\cite{Thouless:1983}. It describes the generic situation of a crystal with filled bands, which is characterized by parameters that can be externally controlled. When these parameters are varied around a closed loop, the number of particles that are transported is quantized. It is thus a robust quantity that is not affected by a small change of the geometry of the loop in parameter space.  Note that at this stage no intuition has been provided on the mechanism at the basis of the transport, nor its direction. For a physical discussion we refer the reader to Sec.\ref{subsubsec:Exp_transport_pumping}, in particular Fig. \ref{fig:pump_JD_fig}. 

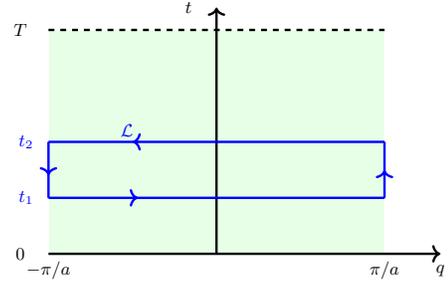
\begin{figure}
\resizebox{0.7\columnwidth}{!}{
\begin{tikzpicture}
\draw[white, fill=green!10!white] (3,4) rectangle (-3,0);

\draw[blue,very thick](3,1)--(3,2);
\node at (3,-0.3) {${\Large \pi/a}$};

\draw[blue,very thick](-3,1)--(-3,2);
\node at (-3,-0.3) {${\Large -\pi/a}$};

\draw[-{>[scale=1]},blue,very thick](-3,1.5)--(-3,1.4);
\draw[-{>[scale=1]},blue,very thick,](3,1.4)--(3,1.5);
\draw[-{>[scale=1]},blue,very thick](-1.5,1)--(-1.4,1);
\draw[-{>[scale=1]},blue,very thick,](-1.4,2)--(-1.5,2);
\node at (-1.6,2.2) {\textcolor{blue}{${\Large {\cal L}}$}};

\draw[->,black,very thick](-3,0)--(4,0);
\node at (4,-0.3) {${\Large q}$};

\draw[->,black,very thick](0,0)--(0,4.4);
\node at (-0.5,4.4) {${\Large t}$};

\draw[black,very thick, dashed](-3,4)--(3,4);

\node at (-3.5,4.0) {${\Large T}$};
\node at (-3.5,0.0) {${\Large 0}$};

\draw[blue,very thick](-3,2)--(3,2);
\node at (-3.4,2) {\textcolor{blue}{${\Large t_2}$}};

\draw[blue,very thick](-3,1)--(3,1);
\node at (-3.4,1) {\textcolor{blue}{${\Large t_1}$}};

\end{tikzpicture}}
\caption{Illustration of the relationship of the change in the Zak phase for a 1D band insulator, with momentum $-\pi/a < q \leq \pi/a$, under adiabatic variation of parameters in time around a cycle of period $T$, \ie with time $0< t \leq T$.
The change in Zak phase between times $t_1$ and $t_2$ is represented by the line integral around a contour ${\cal L}$. When ${\cal L}$ is extended to include the full $q-t$ plane, the periodicity in $q \to q + 2\pi/a$ and $t \to t+ T$ ensures that the line integral must by an integer multiple of $2\pi$. This
integer is the Chern number defined in the $q-t$ plane.
}
\label{fig:pumpingcontour}
\end{figure}

This quantization has the same topological origin as the integer
quantum Hall effect. To see this, consider tracing out the closed loop
in parameter space as time $t$ varies from $t=0$ to $t=T$. The Zak
phase at any given time $t$ is defined by the integral of the Berry connection along the BZ
$-\pi/a< q \leq \pi/a$ at that instant.  The change in 
Zak phase between any two times $t_1$ and $t_2$ can be written as a line integral
\begin{equation}
\phi_{\rm Zak}(t_1) - \phi_{\rm Zak}(t_2) \equiv \oint_{\cal L} \vv{A} \cdot \D\vv{q}
\label{eq:lineint}
\end{equation}
where we label the position in the $q-t$
plane by a two-component vector ${\bm
  q} = (q_1,q_2) = (q,t)$ and the associated Berry connection $A_i
\equiv \I \langle u_\vv{q} | \partial_i | u_\vv{q}\rangle$, with $\partial_1
\equiv \partial/\partial q$ and $\partial_2 \equiv \partial/\partial
t$. The integration contour ${\cal
  L}$ is shown in Fig.~\ref{fig:pumpingcontour}.  The horizontal lines at
fixed $t=t_1$
and $t=t_2$
recover $\phi_{\rm
  Zak}(t_1) - \phi_{\rm
  Zak}(t_2)$, while the integrals on the lines at $q=\pm
\pi/a$ cancel as a result of the periodicity of the BZ. Applying
Stokes's theorem, the line integral in (\ref{eq:lineint}) can be
written as the integral of the Berry curvature, $\Omega = \epsilon_{ij}\partial_iA_j$,  over the area bounded
by ${\cal
  L}$.  The fact that the parameters of the Hamiltonian return to
their original values as $t=0
\to T$ enforces periodicity also in $t$, such that the
$q-t$
plane has the topology of a torus. Thus, extending the contour to
enclose the full region from $t=0$
to $t=T$, thereby computing $\phi_{\rm
  Zak}(0) - \phi_{\rm Zak}(T)$, 
recovers
$2\pi$ times a Chern number. The relevant integral is entirely analogous to Eqn.~(\ref{eq:chern}) with the measure $\D^2 q$ replaced by $\D q \,\D t$.

The link between the winding of $\phi_{\rm Zak}$ and the transported
particle number for a filled band can be established using arguments detailed in
Appendix~\ref{appsubsubsec:pumping}. The transported particle number is determined by computing the change
in the mean particle position $\Delta x$ over one cycle of the pump,
averaged over all states in the band.  The result is that this net
displacement of particles over one cycle, for example the closed loop in Fig.~\ref{fig:zakphaserm1}, is
\begin{equation}
\Delta x= \frac{a}{2\pi} \left[\phi_{\rm Zak}(T) - \phi_{\rm Zak}(0)\right]\,.
\end{equation}
Because $\phi_{\rm Zak}(T)$  equals $\phi_{\rm Zak}(0)$ plus $2\pi$ times the winding number corresponding to the vortex of Fig.~\ref{fig:zakphaserm1},  the displacement is quantized in units of the lattice period $a$. When the band is filled with exactly one particle per state, this entails the quantization of the number of transported particles.

\section{Implementations of Topological Lattices}

\label{sec:implementations}

To this point in this review, we have developed our understanding of lattices, and discussed how topology presents additional ``labels'' tied to individual Bloch bands: providing a new way to categorize band structure.  This section will take the next step and describe the currently implemented techniques by which 1D and 2D band structures with non-trivial topology have been created.

\subsection{Iconic models}

\label{subsec:iconic}

Cold atom experiments are often able to nearly perfectly realize iconic topological models from condensed matter theory.  Here we will briefly describe two such models -- the Harper-Hofstadter model~\cite{Harper1955,Hofstadter1976} and the Haldane model~\cite{Haldane:1988} -- as particularly simple examples of topological lattices and explore what is essential about each of these models.  This will allow us to place experimental approaches in context and to identify what types of new terms must be added to standard optical lattices.

\begin{figure}[tbhp]
\centering
   \includegraphics{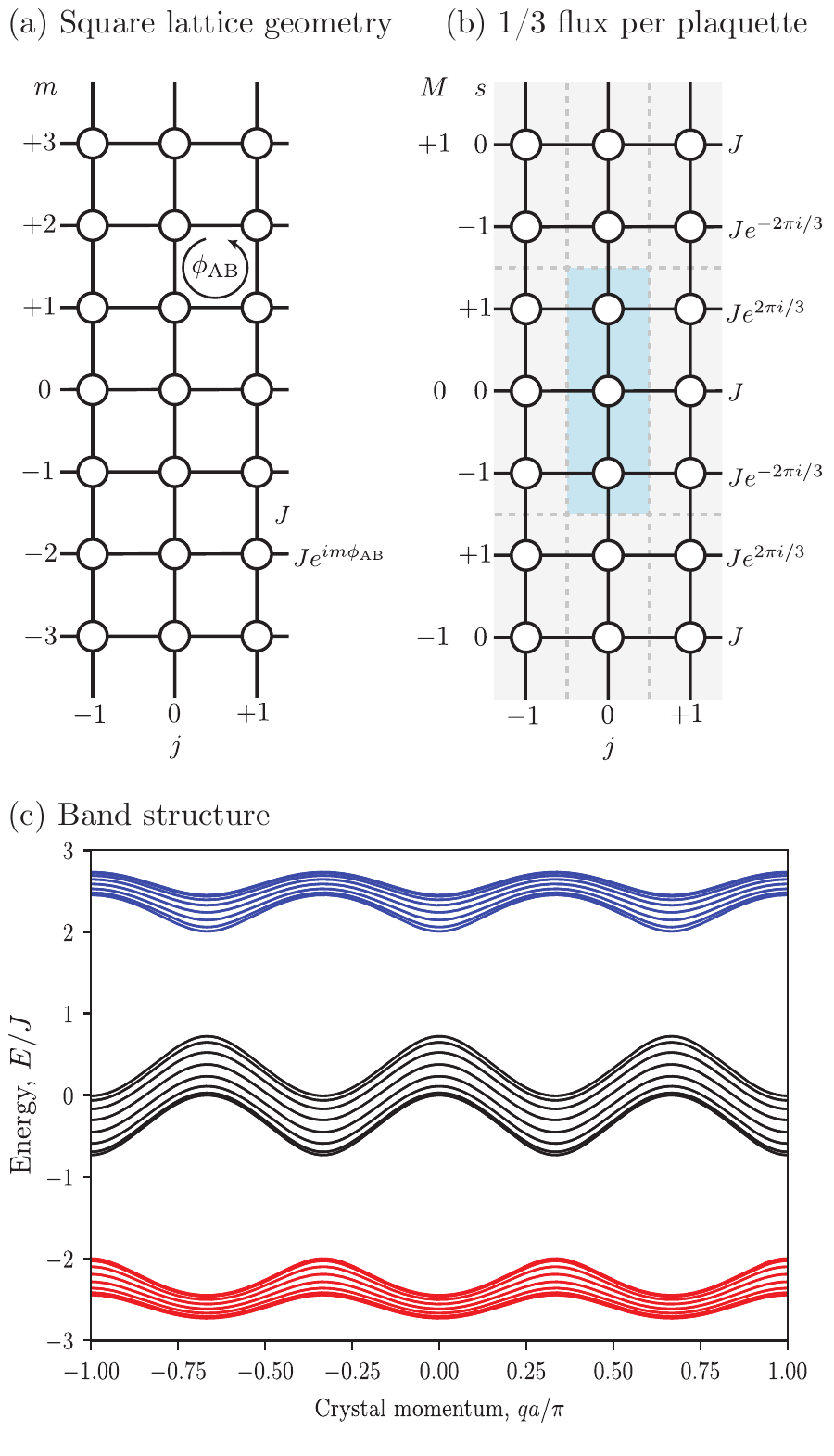}
   \caption{\textbf{Harper-Hofstadter model.} \textbf{a}, Harper-Hofstadter lattice geometry with symmetric hopping $J$ and a flux $\phi_{\rm AB}$ in each plaquette. \textbf{b}, Harper-Hofstadter lattice geometry with flux $\phi_{\rm AB} = 2\pi/3$ per plaquette.  The individual magnetic unit cells are delineated by grey dashed lines with a representative  magnetic unit cell set off in blue for clarity.  \textbf{c}, Computed band structure with $\phi_{\rm AB} = 2\pi/3$ showing the three topological bands, with Chern numbers $+1$, $-2$ and $+1$ built from the three inequivalent sites within the magnetic unit cell.}
\label{fig:hofstadter}
\end{figure}

\subsubsection{Harper-Hofstadter model} The Harper-Hofstadter model -- describing charged particles in a square lattice with a uniform of magnetic field -- derives from the simple 2D tight-binding Hamiltonian 
\begin{align}
\hat H = -J\sum_{j,m}\big( \hat a_{j+1,m}^\dagger \hat a_{j,m} + \hat a_{j,m+1}^\dagger \hat a_{j,m} + \rm{h.c.}\big)\label{eq:tightbinding}
\end{align}
for particles hopping in a square lattice with tunneling strength $J$.  Each individual site of this lattice is labeled a pair of integers $j$ and $m$.  The first term in Eq.~\eqref{eq:tightbinding} denotes tunneling along the $j$ direction (horizontal) and the second term marks tunneling along the $m$ direction (vertical).  This model results in a single, non-topological, cosinusoidal band with full width $8 J$.

As shown in Fig.~\ref{fig:hofstadter}a, our task is to imbue the tunneling matrix elements with nonzero Peierls phase factors so that the phase accrued by a particle encircling a single plaquette is $\phi_{\rm AB}$, equivalent to the Aharonov-Bohm phase $\phi_{\rm AB}= q B A / \hbar$ acquired by a particle with charge $q$ moved around a plaquette of area $A$.  {The resulting complex tunneling matrix elements are required to break time-reversal symmetry and allow nonzero Chern number.} As can be confirmed by the phases depicted in Fig.~\ref{fig:hofstadter}a, the associated Harper-Hofstadter Hamiltonian expressed in the Landau gauge 
\begin{align}
\hat H = -J \sum_{j,m}\big( \E^{\I m \phi_{\rm AB}} \hat a_{j+1,m}^\dagger \hat a_{j,m} + \hat a_{j,m+1}^\dagger \hat a_{j,m} + \rm{h.c.}\big)\label{eq:hofstadter},
\end{align}
gives a phase $\phi_{\rm AB}$ for tunneling around each plaquette.  For rational  $\phi_{\rm AB}/2\pi=p/q$, 
expressed in reduced form, the single band of Eq.~\eqref{eq:tightbinding} fragments into $q$ (generally) topological bands, with zero aggregate Chern number.  

We can focus in on the essential features of this model by considering the special case of one third flux per unit cell, \ie, $\phi_{\rm AB} = 2\pi/3$.  First take note of the tunneling phase $\E^{\I m \phi_{\rm AB}}$ for motion along $j$: As shown in Fig.~\ref{fig:hofstadter}a, this tunneling phase depends on $m$ and has a spatial period of three lattice sites, implying that the lattice's unit cell is enlarged beyond the plaquettes of the underlying square lattice (the unit cells without magnetic flux) to three plaquettes at $\phi_{\rm AB} = 2\pi/3$.  In Fig.~\ref{fig:hofstadter}b, these unit cells are graphically indicated by the gray dashed lines, with a representative unit cell shaded light blue for clarity.  Each of these unit cells is identified by integers $j$ and $M$.  In order to distinguish between the three inequivalent sublattice sites within each unit cell, we introduce also the site index $s=-1,0,+1$, related to the individual plaquette index $m$ via $m=3M + s$.

As a result of this expanded unit cell, the associated Brillouin zone is reduced to $1/3$ of its initial size along the $m$ direction, and the number of bands correspondingly increases from 1 to 3.  Following textbook techniques, we express this Hamiltonian in the Fourier representation, \ie, giving states labeled by their crystal momentum $\vv{q} = (q_j, q_m)$ along with the sublattice index $s$, \ie for $\phi_{\rm AB} = 2\pi/3$ the states are $\left\{\ket{\vv{q}, -1}, \ket{\vv{q}, 0}, \ket{\vv{q}, +1} \right\}$.  For each crystal momentum $\vv{q}$ the Hamiltonian matrix coupling these sublattice sites together is 
\begin{align}
\frac{H_{\vv{q}}}{J} &\deq  -\left(\begin{array}{ccc}2\cos\left(q_ja-2\pi/3\right) & 1 & \exp(\I q_m (3a_s)) \\1 & 2\cos\left(q_ja\right) & 1 \\ \exp(-\I q_m (3a_s)) & 1 & 2\cos\left(q_ja+2\pi/3\right)\end{array}\right);
\end{align}
where $a$ and $a_s$ denote the nearest neighbor lattice spacings in the spatial and synthetic dimensions. (The notion of length $a_s$ in the synthetic dimension is a useful book-keeping device, by which $3 a_s$ is the side of the expanded unit cell.)
The three eigenvalues of this matrix define three separate bands.  Figure~\ref{fig:hofstadter}c shows the resulting band structure, where each of the three bands is endowed with non-zero Chern number.  The expansion of the unit cell to contain three sublattice sites is essential for the formation of topological bands.  Recall that Chern numbers are derived from the integrated Berry curvature over the Brillouin zone.  For a tight-binding model, the Berry curvature can only be non-zero when each Bloch wave function has a spin or pseudo-spin degree of freedom, here provided by the sublattice degree of freedom.

In Sec.~\ref{sec:hopping} we describe how to experimentally imprint these hopping phases using tailored laser fields and will comment on the limitations of different experimental approaches.

\subsubsection{Haldane model} The Haldane model~\cite{Haldane:1988}, an extension of the well-known honeycomb lattice, was an early model of topological band structure but without the presence of an overall magnetic field which requires the expansion of the unit cell size.  Figure \ref{fig:haldane}a plots the Haldane lattice in the conventional honeycomb geometry. We also show its deformation to a ``brick-wall'' geometry most relevant for its experimental realization with cold atoms. The tunneling matrix elements (black lines) with strength $J$ define the underlying honeycomb lattice, along the three nearest-neighbor bonds.
As indicated, even the simple honeycomb lattice describes a dimerized lattice with $\left|{\rm A}\right>$ and $\left|{\rm B}\right>$ sublattice sites (possibly offset in energy by $\pm\Delta$), making it an ideal starting point for realizing topological band structures.  Figure \ref{fig:haldane}b shows the resulting two bands kissing at a pair of Dirac points.  Haldane's addition of next-nearest-neighbor tunneling with strength $J'$ and phase $\phi_{\rm AB}$ (pink lines in Fig.~\ref{fig:haldane}a, along bonds connecting sites of the same sublattice)
renders this model topological. 

\begin{figure*}[tbhp]
\centering
\include{figures/Fig_Haldane-brickwall-new-revised}
   \caption{\textbf{Haldane model.} \textbf{a},  Haldane lattice geometry showing overall honeycomb lattice structure with the addition of next-nearest-neighbor hopping with Aharonov-Bohm tunneling phases $\phi_{\rm AB}$.  The blue box marks the two sites comprising a single unit cell. We also show the deformation of the underlying lattice to the brick-wall geometry, used to plot the dispersions with $q_x a = \vv{q}\cdot\vv{a}_1$ and $q_y a =\vv{q}\cdot \vv{a}_2$. \textbf{b}, Haldane model phase diagram showing the two topological lobes immersed in a non-topological background. \textbf{c}, Band structure computed with only  nearest-neighbor tunneling (black lines in \textbf{a}), showing the familiar pair of Dirac points from this ``brick-wall" lattice.  \textbf{d}, Band structure computed in the topological phase at the marked red point in \textbf{b}, with $\phi_{\rm AB}=\pi/2$ and ``tilt'' $\Delta = 0$, $J'=J/10$. \textbf{e}, Band structure at the topological transition, marked by the  blue point in \textbf{b}, with $\phi_{\rm AB}=\pi/2$ and $\Delta = -3\sqrt{3}J'$, showing the formation of a single Dirac point.}
\label{fig:haldane}
\end{figure*}

This Hamiltonian too can be readily expressed in a crystal momentum dependent matrix, now with two contributions.  Firstly the energy offset and nearest-neighbor tunneling from the underlying honeycomb lattice contribute the matrix
\begin{align}
H_{\vv{q},0} & = \left(\begin{array}{cc} \Delta & -J\sum_i \exp(-\I \vv{q}\cdot \vv{\rho}_i)  \\-J\sum_i \exp(\I \vv{q}\cdot \vv{\rho}_i)   & -\Delta\end{array}\right),
\label{eq:haldaneimplement}
\end{align}
and the next-nearest-neighbor links contribute a second term
\begin{align}
H_{\vv{q},1} &\deq -2J'\left(\begin{array}{cc}\sum_i \cos(\vv{q}\cdot\vv{A}_i\!-\!\phi_{\rm AB}) & 0 \\ 0 & \sum_i \cos(\vv{q}\cdot\vv{A}_i\!+\!\phi_{\rm AB})\end{array}\right) \,.
\end{align}
{We have defined $\vv{A}_1=\vv{a}_1, \vv{A}_2 = - \vv{a}_2, \vv{A}_3 = \vv{a}_2-\vv{a}_1$, with the vectors $\vv{\rho}_i$ and $\vv{a}_i$ labeling nearest-neighbor and next-nearest-neighbor separations as indicated in Fig.~\ref{fig:haldanewrap}a. (For practical realizations of the brick-wall lattice, the next nearest neighbour coupling, along $\vv{A}_3$, is suppressed compared to couplings along $\vv{A}_{1,2}$. However, we use equal strengths for all in 
Fig.~\ref{fig:haldane}.)}

Figure \ref{fig:haldane}b plots the topological phase diagram associated with this model as a function of the Aharonov-Bohm tunneling phases $\phi_{\rm AB}$ and tilt $2\Delta$.  This system supports three distinct topological regions: zones with Chern number $\pm 1$, with the majority of parameter space in the topologically trivial phase with Chern number $0$.

The Haldane model is particularly amenable to experimental study because tuning experimental parameters such as $\phi_{\rm AB}$ can directly drive topological phase transitions.  While for the Harper-Hofstadter lattice, tuning $\phi_{\rm AB}$ does lead to different Chern numbers, the size of the unit cell also changes, leading to more dramatic changes in the band structure.  In Sec.~\ref{sec:force} we show how strongly driving the parameters of a brick-wall lattice 
can break time reversal symmetry, and imbue the lattice's two bands with non-trivial topology.

In these examples of topological band structure, we identified two common elements that experimentalists need to introduce to create non-trivial topology: complex-valued tunneling matrix elements, and unit cells with more than one underlying lattice site, or spin-degree of freedom.

\subsection{Realization of SSH model}\label{subsec:realization_SSH}

\begin{figure}[tbh]
\centering
   \includegraphics{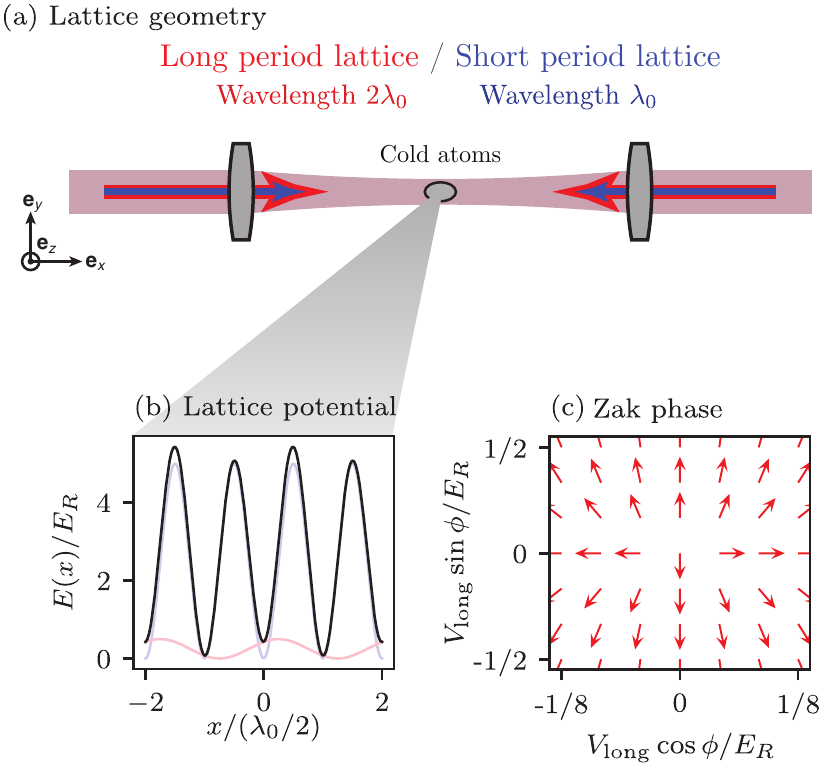}
   \caption{\textbf{ Implementation of the SSH and Rice--Mele models.} \textbf{a}, Representative laser configuration:  A pair of overlapping lasers with wavelengths $\lambda_0$ and $2\lambda_0$ subject a cloud of ultracold atoms to  1D optical lattices, with period $\lambda_0/2$ and $\lambda_0$ respectively, and a spatial relative phase $\phi$. \textbf{b}, Energies of these two lattices and combined potential, showing the long period lattice shifted in position with respect to the short period lattice by a controllable phase shift $\phi$.  The SSH model is realized for $\phi=\pi/2 + n\pi$ (for integer $n$), in which case all minima have the same energy and are separated by potential barriers with staggered height. An example of band structure in this case is shown in Fig. \ref{fig:SSH2}. The choices $\phi=n\pi$ (for integer $n$) lead to equal barrier heights between adjacent sites, and staggered site energies, but with $J'=J$, $\Delta\neq 0$. The displayed data is for an intermediate case of $\phi=\pi/4$ where $J'\neq J$ and $\Delta\neq 0$.
\textbf{c}, Zak phase for the lower pair of bands controlled by tuning the phase shift $\phi$ between the long and short period lattices and the strength of the long-period lattice, as represented by the unit vector $(\cos\phi_{\rm Zak}, \sin\phi_{\rm Zak})$.}
\label{fig:SSH}
\end{figure}

\begin{figure}[tbh]
\includegraphics{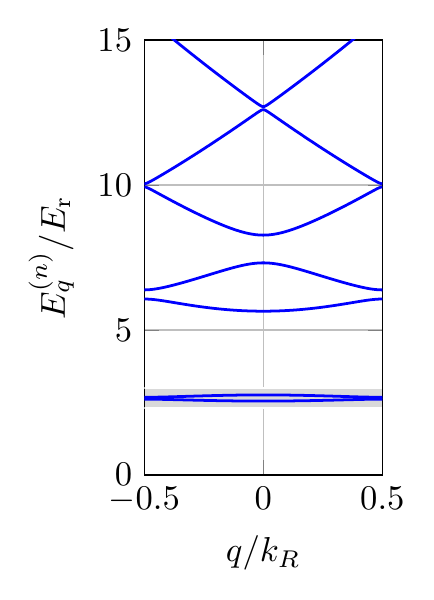}
\includegraphics{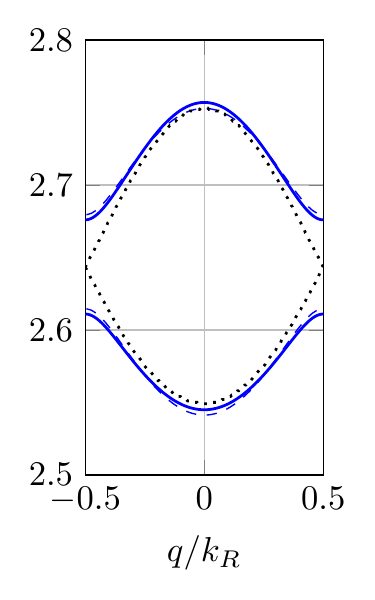}
\caption{{\bf Bipartite lattice band structure compared to the Rice--Mele model.} Left: Band structure computed for the superlattice potential given by Eq.~\eqref{eq:superlattice_potential} with $V_{\rm short}=6\,E_R$, {$V_{\rm long}=E_R$} and a relative phase $\phi=\pi/2$, corresponding to the particular case of the SSH model. Right: Zoom on the lower pair of bands (continuous lines). The energy offset between adjacent minima and a fit to the prediction (\ref{eq:energies_Rice_Mele}) for the band structure of the Rice--Mele model (dashed lines) allow one to extract the practical values of $\Delta$, $J$ and $J'$. (Here $\Delta=0$, $J=0.069\,E_R$ and $J'=0.037\,E_R$.) The dotted lines show the (folded) lowest band for $V_{\rm long}=0$, up to a global energy shift.}
\label{fig:SSH2}
\end{figure}

The 1D-SSH model of polyacetylene, and its generalization the Rice--Mele model are among the most simple topological models to realize.  As described in Sec.~\ref{subsubsec:rm}, the Rice--Mele model, Eqn.~(\ref{eq:hrm1}), consists of a bipartite 1D lattice with tunneling strengths alternating between $J$ and $J^\prime$, and energies of the two sublattice sites, staggered by $\pm\Delta$.

Figure~\ref{fig:SSH}a depicts a typical laser system required to approximate this idealized model, similar to the experimental realization of~\textcite{Atala2013}.  Here a conventional 1D optical lattice with period $\lambda_0/2$ is generated by a pair of counter-propagating lasers each with wavelength $\lambda_0$.  In this lattice, the tunneling is uniform with strength  $J_0$  and the energy minima of the lattice sites are degenerate, as indicated by the pale blue curve in Fig.~\ref{fig:SSH}b.  A second, weaker, lattice with period $\lambda_0$, generated by a laser with wavelength $2\lambda_0$ (pink curve in Fig.~\ref{fig:SSH}b), gives a combined potential (black curve in Fig.~\ref{fig:SSH}b) with generally staggered energy minima and alternating tunneling.  This gives the overall potential
\begin{equation}
V(x) = V_{\rm short} \sin^2(k_R x) + V_{\rm long}\sin^2\left[(k_R x + \phi)/2\right],
\label{eq:superlattice_potential}
\end{equation}
where we have defined: the single photon recoil momentum $\hbar k_R = 2\pi\hbar/\lambda_0$, the associated recoil energy $E_R = \hbar^2 k_R^2 / 2 m$, and the atomic mass $m$ of the atom under study.  

Figure~\ref{fig:SSH}c shows the  Zak phase $\phi_{\rm Zak}$, Eqn.~(\ref{eq:zak}), computed for this physical system in terms of the experimental control parameters.  This figure clearly depicts the singularity expected  when $V_{\rm long} = 0$, which at $\phi = \pi/2$ corresponds to the location of the topological transition in the SSH model when $J=J^\prime$.

Fig. \ref{fig:SSH2} shows the bottom part of the (infinite) band spectrum for $V_{\rm short} = 6 E_R$ and 
{$V_{\rm long}= 1 E_R$}, with the relative phase set to $\phi=\pi/2$. The right part of this figure shows a zoom on the lowest pair of bands, which are the only relevant ones when the temperature and the interaction energies are comparable to or lower than $E_R$. The result for $V_{\rm long}=0$ is indicated for comparison with dotted lines. The BZ is reduced in size by a factor of two as compared to that of the short period lattice only. This is reflected by the $V_{\rm long} = 0$ bands touching at the edge of the BZ; in 1D this marks each such linked-pair as truly being one band in a doubled BZ. 

Then the additional independent control of both $J$, $J^\prime$ and $\Delta$ requires two additional experimental degrees of freedom.  In this realization these parameters are the relative phase $\phi$ of the long and short period lattices (displacing one lattice with respect to the other) and the depth of the long-period lattice $V_{\rm long}$. Fig. \ref{fig:SSH2} has been calculated for $\phi=\pi/2$, in which case all minima of $V(x)$ have the same energy while the barrier heights between adjacent minima alternate between two values. This realizes the SSH model and the dashed lines in the right panel of  Fig.\ref{fig:SSH2} show a fit of the SSH prediction (\ref{eq:energies_SSH}) to the two lowest bands, providing thus the relevant values of $J$ and $J'$.  

\paragraph*{Model parameters.} \label{sec:realization_SSH:parameters}
The $V_{\rm short} = 6 E_R$ short period lattice depth used in these simulations -- a typical laboratory scale -- sets the nominal tunneling strength of $J\approx 0.05 E_R$.  Intuitively, we expect that the energy difference between the minima to be about  $V_{\rm long} \cos(\phi)$.  Similarly, the barriers between the sites differ in height by roughly $V_{\rm long} \sin(\phi)$. The tunneling strength in the effective SSH model has a non-trivial, but monotonically decreasing, exponential-type behavior in the barrier height.  This then begs the question of obtaining the parameters of the Rice--Mele model in Eq.~\eqref{eq:hrm1}, including the two tunneling strengths $J$ and $J^\prime$ along with the energy difference $\Delta$ between sub-lattice sites.

First, recall that the band structure of simple 1D optical lattice potential $V \sin^2(k_R x)$ only approaches that of a tight-binding model with nearest-neighbor tunneling when $V\gg E_R$.  For nearest-neighbor tunneling strength $J$, the resulting dispersion is simply $-2J\cos(\pi q/k_R)$.  As a result the effective nearest-neighbor tunneling strength can be directly obtained from the lowest term in a Fourier expansion of the band structure of the physical 1D optical lattice. (The higher terms in the series describe longer range tunneling, which becomes negligible for deep lattices, see~\textcite{Jimenez-Garcia2013} for an introduction.)
This approach is insufficient for the Rice--Mele band structure
$\pm[\Delta^2 + \delta J^2 + (4\bar J^2 - \delta J^2) \cos^2(\pi q/k_R)]^{1/2}$,
because fits to this dispersion alone cannot effectively disentangle $\Delta$ from $\delta J = J-J^\prime$, nor $\delta J$ from $\bar J = (J+J^\prime)/2$.
One practical resolution to this difficulty is to employ symmetry and evaluate the band structure of the bipartite lattice for two cases: firstly compute the band structure for $\phi=0$ where $J$ and $J^\prime$ are manifestly equal, and then compute the band structure for $\phi=\pi/2$ where $\Delta=0$.  This then allows for the independent determination of  $\Delta$, $J$ and $J^\prime$.  For example, for $V_{\rm short} = 6 E_R$ and $V_{\rm long} = E_R$ this procedure gives $\bar J = 0.053 E_R$, almost independent of $\phi$.  For $\phi = 0$, we further find $\Delta = 0.43 E_R$ and {$\delta J = 0$},  while for $\phi = \pi/2$, this becomes $\Delta = 0$ and $\delta J = 0.032 E_R$.

\subsection{Inertial forces}\label{sec:force}

The common experience of starting water in a pail spinning by moving the bucket in a circular manner -- not rotating the bucket -- suggests that applied inertial forces might produce effects akin to those present in rotating systems: described best by effective Lorentz forces.  We shall see below how these ideas are implemented for ultracold atoms in optical lattices and also come to understand the limitations of these approaches.

The tight-binding model depicted in Eq.~\eqref{eq:tightbinding} is representative of the tunnel-coupling structure present for atoms confined in optical lattices.  It is of particular importance that the tunneling matrix elements $J$ are real-valued (more specifically, transformations between different gauges can introduce ``trivial'' complex amplitudes to the tunneling matrix elements, but in these simple lattices there always exists a gauge choice for which the amplitudes are real valued).  In this section we will develop a simple model illustrating how inertial forces (linear potential gradients or equivalently spatially shaking the lattice potential) can add tunable complex hopping phases to these matrix elements.

\begin{figure}
\centering
   \includegraphics{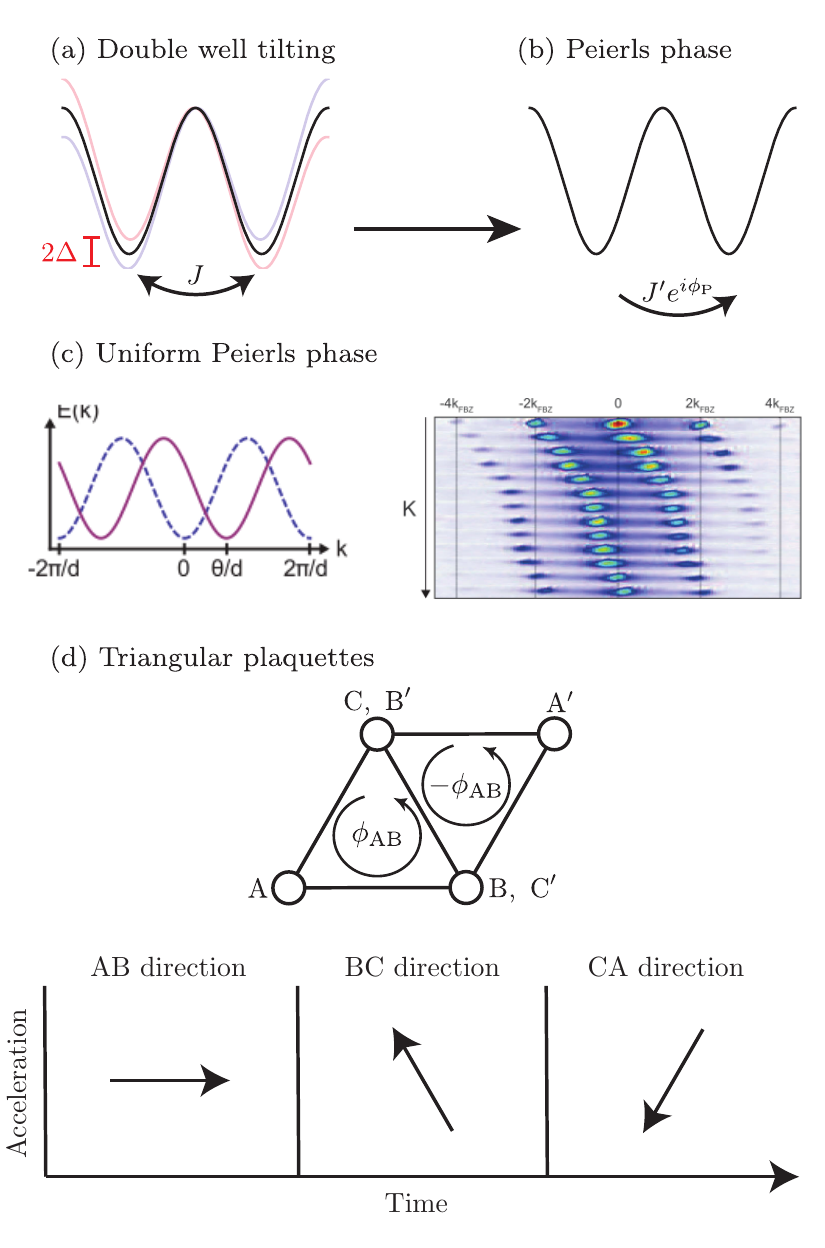}
   \caption{\textbf{Inertial forces.} \textbf{a}, a simple double well lattice subject to modulation, creating \textbf{b} experimentally tunable hopping phases.  \textbf{c} Shaking or tilting in 1D gives rise to a uniform Peierls phase factor that shifts the minima of the tight-binding band structure~\cite{Struck2012}.  \textbf{d}, Shaking in 2d can break time reversal symmetry giving rise to topological lattices.}
\label{fig:force}
\end{figure}

From a quantum mechanical perspective the essential concept is to engineer non-trivial phases acquired by the unitary evolution of a time-periodic Hamiltonian which can be cast as complex hopping amplitudes in an effective time-independent Hamiltonian.  This physics is minimally captured by the tunnel coupled pair of lattices sites shown in Fig.~\ref{fig:force}, essentially comprising a single unit cell of the Rice--Mele model.  We aim for a two-site model described by the Hamiltonian
\begin{align}
\hat H &= -J \left(\ket{r}\bra{l} \E^{\I\phi_{\rm P}} + \ket{l}\bra{r} \E^{-\I\phi_{\rm P}}\right) + \Delta\left(\ket{r}\bra{r} - \ket{l}\bra{l}\right)\nonumber\\ 
&= -J \cos(\phi_{\rm P})\hat \sigma_x + J \sin(\phi_{\rm P})\hat \sigma_y + \Delta\hat \sigma_z, \label{eq:peierls}
\end{align}
including a laboratory-controllable tunneling phase $\phi_{\rm P}$, alas, our lattice is born with $\phi_{\rm P}=0$.  In the second line we expressed this Hamiltonian in terms of the Pauli operators $\hat \sigma_{x,y,z}$, allowing us to follow a simple analysis of a spin-1/2 system~\cite{Haroche1970}.

We earlier noted that gauge transformations can introduce complex-valued tunneling phases.  A gauge transformation is simply a position-dependent unitary transformation that adjusts the local phase of the wavefunction and compensates the Hamiltonian accordingly; in our double-well model, a gauge transformation in the spatial picture becomes a $\hat \sigma_z$-rotation in the spin picture.  Evidently, the Peierls phase factor $\phi_{\rm P}$ can be fixed to a non-zero value by the choice of gauge.  Since such a choice is of no physical consequence, it is instead the ability to change $\phi_{\rm P}$ (either spatially to induce Aharonov-Bohm fluxes, or temporally to induce artificial electric fields) that is the essential content of this discussion.  Here we adopt the most straightforward gauge choice that sets $\phi_{\rm P}=0$ for unadulterated lattices.

In the following, we show how a time-periodic linear gradient provides control over both the tunneling amplitude $J$ and phase $\phi_{\rm P}$.  In our two-site model, this modulation (\ie, detuning) is described by $\Delta(t) = \Delta(t+T)$, with period $T$, angular frequency $\omega = 2\pi/T$, and with zero per-cycle average $\langle\Delta(t)\rangle_T=0$.

It is straightforward to eliminate the time-dependent $\Delta(t)$ term in Eq.~\eqref{eq:peierls} (initially, with $\phi_{\rm P} = 0$) by making the unitary transformation
\begin{align}
\ket{\psi^\prime(t)} &= \exp\left(\frac{\I}{\hbar} \int_0^t  \Delta(t^\prime) \hat\sigma_z \,\D t^\prime\right)\ket{\psi(t)};\label{eq:force:gauge}
\end{align}
in the language of quantum optics this is akin to the transformation into the time-dependent ``interaction'' picture.  Since this transformation is a $\hat \sigma_z$-rotation it is equivalent to a time-dependent gauge transformation, leading to a {\it non-zero}, time-dependent Peierls phase factor
\begin{align}
\phi_{\rm P}(t) &= \frac{2}{\hbar}\int_0^t  \Delta(t^\prime)\,\D t^\prime.
\label{eq:modulated_phase}
\end{align} 
When the modulation frequency's associated energy $\hbar \omega$ is greatly in excess of the tunneling $J$, we make a rotating wave approximation to replace the time-dependent terms introduced by this rotation by their time averages, giving the time-averaged interaction picture Hamiltonian
\begin{align}
\hat H &= -J \langle \cos \phi_{\rm P}(t)\rangle_T \,\hat \sigma_x + J \langle\sin \phi_{\rm P}(t)\rangle_T\, \hat \sigma_y,
\end{align}
with a potentially non-zero dc Peierls phase factor
\begin{equation}
\tan(\phi_{\rm P, dc}) =  \frac{\langle \sin \phi_{\rm P}(t) \rangle_T}{\langle \cos \phi_{\rm P}(t) \rangle_T}\,.
\end{equation}
Physically, the time-dependent gauge transformation in Eq.~\eqref{eq:force:gauge} allows the system to sample a range of Peierls phase factors, and retain a non-zero average.  In effect, our task is to make $\langle \sin \phi_{\rm P}(t) \rangle_T$ non-zero; and because $\sin$ is an odd function we seek a waveform $\phi_{\rm P}(t)$ that takes on positive and negative values in an ``imbalanced'' manner.

The most simple example to deploy in the laboratory is a monochromatic sinusoidal modulation $\Delta(t)=\Delta\cos(\omega t)$ of the tilt.  In this case, the Bessel series expansion gives $\langle \cos \phi_{\rm P}(t) \rangle_T = {\mathcal J}_0(2\Delta/\hbar\omega)$ and $\langle \sin \phi_{\rm P}(t) \rangle_T = 0$.  Because the integrated sinusoidal waveform takes on positive and negative values with equal frequency the average tunneling phase is zero, however, this modulation does renormalize the tunneling strength $J\rightarrow J \times {\mathcal J}_0(2\Delta/\hbar\omega)$, as was observed experimentally~\cite{Lignier2007}.  While monochromatic sinusoidal modulation is simple to deploy, it can obscure the underlying physics, rapidly becoming a tangle of Bessel functions.

Instead consider a waveform consisting of two delta-function ``kicks'' in each drive cycle~\cite{Sorensen:2005,Anderson2013a}, the first with strength $\Delta/\omega$ at time $t = 0$ and the second with strength $-\Delta/\omega$ at time $f\times T$, a fraction $f$ through the drive period $T$.  Integrating this waveform gives the time-dependent Peierls phase factor
\begin{align}
\phi_{\rm P}(t) &= \frac{2\Delta}{\hbar\omega}\left\{\begin{array}{cc} 1-f & {\rm for}\ 0\leq t \leq f T \\
-f & {\rm for}\ f T < t \leq T \end{array}\right.,
\end{align}
a pulse-width modulated waveform with zero average, with duty cycle $f\in[0,1)$.  For $f\notin\left\{0,1/2\right\}$, the resulting asymmetric and skewed waveform leads to a non-zero average of $\langle \sin \phi_{\rm P}(t) \rangle_T$.  

For the special case $\Delta/\hbar\omega = \pi$ , the time-averaged Peierls phase factor is $\phi_{\rm P, dc} = - 2\pi (f-1/2)$ with tunneling strength unchanged at $J$.  The basic physical picture is that after an atom tunnels between sites it acquires a phase different from what it would have acquired on its initial site, and $\phi_{\rm P, dc}$ expresses the differential phase acquired upon returning to its initial site.  The time-dependent phase $\phi_{\rm P}(t)$ must break time-reversal symmetry to give a non-zero average of $\sin \phi_{\rm P}(t)$.  
Figure~\ref{fig:force}c depicts the first experimental realization of a non-zero Peierls phase factor imprinted using inertial forces~\cite{Struck2013}.  Rather than tilting the lattice potential, ~\onlinecite{Struck2013} found it more convenient to spatially shake the lattice potential by modulating the phase of the lasers creating the optical standing wave, giving the potential $V(\hat x,t) = V_0 \cos[2 k_{\rm R} (\hat x - \delta x(t))]/2$.  Here $V_0$ is the lattice depth and $k_{\rm R}$ is the two photon recoil momentum from the wavelength $\lambda$ of the lasers creating the lattice potential.  

Although this shaking process is physically quite different from applying a time-dependent gradient, they are functionally equivalent. The connection between the two can be seen clearly in terms of a pair of time-dependent transformations.  We begin by using the spatial displacement operator $\hat D_x[-\delta x(t)] = \exp[\I \hat k \delta x(t)]$ to transform to the non-inertial frame co-moving with the lattice, \ie $\hat D_x[-\delta x(t)] \hat x \hat D_x[-\delta x(t)]^\dagger = \hat x + \delta x(t)$.  This exchanges the lattice's motion for a new time-dependent contribution to the Hamiltonian $-\hbar \hat k \partial_t \delta x(t)$. The once-transformed Hamiltonian
\begin{align}
\hat H^{(1)}(t) =& \frac{\hbar^2}{2m} \left[\hat k -\frac{m}{\hbar}\partial_t \delta x(t)\right]^2 + \frac{V_0}{2} \cos(2 k_{\rm R} \hat x)\\
&-\frac{m}{2}\left[\partial_t \delta x(t)\right]^2,\nonumber
\end{align}
contains a new time-dependent vector potential and a global time-dependent energy shift that does not impact the system's dynamics.  Evoking Hamilton's equation $\dot x = \partial_{\hbar k} \hat H$, we see that the appearance of this vector potential simply describes the fact that in the moving frame the velocity of an object differs from that in the lab frame by the instantaneous velocity of the moving frame, $-\partial_t \delta x$.

We complete our argument using the time-dependent momentum displacement operator $\hat D_k[-\delta k(t)] = \exp[-i\delta k(t) \hat x]$ -- a gauge transformation -- that converts the time-dependent vector potential into a potential gradient $\hat V = \left[m \partial^2_t \delta x(t)\right] \hat x$.  This reminds us that inertial forces are present in accelerating frames, and informs us that experimenters are free to either use shaken lattices or potential gradients equivalently to produce the inertial forces required to induce $\phi_{\rm P}$.

While this non-zero and uniform Peierls phase factor is an essential first step for emulating Aharonov-Bohm fluxes, a uniform Peierls phase factor in 1D can be eliminated via a gauge transformation (although any temporal change can still lead to effective electric fields). In contrast by moving to 2D systems such as in modulated and shaken honeycomb-geometry lattices, this technique has non-trivial alterations to band structure~\cite{Struck2013} and including those topologically equivalent to the Haldane model~\cite{Jotzu:2014}.

We now extend our discussion to 2D to understand Aharonov-Bohm fluxes.  Figure~\ref{fig:force}d depicts a minimal model of shaking in 2D; the top panel illustrates the two triangular plaquettes that make up a single unit cell of a triangular lattice, while the bottom panel graphs a shaking protocol that first accelerates parallel to the A-B link of the left plaquette (\ie, along $\vv{e}_x$), then accelerates parallel to B-C, and finally accelerates parallel to C-A.  For each link, this protocol leads to the same time-dependent Peierls phase factor given in (\ref{eq:modulated_phase}), with each phase shifted in time by $2\pi/3$, giving the same non-zero tunneling phase $\phi_{\rm P, dc}$ to each side of the plaquette.   This then leads to an overall Aharonov-Bohm phase $\Phi_{\rm AB} = \phi_{\rm P, dc}$.  

It might appear that our task is complete, but have we truly created a uniform Aharonov-Bohm phase over all plaquettes?  Unfortunately~\textcite{Struck2013} demonstrated that this is not the case.  Following the same argument for the second (inverted) plaquette that completes a single unit cell shows that accumulated phases give a negative flux $\Phi_{\rm AB} = -\phi_{\rm P, dc}$: leading to a staggered flux.  Therefore on average the Aharonov-Bohm flux through this lattice is zero.  One way to understand this is that the Peierls phase factors are created with uniform amplitude throughout the lattice, rather than with the linear dependence on position as for our Landau-gauge example of the Harper-Hofstadter Hamiltonian.  \textcite{Sorensen:2005} proposed remedying this by applying a potential whose gradient itself increased away from the systems center, which they termed a quadrupole potential.

Evidently shaking of this type does not provide a straightforward route for realizing Hamiltonians such as the Harper-Hofstadter model with uniform fields, but it has proven a successful route for creating a Haldane-type Hamiltonian that has a zero average flux but still with topological bands~\cite{Jotzu:2014}. Important to this realization is the fact that the effective Hamiltonian for the shaken lattice acquires a next-nearest-neighbor hopping with a non-zero Peierls phase factor. 
Beyond the time-averaged Hamiltonian discussed above,
perturbative corrections from the  time-varying part of the nearest-neighbor tunneling lead to next-nearest-neighbor tunneling of order $J' \sim J^2/(\hbar\omega)$, arising from a second-order process through an intermediate virtual state detuned by $\hbar\omega$. Such terms are conveniently obtained from the Magnus expansion of this tight-binding model is described in  Appendix~\ref{app:floquet}. {An analysis of the effective model for the shaken lattice that goes beyond this Magnus expansion approach is provided by \textcite{modugno2017}.}

\subsection{Resonant coupling: laser-assisted tunneling}\label{sec:hopping}

The previous section outlined the broad range of engineered tunnel couplings possible via temporal modulation of the parameters of the lasers underlying the lattice potential.  While it was possible to create complex-valued tunneling, it was not possible to independently control the phase and amplitude of tunneling on each lattice-link: more control is required.  Following the ideas in~\textcite{Jaksch2003}, we describe how such a fine-grained control is in principle possible using laser-assisted tunneling, and how experimental implementations have approached this task.  As we shall see, although this laser-assisted tunneling is effected by temporal modulation, the modulation results from additional lasers, rather the lasers from which the underlying lattice is assembled.

\begin{figure*}[tbhp]
\centering
   \includegraphics{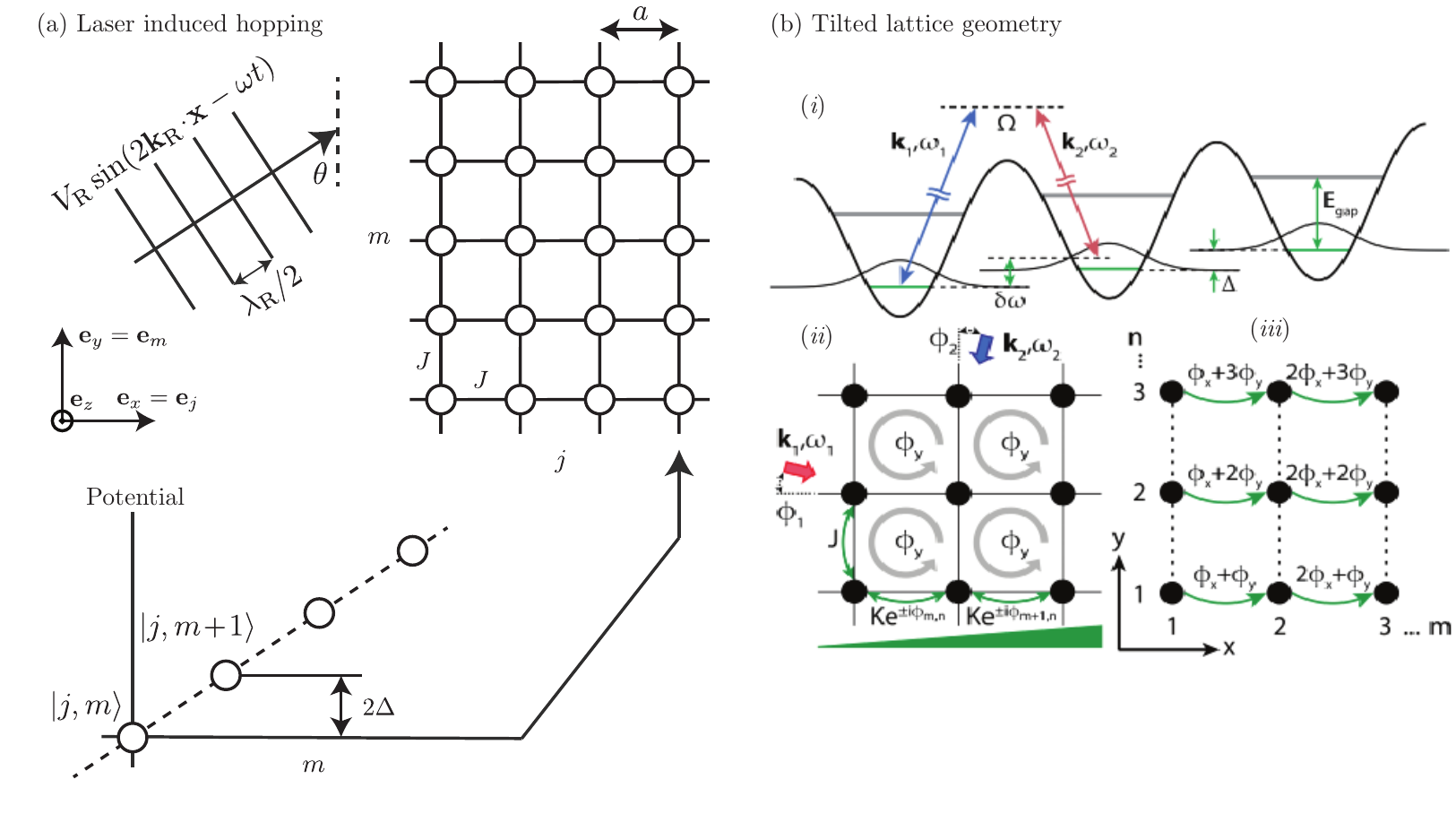}
   \caption{\textbf{Laser induced hopping.} \textbf{a} 2D square lattice (right) with a potential gradient along $\vv{e}_m$ (vertical) illuminated by a traveling wave potential.  The coupling of any pair sites of this lattice $\ket{j,m}$ and $\ket{j,m+1}$ is qualitatively described as a two-level system with detuning $2\Delta$ coupled by the traveling wave.  \textbf{b} Technique for creating the half-flux Harper-Hofstadter Hamiltonian in tilted spin-dependent lattices as implemented in MIT~\cite{Miyake2013} similar to~\textcite{Aidelsburger2013}.}
   \label{fig:laser_hopping}
\end{figure*}

The essential concept of this technique is straightforwardly illustrated in the 2D square lattice depicted in Fig.~\ref{fig:laser_hopping}a: the native tunneling along the vertical direction 
is first eliminated by applying a potential gradient (\ie tilting the lattice), then coupling between neighboring lattices sites is re-established with a traveling wave potential.  Here, the spatially non-uniform phase of the traveling wave is imprinted upon atoms as they are moved from site-to-site: described by complex-valued tunneling amplitudes.  Because local optical phases are relatively easy to control (for example by creating higher order optical modes such as Laguerre-Gauss modes, as has  been done in~\textcite{Chen2018}), these techniques in principle allow for more subtle engineering of the local Aharonov-Bohm phases than is possible with whole-scale modulation of lattice parameters.  Still, current implementations~\cite{Aidelsburger2013,Miyake2013} rely only on the uniformly changing phase from plane waves to generate homogeneous fields.

The basic principle can be understood in terms of the same sort of two-level system discussed in Sec.~\ref{sec:force}, but from a perspective in which the rotating wave approximation (RWA) is valid.  In the present case, we focus on two neighboring lattice sites in Fig.~\ref{fig:laser_hopping}a, labeled by $\ket{j,m}$ and $\ket{j,m+1}$ coupled by the traveling wave potential $V_{\rm R}(\vv{x}) = V_{\rm R} \sin(2 \vv{k}_{\rm R}\!\cdot\!\vv{x} - \omega t)$ that locally modulates potential intersecting $\vv{e}_m$ with angle $\theta$.  Physically this is directly realized~\cite{Aidelsburger2013,Miyake2013} by a pair of interfering lasers giving rise to a moving standing wave with periodicity $\lambda_{\rm R}/2$, and recoil wave-vector $|\vv{k}_{\rm R}| = 2\pi / \lambda_{\rm R}$ (the wavelength $\lambda_{\rm R}$ incorporates all geometric factors present from the intersection angle between these lasers).

Midway between these two sites, at position $\vv{x}_0 = a(j, m+1/2)$, this potential is
\begin{align}
V_{\rm R}(\vv{x}) &\approx \frac{V_{\rm R}}{2\I} \left\{\E^{\I (2\vv{k}_{\rm R}\!\cdot\!\vv{x}_0-\omega t)}\left[1 + 2\I \vv{k}_{\rm R}\!\cdot\! \left(\vv{x}-\vv{x}_0\right)\right] - {\rm c.c.}\right\},
\end{align}
to first order in position.  In this expression: (1) the first term describes a modulated, but spatially uniform shift in the potential with no physical consequence, that may therefore be neglected; and (2) the horizontal, $\vv{e}_j$ dependence drops out at this order because the localized wavefunctions, both centered at $a j$, are symmetric and compact in square geometry optical lattices. The remaining terms add a modulated contribution to the detuning
\begin{align}
\Delta(t) &\approx [k_{\rm R} a \cos(\theta)] V_{\rm R} \cos(2 \vv{k}_{\rm R}\!\cdot\! \vv{x}_0-\omega t):
\end{align}
the same sort of shaking potential we studied in Sec.~\ref{sec:force}, now with an overall phase dependent on the center position $\vv{x}_0$.  Here we focus on the limit in which $\hbar\omega=2\Delta \gg J$, that leads to the time-independent RWA Hamiltonian
\begin{align}
\hat H_{\rm RWA} &= - J_{\rm RWA} \left[\ket{j,m+1}\bra{j,m} \E^{\I\phi_{\rm RWA}(m,j)} + {\rm H.c}\right],
\end{align}
with {double-well} tunneling strength
\begin{align}
J_{\rm RWA} &= -J\times\left(\frac{V_{\rm R}}{2\Delta}\right) k_{\rm R} a \cos(\theta)
\end{align}
and phase
\begin{align}
\phi_{\rm RWA}(m,j) &= 2 k_{\rm R} a \left[j\sin\theta + \left(m+\frac{1}{2}\right)\cos\theta\right].\label{eq:phi_RWA}
\end{align}
Here the local phase of the traveling wave potential $2 \vv{k}_{\rm R}\!\cdot\! \vv{x}_0$ at the double well is directly imprinted onto the atoms as they tunnel in the $m$ direction, but not when they tunnel in the $j$ direction.  The result of this double well analysis can be extended to the whole lattice, where the expression for $\phi_{\rm RWA}$ is unchanged, and $J_{\rm RWA}$ is qualitatively the same but quantitatively altered.

Summing the tunneling phase around any plaquette gives an Aharonov-Bohm flux
\begin{align}
\Phi_{\rm AB} &= 2 k_{\rm R} a \sin\theta,
\end{align}
with no spatial dependence.  As a result, the RWA Hamiltonian is gauge equivalent to the Landau-gauge Harper-Hofstadter Hamiltonian
\begin{align}
H = -\sum_{j,m}J \hat a_{j+1,m}^\dagger \hat a_{j,m} + J_{\rm RWA} \E^{\I j\phi_{\rm AB}} \hat a_{j,m+1}^\dagger \hat a_{j,m} + \rm{H.c.}\ . 
\end{align}
This Hamiltonian was realized in the manner described above both by the Munich and MIT groups~\cite{Aidelsburger2013,Miyake2013}, illustrated in Fig.~\ref{fig:laser_hopping}.  The MIT group used a lattice derived from a $1064\ {\rm nm}$ laser, with a traveling wave generated by beams from the same laser intersecting the $\vv{e}_m$ axis at $\theta = \pi/4$.  This geometry gives a flux $\Phi_{\rm AB}/2\pi = 1/2$ per plaquette, and illustrates an important practical point of this technique.  In a similar manner, but with a different laser geometry the Munich group realized 1/4 flux per plaquette.  In both cases, the laser induced tunneling strength is proportional to $\cos\theta$, while Aharonov-Bohm phase is proportional to $\sin\theta$, requiring a compromise dependent on the experimental goals.  Following this initial experiment, the Munich group retooled their technique as pictured in Fig.~ \ref{fig:laser_hopping}c  by using the staggered potential inside individual four-site plaquettes and laser induced hopping to establish tunneling along all the lattice directions, enabling the measurement of the Chern number~\cite{munichchern}; see Sec.~\ref{subsubsec:cofm}.

\subsection{Synthetic dimensions}

The concept of synthetic dimensions is rooted in the fact that a lattice is no more than a set of states labeled by integers, \eg $j$ and $m$ in the preceding discussions labeled the atoms at sites described by wavefunctions $\ket{j,m}$.  This insight allows the creation of lattices that use the atoms' internal or `spin' degrees of freedom as additional synthetic dimensions.  \textcite{Boada2012} and \textcite{Celi:2014} describe how the techniques discussed in Sec.~\ref{sec:hopping} can be used to create a lattice with one spatial dimension (denoted by $j$) and one synthetic dimension (denoted by $m$ to evoke the atomic $m_F$ states from which it is built).  Large artificial magnetic fields using synthetic dimensions were simultaneously realized at NIST and LENS (\textcite{Stuhl:2015} and \textcite{Mancini:2015}) using hyperfine ground states of bosonic $^{87}{\rm Rb}$ and fermionic $^{173}{\rm Yb}$, respectively.

Both synthetic dimension experiments then replaced photon assisted tunneling with two-photon Raman transitions.  Physically, these transitions simultaneously change the internal atomic state and impart the two-photon recoil momentum.  For a one-dimensional (1D) optical lattice -- essentially one chain along $\vv{e}_j$ of the 1D lattice in Fig.~\ref{fig:laser_hopping} -- the spatially imprinted phase is
$\phi_{\rm syn} = 2 k_{\rm R} a j \sin \theta$. This expression is morally equivalent to Eq.~\eqref{eq:phi_RWA} derived for photon assisted tunneling, but without any dependence on $m$ since the Raman laser's $\vv{k}$ vector is always ``perpendicular'' to the synthetic $m$ direction, rendering $\cos\theta\rightarrow 0$.  This thereby eliminates the geometric compromise required to maximize the laser assisted tunneling strength at simultaneously large flux.

Although synthetic dimension and photon-assisted tunneling experiments can produce the same sort of magnetic lattice geometrics, the techniques have important practical differences.  For example, spin selective measurements allow the synthetic dimension lattice site to be measured with near-perfect ``spatial'' resolution.  In addition the limited number of spin states (typically 3 to 5) produce synthetic dimension lattices with strip-like geometrics with perfect hard-wall boundary conditions, rather than extended planes as for conventional 2D optical lattices.  In addition synthetic dimension lattice sites with the same spatial index $j$ but with different internal index $m$ are in reality spatially overlapping, so that the spatially local atom-atom interactions become anisotropic: long ranged in $m$ and short ranged in $j$.

Both experimental synthetic dimensions realizations ($^{87}{\rm Rb}$ and $^{173}{\rm Yb}$ in~\textcite{Stuhl:2015} and \textcite{Mancini:2015}, respectively) created large flux, highly elongated strips along $j$, with just three sites in width along $m$.  More recently two-leg ladder implementations of the synthetic dimension concept have been realized~\cite{livi2016synthetic,kolkowitz2017spin} using the optical clock transition of both $^{173}{\rm Yb}$ and $^{87}{\rm Sr}$, and synthetic dimensions have even been constructed using momentum states in lieu of spin states~\cite{Meier2016}.

\subsection{Flux lattices: intrinsic topology}

\label{subsec:fluxlattice}

Each of the topological lattices discussed in the preceding sections was engineered beginning with a non-topological lattice, to which modulation or light-assisted tunneling was added to engineer a desired topological model.  In this section we will focus on a different approach for generating a topological lattice without the need for modulation or light assisted tunneling. This approach applies to an atom with several internal states, and subjected to a combination of three light-matter interactions: those that are independent of internal atomic states; those that depend on internal atomic states; and  those couple between the different internal atomic states. In particular, 
as we show in Appendix~\ref{app:alkali}, the far-detuned  light-matter interaction for alkali atoms takes the form
\begin{align}
\hat H_{\rm RWA} &= U(\vv{r}) \hat 1 + {\boldsymbol \kappa}(\vv{r})\!\cdot\!\hat{\vv{F}},
\label{eq:scalarvectorls}
\end{align}
where $U(\vv{r})$ and ${\boldsymbol \kappa}(\vv{r})\!\cdot\!\hat{\vv{F}}$ describe the rank-0 (scalar) and rank-1 (vector) light shifts acting on an atom with internal angular momentum operator $\hat{\vv{F}}$. The possible lattices formed from (\ref{eq:scalarvectorls}) have a very rich range of structures, characterized by the spatial variations of $U(\vv{r})$ and the three components of ${\boldsymbol \kappa}(\vv{r})$. Because this approach is not tied to a pre-existing tight-binding lattice, it is not limited to deep optical lattices and can have topological properties even for very shallow or weak optical coupling strengths.

We will discuss the essence of this approach in two ways: (1) we will explore ``flux lattices'' in which the Aharonov-Bohm flux emulated by a Berry phase in real space leads to lattices with nonzero effective magnetic fields, and (2) we will explore the connection between spin-dependent lattices and established topological models.  Indeed in our initial discussion of the Haldane model, we identified the two sublattice sites in each unit cell with a pseudospin degree of freedom and arrived at a spin-dependent band structure, Eq.~\eqref{eq:haldaneimplement}, we now make this literal.

\subsubsection{Flux lattices}\label{sec:flux_lattice}

\begin{figure*}[htbp]
\begin{center}
\includegraphics{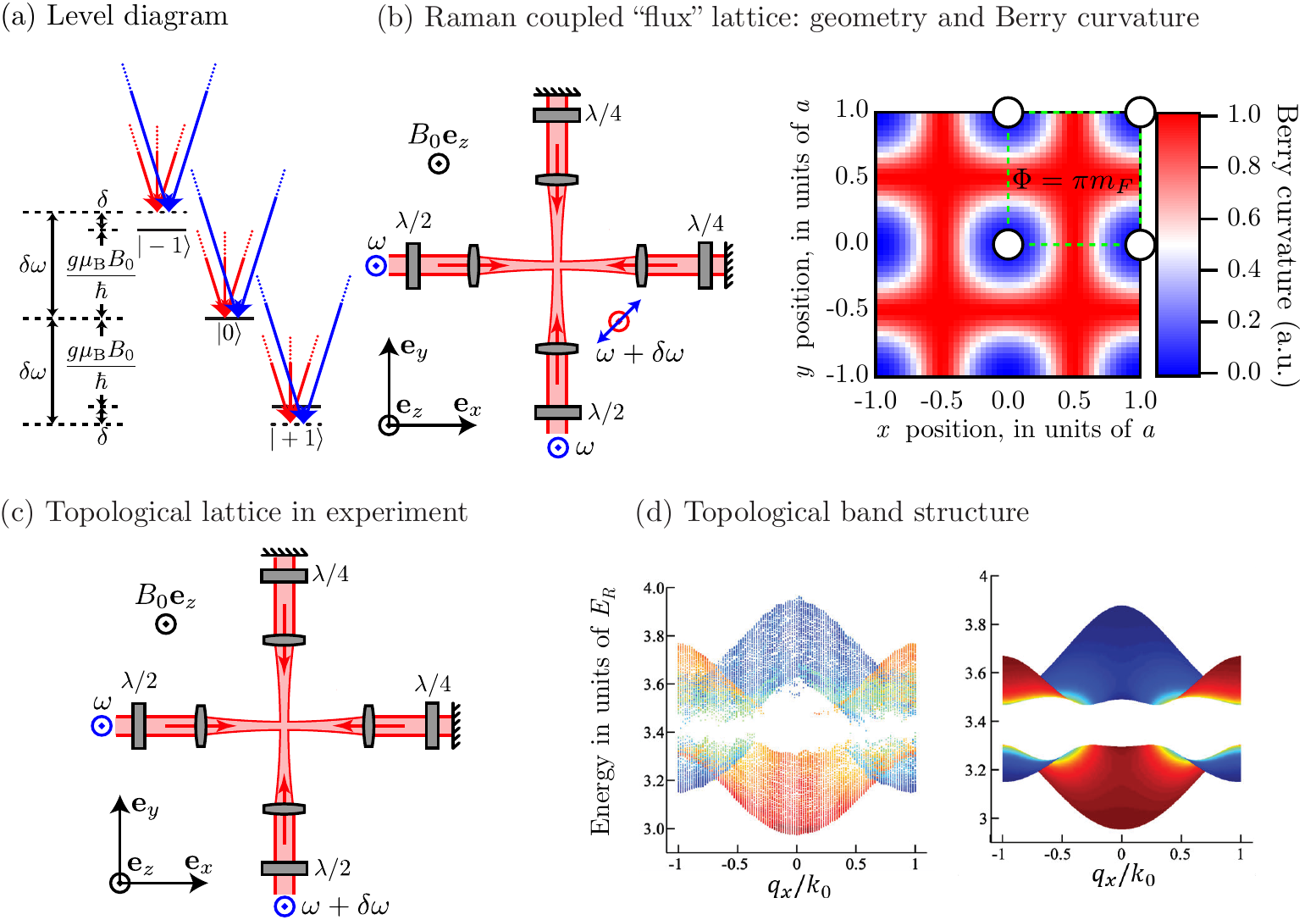}
\end{center}
\caption[Proposed experimental geometry]{\textbf{Spin dependent topological lattices.}  {\bf a} Level diagram for three-level total angular momentum $f=1$ case with $m_F$ states labeled, as is applicable for the common alkali atoms$^{7}{\rm Li}$, $^{23}{\rm Na}$, $^{39}{\rm K}$, $^{41}{\rm K}$, and $^{87}{\rm Rb}$.  For reference, the diagram shows the decomposition of these optical fields into $\sigma_{\pm}$ and $\pi$, but as discussed in the text, this is not an overly useful way of considering this problem.  \textbf{b},  Laser geometry for a typical optical flux lattice (left) producing a real-space Berry curvature with non-zero average (right). \textbf{c}, Experimental geometry for spin-dependent topological lattice~\cite{Sun2017}. \textbf{d}, Directly observed topological band structure (left) and computed (right) also from Ref.~\textbf{c}.
}
\label{fig:fluxlattice}
\end{figure*}

Before turning to our discussion of flux lattices, we shall pause to reflect on our discussion of topological lattices to this point.  Section~\ref{sec:topology} introduced the concept of topological invariants in terms of the Berry connection (Zak phase) or Berry curvature (Chern number) integrated over the BZ.  In the latter case, the Chern number can be cast as a momentum-space statement of Gauss's law, in which the Berry curvature integrated over the toroidal BZ counts an integer number of topological ``charges'' in the inside of the torus yielding the Chern number.  

A ``flux lattice'' is an optical lattice potential that instead is defined as a lattice in which the integrated Berry curvature in each {\it spatial} unit cell is non-zero, suggesting that the atoms might behave as if large magnetic fields are present.  This provides an intuitive framework in which to link the physics of Landau levels and lattice bands. Indeed, in general, the atom experiences a combination of a periodic magnetic field (with non-zero average) and a periodic scalar potential. 
Figure~\ref{fig:fluxlattice}b depicts a bichromatic laser configuration that gives the same $\vv{\kappa}$ used in the initial flux lattice proposals~\cite{Cooper2011,Cooper2011a}, with four frequency degenerate in-plane lasers and with a single down-going laser at a different frequency.   
As shown by~\textcite{Juzeliunas2012}, this geometry can be tuned to produce the desired effective magnetic field with vector strength
\begin{align}
\vv{\kappa} =& \kappa_\perp \left[\cos \left(\frac{\pi x}{a}\right)\vv{e}_x + \cos\left(\frac{\pi y}{a}\right) \vv{e}_y\right]\nonumber\\ 
&+ \kappa_\parallel \sin \left(\frac{\pi x}{a}\right) \sin \left(\frac{\pi y}{a}\right)  \vv{e}_z,
\end{align}
where $\kappa_\perp$ and $\kappa_\parallel$ are set by the intensity and polarization of the laser fields, and $a=\lambda$ is equal to the laser wavelength $\lambda$.  This result follows directly from the expressions in Appendix~\ref{app:alkali}.
The right panel of Fig.~\ref{fig:fluxlattice}b shows the spatial distribution of the Berry curvature in a single unit cell with spatial extent $a$, with a clear non-negative mean, evaluated for $\kappa_\perp = \kappa_\parallel$.  Thus, it achieves the goal of having a net non-zero effective magnetic field piercing the unit cell. 
In both the square geometry and a similar three-beam setup with $2\pi/3$ intersection, the resulting band structure can have topological bands~\cite{Cooper2011,Cooper2011a}.
 
This discussion hides one subtle point: the integrated curvature over each unit cell must be a multiple of $2\pi$.  In the present case the integrated curvature is $8\pi m_F$ over the complete unit cell. The circles in Fig.~\ref{fig:fluxlattice}b mark the locations of the minima of the adiabatic potential $\hbar m_F \left|\vv{\kappa}\right|$. For a deep lattice, in which this  potential is large compared to the recoil energy, the atoms would be strongly confined close to these minima. Their locations would define effective lattice sites of a tight-binding description, with the unit cell containing four such sites and therefore divided into four plaquettes.  For a spin-1/2 system, with $m_F = \pm1/2$, each of these plaquettes will have a flux $\Phi = \pi$,
while for bosonic alkali atoms, $m_F$ takes on integer values implying $\Phi=2\pi m_F$ is a multiple of $2\pi$.  In either case, nearest-neighbor hopping on this square lattice geometry would not break time-reversal symmetry, so the Berry curvature of the bands must vanish. As a result, for topological band structure to emerge in a straightforward way, longer range hopping as in the Haldane model is required. These considerations indicate that, for this square geometry, the flux lattice would not lead to topological bands in the deep-lattice limit. However, we reiterate that the flux lattice approach is not restricted to deep lattices, but applies also for shallower lattices in which the atoms can move throughout the unit cell and a restriction to nearest-neighbor hopping is inappropriate. 

We have demonstrated that flux lattices can give rise to a net non-zero magnetic field piercing the real space unit cell. The comparison with free particles in a uniform magnetic field -- which form Landau levels -- suggests the appearance of topological Chern bands. However, 
this is not guaranteed: such lattices may or may not be topological, as defined by the usual Chern number computed in momentum space. (The deep lattice limit of the square flux lattice described above provides an example in which the bands are not topological despite the nonzero effective magnetic flux.)  
Moreover, as we now discuss, there can be cases in which the net flux through the unit cell vanishes, yet the bands are topological~\cite{coopermoessner}. In general, to determine the band topology requires a full calculation of the band structure, including  the atom's kinetic energy. 

\subsubsection{Fluxless lattices}\label{sec:fluxless_lattice}

A key insight from the Haldane model is that a net magnetic field is not a prerequisite for topological band structure.  It is sufficient to break time-reversal symmetry, a condition that can readily be achieved for lattices that couple internal states. The appearance of topological bands for zero net flux is readily established for shallow lattices~\cite{coopermoessner}. Here we focus on the regime of deep lattices, where an effective (spin-dependent) tight-binding model can be developed.
As we made explicit in our discussion of the Haldane model, the two basis sites consisting of a single unit cell can be assigned a pseudospin label, and these pseudospins are then arrayed to form a honeycomb or brick-wall lattice.  Building from this understanding, we conclude by discussing the topology of state-dependent lattices, without any explicit reference to Berry phases.

Figure~\ref{fig:fluxlattice}c displays the phase-stable lattice geometry realized in~\textcite{Sun2017}, which is closely related to the flux lattice geometry, absent the vertical beam and with the in-plane beams driving Raman transitions.  A predecessor of this setup described by~\textcite{Wu:2015} generated a 2D optical lattice with spin-orbit coupling with $^{87}$Rb atoms (a first realization of 2D spin-orbit coupling was reported by \textcite{Huang:2016} for a bulk geometry).  The two pseudospin states are the Zeeman levels $|F=1,m=-1\rangle$ and $|F=1,m=0\rangle$ of the lowest hyperfine state of the ground state manifold.  The essential concept of this lattice is to first use the scalar light shift from the detuned, counter-propagating beams to create a conventional 2D optical lattice operating in the tight-binding regime, and then to use the vector contribution of the light shift to give a combination of local effective magnetic fields and spin-dependent tunneling.  The resulting tight-binding model is very closely related to that of the Haldane model and shares its topological properties.  The left panel of Fig.~\ref{fig:fluxlattice}d shows the experimentally measured band structure in good agreement with the predictions of theory displayed in the right panel. 

\section{Experimental Consequences}

\label{sec:experimental}
\newcommand{\parencite}{\cite}

This section presents recent experimental investigations of non-trivial (global or local) topological properties of energy bands, either in 1D or 2D geometries. Interactions play a non-essential role for the experiments described below, hence phenomena addressed here correspond to single-particle (or ideal gas) physics. 

This section is divided into three parts. In the first one we describe measurements that are performed on an atomic system at equilibrium, using local probes in momentum space that allow one to reconstruct the topology of the occupied band(s). In the second part we present analyses performed by looking at the dynamics of wave-packets. These wave-packets are well localized  at the scale of the Brillouin zone  and one can bring them  close to some points of specific interest, Dirac points for example, using an external force. The last part is devoted to transport measurements, which are closer in spirit to the techniques that are commonly used in condensed matter physics.

\subsection{Characterization of equilibrium properties}

\subsubsection{Time-of-flight measurements}

Before entering into a discussion of specific measurements, we briefly comment on implications of the time-of-flight (TOF) measurements commonly used in experiment.  In the vast majority of cold atom experiments, the measurement procedure begins with the rapid removal of all applied  fields (both those involved in trapping the ensemble and those required to create the topological lattice of interest).  After this abrupt -- projective -- turn off, the atoms then undergo a period of ballistic expansion followed by a measurement of their density (often in a spin-resolved manner).  In many experiments, this procedure gives a direct measurement of the momentum distribution as it was when the applied fields were just removed.

While this sounds simple in principle, this procedure can appear to give measurement results that are {\it gauge dependent}.  Fortunately {any supposed contradiction with general principles of local gauge invariance} is illusory.  For example, two different experiments might well create a Harper-Hofstadter lattice with the same flux using very different laser geometries, {which naturally define the synthetic vector potential in two different gauges. The observed momentum distribution will in general differ in these two cases\cite{Kennedy2015}. As discussed in~\textcite{mollercooper2010} and~\textcite{LeBlanc2015} these differences arise because the synthetic vector potential vanishes during TOF when the laser fields are removed: there are physical differences in how exactly the synthetic vector potential} returns to zero when TOF begins, and physically different effective electric fields present during this turnoff. {All physical observables remain invariant to local gauge transformations, provided these are applied consistently, \ie changing the vector potential both before and during TOF.}

\subsubsection{Local measurement of the Berry curvature}

\label{subsubsec:localmeasureberry}

We consider in this paragraph a cold atomic gas at equilibrium in a 2D optical lattice. With the so-called band-mapping technique one can measure precisely the distribution ${\cal N}(\vv{q})$ of the quasi-momentum $\vv{q}$ for each energy band. In this technique the laser beams forming the lattice are turned off in a controlled manner, so that the populations of the Bloch states forming the various energy bands in the presence of the lattice are transferred to states with a well-defined momentum in the absence of the lattice~\cite{Greiner:2001}. The measurement of ${\cal N}(\vv{q})$ in combination with a suitable temporal variation of the lattice parameters before the complete turn-off, allows one to characterize the band topology, \ie to access not only the energies, but also the Berry curvature ${\vv{\Omega}}(\vv{q})$. 

To illustrate this point we consider a 2D lattice and use the same two-band model as in section~\ref{sec:topology}, assuming a unit cell with two non-equivalent sites labeled A and B. The generic Hamiltonian in reciprocal space is $\hat H_{\vv{q}}=h_0(\vv{q}) \hat 1- \vv{h}(\vv{q})\cdot \hat{\vv{\sigma}}$ where $(h_0,\vv{h})$ is a 4-vector with real components that are periodic over the BZ. The energies of the two bands are $E_{\vv{q}}^{\pm}=h_0({\vv{q}})\pm |\vv{h}({\vv{q}})|$  and the Bloch states  $|u_{\vv{q}}^\pm\rangle$ can be written as linear combinations of $|\vv{q}, {\rm A}\rangle$ and $|\vv{q}, {\rm B}\rangle $. Using the expression (\ref{eq:2dwavefunction}) for these Bloch states, one finds:
\begin{eqnarray}
{\vv{\Omega}}^\pm(\vv{q})&=&\vv{\nabla}_{\vv{q}} \left(\langle  u_{\vv{q}}^\pm|\right) \times \vv{\nabla}_{\vv{q}} \left(|  u_{\vv{q}}^\pm \rangle\right) \nonumber \\
&=& 
\pm\frac{1}{2}\vv{\nabla}_{\vv{q}}(\cos \theta_{\vv{q}})\times \vv{\nabla}_{\vv{q}} \phi_{\vv{q}} \,,
\label{eq:Berry_curvature}
\end{eqnarray}
where $(\theta_{\vv{q}},\phi_{\vv{q}})$ defines the direction of $\vv{h}({\vv{q}})$ in spherical co-ordinates.

A procedure to determine $\theta_{\vv{q}}$ and $\phi_{\vv{q}}$, hence the curvature  ${\vv{\Omega}}$, in such a two-band situation has been proposed by \textcite{Hauke:2014} and implemented by \textcite{Flaschner:2016}. The starting point is the momentum distribution ${\cal N}_{\bs q}(\bs k)$ associated with a Bloch state
\begin{equation}
\psi_{\vv{q}}^{(-)}(\vv{r})=\sum_{s={\rm A},{\rm B}}\sum_{\vv{R}_s} \alpha_{\vv{q},s}\; w(\vv{r}-\vv{R}_s) \;\E^{\I \bs q\cdot\bs R_s}.
\end{equation}
 Here $w(\vv{r})$ is the Wannier function of the band, supposed to be identical for the two sublattices $s=A,B$, and $(\alpha_{\vv{q},{\rm A}},\alpha_{\vv{q},{\rm B}})=(\cos(\theta_{\vv{q}}/2),\E^{\I \phi_{\vv{q}}}\sin(\theta_{\vv{q}}/2))$. The momentum distribution ${\cal N}_{\bs q}(\bs k)$ is the square of the Fourier transform of $\psi_{\vv{q}}^{(-)}(\vv{r})$. For $\vv{k}$ inside the first BZ, it is peaked around $\vv{k}=\vv{q}$ and given by $|\tilde w(\vv{k})|^2\;  \left| \alpha_{\vv{k},{\rm A}}+\alpha_{\vv{k},{\rm B}}\right|^2$, where 
 $\tilde w(\vv{k})$ is the Fourier transform of $w(\vv{r})$. When the lowest band is uniformly filled with independent fermions, the momentum distribution of the gas is obtained by summing the contributions ${\cal N}_{\bs q}(\bs k)$ over all quasi-momenta $\vv{q}$ of the BZ: 
 \begin{eqnarray}
{\cal N}(\vv{k}) = |\tilde w(\vv{k})|^2  \left[ 1-\sin \theta_{\vv{k}}\cos \phi_{\vv{k}} \right].
\label{eq:Hauke_result}
\end{eqnarray}
This distribution can be measured using a ballistic expansion after a sudden switch-off of the lattice.
The key point of (\ref{eq:Hauke_result}) is that the measured distribution ${\cal N}(\vv{k})$ is sensitive to  the relative phase of the contributions $\alpha_{\vv{q},s}$ of the two sites $s={\rm A},{\rm B}$ in the expression of each Bloch state $|\psi_{\vv{q}}^-\rangle$. More precisely, although not sufficient to determine unambiguously the angles $\theta$ and $\phi$, this measurement already provides the value of the product $\sin \theta\cos \phi$. To go one step further, 
\textcite{Hauke:2014} suggested to apply an abrupt quench to the lattice parameters so that the Hamiltonian becomes 
$\hat H'_{\vv{q}}=({\hbar \omega_0}/{2})\;\hat \sigma_z$.
One then lets the gas evolve in the lattice during a time interval $t$ before measuring the momentum distribution. Since the evolution during this time simply consists of adding the phase  $\pm \omega_0 t/2$ to $\alpha_{\vv{q},s}$, the momentum distribution at time $t$ reads:
\begin{equation}
{\cal N}(\vv{k},t)=|\tilde w(\vv{k})|^2  \left[ 1-\sin \theta_{\vv{k}}\cos (\phi_{\vv{k}}+\omega_0 t) \right].
\label{eq:evol_Hauke}
\end{equation}
By repeating this procedure for various times $t$ and measuring the amplitude and the phase of the time-oscillating signal, one can determine simultaneously $\phi$ and $\theta$ at any point in the BZ. 
 
\begin{figure}[t]
\begin{center}
\includegraphics[width=8cm]{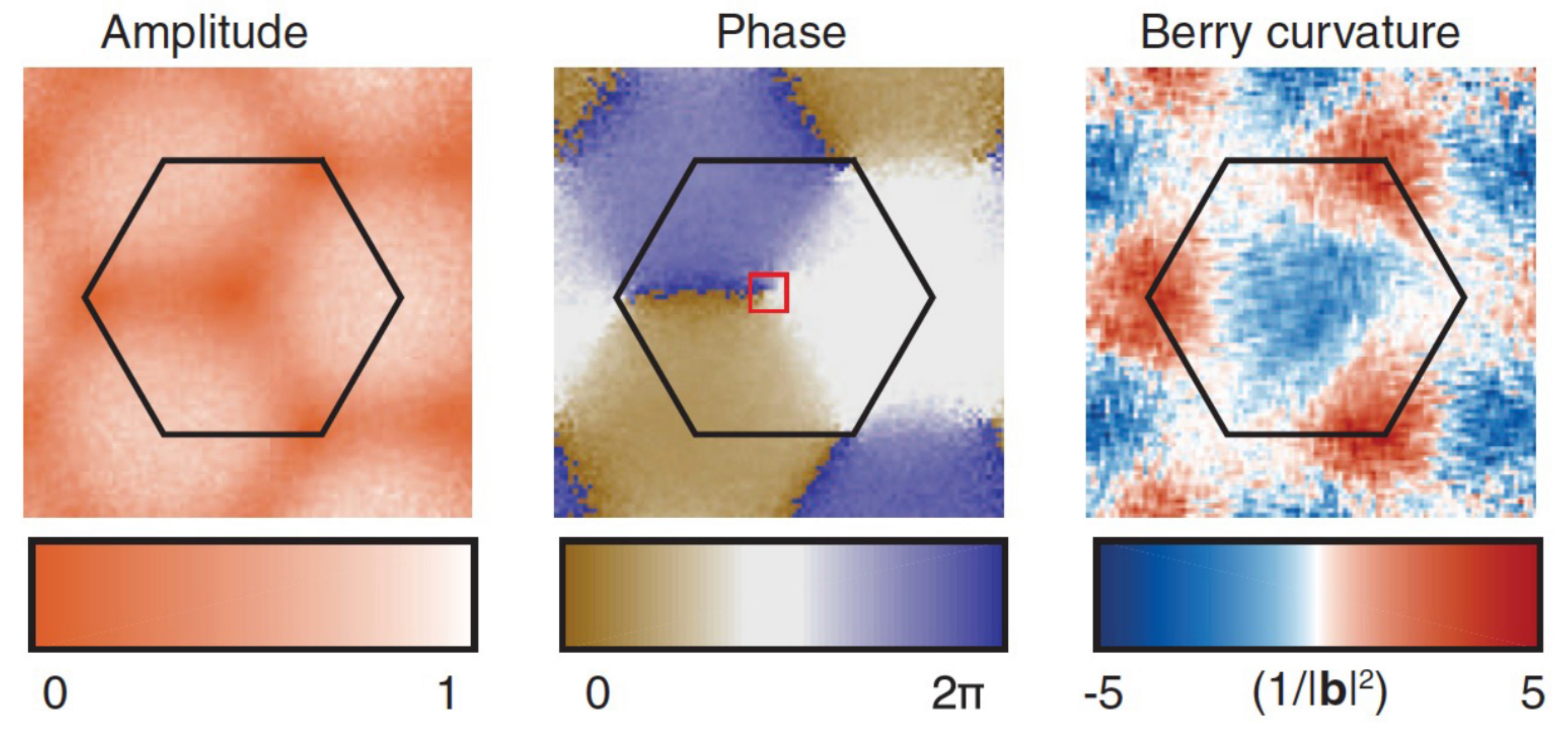}
\end{center}
\caption{Amplitude $\propto \sin \theta_{\vv{q}}$ (left column) and phase $\phi_{\vv{q}}$ (middle column) obtained from the fits to the oscillation (\ref{eq:evol_Hauke}) of the momentum distribution.  From those fit results, one can reconstruct the momentum-resolved Berry curvature (right column) given in units of the inverse reciprocal lattice vector
length $|\vv{b}|$ squared. (Courtesy of \textcite{Flaschner:2016}.)}
\label{fig:2016_Flaschner_Berry_curv}
\end{figure}

This procedure was implemented by \textcite{Flaschner:2016} using an hexagonal lattice of tubes filled with fermionic $^{40}$K atoms. The unit cell for this graphene-like geometry contains two sites and the lattice parameters are chosen such that there is initially a large energy offset $\hbar \omega_{AB}$ between the $A$ and $B$ sites, corresponding to essentially flat bands with no tunnelling. As explained in Sec. \ref{sec:hopping}, the dynamics in the lattice can be restored by a resonant, circular shaking of the lattice at a frequency $\Omega \approx \omega_{AB}$. In the experiment of \textcite{Flaschner:2016}, the shaking was produced by a phase modulation of the three laser beams forming the lattice, and it also resulted in a non-negligible value for the Berry curvature. Once the atoms equilibrated in this lattice, the abrupt quench needed for the procedure of \textcite{Hauke:2014}  was obtained by simply switching off the modulation. Measured amplitudes and phases of the time oscillation of ${\cal N}(\vv{k},t)$ are plotted in Figure \ref{fig:2016_Flaschner_Berry_curv}, together with the Berry curvature reconstructed from (\ref{eq:Berry_curvature}). One then expects the integral of ${\Omega}_{\vv{q}}$ to be an integer ${\cal C}$ times $2\pi$, where ${\cal C}$ is the Chern index of the populated band. Here the reconstructed Berry curvature leads to a value of ${\cal C}$ compatible with 0~\cite{Flaschner:2016}. This is in agreement  with the expected band topology in this case. 

The method outlined by  \textcite{Hauke:2014} is reminiscent of a previous proposal by  \textcite{Alba:2011}. There, the two lattice sites $A$ and $B$ are supposed to be occupied by two different internal (pseudospin) states $|a\rangle$ and $|b\rangle$. The momentum distribution measurement  can be done in a spin resolved way, which provides the local spin polarization for the lowest band [see (\ref{eq:2dwavefunction})]:
\begin{equation}
\frac{h_z(\vv{q})}{|\vv{h}(\vv{q})|}=\cos \theta_{\vv{q}}=\frac{{\cal N}_b(\vv{q})-{\cal N}_a(\vv{q})}{{\cal N}_a(\vv{q})+{\cal N}_b(\vv{q})}\,.
\label{eq:polarization}
\end{equation}
The two other components of $\vv{h}/|\vv{h}|$ can be obtained by inducing a coherent transition  between $|a\rangle$ and $|b\rangle$ (Raman pulse) with an adjustable phase and duration, in order to rotate the pseudospin  during the time-of-flight. Once the direction of $\vv{h}({\vv{q}})$ is known at all points of the Brillouin zone, the value of the Berry curvature follows from Eq.~(\ref{eq:Berry_curvature}).

\subsubsection{Topological bands and spin-orbit coupling}

When the considered lattice has some specific geometrical symmetries, the assessment of the topological nature of a band can be notably simplified with respect to the procedure outlined above.  One does not need to characterize the eigenstates of the Hamiltonian at all points of the Brillouin zone to determine if the integral of the Berry curvature (\ref{eq:Berry_curvature}) over this zone is non-zero, and it is sufficient to concentrate on some highly symmetric points.  The basis of this simplification, which is discussed in  Appendix~\ref{app:1dmodels} for the case of a 1D lattice (see in particular Eq.~(\ref{eq:halfBZwinding})),  was outlined by \textcite{Liu:2013} for a 2D square optical lattice for pseudo-spin 1/2 particles, in the presence of spin-orbit coupling terms. 

Let us briefly outline the main result of \textcite{Liu:2013}. Suppose that the Hamiltonian is invariant under the combined action of the spin operator $\hat \sigma_z$ and the spatial operator transforming a Bravais lattice vector  $\vv{R} $ into $ -\vv{R}$. Consider the energy eigenstates (Bloch states) at the four points of the BZ: $\{\vv{\Lambda_i}\}=\{ (0,0), (0,\pi), (\pi,0), (\pi,\pi)  \}$, $i=1,\ldots,4$. The two Bloch states $|\psi_\pm(\vv{\Lambda}_i)\rangle$ at each of these locations are also eigenstates of $\hat \sigma_z$,  and the corresponding eigenvalues $\xi_{i,\pm}$  can only take the values $+1$ or $-1$. Now one can show that the sign of the product $P_\eta=\prod_i \xi_{i,\eta}$ of the four $\xi_{i,\eta}$ of a given subband $\eta=\pm$ 
is directly related to the Chern number of this subband. More specifically, if the sign of $P_\eta$ is negative, the Chern number of the subband $\eta$ is odd, hence non-zero: this unambiguously signals a topological band. If $P_\eta$ has a positive sign, the Chern number of the subband $\eta$ is even, and this most likely signals a non-topological character for this subband (zero Chern number). 

This procedure was realized experimentally by the type of lattice we described in Sec.~\ref{sec:flux_lattice} involving two pairs of retroreflected Raman lasers intersecting at right angles.  This configuration can produce the reciprocal space tight-binding Hamiltonian with the generic form (\ref{eq:haldaneham}):
\begin{eqnarray}
\hat H_\vv{q} &=& 2t_{\rm SO}\left[ \sin(aq_y)\, \hat \sigma_x +  \sin(aq_x) \,\hat \sigma_y\right]\nonumber \\
&+& 
\left[ m_z-2t_0 \left( \cos(aq_x)+\cos(aq_y) \right)\right]\hat \sigma_z\,,
\label{eq:solattice}
\end{eqnarray}
where $m_z$ is proportional to the detuning from Raman resonance. The two  amplitudes $t_0$ and $t_{\rm SO}$ characterize the tunneling amplitudes without and with spin flip, respectively. They can be controlled independently by varying the intensities of the two pairs of laser beams. In particular the term proportional to $t_{\rm SO}$ can be simplified in the limit of low momenta into $\propto k_y\hat \sigma_x+k_x \hat \sigma_y$, corresponding to the usual form of the Rashba-Dresselhaus spin-orbit coupling for a bulk material~\cite{Galitski:2013}.

An interesting feature of this Hamiltonian is the possibility to control the topology of the lowest band:   It is non trivial if and only if $|m_z| < 4 t_0$. \textcite{Wu:2015} tested this prediction by loading the $^{87}$Rb gas at a temperature $T \sim 100\;$nK such that the lowest band was quasi-uniformly filled, whereas the population of higher bands remains small. The polarization defined in Eq.~(\ref{eq:polarization}) was measured by a spin-resolved imaging of the atomic cloud after time-of-flight, and the product  $P_-$ for the lowest band was found to be negative in the expected range of values of $m_z$.  


\subsubsection{Momentum distribution and edges states}

It is well known from integer quantum Hall physics that the non-trivial topology of a band can give rise to a quantized Hall conductance $\sigma_{xy}$. Applying a voltage difference $V_y$ between the two opposite edges of a rectangular 2D sample gives rise to a global current $I_x=\sigma_{xy}V_y$ along the $x$ direction. This current flows on the edges of the sample in a chiral way, with (for example) a positive value $I_x^{(+)}$ on the  $y>0$ side of the sample and a negative value $I_x^{(-)}$ on the $y<0$ side~\cite{Hatsugai:1993}. In the absence of an applied voltage ($V_y=0$), the edge currents are still non-zero but they exactly compensate each other: $I_x^{(-)}=-I_x^{(+)}$ .

The possibility to engineer  a 2D atomic gas with one real dimension and one synthetic dimension that we discussed in Sec.~\ref{subsubsec:magnetoplasmons} offers a way to access directly these edge currents in a cold atom experiment. Indeed when the second dimension (labeled $y$ above) is synthetic, \ie associated to an internal degree of freedom (pseudo-spin), one expects a given sign of $I_x$ for the largest  value of the pseudo-spin, and the opposite sign for its smallest value. 

The advantage of a synthetic dimension for observing these edge states is clear: It provides a sharp boundary to the sample, whereas a standard 2D optical lattice would lead to edge states that would be smeared over several lattice sites, hence much more difficult to observe. In addition, the momentum distribution can be measured individually for each pseudo-spin value. One thus obtains the value of the current on each ``site" of the synthetic direction. The transposition of such a measurement to a real direction $y$ would imply a single-site resolving imaging, which would be much more demanding from an experimental point of view.  However, the use of spin states for synthetic dimensions greatly limits the potential extent of the synthetic dimension: in the extreme limit of just two spin states, the system can even behave as if it were all edge with no bulk~\cite{Hugel2014}.  For narrow systems with synthetic extent less than $q$ for flux $p/q$, the states behave more like those in a continuum (with a single guiding center, see the supplementary material in~\textcite{Stuhl:2015}, and as the width further increases, the familiar topological edge modes emerge, but remain slightly gapped at the edge of the 1D Brillouin zone from the bulk bands.

In addition, recent experiments have instead turned to using momentum states for a synthetic dimension, in principle allowing for more extended systems.  Indeed the edge states associated with the 1D SSH model have already been observed using a momentum-space lattice~\cite{Meier2016}.  

\begin{figure}[t]
\begin{center}
\includegraphics[width=90mm]{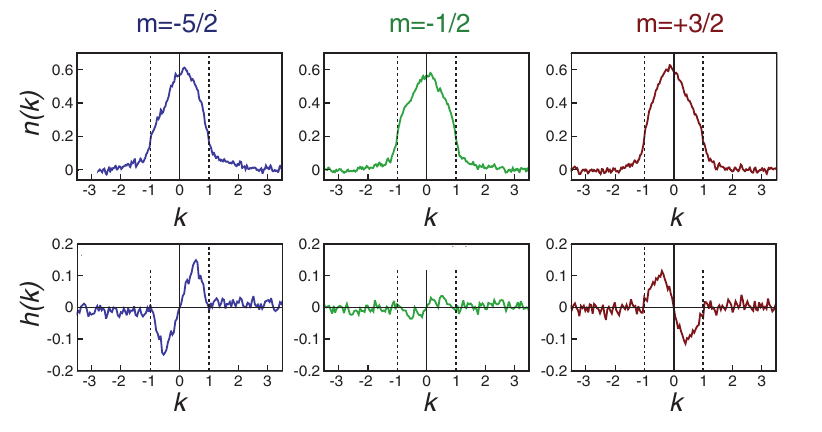}
\end{center}
\caption{Direct visualization of an edge-state current for a two-dimension (one real, one synthetic) lattice in the presence of an artificial magnetic field. Upper row: Momentum distributions $n_m(k)$ along the real direction $x$ for the three values of the pseudo-spin $m$ corresponding to the synthetic direction. The value $m=-1/2$ can be viewed as the bulk, whereas $m=+3/2$ and $m=-5/2$ corresponding to the opposite edges of the sample along the synthetic dimension. Lower row: Function $h_m(k)=n_m(k)-n_m(-k)$. Courtesy of \textcite{Mancini:2015}.}
\label{fig:2015_Mancini_edges}
\end{figure}

We show in  Figure \ref{fig:2015_Mancini_edges} the results of an experiment performed by \textcite{Mancini:2015}. A gas of fermionic $^{173}$Yb atoms can tunnel between the sites of an optical lattice along the $x$ direction, with an essentially frozen motion along the two other (real) $y$ and $z$ directions. The synthetic direction consists of 3 Zeeman substates $m=-5/2,-1/2,+3/2$ selected among the 6 Zeeman states of the ground level. The ``tunneling" along this synthetic direction is provided by a pair of light beams. These beams induce  stimulated Raman processes between the Zeeman states, hence the desired laser-induced hopping. These beams also provide an artificial gauge field thanks to the space-dependent phase $\pm \varphi(x)$ printed on the atomic state in a $\Delta m=\pm 2$ transition. 

The atoms are prepared in a metallic state (less than one atom/lattice site) in the lowest band of the single particle Hamiltonian. Figure \ref{fig:2015_Mancini_edges} shows the momentum distributions $n_m(k)$ along the $x$ direction for the three values $m$ of the pseudo-spin. Here the central value of the pseudo-spin ($m=-1/2$) plays the role of the bulk of the material. The momentum distribution in this internal state is thus symmetric around 0, corresponding to a null net current. In contrast, a non-zero current is associated to  the side values of the pseudo-spin: The distribution for the largest (smallest) value of $m$  is displaced towards positive (resp. negative) values. The displacement is made even clearer in the second row of Figure \ref{fig:2015_Mancini_edges} where the function
$h_m(k)=n_m(k)-n_m(-k)$ is plotted. \textcite{Mancini:2015} also checked that the sign of the edge current is reversed when the sign of the artificial magnetic field is changed.

\subsection{Wave-packet analysis of the BZ topology}
\label{subsec:wave_packet}

Cold atom experiments offer the possibility to prepare the particles in a state described by a wave-packet that is well-localized in momentum space, in comparison with the size of the Brillouin zone. In this case, the dynamics of the wave-packet directly reveals the local properties in the BZ: energy landscape $E(\vv{q})$ and Berry curvature ${\Omega}(\vv{q})$. These properties can be encoded either on the dynamics of the center of the wave-packet, or on the interference pattern that occurs when several paths can be simultaneously followed.

\subsubsection{Bloch oscillations and Zak phase in 1D} 

Conceptually the simplest example of a wave-packet analysis of the band topology is found in a 1D lattice of period $a$, for which the BZ extends between $q=-\pi/a$ and $q=\pi/a$. When an atom initially prepared with the quasi-momentum $ q_0$ is submitted to an additional uniform force $F$, its quasi-momentum $ q(t)$ periodically spans the Brillouin zone (Bloch oscillation) at a frequency $aF/h$:
$q(t)=q_0+Ft/\hbar$ (mod. $2\pi/a$). The phase that is accumulated in this periodic motion contains relevant information about the topology of the energy bands.

Since Bloch oscillations will play a key role in several instances in the following, we briefly outline the principle of their theoretical description. The Hamiltonian of the particle in the presence of the periodic lattice potential $V(x)$ and of the force $F$ reads $\hat{H}=\hat p^2/2m+ V(\hat x)- F\hat x$. Let us write the initial state of the particle $\psi(x,0)=\E^{\I xq_0}u_{q_0}^{(n)}(x)$, where $u_{q_0}^{(n)}(x)$ is the periodic part of the Bloch function associated to the $n$-th band and to the quasi-momentum $q_0$.  One can look for a solution of the time-dependent Schr\"odinger equation under the usual Bloch form $\psi(x,t)=\E^{\I x q(t)}\;u(x,t)$. The evolution of $u$ is governed by the periodic Hamiltonian $\hat H_{q(t)}=[\hat p+\hbar q(t)]^2/2m +V(\hat x)$, therefore the solution $u(x,t)$ remains spatially periodic at any time. If we add the assumption that the force $F$ is weak enough so that interband transitions play a negligible role, 
the state of the particle will follow adiabatically the corresponding Bloch state of the $n$th band, \ie   $u(x,t)=\E^{\I \varphi(t)}\;u_{q(t)}^{(n)}(x)$, and the relevant information is encoded in the phase $\varphi(t)$. Let us focus on the state of the particle after one Bloch period. At this moment, the quasi-momentum  is back to its initial value $q_0$ and the phase is the sum of two contributions: (i) the dynamical phase $-\int E[q(t)]\;\D t/\hbar$ and (ii) the Zak phase of the band, Eq.~(\ref{eq:zak}). 

\begin{figure}[t]
\begin{center}
\includegraphics[width=88mm]{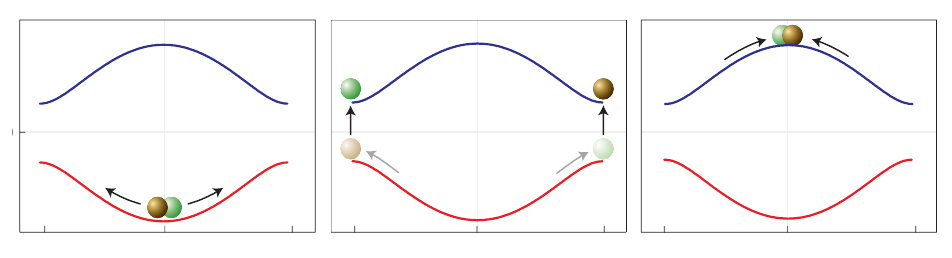}
\end{center}
\caption{ Two-band spectrum of the SSH Hamiltonian and Zak phase measurement. An initial wave-packet is prepared at the bottom of the lowest band (left image). Each atom is in the superposition of the $|\pm\rangle$ spin states, which experience opposite forces in the presence of a magnetic field gradient. After a combination of  exchanges of the atomic internal state and of the dimerization (middle image), the two wave-packets recombine at the top of the highest band (right image). The measurement of the accumulated phase provides the value of the Zak phase. Courtesy of \textcite{Atala2013}.}
\label{fig:2013_Atala_Zak}
\end{figure}

The first measurement of the Zak phase in a cold atom context was performed by \textcite{Atala2013}.  A 1D superlattice was made out of two standing waves with a long and a short period (see~\textcite{Sebby-Strabley2006} and~\textcite{Trotzky2008} for two methods for creating double-well superlattice potentials), generating the potential (\ref{eq:superlattice_potential})
where both the relative phase $\phi$ and the amplitudes $V_{\rm long,short}$ are control parameters. As detailed in section~\ref{subsec:realization_SSH}, this constitutes a realization of the Hamiltonian proposed by \textcite{Rice:1982}, reducing to the SSH model~\cite{Su:1979} for $\phi=\pi/2$. Here we restrict for simplicity to the SSH case and we refer the reader to \textcite{Atala2013} for the discussion of the general case. First we recall that as discussed in Sec. \ref{sec:zak_phase}, the Zak phase itself is not 
invariant under gauge transformations in momentum space. Indeed it depends on the choice of the relative phases between Wannier states, \ie the arbitrariness in deciding if the unit cell is formed by a pair $\{A_j,B_j\}$ ($N_{\rm A}=N_{\rm B}$ in  Eq.~(\ref{eq:Zak_gauge_change1})) or by a pair  $\{B_{j-1},A_j\}$ ($N_{\rm A}=N_{\rm B}+1$ in  Eq.~(\ref{eq:Zak_gauge_change1})). However, once this choice is made, the gauge-invariant quantity $\delta \phi_{\rm Zak} \equiv\phi_{\rm Zak}^{(D_1)}- \phi_{\rm Zak}^{(D_2)}=\pi$, where $D_1$ and $D_2$ correspond to the two possible dimerizations of the system, obtained by choosing either $\phi=0$ or $\phi=\pi$. 

 In order to measure $\delta \phi_{\rm Zak}$, \textcite{Atala2013} first prepared the atoms in a wave-packet localized at the bottom of the lowest band. Then using a $\pi/2$ microwave pulse the atoms were placed in a superposition of two spin states $|\uparrow\rangle$ and $|\downarrow\rangle$, which underwent Bloch oscillations in opposite directions in the presence of a magnetic gradient (Figure~\ref{fig:2013_Atala_Zak}). At the moment when the two wave-packets reach the edges of the BZ, the Zak phase is encoded in the relative phase between these wave packets.  In principle it could thus be read using a second $\pi/2$  microwave pulse, closing the interferometer in quasi-momentum space. However the magnetic field fluctuations in the lab cause a random dephasing between the two arms of the interferometer, and prevent one from performing this direct measurement. To circumvent this problem, \textcite{Atala2013} used a spin echo technique. A second microwave pulse flipped the spins as the atoms reached the edges of the BZ and at the same moment, the dimerization was changed by switching $\phi$ from 0 to $\pi$, corresponding to an exchange of states between the lower and the upper bands. Finally the two paths  were recombined when  the wave-packets reached the top of the upper band, and the accumulated phase revealed the value of $\delta \phi_{\rm Zak}$. The experimental result  $\delta \phi_{\rm Zak}/\pi=0.97\,(2)$ was in excellent agreement with the expected value.


\subsubsection{Measurement of the anomalous velocity}

\label{subsubsec:anomalousvelocity}

We now turn to the case of a 2D lattice and we investigate how the semi-classical dynamics of a wave-packet can reveal the topological features of a given energy band. The starting point is the set of equations that govern the evolution of the  average quasi-momentum  $\vv{q}$ and average position $\vv{r}$   when a constant force $\vv{F}$ is superimposed to the lattice potential
\begin{eqnarray}
 \hbar\,\dot{\vv{q}}&=& \vv{F}, \\
\hbar\, \dot{\vv{r}}&=&\vv{\nabla}_{\vv{q}}E(\vv{q}) +  {\vv{\Omega}}(\vv{q})\times \vv{F} \, .
\label{eq:anomalous}
\end{eqnarray}
This set of equations is valid when the applied force $\vv{F}$ is weak enough, so that transitions to other  bands can be neglected. With the first equation we recover the Bloch oscillation phenomenon: The momentum drifts linearly in time in response to the applied force $\vv{F}$. 
The second equation provides the value of the velocity of the wave-packet at a given position in the BZ. It contains two contributions: The first one is the well-known expression for the group velocity (see \eg \textcite{Ashcroft:1976} for the case of a periodic potential). The second contribution,  which is sometimes called  the ``anomalous velocity"~\cite{Xiao:2010}, couples the wave-packet dynamics to the local value of the Berry curvature  ${\vv{\Omega}}(\vv{q})$. As a result of this contribution, the recording of a given trajectory  inside the Brillouin zone allows one to reconstruct the Berry curvature along this trajectory~\cite{Price:2012}.

\begin{figure}[t]
\begin{center}
\includegraphics[width=44mm]{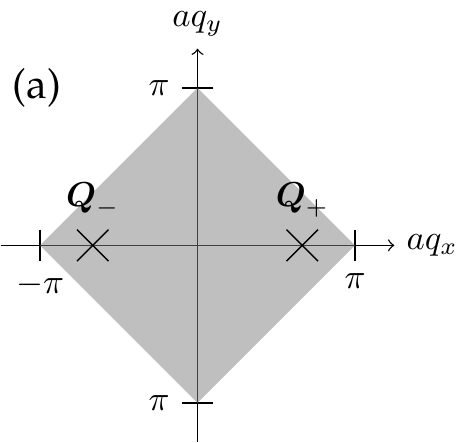}\\
\includegraphics[width=44mm]{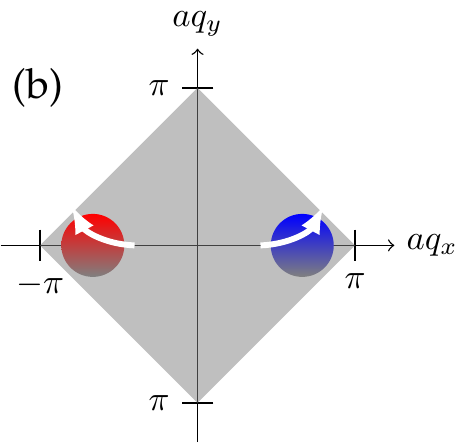}\includegraphics[width=44mm]{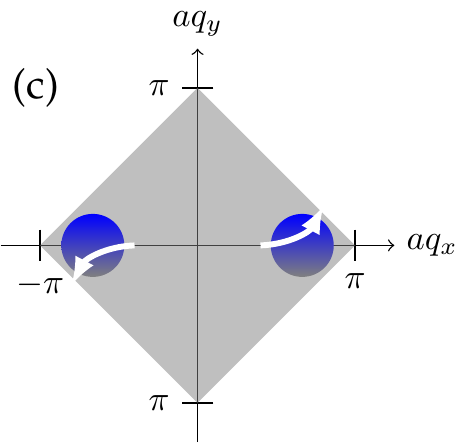}
\end{center}
\caption{ (a) Brillouin zone and Dirac points $\bs Q_\pm$ for the brick-wall lattice of Fig. \ref{fig:haldane}. (b) When the degeneracy is lifted by introducing an energy offset between the A and B sites, the subbands are topologically trivial. For the lowest band, the Berry curvature has opposite signs (marked with blue and red colors) in the vicinity of $\bs Q_\pm$. The anomalous velocity (white arrow) thus has opposite chirality at these points. (c) Lifting of degeneracy obtained by adding complex NNN couplings~\cite{Haldane:1988}. The Berry curvature then keeps a constant sign over the BZ, and the anomalous velocity has the same chirality at both points.
Figure adapted from \textcite{Jotzu:2014}.}
\label{fig:2014_Jotzu_Haldane_v3}
\end{figure}

This method was implemented by \textcite{Jotzu:2014} in order to analyze the topology of a 2D optical lattice in the vicinity of Dirac points. The experiment was performed with  a brick-wall lattice (Figure~\ref{fig:haldane}a), which is topologically equivalent to the hexagonal lattice of graphene, with two sites A and B  per unit cell~\cite{Tarruell:2012}. In such a lattice, when only nearest-neighbor couplings ${\rm A}\to {\rm B}$ and ${\rm B} \to {\rm A}$ are taken into account, the spectrum consists of two bands touching at two Dirac points $\vv{Q}_+$ and $\vv{Q}_-$ in the BZ (Figure~\ref{fig:2014_Jotzu_Haldane_v3}a). As explained in Sec. \ref{sec:force}, an additional circular shaking of the lattice breaks the time-reversal symmetry of the system and allows one to lift the degeneracy at the Dirac points, with the two subbands acquiring a non-trivial topology. Another possibility to lift this degeneracy consists of simply introducing an energy offset between sites A and B. However in this case each subband is topologically trivial (Sec. \ref{subsubsec:The Chern number}). With this setup one can thus explore the phase diagram of the Haldane model shown in Fig. \ref{fig:haldane}b. 

Using an analysis of wave-packet dynamics close to the points $\vv{Q}_{\pm}$, \textcite{Jotzu:2014} studied the transition between the topological and trivial cases. Starting from a wave-packet at the center of the BZ, they dragged it close to $\vv{Q}_+$ or $\vv{Q}_-$ using the force $\vv{F}$ created by a magnetic gradient. The curvature of the motion of the wave-packet in the BZ revealed the sign of the anomalous velocity, hence of the Berry curvature. In the trivial case, ${\vv{\Omega}}(\vv{Q}_{+})$ and ${\vv{\Omega}}(\vv{Q}_{-})$ have opposite signs, hence the Chern number which is proportional to the integral of $\vv{\Omega}(\vv{q})$ over the BZ is zero (Fig.~\ref{fig:2014_Jotzu_Haldane_v3}b). In the topologically non-trivial case, ${\vv{\Omega}}(\vv{Q}_{\pm})$ have the same sign (Fig.~\ref{fig:2014_Jotzu_Haldane_v3}c). The measurements of the drift of the wave-packet performed by \textcite{Jotzu:2014} confirmed quantitatively this scenario.


\subsubsection{Interferometry in the BZ}

We now come back to the simple case of a graphene-like lattice with only NN couplings, in which case the two subbands touch at two Dirac points $\vv{Q}_{\pm}$. In this case it is not possible to calculate the Berry curvature $\vv{\Omega}$ for each subband because the definition (\ref{eq:Berry_curvature}) is singular in $\vv{Q}_{\pm}$. The situation can be viewed as an equivalent (for momentum space)  of an infinitely narrow solenoid (in real space) providing a finite magnetic flux. In the latter case it is known that the presence of the solenoid can be probed by interferometric means. This is indeed a paradigmatic example for the Aharonov--Bohm effect: Using a two-path interferometer such that the solenoid passes through its enclosed area, there exists  between the two paths  a phase difference proportional to the magnetic flux. In the case of a two-path interferometer enclosing a Dirac point in momentum space, the phase difference between the two paths is $\pi$~\cite{Mikitik:1999}.

\begin{figure}[t]
\begin{center}
\includegraphics[width=80mm]{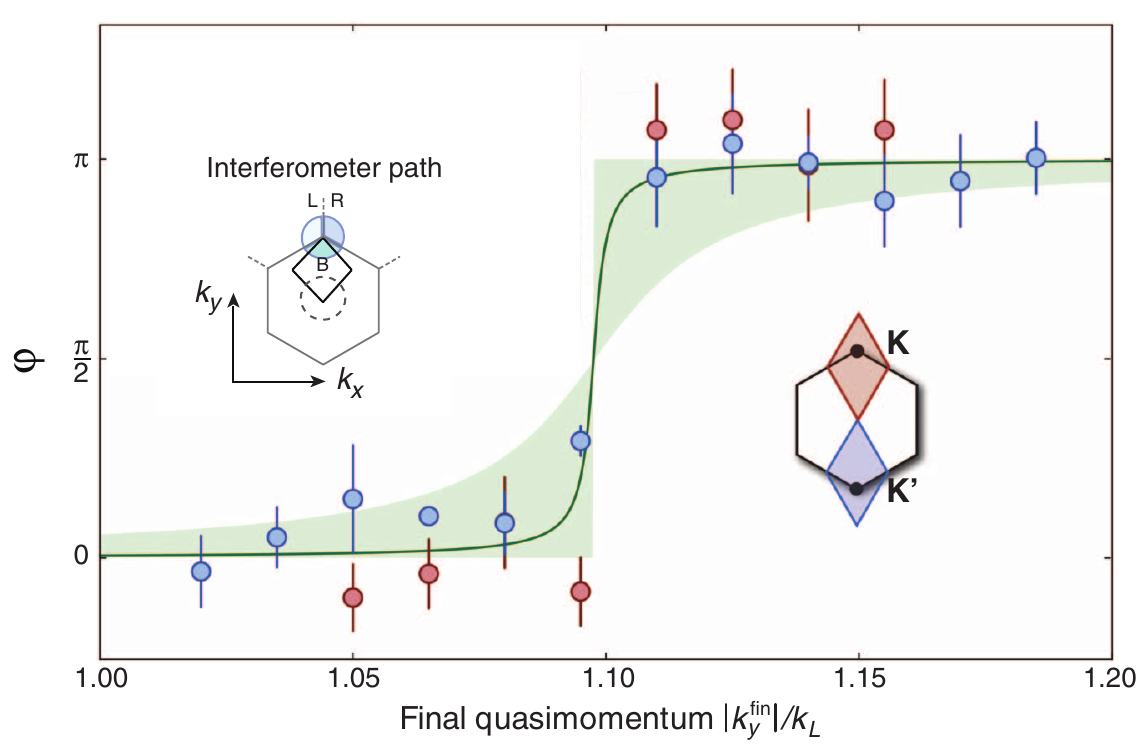}
\end{center}
\vskip-5mm
\caption{Aharonov--Bohm interferometer around a Dirac point. Starting from a wave-packet localized at the center of the BZ of a graphene-like optical lattice, one measures the phase difference $\varphi$ between the  two paths of an interferometer ending at the quasi-momentum $(k_x=0,k_y)$. When one of the Dirac points $K$ or $K'$ is inside the enclosed area of the interferometer, the measured value for $\varphi$ is in good agreement with the prediction $\varphi=\pi$. Courtesy of  \textcite{Duca:2015}. }
\label{fig:2015_Duca_aharonov}
\end{figure}

This phase shift has been measured in the cold atom context by \textcite{Duca:2015}. The graphene-like structure was generated using an optical lattice with three beams at 120$^\circ$ angles. Initially the external state of the $^{87}$Rb atoms is a wave-packet located at the center of the BZ, and their internal state is a given Zeeman state $|\uparrow\rangle$. A microwave $\pi/2$ pulse prepares a coherent superposition of  $|\uparrow\rangle$ and $ |\downarrow\rangle$, where the magnetic moment of $|\downarrow\rangle$ is opposite to that of $|\uparrow\rangle$. Then the displacements of the two corresponding wave-packets are controlled using simultaneously: (i) A lattice acceleration along $y$ which creates the same inertial force on the two Zeeman states; (ii) A magnetic gradient which creates opposite forces on them along $x$; (iii) A $\pi$ microwave pulse which exchanges the atomic spins in the middle of the trajectories, in order to close the interferometer in momentum space. The result obtained by \textcite{Duca:2015}, shown in Figure~\ref{fig:2015_Duca_aharonov}, shows clearly that a phase  difference of $\sim \pi$ between the two paths appears if and only if the enclosed area contains one of the two Dirac points. Using a similar technique \textcite{Li:2016}  could also access the Wilson line regime, which generalizes the Berry phase concept to the case where the state of the system belongs to a (quasi-)\-degenerate manifold. \textcite{Li:2016} investigated the case where the transport from one place to another in the BZ is done in a time much shorter than the inverse of the frequency width of the bands. In that experiment the transport was characterized by a single phase factor and could be analyzed within the framework of St\"uckelberg interferometry \cite{Lim2014,Lim2015}. However in more complex situations,  interferometry can  also reveal non-Abelian features  of the transport (see \eg \textcite{Alexandradinata:2014}).

\subsubsection{Direct imaging of edge magneto-plasmons}

\label{subsubsec:magnetoplasmons}

\begin{figure}[t]
\begin{center}
\includegraphics[width=82mm]{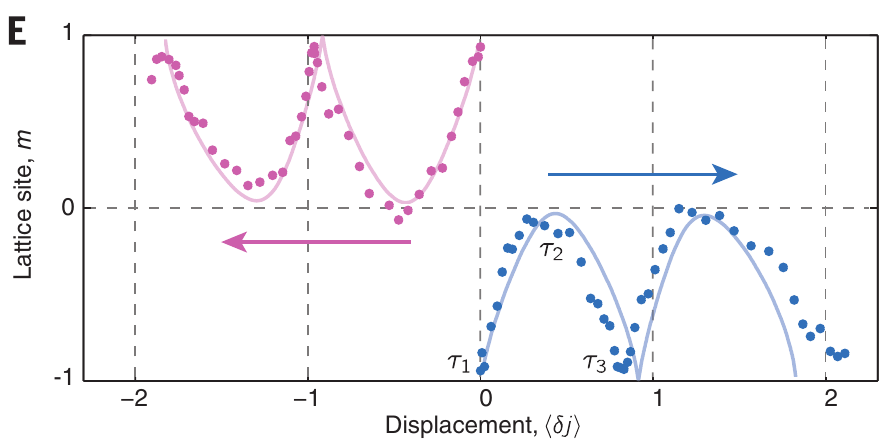}
\end{center}
\caption{Edge-magnetoplasmon trajectories, where the displacement was obtained by integrating the velocity.}
\label{fig:2015_Stuhl_edges}
\end{figure}

Although the synthetic-dimension systems can be described by the same Harper-Hofstadter Hamiltonian as with real-space laser-assisted tunneling approaches, using the spin degree of freedom to encode one spatial direction enables new preparation, control and measurement opportunities.  For example: the finite number of spin states in effect generates infinitely sharp hard-wall boundaries in the synthetic dimension; the synthetic-dimension tunneling can be applied and removed on any experimental time-scale; atoms can be initially prepared in any initial synthetic dimension site; and conventional time-of-flight measurements, along with Stern-Gerlach techniques allow the near-perfect resolution of ``position'' in the spin direction.

Recent synthetic dimension experiments~\cite{Mancini:2015,Stuhl:2015} used these techniques to directly image the evolution of edge magneto-plasmons.  In a quantum-Hall system, an edge magneto-plasmon is the quantum analogue to the skipping cyclotron orbits which ``bounce" down the edge of a system in a chiral manner, essentially following cyclotron orbits that are interrupted by the system's edge (see Fig.~\ref{fig:skipping}).  These dynamical excitations evolve with a characteristic frequency given by the cyclotron frequency and are  superposition states  between different Landau levels.  These should not be confused with the chiral edge modes that underly quantized conductance: these modes are built from states all within the same Landau level.  These modes have not been observed in two-dimensional experiments, however, their analog has been observed in 1D experiments, also using synthetic dimensions, where the end-modes of 1D systems have been observed~\cite{Meier2016}.

In condensed matter systems, edge magneto-plasmons certainly have been launched and detected.  Both in steady state, in magnetic focusing experiments~\cite{Houten1989} and directly in the time-domain using electrostatic gates~\cite{Ashoori1992}.  Cold atom experiments complete the picture by allowing for direct space and time resolution of these skipping orbits.

In these experiments~\cite{Mancini:2015,Stuhl:2015}, the system was initialized with no hopping along the synthetic dimension, and with all atoms either in one or the other edge along the synthetic dimension.  Once this initial state was prepared, tunneling was instantly turned on.  The highly localized initial state, described by a superposition of different Landau-levels, then began to evolve, skipping down the system's edge.  Following a tunable period of evolution time-of-flight measurements directly imaged the position along the synthetic direction along with velocity in the spatial direction~\footnote{Because TOF measurements yield the momentum distribution, some effort is required to derive the velocity in the lattice from this information.}, which gives position by direct imaging.  As shown in Fig.~\ref{fig:2015_Stuhl_edges}, this prescription allows for direct imaging of edge magneto-plasmons.

\subsection{Transport measurements}

In the last part of this section, we turn to experimental procedures that are closer in spirit to well-established condensed matter techniques. Starting from a uniformly filled band, one can perform a transport-like experiment, measure the displacement of the whole cloud  in the lattice after a certain duration, and infer non-trivial topological aspects from this dynamics~\cite{Dauphin2013}.


\subsubsection{Adiabatic pumping}
\label{subsubsec:Exp_transport_pumping}

The concept of a quantized pump introduced by \textcite{Thouless:1983} has been described in general terms in Sec.~\ref{subsec:pumping}. It requires a lattice whose shape is controlled by at least two effective parameters, such as $\Delta$ and $J'-J$ for the Rice--Mele model (see Eq.~\ref{eq:hrm1}). One starts with a gas that  uniformly fills a band of the lattice. Then one slowly modifies the lattice shape in a way that corresponds to a closed loop in parameter space. As shown in the Appendix~\ref{app:1dmodels},  the resulting displacement of the center-of-mass of the gas is then quantized in units of the lattice spacing.

\begin{figure}[t]
\begin{center}
\includegraphics[width=85mm]{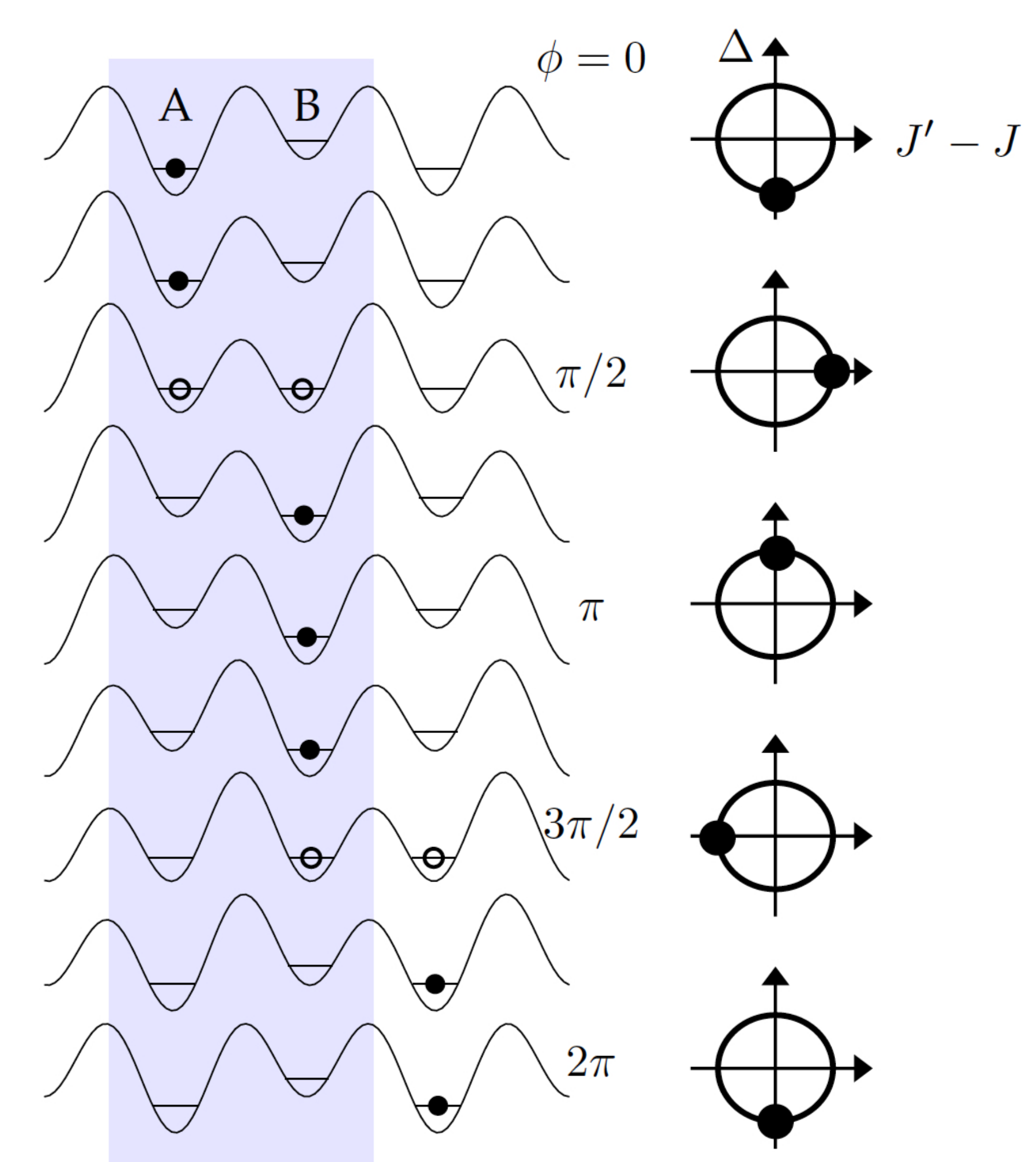}
\end{center}
\caption{Illustration of topological pumping with a 1D superlattice described by the potential (\ref{eq:superlattice_potential}). Left: The phase $\phi$ of the long lattice (see (\ref{eq:superlattice_potential})) is varied from 0 to $2\pi$ from top to bottom. Initially the phase $\phi=0$ and the particle is  supposed to be localized on the site $A_j$ of a given lattice cell $j$. In the limiting case where the energy difference between A and B sites is large compared to the tunnel matrix elements, this state would be stationary if $\phi$ was kept at the value 0. When $\phi$ is increased up to $\pi/2$, the  sites $A_j$ and $B_j$ have the same energy and the particle is adiabatically transferred to $B_j$. Note that we neglect here the tunneling of the particle from $A_j$ to $B_{j-1}$, assuming that it is inhibited by the large barrier between these two sites. The particle then remains in $B_j$ until the phase reaches the value $3\pi/2$, when the particle again undergoes an adiabatic transfer, now from $B_j$ to $A_{j+1}$. (Here again we neglect tunneling across the large barrier now present between $B_j$ and $A_j$.) When the phase $\phi=2\pi$ the potential is back to its initial value and the particle has moved by one lattice site. Note that a motion in the opposite direction occurs if the particle starts for the site $B_j$ when $\phi=0$. Right: In the two-band approximation corresponding to the Rice--Mele model, the system performs a closed loop around the origin in the parameter space $(J'-J,\Delta)$.}
\label{fig:pump_JD_fig}
\end{figure}

This concept can be addressed in a 1D geometry, using a superlattice with the potential (\ref{eq:superlattice_potential}). As explained in Sec.~\ref{subsec:realization_SSH}, the motion of atoms in the two lowest bands of this potential for a deep enough lattice can be described by the Rice--Mele Hamiltonian. Suppose that the relative phase $\phi$ of the long-period lattice is scanned from 0 to $2\pi$, while the lattice depths $V_{\rm long}$ and $V_{\rm short}$ are kept fixed (Figure~\ref{fig:pump_JD_fig}).  Initially when $\phi= 0$, the energy of a A site is below that of a B site, \ie $\Delta<0$ in the framework of the Rice--Mele model. At this moment the superlattice is symmetric, hence the tunnel matrix elements $J'$ and $J$ from ${\rm B}_j$ to ${\rm A}_{j}$ and ${\rm A}_{j+1}$ are equal. When $0<\phi<\pi $, the energy barrier between ${\rm A}_j$ and ${\rm B}_j$ is smaller than that between ${\rm B}_j$ and ${\rm A}_{j+1}$, hence $J'>J$. During this time period, $\Delta$ increases and changes sign when $\phi=\pi/2$. During the second half of the cycle, $\pi<\phi<2\pi$, one now finds $J'<J$ and $\Delta$ again changes sign. When $\phi=2\pi$, the system has performed a closed loop in the parameter space $(J'-J,\Delta)$, encircling the vortex localized in $J'-J=0$, $\Delta=0$. One can therefore conclude that the center-of-mass of the cloud  has moved by one lattice cell. In the limiting case of a particle initially localized in one of the lattice sites, one can recover this result by a simple reasoning based on the adiabatic following of the instantaneous energy levels (see  Fig. \ref{fig:pump_JD_fig}).

\begin{figure}[t]
\begin{center}
\includegraphics[width=85mm]{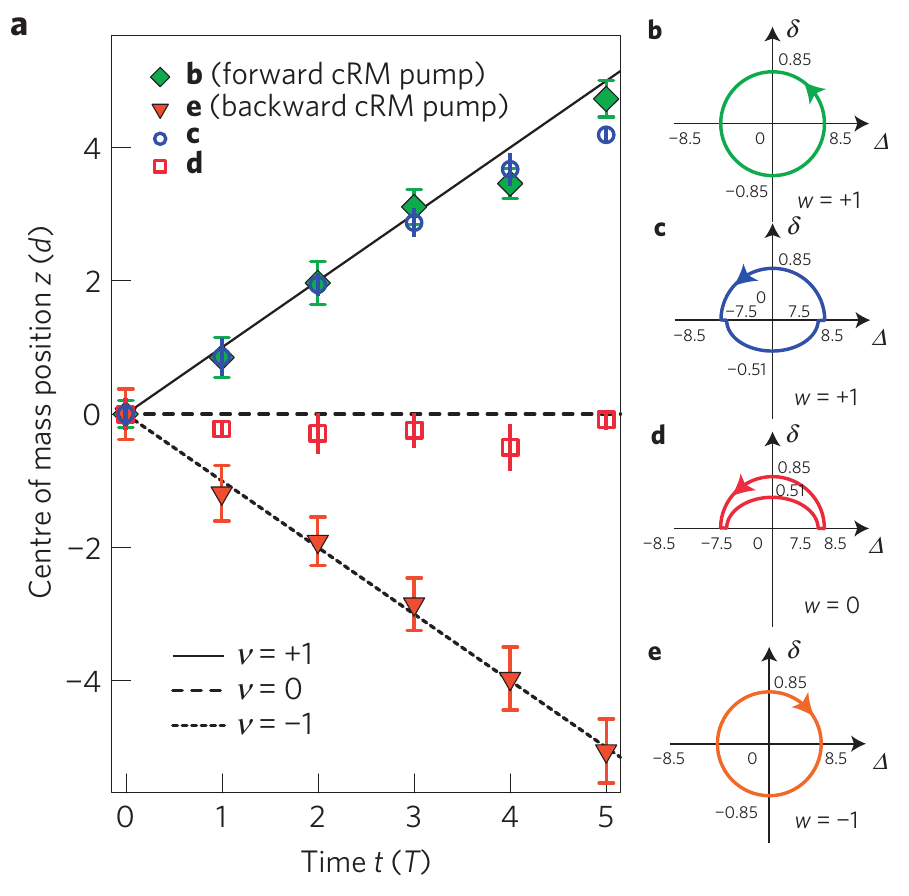}
\end{center}
\caption{Quantization of the displacement of a cloud of fermionic $^{171}$Yb atoms placed in a 1D optical superlattice described by the Rice--Mele Hamiltonian. Depending on the closed loop in parameter space $(J'-J\equiv 2\delta,\Delta)$, the displacement can be positive, zero or negative. The time $T$ represents the duration of a pump cycle. Courtesy of \textcite{Nakajima:2016}.}
\label{fig:2016_Nakajima_pump}
\end{figure}

A Thouless pump has been implemented in cold atom setups~\cite{Lu:2016,Nakajima:2016,Lohse:2016}. The experiment by \textcite{Nakajima:2016} uses a superlattice potential similar to that represented in Fig. \ref{fig:pump_JD_fig}, with $V_{\rm long}=30\,E_r$ and $V_{\rm short}=20\,E_r$, respectively, with $E_r=\hbar^2/(8ma^2)$. Using a gas of non interacting $^{171}$Yb atoms (fermions), \textcite{Nakajima:2016} verified that the displacement of the cloud when $\phi$ varies from 0 to $2\pi$ is equal to one lattice period, as expected (Figure~\ref{fig:2016_Nakajima_pump}). They also checked that it is topologically robust, {\it \ie} it does not change if one slightly deforms the path in parameter space by adding a time variation modulation of $V_{\rm long}$ and $V_{\rm short}$. However if the modification is such that the closed trajectory in parameter space $(J'-J,\Delta)$ does not encircle the origin point anymore, the displacement per pump cycle drops to zero. 

%


\subsubsection{Center-of-mass dynamics in 2D}
\label{subsubsec:cofm}
\begin{figure}
\begin{center}
\includegraphics[width=85mm]{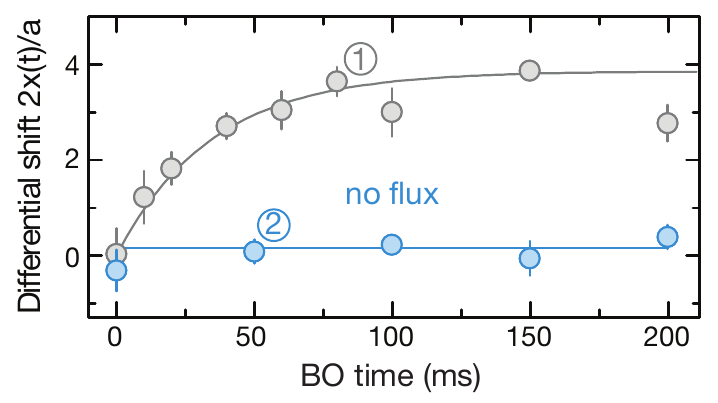}
\end{center}
\caption{Transport measurement in a square lattice with  a flux/plaquette $\phi_{\rm AB}= \pi/2$ (data 1). A Hall-type current is observed through the displacement of the center-of-mass of the cloud along the $x$ direction, when a uniform force inducing Bloch oscillations (BO) is applied along $y$ for an adjustable time. For short BO times, only the lowest subband is populated, resulting in a linear variation of $x(t)$ with time, in agreement with the expected Chern index of this subband. At longer times, heating  processes equalize the populations of the subbands and the Hall drift stops. When the flux is zero, no Hall current is observed (data 2). {Courtesy of Monika Aidelsburger, adapted from \textcite{munichchern}.}}
\label{fig:2014_Aidelsburger_Chern}
\end{figure}

As a last example of a cold atom probe of band topology, let us briefly describe the analysis of the dynamics associated with the Harper-Hofstadter Hamiltonian (\ref{eq:hofstadter}) by \textcite{munichchern}. This Hamiltonian was implemented using the laser induced hopping method explained in Sec.~\ref{sec:hopping}, producing the flux per plaquette $\phi_{\rm AB}=\pi/2$. As a result of this applied gauge field, the lowest band of the lattice was split in 4 subbands, with Chern numbers $(+1,-1,-1,+1)$, with the two intermediate subbands touching at Dirac points. Bosonic $^{87}$Rb atoms were loaded in majority in the lowest subband of the lattice, at a temperature such that this subband was filled quasi-uniformly. Then, a weak uniform force along the $y$ axis originating from a gradient of the light intensity of an auxiliary beam was applied to the atoms for an adjustable duration $t$. The position of the center-of-mass of the atom cloud was monitored as a function of time and it revealed the topology of the bands. At short times (typically less than 50 ms), the drift of the cloud along the $x$ direction, similar to a Hall current, was found to be linear with $t$ (Figure~\ref{fig:2014_Aidelsburger_Chern}). It provided a measurement of the anomalous velocity in good agreement with the expected value for the Chern number of the lowest subband.  For longer times $t$, heating originating from the resonant modulation applied to the lattice induced  transitions between the various subbands, which eventually get equally populated. Since the sum of all four Chern number is zero, the drift of the center-of-mass then stopped. A careful analysis of this dynamics, associated with an independent measurement of the population of each subband provided a measurement of the individual Chern numbers  with a 1\% precision.

\section{Interaction Effects}

\label{sec:interactions}

Some of the most interesting directions for future work on cold atoms
in topological optical lattices involve studies of collective effects
that arise from interparticle interactions.  Such studies hold promise
for the exploration of novel phases of matter, and to elucidate the
role of topology in strongly correlated many body systems.

\subsection{Two-body interactions}

The methods described above for generating topological lattices consist either of periodic modulation of site energies, forming a so-called ``Floquet" system (see Appendix~\ref{app:floquet}), or of Raman coupling of internal spin states, leading to optically ``dressed states" of the atoms. Both of these methods lead to effective interactions between particles in the resulting energy bands that have some novel features.

\subsubsection{Beyond contact interactions}

\label{subsec:nonlocal}

For ultracold atoms in the continuum, the typical two-body
interactions are dominated by the $s$-wave scattering, which can be
represented by a short-range (contact) interaction $g\,\delta(\bs r_1-\bs r_2)$, where the Dirac $\delta$ distribution is supposed to be properly regularized. 
However, for
particles ``dressed" by a laser field like in the optical lattices described above, the interactions
typically acquire non-local character.

\paragraph{Continuum models.} For the continuum setting of optical flux lattices of Sec.~\ref{subsec:fluxlattice} with laser plane waves inducing Raman couplings between internal spin states, non-local interactions can arise from the momentum-dependence of the dressed-state wavefunctions.  For the two-state system described by the Hamiltonian (\ref{eq:solattice}), an energy eigenstate in a given band can be written
\be
|\Psi_\vv{q}\rangle = |\psi_{\bs q}\rangle\otimes |\Sigma_{\bs q}\rangle,
\label{eq:solatticewf}
\ee
\ie the product of an orbital Bloch state $|\psi_{\bs q}\rangle$ of quasi-momentum $\bs q$ and of a $\bs q$-dependent  spin state $|\Sigma_{\bs q}\rangle$ written in the basis $|\pm\rangle_z$. 
The dependence of the spin component $|\Sigma_\vv{q}\rangle$ on wavevector has direct consequences on the two-body scattering matrix elements within this band.

Let us first recall the simple case of spinless or fully polarized particles. 
For the contact interaction $\hat V=g\,\delta(\hat{\bs r})$, the matrix element of $\hat V$ for a pair of free distinguishable particles transitioning from $(\vv{q}_1,\vv{q}_2) \to (\vv{q}_1+\vv{Q},\vv{q}_2-\vv{Q})$ is equal to $g$ (up to a normalization factor) and it is independent of the momentum transfer $\bs Q$. Here we neglect for simplicity energy-dependent corrections of the $s$-wave scattering amplitude, which is valid when $Q^{-1}$ is larger than the $s$-wave scattering length, a criterion usually satisfied in quantum gases.
For two indistinguishable particles, physical observables involve the sum
over the two permutations 
\begin{equation}
(\vv{q}_1,\vv{q}_2) \to (\vv{q}_1+\vv{Q},\vv{q}_2-\vv{Q}) \mbox{ or }  (\vv{q}_2-\vv{Q},\vv{q}_1+\vv{Q})
\label{eq:two_paths}
\end{equation}
with a relative sign of $\varepsilon=\pm 1$ for
bosons/fermions. The matrix element of $\hat V$ is doubled for polarized 
bosons and vanishes for polarized fermions: the latter result simply
reflects the fact that single-component fermions are insensitive to
the contact interaction, since the Pauli principle precludes them from
having the same spatial position.

Consider now the scattering of atoms in the dressed state band with wavefunctions  (\ref{eq:solatticewf}) and assume for simplicity that the contact interaction is independent of the internal states $|\pm\rangle$. For bosons/fermions ($\varepsilon=\pm 1$), the transition matrix element of $\hat V$ for the two-path process (\ref{eq:two_paths})
is now
\begin{eqnarray}
&&\langle \psi_{\bs q_1+\bs Q},\psi_{\bs q_2-\bs Q}|\hat V|\psi_{\bs q_1},\psi_{\bs q_2}\rangle \;
\langle \Sigma_{\bs q_1+\bs Q}|\Sigma_{\bs q_1}\rangle\;\langle \Sigma_{\bs q_2-\bs Q}|\Sigma_{\bs q_2}\rangle\nonumber\\
&+\varepsilon&\langle \psi_{\bs q_2-\bs Q},\psi_{\bs q_1+\bs Q}|\hat V|\psi_{\bs q_1},\psi_{\bs q_2}\rangle \;
\langle \Sigma_{\bs q_2-\bs Q}|\Sigma_{\bs q_1}\rangle\;\langle \Sigma_{\bs q_1+\bs Q}|\Sigma_{\bs q_2}\rangle
\nonumber
\end{eqnarray}
For fermions, this matrix element no longer vanishes in general. For
example, for $\vv{Q}=0$, it is
$\propto g \left( 1 - \left|\langle
  \Sigma_{\vv{q}_2}|\Sigma_{\vv{q}_1} \rangle\right|^2\right)$.  
This is an important result that shows that starting from
contact-interacting fermions occupying a non-degenerate band,
the optical dressing leads to an {\it interacting} one-component Fermi gas, 
more precisely to effective $p$-wave interactions
\cite{zhang_prl_2008}.  The possibility of interaction between two identical
fermions can be viewed as arising from non-adiabatic corrections to the optical dressing, allowing two particles to coincide in real space by being
in different internal dressed states, dependent on their
momentum~\cite{Cooper2011a}. The momentum dependence of the
dressed-state wavefunctions converts the contact interactions into a
momentum-dependent interaction for atoms,
thereby allowing effective $p$-wave scattering even at ultralow
temperatures.  Similarly, dressed-state bosons can acquire a momentum-dependent
interaction allowing the appearance of $d$-wave and higher angular-momentum channels in the scattering~\cite{Williams20012012}.

\paragraph{Tight-binding models.}  For tight-binding lattice systems, we have described in Sec.~\ref{sec:implementations} how
periodic driving at frequency $\omega$ can be used to tailor the
amplitudes and phases of the hopping matrix elements between the
sites.  We demonstrated
how, at rapid driving frequencies, an effective Hamiltonian with (potentially complex) modified tunneling
amplitudes arises. This is the effective Floquet Hamiltonian for the modulated system, which governs the dynamics of the particles on
timescales large compared to the period of the modulation,
$T = 2\pi/\omega$, as described in Appendix~\ref{app:floquet}.  In the presence of interparticle interactions --
\eg Hubbard interaction $U$ between particles on the same lattice site --
new terms appear in this effective Floquet Hamiltonian,  including non-local interactions~\cite{eckardtreview}. 
To understand the origin of the  non-local interactions, consider the modulated two-site system described in \ref{sec:force} under a harmonic drive of the energy offset between the sites, $\Delta(t) = \Delta \cos\omega t$. The unitary transformation Eq.~\ref{eq:force:gauge} leads to a Hamiltonian
\begin{eqnarray}
\hat{H}' & = & -J\left[ \E^{\I\phi_{\rm P}(t)} |r \rangle \langle l | 
+ \E^{-\I\phi_{\rm P}(t)}  |l \rangle \langle r | \right]\,, \\
\label{eq:tunneloscillate}
\phi_{\rm P}(t) & = & \frac{\Delta}{\hbar\omega}\sin(\omega t)\,.
\end{eqnarray}
It is convenient to expand this Hamiltonian in its harmonics
\begin{eqnarray}
\hat{H}' & = & -J \sum_{m=-\infty}^\infty {\mathcal J}_m\left(\frac{\Delta}{\hbar\omega}\right) \E^{\I m\omega t}
 |r \rangle \langle l |  + \mbox{H.c}
\end{eqnarray}
where ${\mathcal J}_m(z)$ are Bessel functions of the first kind. The $m=0$ term gives the
time-averaged Hamiltonian discussed in Sec.~\ref{sec:implementations}: it describes intersite tunneling with an
effective tunneling rate  that is modified from the bare rate $J$ 
by ${\mathcal J}_0(\Delta/\hbar\omega)$.
For $\hbar\omega$ large compared to all other energy scales (tunneling
$J$ and any onsite-interaction energy $U$) the higher-order terms
$|m|>0$ are far off-resonant, so are rapidly oscillating and have
small effects on the wavefunction. Still these terms do 
perturb the atomic wavefunction, leading to the 
new features we
are interested in here. For a particle that is initially in the right
well, $|r\rangle$, a
rapidly oscillating perturbation $\hat{V}_m \E^{\I m\omega t}$  (\ie $m\neq 0$)
 causes a first-order correction
to the wavefunction of
\begin{eqnarray}
|\psi'\rangle & \approx & |r\rangle + |l\rangle \langle l | \hat{V}_m |r \rangle \E^{\I m\omega t/2} \frac{\sin (m\omega t/2)}{m\hbar\omega}\\
 &  \approx & |r\rangle 
-J {\mathcal J}_m(\Delta/\hbar\omega)\E^{\I m\omega t/2} \frac{\sin (m\omega t/2)}{m\hbar\omega}
|l\rangle
\end{eqnarray}
Thus, these terms induce a (small) nonzero probability for the particle to be in the left well,
\begin{equation}
p_l = \sum_{m\neq 0} \frac{J^2}{2(m\hbar\omega)^2} \left[{\mathcal J}_m\left(\frac{\Delta}{\hbar\omega}\right)\right]^2\,,
\end{equation}
after averaging over the rapid oscillations and summing over all such terms. If in addition to the tunnelling (\ref{eq:tunneloscillate}) the static Hamiltonian has an on-site interaction between particles in the left well, which we consider to be a Hubbard interaction $U \hat{n}_{l}(\hat{n}_{l}-1)/2$ (with $\hat{n}_{l}$ the number operator in the left well), then these rapidly oscillating terms that move particles from right to left well will give rise to effective 
non-local interactions $U^{\rm eff}_{lr} \hat{n}_{l} \hat{n}_{r}$, with
\begin{equation}
U^{\rm eff}_{lr} = U p_l \propto \frac{U J^2}{(\hbar\omega)^2}\,.
\label{eq:nonlocalsite}
\end{equation}
A similar nonlocal term will arise from interactions in the right well.
A full analysis of such effects is best achieved through 
a construction of the Floquet Hamiltonian at high drive frequency via the Magnus expansion~\cite{goldmandalibard} or other systematic approach~\cite{eckardt:2015}. An overview of the Magnus expansion is given in Appendix~\ref{app:floquet}.

\paragraph{Synthetic dimensions.} 
An extreme example of non-local interactions arises for systems
involving synthetic
dimensions~\cite{Celi:2014,Stuhl:2015,Mancini:2015}.  There, interactions are typically
long-ranged among the set of $s=1,N_s$ internal states which form the
synthetic dimension, and short ranged in the $d$ spatial
co-ordinates. For a synthetic dimension formed from an internal spin
degree of freedom,  the interactions are of infinite range in the
synthetic dimension, and of conventional short range in position. {For example spin-exchange interactions can be viewed as a correlated tunneling, such as $(m,m')\to (m-1,m'+1)$.}
Such
systems realize interesting intermediate situations in which the
single-particle physics can be viewed as $d+1$ dimensional (for $N_s$
is large) while the interactions remain $d$-dimensional. For a
synthetic dimension formed from simple harmonic oscillator subband
states~\cite{price_ozawa_goldman} the interactions fall off with
increasing spacing along the synthetic dimension. {Interactions in an effective two-leg ladder, formed from spin-orbit coupled Strontium atoms, have been studied experimentally by \textcite{Bromley2018}.}

\paragraph{Current-density coupling.}
Finally, we note that the novel two-body interactions that are induced
by Raman coupling or Floquet modulation can also include terms that
are not just density-density interactions, but that couple the
particle motion to particle density.
These effects arise naturally in photon-assisted tunneling between
lattice sites as a consequence of interaction-induced energy
shifts. The imposed energy offset between two adjacent lattice sites
is modified by the onsite interactions, $U$, in such a way that the photon-assisted tunneling can be brought into or out-of resonance depending on the
occupations of the sites by other particles. This leads to a
density-dependent tunneling term of the form
\be
\sum_{i} \left[J(\hat{n}_i,\hat{n}_{i+1}) \hat{b}^\dag_i\hat{b}_{i+1} + \mbox{h.c.}\right]\,,
\ee
shown here for spinless bosons. Such situations have been analysed in Refs.~\cite{munichchern,PhysRevA.93.043618,bermudezporras}. The density-dependent corrections are typically a small modification of the photon-assisted hopping in current experimental setups of gauge fields on optical lattices [see supplementary information in~\textcite{munichchern}]. However, they can be made to dominate in regimes where $U$ is large, such that the photon drive frequency $\omega$ is not resonant with the bare detuning $\Delta$ but with $\Delta\pm U$. Such situations have been explored in experiments on two-component Fermi  systems~\cite{goerg2017,2017arXiv171102061X}.
Analogous coupling between current and density  arises also in the continuum if interactions lead to detunings that influence the local dressed states, thus causing the induced vector potential $\vv{A}$ to be density-dependent~\cite{edmonds2013}.
Density-mediated hopping can also arise in settings where the single particle energy bands are flat (dispersionless). Then individual particles do not move, being localized to a region of size of the Wannier orbital, and particle motion arises only through interparticle interactions. Such settings include the flat energy  bands of frustrated lattices, such as the Creutz ladder~\cite{Creutz:1999,Mazza:2012,PhysRevX.7.031057} in 1D, and the  Kagom{\' e}~\cite{PhysRevB.82.184502} and dice lattices~\cite{mollercooperdice} in 2D, as well as for particles in dressed states of internal spin states in which the direct nearest-neighbor tunneling can be made to vanish~\cite{bilitewskicoopersd}.

\subsubsection{Floquet heating}

The explicit time-dependence of the Hamiltonian in periodically driven
systems relaxes energy conservation, and leads to forms of
inelastic scattering and heating not present in the time-independent
case. For a Floquet system at frequency $\omega$, these correspond to
the absorption (or emission) of an integer number of ``photons" of energy
$\hbar\omega$ from the external drive. For the ``Floquet-Bloch" waves
of a spatial- and time-periodic potential, such inelastic scattering
could occur for even for a single particle that scatters from a defect
in the lattice, which allows a change in momentum transfer. However, an important source of
potential inelastic scattering is for pairs of particles via the
interparticle interactions, \ie  inelastic two-body collisions.

A general description of the inelastic scattering of Floquet-Bloch
waves was provided in~\textcite{bilitewski2015}.  Consider a time-periodic
Hamiltonian of frequency $\omega$, and denote the single-particle
energy band, with band index $\nu$, by the energies
$\epsilon_\nu(\vv{q})$ defined as continuous functions of $\vv{q}$
over the Brillouin zone.  Owing to the time-periodicity, this can be
viewed as one member of a sequence of Floquet energy bands
$\epsilon^{(m)}_\nu(\vv{q}) = \epsilon_\nu(\vv{q}) +
m\hbar\omega$~\cite{eckardtreview}.
We
define inelastic scattering to be those processes in which $m$, or
$\nu$, or both, change under a scattering event.  For weak scattering
from state ${\rm i}$ to state ${\rm f}$, the inelastic rate can be
computed through a ``Floquet Fermi Golden Rule"~\cite{PhysRevB.84.235108}
\be
\gamma_{\rm{i} \to {\rm f}}  = \frac{2\pi}{\hbar}\sum_m |\langle \langle \Phi_{\rm f}^{(m)}|\hat{V}\Phi_{\rm i}^{(0)}\rangle\rangle|^2\delta(E_{\rm i}-E_{\rm f}-m\hbar\omega)\,,
\ee
where $|\Phi_{\rm f,i}^{(m)}(t)\rangle \equiv \E^{-\I m\omega t}|\Phi_{\rm f,i}^{(0)}(t)\rangle$ and $\langle\langle \Phi_1|\Phi_2\rangle\rangle \equiv \frac{1}{T}\int_0^T \langle\Phi_1|\Phi_2\rangle \D t$. Here $\hat{V}$ can denote a one-body potential (e.g. a lattice defect) or a two-body interaction. Processes with nonzero $\Delta m$ correspond to the exchange of $\Delta m$ quanta from the drive field, changing the energy of the atom by $m\hbar\omega$.
This provides a simple prescription by which to calculate the inelastic ($\Delta m\neq 0$) contribution to the two-body scattering of Floquet-Bloch states. 
Stepping beyond the simple two-body calculation to a many-body setting can be achieved by analysing instabilities within the Gross-Pitaevskii description~\cite{PhysRevA.90.013621,PhysRevA.91.023624,PhysRevX.7.021015}. 

Calculations for the parameters used in the experimental studies of weakly interacting bosons in the Harper-Hofstadter model~\cite{munichchern} show that the heating rates observed in those experiments are consistent with the expected inelastic two-body scattering processes dominated by single-photon absorption~\cite{bilitewskiharper2015}. This analysis emphasizes the role played by the motion transverse to the 2D plane along the (weakly confined) third dimension. The application of an optical lattice to confine this motion and open up gaps at the one-photon resonance is expected to significantly reduce the heating rate in such experiments~\cite{bilitewskiharper2015,PhysRevA.91.023624}. The excitation of motion along tubes, transverse to the optical lattice, has been argued to be responsible for heating rates in a recent experimental study of a periodically modulated 1D lattice~\cite{schneiderdata}.
In strongly driven systems~\cite{PhysRevA.92.043621} particle transfer to higher bands can arise from multiphoton resonances at the single-particle level~\cite{2016arXiv160400850S}.

Floquet heating in the full, many-body, system presents an interesting theoretical issue which remains an active area of investigation. That energy is not conserved leads to the expectation that the system will be driven to an infinite temperature state at long times~\cite{PhysRevE.90.012110}. This expectation relies on the assumption that energy is redistributed between all degrees of freedom through the interparticle interactions. There can, however, arise situations in which many-body systems show steady states that are not at infinite temperature~\cite{PhysRevB.93.174305,dalessio}, or in which there form prethermalized states on intermediate timescales~\cite{PhysRevLett.115.205301,PhysRevE.93.012130}.
Furthermore, in settings involving disorder,  it has been shown that many-body localized (MBL) phases can be robust to Floquet modulations, allowing the existence of non-thermal steady states to arbitrarily long times~\cite{abanin2016}. A striking example of such a ``Floquet MBL" phase is the Floquet time crystal~\cite{moessnersondhi}.

\subsection{Many-body phases}

The topological optical lattices described above support an array of
many-body phases. Novel features arise for weakly interacting gases through the
geometrical and topological characters of the underlying band structure. Furthermore, there can arise 
interesting strongly correlated phases, driven by strong interparticle interactions.

\subsubsection{Bose-Einstein condensates}

For an optical lattice loaded with a gas of non-interacting bosons one
expects the ground state to be a Bose-Einstein condensate (BEC), in
which all particles condense at minimum of the lowest energy
band. This expectation applies just as well to the topological optical
lattices as to regular optical lattices.
However, topological optical lattices can bring several novel features.

\begin{figure}
\includegraphics[width=0.45\columnwidth]{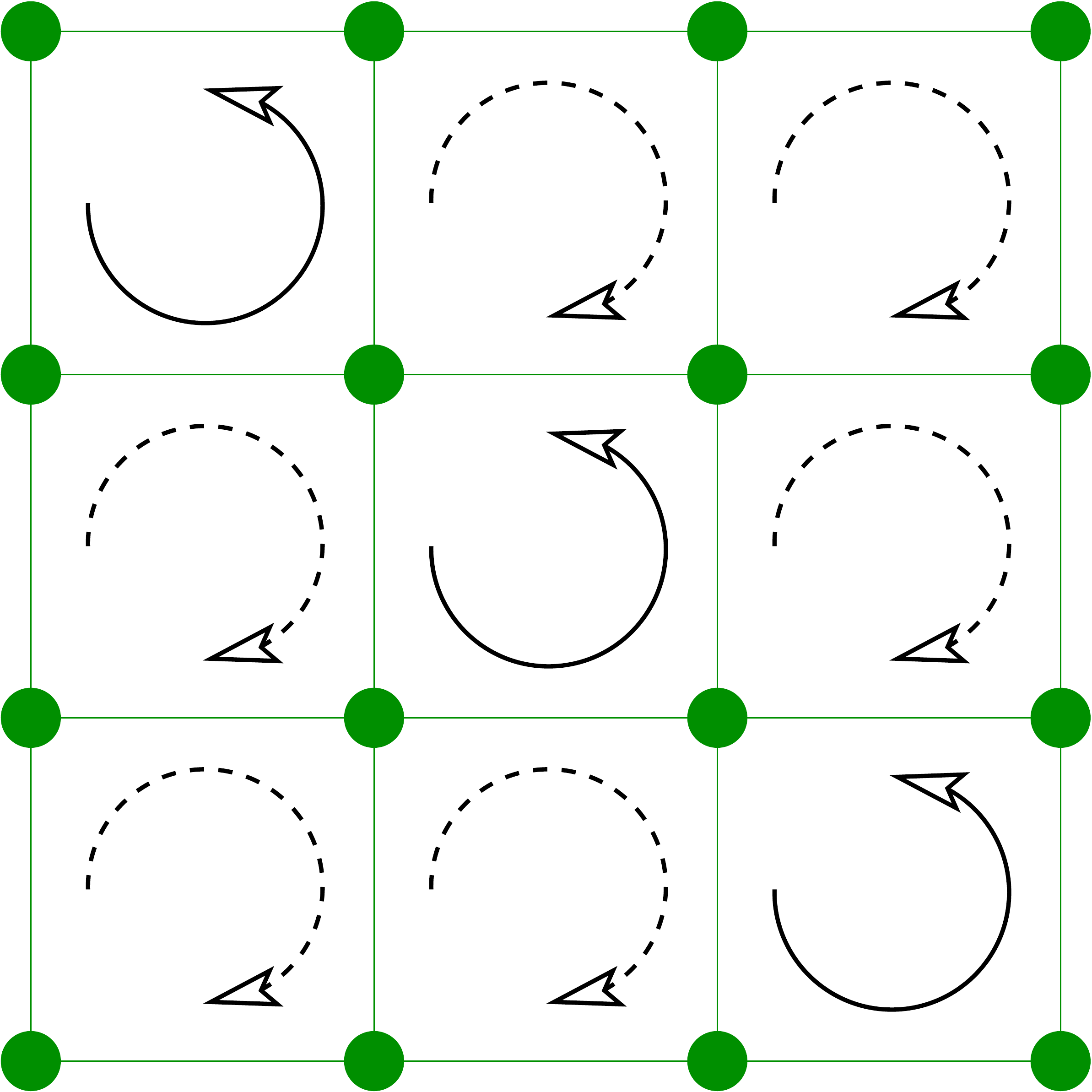}
\caption{Illustration of a vortex lattice configuration for weakly interacting bosons in the Harper-Hofstadter model at flux  $\phi_{\rm AB}=2\pi/3$.
The gauge-invariant currents flow on the links between the sites of the square lattice.
The pattern of current breaks the translational invariance of the system. The strongly circulating currents (shown in solid lines and arrows) are around plaquettes that lie along diagonal lines. These are the lattice-equivalents of the ``vortex cores'', now distorted from the triangular lattice expected in the continuum models, to be pinned to the 
plaquettes of the square lattice. There are weak counter-circulating currents around other plaquettes, such that the net particle flow vanishes, when coarse grained on scales large compared to the lattice spacing}
\label{fig:vortexlattice}
\end{figure}

(1) For systems without time-reversal symmetry (as required to generate
a Chern band), the individual Bloch wavefunctions have phase
variations which in general give rise to nonzero local current
density. Since the Bloch wavefunctions are stationary states, these currents must be divergenceless, but this still allows the
BEC to support circulating currents in dimensions $d>1$.  These
currents take the form of the local current density of a vortex
lattice. An example is shown in Fig.~\ref{fig:vortexlattice}.

(2) For topological optical lattices in $d>1$ the energy minimum in
which the BEC forms will, in general, be characterized by a non-zero
Berry curvature. As described in
Sec.~\ref{subsec:intexpt} this Berry curvature affects the
collective modes of the BEC.

(3) There can arise situations in which there are a set of degenerate
energy minima. An important example is provided by the Harper-Hofstadter model,
for which the energy band typically has multiple degenerate
minima. For example, at flux $\phi_{\rm AB} = 2\pi/s$ the band has $s$ degenerate
minima in the magnetic Brillouin zone. For non-interacting particles, there is a macroscopic degeneracy associated with the occupation of these $s$ degenerate single particle states: a BEC could form in any linear
superposition of these states; or indeed a ``fragmented'' condensate
could form~\cite{mueller2006}.  The inclusion of interactions, $U\neq 0$, is required to
resolve this macroscopic degeneracy.
The simplest way to include interactions is via Gross-Pitaevskii 
mean-field theory.  This is valid for sufficiently weak interactions
$U\ll J$ and at high mean particle density $\bar{n}\gg 1$. Studies of
the mean-field ground state (in regimes of both weak and strong
interactions, $U \bar{n}\ll J$ and $U \bar{n}\gg J$) show that
repulsive interactions stabilize a simple BEC with the form of a
vortex lattice~\cite{PhysRevB.48.3309,PhysRevLett.105.155302,PhysRevA.83.013612}. This
state breaks the underlying translational symmetry of the lattice, so
there are several discrete (symmetry-related) states that are
degenerate.  For example, for flux $\phi_{\rm AB}= 2\pi/3$ it has the
form of vortex lines along diagonals of the lattice,
Fig.~\ref{fig:vortexlattice}.  Numerical studies of the vortex lattice
ground states for $U\bar{n}\gg J$ have been conducted for a wide range
of flux $\phi$: these show simple ordered lattices for
$|\phi_{\rm AB}| > \pi/2$, but complex behavior, with many local
energy minima, for $|\phi_{\rm AB}| < \pi/2$~\cite{PhysRevB.48.3309}.

Similar vortex lattice states appear in ladder systems.
In experiments on weakly interacting bosons on a two-leg ladder with flux, \textcite{Atala2014} demonstrated a transition between a uniform superfluid phase with Meissner-like chiral currents and a vortex phase (with broken translational symmetry along the ladder) as a function of the tunneling strength across the rungs of the ladder, akin to the Meissner and vortex lattice phases of a  type-II superconductor.

\subsubsection{Topological superfluids}

\label{sec:topologicalsuperfluids}

In both the tight-binding lattices and the Raman-dressed flux lattices,
there can arise situations in which there are non-local interactions
between fermions in a single band. This allows attractive $p$-wave
pairing and can lead to interesting forms of topological
superfluidity with ``Majorana'' excitations.  Here
we discuss the general features of these topological superfluids.

Their properties are understood within mean-field theory, as described by the
Bogoliubov-de Gennes (BdG) Hamiltonian. This takes the form
\begin{align}
\hat{H}-\mu\hat{N}  =   \sum_{\alpha,\beta} 
\left[\varepsilon_{\alpha\beta} \hat{c}^\dag_\alpha \hat{c}^{\phantom{\dag}}_\beta + \frac{1}{2}\left( \Delta_{\alpha\beta} \hat{c}^\dag_\alpha \hat{c}^{\dag}_\beta
+\Delta^*_{\alpha\beta} \hat{c}^{\phantom{\dag}}_\beta \hat{c}^{\phantom{\dag}}_\alpha
\right) \right] \,,\label{eq:hambdg}
\end{align}where $\hat{c}_\alpha^{(\dag)}$ are fermionic creation/annihilation operators for the single-particle states labeled by $\alpha=1,\ldots N_s$, which encodes both positional and internal (spin) degrees of freedom.  They obey the fermionic  anticommutation relations, $\{\hat{c}^{\phantom{\dag}}_\alpha,\hat{c}^{\phantom{\dag}}_\beta\}=
\{\hat{c}^{\dag}_\alpha,\hat{c}^{\dag}_\beta\}=0$,
$\{\hat{c}^{\phantom{\dag}}_\alpha,\hat{c}^\dag_\beta\}=\delta_{\alpha\beta}$.
The $N_s\times N_s$ matrix $\underline{\underline{\varepsilon}}$ describes  the conventional particle motion and potentials, and must be Hermitian, $\varepsilon_{\alpha\beta}^* = \varepsilon^{\phantom{*}}_{\beta\alpha}$;  the $N_s\times N_s$ matrix $\underline{\underline{\Delta}}$ represents the superconducting pairing,  and is antisymmetric $\Delta_{\alpha\beta}=-\Delta_{\beta\alpha}$.

It is convenient to express the Bogoliubov-de Gennes Hamiltonian 
as (up to a constant shift)
\begin{eqnarray}
\hat{H}-\mu\hat{N}  & = & \frac{1}{2}
\left(\underline{\hat{c}}^\dag \; \underline{\hat{c}}^{\phantom{\dag}}\right) 
{\mathcal H}^{\rm BdG} 
 \left(\begin{array}{cc} \underline{\hat{c}}^{\phantom{\dag}}  \\
\underline{\hat{c}}^\dag \end{array}\right) \\
 {\mathcal H}^{\rm BdG} & = &  \left(\begin{array}{cc} \underline{\underline{\varepsilon}} & \underline{\underline{\Delta}} \\
 -\underline{\underline{\Delta}}^* & -\underline{\underline{\varepsilon}}^*\end{array}
\right)\,,\label{eq:hbdg}
\end{eqnarray}
where $(\underline{\hat{c}},\underline{\hat{c}}^\dag)^{\rm T}$ is a $2N_s$-component column vector formed by listing all fermionic destruction $\hat{c}^{\phantom{\dag}}_{\alpha  = 1, \ldots N_s}$
and  creation $\hat{c}^\dag_{\alpha  = 1, \ldots N_s}$ operators.
The  quasiparticle excitations are determined by the spectrum of the $2N_s\times 2N_s$ matrix ${\mathcal H}^{\rm BdG}$ 
\begin{equation}
E_{\lambda}
\left(\begin{array}{c} \underline{u}^{(\lambda)} \\
\underline{v}^{(\lambda)} \end{array}\right) = 
 {\mathcal H}^{\rm BdG} \left(\begin{array}{c} \underline{u}^{(\lambda)} \\
\underline{v}^{(\lambda)} \end{array}\right)
\end{equation}
in terms of the $2N_s$-component vector of amplitudes $u^{(\lambda)}_{\alpha=1,\ldots N_s}$ and $v^{(\lambda)}_{\alpha=1,\ldots N_s}$, where
$\lambda$ labels the $2N_s$ eigenvalues.
The matrix BdG Hamiltonian (\ref{eq:hbdg}) has a special symmetry:
\begin{eqnarray}
 {\mathcal H}^{\rm BdG}  & = & - 
\left(\begin{array}{cc} \underline{\underline{0}}  & \underline{\underline{I}} \\
 \underline{\underline{I}} &  \underline{\underline{0}} \end{array}\right)
 \left[ {\mathcal H}^{\rm BdG}\right]^* 
\left(\begin{array}{cc} \underline{\underline{0}}  & \underline{\underline{I}} \\
 \underline{\underline{I}} &  \underline{\underline{0}} \end{array}\right) 
 \label{eq:bdgphsymmetry}
 \end{eqnarray}
 where $\underline{\underline{0}}$ and $\underline{\underline{I}}$ are the $N_s\times N_s$ null and  identity matrices respectively.  This has the consequence that for any eigenstate  $(\underline{u}^{(\lambda)},\underline{v}^{(\lambda)})^{\rm T}$ 
with eigenvalue $E_{\lambda}$ there is another eigenstate
\begin{equation}
 \left(\begin{array}{c} \underline{u}^{(\bar{\lambda})} \\
\underline{v}^{(\bar{\lambda})} \end{array}\right) = 
 \left(\begin{array}{c} \underline{v}^{(\lambda)*} \\
\underline{u}^{(\lambda)*} \end{array}\right)
\label{eq:bdgsymmetry}
\end{equation}
with eigenvalue $E_{\bar{\lambda}} = -E_{\lambda}$: \ie the spectrum is symmetric in energy around $E=0$. (Note that energies have been defined relative to the chemical potential, $\mu$.) This intrinsic particle-hole symmetry in fact represents an inherent redundancy in the theory, by which the eigenstates of $ {\mathcal H}^{\rm BdG}$ must include both the destruction and the creation operator of any quasiparticle state~\cite{bernevigbook}. 
Specifically, defining the operator associated with each eigenvector $\lambda=1\ldots 2N_s$
\begin{equation}\hat{C}_\lambda =   \sum_{\alpha} u^{(\lambda)}_\alpha \hat{c}_\alpha 
+v^{(\lambda)}_\alpha \hat{c}^\dag_\alpha \, 
\label{eq:qpoperator}
\end{equation}
 the Hamiltonian may be expressed in the diagonalized form 
\begin{equation}
\hat{H}-\mu\hat{N}  = \frac{1}{2}\sum_{\lambda=1}^{2N_s}E_{\lambda} \hat{C}_\lambda^\dag\hat{C}_\lambda \, .
\end{equation}
 The above symmetry (\ref{eq:bdgsymmetry}) allows this to be written as
\begin{eqnarray}
\hat{H}-\mu\hat{N}  & = & \frac{1}{2}
\sum_{\lambda=1}^{N_s} E_{\lambda} 
\hat{C}_\lambda^\dag\hat{C}_\lambda^{\phantom{\dag}} +E_{\bar{\lambda}} \hat{C}_{\bar\lambda}^\dag\hat{C}_{\bar\lambda}^{\phantom{\dag}} \\
& = & \frac{1}{2}
\sum_{\lambda=1}^{N_s}
E_{\lambda} \left(\hat{C}_\lambda^\dag\hat{C}_\lambda^{\phantom{\dag}} -\hat{C}_{\lambda}^{\phantom{\dag}}\hat{C}^\dag_{\lambda} \right)\\
& = & \sum_{\lambda=1}^{N_s} E_{\lambda} \left(\hat{C}_\lambda^\dag \hat{C}^{\phantom{\dag}}_\lambda
-\frac{1}{2}\right) \,.
\label{eq:hbdgdiag}
\end{eqnarray}
where the sum is now over just $N_s$ eigenstates. These are conventionally chosen to be states for which $E_\lambda\geq 0$, such that the ground state $|0\rangle$ is defined by $\hat{C}^{\phantom{\dag}}_\lambda |0\rangle = 0$, and 
$\hat{C}^\dag_\lambda$ are the quasiparticle creation operators.

For spinless fermions in translationally invariant settings we can use the plane-wave operators $\hat{\tilde{c}}_\vv{q} \propto \sum_\alpha \E^{-\I\vv{q}\cdot \vv{r}_\alpha} \hat{c}_\alpha$ to write
\begin{equation}
\label{eq:bdgk}
\hat{H}-\mu\hat{N} = \frac{1}{2} \sum_{\vv{q}} 
\left(\hat{\tilde c}^\dag_\vv{q}\; \hat{\tilde c}^{\phantom{\dag}}_{-\vv{q}}\right)
\left(\begin{array}{cc} \varepsilon_\vv{q} & \Delta_\vv{q} \\
\Delta^*_{ \vv{q}} & -\varepsilon_\vv{q}\end{array}
\right)
\left(\begin{array}{cc} \hat{\tilde c}^{\phantom{\dag}}_\vv{q} \\
\hat{\tilde c}^\dag_{- \vv{q}} \end{array} \,.
\right)\end{equation}
with the superconducting gap required to satisfy $\Delta_{-\vv{q}} =  -\Delta_{\vv{q}}$. 
The BdG matrix (\ref{eq:hbdg}) reduces to a momentum-dependent $2\times 2$ matrix, ${\mathcal H}_\vv{q}^{\rm BdG}$.
An important example is the Kitaev model of a $p$-wave superfluid in 1D, described in detail in Appendix~\ref{app:1dmodels},  for which 
\begin{eqnarray}
{\mathcal H}_q^{\rm BdG} & =  &
\left(\begin{array}{cc} -2J\cos(qa) -\mu & -2\I\Delta\sin(qa) \\
2\I \Delta\sin(qa) & \mu+2J\cos(qa) \end{array}
\right) \\
 & \equiv &  -{\vv{h}}({q}) \cdot {\vv{\sigma}} \,.
\end{eqnarray}
The energy eigenvalues show the above particle-hole symmetry,
$E^\pm_q = \pm |{\vv{h}}({q})|$. Furthermore,  
as shown in the Appendix~\ref{app:1dmodels}, the bands are characterized by a winding number relating to how ${\vv{h}}({q})$ encircles the origin, analogous to the winding number of the SSH model. This winding number can be used to identify a topological superconducting phase.

On a finite geometry this topological phase hosts an exact zero energy state on its boundary, \ie with 
$E_{\lambda_0}=0$ in (\ref{eq:hbdgdiag}). The prescription that we used to define the form (\ref{eq:hbdgdiag}) is ambiguous for such zero modes. Consider the operator $\hat{C}_{\lambda_0}$ constructed via (\ref{eq:qpoperator})  from the eigenvector of such a
zero mode, $\lambda_0$. The particle-hole symmetry implies that there is another zero-energy state with label $\bar{\lambda}_0$, which, using Eqn.~(\ref{eq:bdgsymmetry}), would lead to an operator $\hat{C}_{\bar{\lambda}_0}= \hat{C}^\dag_{\lambda_0}$. Together, these two operators $\hat{C}^\dag_{\lambda_0}$ and
$\hat{C}_{\bar{\lambda}_0}$
describe the destruction/creation of a fermionic
quasiparticle at $E_{\lambda_0}=E_{\bar{\lambda}_0}=0$. Since this excitation has zero energy, it describes a ground state that is (two-fold) degenerate depending on whether this zero energy mode is occupied or filled.  It is therefore immaterial which of these operators we choose to view as a particle creation / destruction. Indeed, we could also choose to work in terms of operators that are linear superpositions. One particular choice is to define the {\it Majorana} operators
\begin{eqnarray}
\hat{\gamma}_{1} & \equiv & \left(\hat{C}^{\phantom{\dag}}_{{\lambda_0}}
+ \hat{C}^\dag_{{\lambda_0}}\right)
\label{eq:majorana1}
\\
\hat{\gamma}_{2} & \equiv & {\I}\left(\hat{C}^{\phantom{\dag}}_{{\lambda_0}}
- \hat{C}^\dag_{{\lambda_0}}\right)\,,
\label{eq:majorana2}
\end{eqnarray}
which obey anti-commutation relations $\left\{\hat{\gamma}_i,\hat{\gamma}_j\right\} = 2\delta_{ij}$. 
Since the Majorana operators are self-adjoint $\gamma_i^\dag = \gamma_i$, they can be viewed as describing particles that are their own antiparticles.

The above transformation to Majorana operators appears rather arbitrary: from a mathematical perspective one can always choose either to work in terms of the two Majorana operators $\gamma_{1},\gamma_2$ or in terms of $\hat{C}^{\phantom{\dag}}_{{\lambda_0}}$ and its adjoint
$\hat{C}^\dag_{{\lambda}_0}$. For topological superconductors there is a clear physical reason to prefer the Majorana operators: the Majoranas are {\it spatially localized},  each one tied to a single boundary or  defect. 
Thus, local probes couple directly to the Majorana operators. In contrast, the  quasiparticle operators $\hat{C}^{(\dag)}_{{\lambda_0}}$ are non-local.  The locality of the Majorana operators 
is shown explicitly in the  Appendix~\ref{app:1dmodels} for the Kitaev model, where one Majorana operator acts on the right boundary and one acts on the left boundary.

That the Majorana modes are spatially localized causes the system to have properties that are robust to external perturbations, including disorder potentials: the exact particle-hole symmetry enforces the mode to have $E=0$. Departures from $E=0$ can only arise from mixing with other Majorana modes; since the intervening superconducting state is gapped these corrections are suppressed exponentially in the distance between the local Majorana modes, \eg the length of the 1D Kitaev chain. Note that the two Majorana modes --  one at each end of the 1D Kitaev chain -- describe a single fermionic excitation, such that one should view the Majorana as a fractionalized quasiparticle. In contrast, the zero energy edge states of the SSH model correspond to two separate fermionic excitations, one at each end of the chain. As discussed in~\ref{subsubsec:discretesymmetries} the edge states of the SSH model are sensitive even to local perturbations which break the chiral symmetry.

The above example for the 1D topological superconductor has a very natural generalization to 2D.  There, the bulk spectrum is of the form
\begin{eqnarray}
{\mathcal H}_{\vv{q}}^{\rm BdG} & =  &
\left(\begin{array}{cc} \epsilon_\vv{q} -\mu & \Delta_\vv{q} \\
\Delta^*_{ \vv{q}} & \mu-\epsilon_\vv{q}\end{array}
\right) \equiv - {\vv{h}}(\vv{q}) \cdot {\vv{\sigma}}
\end{eqnarray}
where $\vv{q}$ now runs over a 2D Brillouin zone. Topological superconducting states can appear in situations where the
${\vv{h}}(\vv{q})$ acquires all three components, such as in the continuum  $p_x + i p_y$ superfluid, for which 
\begin{equation}h_z = \hbar ^2 |\vv{q}|^2/2m - \mu \quad , \quad h_x+ \I h_y = \Delta_0 (q_x+\I q_y)\,.\end{equation}
For $\mu >0$, the quasiparticle spectrum is fully gapped, and the unit vector ${\vv{h}}(\vv{q})/|{\vv{h}}(\vv{q})|$ wraps the sphere as $\vv{q}$ runs over all values, indicating that this is a topological phase~\cite{ReadG00}. At $\mu =0$ there is a gap-closing transition to a superconducting phase at $\mu<0$ which is non-topological.
The 1D surface of this topological 2D superconductor has an edge mode that has Majorana character, albeit in a setting in which there is a continuum of edge modes.
In this 2D setting localized Majorana modes arise as bound states on the cores of quantized vortices -- \ie point
 defects localized in the bulk of the system (introduced by rotation or other external means).

It is natural to search for topological superfluids using cold
atoms. Many routes to $p$-wave pairing of single-component fermions
have been suggested. These include methods involving a $p$-wave
Feshbach resonance~\cite{gurarie2007resonantly}, or long-range dipolar interactions~\cite{baranov2012condensed}, or induced
interactions via a background BEC~\cite{Wu2016}. 
In the context of this review, the
connections arise through the use of spin-orbit coupling (optical
dressing) to allow contact interactions (\eg $s$-wave pairing) to
lead to effective $p$-wave interactions between single-component
fermions. Proposals of this kind have been presented in 3D,
2D~\cite{zhang_prl_2008} and 1D~\cite{nascimbene_toposf,Yan2015}.

\subsubsection{Fractional quantum Hall states}

The conditions for realizing fractional quantum Hall (FQH) states in
2D semiconductor systems are well-understood~\cite{prangeandgirvin}.  The application of a
strong magnetic field breaks the single-particle energy spectrum into
degenerate Landau levels. When a Landau level is partially filled with
electrons, the large number of ways in which the electrons can occupy
the single-particle states gives a very
high degeneracy.
This degeneracy is lifted by repulsive inter-particle
interactions, leading to strongly correlated FQH ground states at
certain ratios of particle density to flux density, $\nu = n/n_\phi$.
FQH states are characterized by a non-zero energy gap to making
density excitations in the bulk. For temperatures below
this gap they behave as incompressible liquids: with a bulk energy
gap, but carrying gapless edge modes. In this sense they
resemble integer quantum Hall states. However the edge modes are not
simply described as single-particle states but involve fractionalized
quasiparticles~\cite{wenedge}.  Similarly, the (gapped) particle-like excitations in
the bulk of the system have fractional charge, and are predicted to
have fractional quantum exchange statistics~\cite{adysternreview}.

The achievement of
similar regimes with ultracold atoms would allow
the exploration of several novel variations of FQH physics.

{\it Bosons.} Cold atom experiments would allow the first exploration
of FQH states for bosons.  Theory shows that contact interacting bosons
in the lowest Landau level exhibit FQH states provided the filling
factor is not too large. These states include robust variants of
interesting phases -- the Moore-Read and Read-Rezayi phases -- which
are expected to exhibit non-Abelian particle exchange
statistics~\cite{cooperadvancesreview}.  The stabilization, and
exploration, of non-Abelian phases is a much sought after goal, in
part to find the first evidence that nature does exhibit this exotic
possibility of many-body quantum theory and in part in connection with
the possible relevance for quantum information
processing~\cite{nayakrmp}

The physics of interacting bosons in the lowest Landau level can be
accessed in the Harper-Hofstadter model at relatively low flux
density, $|\phi_{\rm AB}|\lesssim 2\pi/3$, for which the lowest band is
similar to the continuum Landau level, and where similar FQH states
appear for
bosons~\cite{Sorensen:2005,palmer,hafezi-2007,mollercooper2009}.
Optical flux lattices, involving spin-orbit coupling, can lead to 2D
energy bands that are very similar to the lowest Landau level:
topological bands with unit Chern number and with very narrow energy
dispersion.  An advantage of this approach is that the flux density
$n_\phi$ can be high, so FQH states are expected at high particle
density $n$ where interactions are strong. Exact diagonalization
studies have established stable FQH states of bosons
including exotic non-Abelian
phases~\cite{cooperdalibard2013,sterdyniak_2015}.

{\it Lattice effects.} The use of optical lattices to generate
topological energy bands naturally causes cold atomic gases to explore
new aspects of FQH physics.  The underlying lattice makes the
single-particle states differ from those of the continuum Landau
level, and can influence the nature of the many-body ground states.
For the Harper-Hofstadter model, the FQH states found at low flux
density, where the bands resemble the continuum Landau
level,
are replaced by other strongly correlated phases at
high flux
densities where lattice effects become relevant.

It has been shown that states of the same form as FQH states
of the continuum Landau level -- so-called ``fractional Chern
insulator'' (FCI) states -- can be formed, for bosons and fermions, in
a variety of topological energy bands starting from models (such as
the Haldane model) which are far-removed from the continuum
Landau level~\cite{Parameswaran2013816}.
Typically such models require the introduction of further-neighbor
tunneling terms to flatten the lowest energy band, such that one can
enter a regime of strong correlations (mean interaction energy larger
than bandwidth) without mixing with higher bands. Indeed, it has been
shown that by tailoring the tunneling Hamiltonian, one can construct
models for which the {\it exact} many body ground states are FQH
states~\cite{PhysRevLett.105.215303} or FCI
states~\cite{PhysRevLett.116.216802}. No general theorem exists concerning the nature of the many-body ground state in a given Chern band. However, beyond flatness of energy band, it is believed that flatness of the geometry of the states, as measured by the Berry curvature and by a quantity know as the Fubini-Study metric, is advantageous in stabilizing topological many-body phases~\cite{Parameswaran2013816}.

A particularly interesting aspect of the topological energy bands in
lattices is that these can differ qualitatively from a continuum
Landau level, specifically if their Chern number differs from unity. Indeed, for the
Harper-Hofstadter model at flux $\phi_{\rm AB} = \pi + \epsilon$, with $\epsilon$ small, the
lowest energy band has Chern number of $2$, so is topologically
distinct from the lowest Landau level.  Numerical
calculations show the appearance of FQH phases
for particles occupying this energy band.
These are examples of FQH states that have no counterpart in the continuum Landau level but that are stabilized by the lattice
itself~\cite{mollercooper2015,mollercooper2009,2017arXiv170300430H}.

{\it Symmetry-protected topological phases.}  
The topological states
of non-interacting Fermi systems (topological insulators and
superconductors) are now viewed as an example of a
more general class of symmetry-protected topological (SPT) phase,
which allows for interparticle interactions and bosonic
statistics. SPT phases are defined by the conditions that they are
gapped phases, with gapless edge modes, but unlike the FQHE without particle fractionalization. Hence these states have a
unique ground state on a periodic geometry, and the many-body ground state has only short-range entanglement~\cite{senthilreview}.
The phases of strongly
interacting bosons in Chern bands include cases of ``integer'' quantum
Hall states of the bosons, which provide one of the cleanest
realizations of symmetry-protected topological (SPT)
phases~\cite{mollercooper2015,mollercooper2009,2017arXiv170300430H}.
Here ``integer'' refers to the fact that the Hall conductance is
an integer, as opposed to fractional. Such phases still arise from strong
inter-particle interactions, albeit without fractionalized quasiparticles.

{\it Ladders.} Cold atomic gases provide ways in which to study
ladder-like systems which are quasi-1D variants of FQH systems. These
arise either in Harper-Hofstadter models with superlattices to create
ladder geometries, or in systems involving a synthetic dimension for
which there is naturally tight confinement. Both situations have been
shown theoretically to support strongly correlated states that are
closely related to FQH
states~\cite{PhysRevB.91.054520,PhysRevB.92.115446,2016arXiv161206682C}. A
precise connection between FQH states on an infinite 2D system and the
states in these quasi-1D settings can be made by considering the
quantum Hall wavefunctions on a cylindrical geometry which is infinite
in one direction but has finite circumference $L$ in the
other. Studies of the evolution of the ground state from 2D
($L\gg \bar{a}$, with $\bar{a}$ the mean interparticle spacing) into
the ``squeezed geometry" (small $L\lesssim \bar{a}$) show that many
FQH states of the 2D systems evolve smoothly into charge-density wave
states of the quasi-1D geometry. For example the bosonic Laughlin
state, at filling factor $\nu=1/2$, evolves into a CDW state which
breaks translational symmetry to double the unit cell to give two
degenerate ground states of the CDW.  The fractionally charged
quasiparticles of the FQH states map onto domain walls between the
different symmetry-related CDW states.

\subsubsection{Other strongly correlated phases}

{\it Chiral Mott insulator.} For the Harper-Hofstadter-Hubbard model at strong interactions $U\gtrsim J$ the Gross-Pitaevskii approach fails. Just as in the
case of vanishing magnetic field, there can be incompressible Mott
phases, with $n_i =$ integer.  However, other strongly correlated phases are predicted to appear.
One striking example is the chiral Mott insulator: an incompressible phase at integer filling (like the Mott insulator), but which carries a nonzero local current in the ground state (as does a vortex lattice).
Numerical calculations indicate that, within a region of stability between Mott insulator and superfluid, the chiral Mott insulator can form the ground state of the Bose-Hubbard model with external flux on a variety of 2D lattices~\cite{PhysRevB.87.174501,PhysRevB.89.155142,PhysRevB.91.094502}. In ladder systems
the Mott-vortex phase found in numerical calculations~\cite{PhysRevB.91.140406,PhysRevB.91.054520,PhysRevLett.115.190402}  
can be viewed as the 1D analogue of the chiral Mott insulator.

{\it Chiral spin states.}  For hard-core interactions, $U\to \infty$, interesting many-body states can still arise in the Harper-Hofstadter-Hubbard model for bosons provided $n_i\neq$ integer
so the hard-core bosons are mobile. These bosons can be viewed as a
spin-1/2 quantum magnet, with the phases of the tunneling matrix
elements introducing frustrated magnetic
interactions~\cite{mollercooper2010}. Thus, the FQH (or FCI) phases of
these hard-core bosons can be viewed as quantum spin
liquids~\cite{PhysRevLett.59.2095}. Other predicted quantum spin
liquids may find realization with these Bose-Hubbard
models~\cite{2015arXiv150404380L}.
The phases of interacting {\it fermions} in
Hubbard-like models with tunnelling phases have also been studied
theoretically. Much of the focus has been on the Haldane model for
spin-1/2 fermions with onsite repulsion $U$. For systems close to half
filling there is a competition between the band insulator, the Mott
insulator, and various spin-symmetry-broken
phases~\cite{PhysRevLett.116.225305,PhysRevB.94.035109,PhysRevB.91.161107}.
The Mott insulating state naturally leads to frustrated spin models,
with possible unconventional forms of ordering at very low
temperatures, at the superexchange energy scale $J^2/U$ or below, including
chiral spin states~\cite{PhysRevB.93.115110} and chiral
spin-liquids~\cite{PhysRevLett.116.137202}.

{\it Number conserving topological superfluids.} 
The theory of topological superfluids,  Sec.~\ref{sec:topologicalsuperfluids}, is based on the mean-field treatment of pairing described by the Bogoliubov--de Gennes theory.  In this theory, the number of pairs of fermions is not conserved. For a 1D system, such a mean-field theory can be appropriate in settings in which the superfluid pairing is proximity-induced, \eg by coupling to a bulk superfluid of fermion pairs with which pairs of particles can be exchanged and which can impose a fixed superconducting order parameter. However, in the absence of such a proximitizing medium the quantum fluctuations in 1D systems preclude the existence of any long-range order in the superconducting pairing and the structure of the mean-field theory should fail. Theoretical work has shown how 1D topological superfluids can still arise in such number-conserving settings. Models with microscopic symmetry can give rise to zero-energy modes, arising as exact degeneracies in the spectrum of an open chain in the topological phase~\cite{Sau2011,Fidkowski2011,Iemini:2015,Lang:2015,Iemini:2017}. This degeneracy is the many-body counterpart of the Majorana modes of the mean-field 1D topological superconductor.
In settings without such a symmetry, topological degeneracies can arise in geometries in which modulation of the parameters along the chain lead to multiple interfaces between topological and non-topological phases~\cite{Ruhman:2015,Ruhman:2017}.

\subsection{Experimental perspectives}

\label{subsec:intexpt}

Many of the experimental consequences of topological bands described
in Section~\ref{sec:experimental} rely on interactions between the
particles being sufficiently weak that they can be treated as
non-interacting. It is therefore important to consider what
experimental observables can be used to characterize the properties of
atoms in topological bands when interactions cannot be neglected.

{\it Equilibrium observables.} Some of the most important observables for characterizing the
properties of the atoms are well-established from studies of atomic
gases in other settings without topological character.
Measurements of the equation of state, and observations of
density-density correlations in time-of-flight imaging will be crucial
for establishing the existence and forms of strongly correlated phases
such as fractional quantum Hall states. Similarly, the possibility to
image at the single site level in quantum gas microscopes~\cite{Bakr2009, Sherson2010} will allow
precise characterization of microscopic structure in these phases,
including of the local currents for example in the chiral Mott
insulator phase. A method for detecting local currents was presented in
experiments from the Munich group \cite{Atala2014}.

{\it Collective Modes.} Observations of the collective mode frequencies
provides a sensitive way to detect properties of many-body systems~\cite{Dalfovo:1999}. 
For BECs formed in bands with geometrical character,
the collective mode frequencies are sensitive to 
the Berry curvature at the band minimum~\cite{PhysRevLett.111.220407}.
Consider a weakly interacting BEC in a single band minimum. We take minimum to have isotropic effective mass $M^*$, and a local Berry curvature $\Omega \vv{e}_z$. The effect of Berry curvature on the collective modes can be readily determined by adapting the standard hydrodynamics approach~\cite{pethick2002bose}
 to include the anomalous velocity from the Berry curvature. 
For a spherical harmonic trapping potential, 
$V(\vv{r})=\frac{1}{2} \Lambda |\vv{r}|^2$, the collective modes have the same spatial structure as for $\Omega=0$~\cite{PhysRevLett.77.2360}, but  the frequencies depend on the Berry curvature. For example, for small $\Omega$ the three angular momentum components of the dipole mode are split by $\Delta \omega = \Lambda\Omega/2\hbar$, leading to a precessional motion of the center-of-mass oscillation of the cloud at this frequency.

{\it Edge states.}
For incompressible fluids that are topological -- \eg a topological
insulator of non-interacting fermions, or a FQH
fluid -- there exists a special class of low energy collective modes, which are the gapless {\it edge
  states} of the fluid.  For non-interacting particles in Chern bands,
these are the edge states discussed in
Section~\ref{subsec:edge}.  For interacting systems these
may not be easily described as single-particle excitations, but can still appear as long-lived
surface excitations.
Measurements of the propagation of surface waves
can directly probe the edge state structure, for example allowing one
to detect the number of edge channels and their respective
velocities~\cite{cazalilla:121303}.
The edge states also lead to highly characteristic dynamics
in far-from-equilibrium situations in which the confining potential is removed~\cite{Goldman:2013}. Edge states appear also
at interfaces between bulk regions of differing topologies, prepared for example by spatial modulation of the lattice potential~\cite{Goldman:2016}, and have been probed experimentally for the SSH model~\cite{Leder:2016}.
Experiments using a quantum gas microscope have shown the influence of strong interparticle interactions on the chiral edge states in Harper-Hofstadter ladders~\cite{tai2017}.

{\it RF excitation.} Another natural probe of the edge states is to
measure the spectrum for the removal of particles via RF excitation,
ideally performed with single site resolution~\cite{Bakr2009, Sherson2010} to focus on the boundary. Such
spectra have been proposed as a means to detect localized Majorana modes
in topological superfluids, appearing as a (near) zero-energy
contribution inside the spectral
gap~\cite{grosfeld:104516,nascimbene_toposf}. 
A separate definition of fractional statistics is provided by Haldane's exclusion statistics~\cite{haldaneexclusion}: a generalized version of Pauli blocking, by which the fractional quasiparticles reduce the  number of available states for other quasiparticles in a well-defined, but fractional, manner.  An observation of this effect requires counting many-body states, and may be possible in precision spectroscopy of small FQH clusters~\cite{coopersimonexclusion} in which the exclusion statistics reveal themselves in the count of spectral lines.

{\it Adiabatic pumping.}
For topologically ordered systems, the existence of fractional low energy particle-like excitations allows for new features in  Thouless pumping. Specifically, under one full cycle of the adiabatic evolution of the pump, it is possible to transfer a fractional particle number across the system. (In the examples discussed above, the number of particles transferred was constrained to be an integer, set by the Chern number.) This fractional pumping is related to the existence of a ground state degeneracy:  one cycle of the adiabatic pump converts one ground state into another degenerate ground state, and multiple cycles (transferring multiple fractional particles) are required before the system returns to its starting state. This picture forms the basis for understanding of the quantization of the Hall conductance in the fractional quantum Hall effect of 2D systems~\cite{laughlingauge,halperingauge}. For narrow strips of quantum Hall systems, this physics smoothly evolves into the pumping of fractional charges in commensurate charge density waves. The manifestation of such pumping for 1D fermionic systems with synthetic dimension has been described in~\textcite{PhysRevLett.115.095302,1367-2630-17-10-105001} and~\textcite{Taddia:2017}.

{\it Hall conductivity from ``heating''.} An interesting way in which to measure the Hall conductivity, $\sigma_{xy}$, of a system -- \ie the transverse current density in response to a uniform force -- is to measure the {\it heating} rate, as set by the rate of power absorption, $P_\pm(\omega)$, caused by the application of a circularly polarized force $\vv{F}_\pm(t) = F_0 (\cos\omega t,\pm \sin\omega t)$.  The d.c. Hall conductivity is found to be set by 
\begin{equation}
\sigma_{xy} = \frac{2}{\pi F_0^2 A_{\rm syst}}\int_0^\infty \D\omega \frac{1}{\omega}[P_+(\omega)-P_-(\omega)]\,,
\label{eq:sigmaxysumrule}
\end{equation}
\ie the difference of the power absorption rates divided by frequency and integrated over all drive frequencies~\cite{Tran:2017}. For non-interacting particles the power absorption can be found from measurements of the rates of depletion of (initially) occupied states, $\Gamma_\pm$, via $P_\pm = (\hbar\omega)\Gamma_\pm$, allowing the Hall conductivity to be determined via measurements of this depletion rate. For a filled Chern band, the intergrated difference of depletion rates (\ref{eq:sigmaxysumrule}) is therefore expected to recover the quantized Hall conductivity of the filled bands.
The formula (\ref{eq:sigmaxysumrule}) is however valid even for interacting particles, arising from a general sum-rule for the linear response functions, so provides a possible way to measure the Hall conductivity also in strongly correlated phases.

\section{Outlook}

\label{sec:outlook}

\subsection{Turning to atomic species from the Lanthanide family} 
In this review we explored several classes of lattice schemes for which a nontrivial topology 
originates from a two-photon Raman coupling between various sublevels of the electronic ground state manifold. So far most experiments of this type were performed with atoms from the alkali-metal family, which are (relatively) easy to manipulate and cool down to quantum degeneracy. However for such atoms, recoil heating due to spontaneous emission of photons may cause severe problems. Indeed the desired Raman coupling is significant only when the laser is detuned from the resonance by less than the fine-structure splitting $\Delta_{\rm f.s.}$ of the resonance line (see Appendix~\ref{app:alkali}). Because of this relatively small detuning, spontaneous emission of photons occurs with a non-negligible rate $\gamma$. More precisely, one can define the merit factor ${\cal M}=\kappa/\gamma$,
where $\kappa$ is the desired Raman matrix element. Taking as an example the case of the fermionic alkali-metal atom $^{40}$K, one finds after optimization ${\cal M} \sim \Delta_{\rm f.s.}/\Gamma \sim 10^5$, where $\Gamma$ stands for the natural width of the electronic excited state (see e.g. \textcite{Dalibard:2016} for details). If one takes as a typical value $\hbar \kappa$ equal to the recoil energy, the photon scattering rate is $\gamma \sim 0.3\,$s$^{-1}$, leading to the heating rate $\dot E \sim k_{\rm B} \times100\,$nK/s. This may be too large for a  reliable production of strongly correlated topological states. 

A more favorable class of atoms is the lanthanide family, with species like erbium or dysprosium. These atoms have two outer electrons  and an incomplete inner shell (6s$^2$ and 4f$^{10}$ for Dy). Because of this inner shell,  the electronic ground state has a non-zero orbital angular momentum ($L=6$ for Dy). The lower part of the atomic spectrum contains lines corresponding either to the excitation of one of the outer electrons or of one electron of the inner shell. By choosing a laser excitation close to a narrow line resonance and thus with a large detuning $\Delta_b$ from the closest broad line, one reaches after optimization ${\cal M}\sim \Delta_b/\Gamma_b$. Because $\Delta_b$ is now of the order of an optical frequency, the merit factor is ${\cal M}\sim 10^7$, leading to $\gamma \sim 10^{-3}$\,s$^{-1}$ and to the residual heating $\dot E \sim k_{\rm B} \times 0.1 $\,nK/s.

\subsection{Topological lattices without light}

Laser fields constitute a common element for all of the techniques described herein, and in virtually all of these cases these fields lead to unwanted off-resonant scattering leading to heating, atom loss or both.   As discussed in the previous section, these scattering processes might be mitigated or eliminated by suitable selection of different transitions or working with different atoms, however, fundamentally different approaches are also possible.

Of particular interest is the possibility to replace the laser fields with RF or microwave magnetic fields generated by a micro-fabricated atom chip.  Unlike optical fields, these fields have practically no off-resonant emission, solving the in-principle atomic physics limitation, in exchange for the technical complexity of working in the vicinity of an atom chip.  

In one category of proposals, the atom chip serves simply as the source of a large time-modulated gradient magnetic field~\cite{Anderson2013a}. This produces in effect a series of pulses that generates spin-dependent gauge fields; the most simple implementation of this technique was realized in the lab~\cite{Luo2016}, and although spontaneous scattering was eliminated, the fairly low frequency of their drive lead to significant micro-motion induced heating effects.

In a second category of proposals~\cite{PhysRevLett.105.255302}, the atom chip consists of a large array of micro-fabricated parallel wires which give a near-field radio-frequency magnetic field that drives transitions in the same way as Raman lasers, as described here in the context of synthetic dimensions or intrinsic spin-dependent lattices.  Because these fields are structured at the micrometer scale -- far below the free-space wavelength of RF fields -- these are near-field structures and the atomic ensemble must be on the micrometer scale from the chip's surface. 

\subsection{Other topological insulators and topological metals}

We have focused on specific recent experimental realizations of
topological energy bands using cold gases: in 2D systems, and in 1D systems with chiral symmetry. %
Routes to achieving topological superfluid phases in cold gases, arising from  BCS-pairing of fermions, have also been described in Sec.~\ref{sec:topologicalsuperfluids}.

As discussed in Sec.~\ref{subsec:topoinsulator} other
forms of topological energy bands can arise depending on
dimensionality and on the global symmetries that are imposed,
according to the ten-fold way~\cite{RevModPhys.88.035005}.
Important cases in solid state systems are the $\mathbb{Z}_2$ topological
insulators that arise in spin-orbit coupled systems with time-reversal
symmetry (TRS), in 2D and 3D. 
There are proposals for how to realize bands
with this topology for cold gases. The required TRS can be
implemented by fine-tuned engineering of
the relevant terms in the Hamiltonian~\cite{PhysRevLett.105.255302}. TRS
can also be established as an intrinsic property for cold gases: in
the absence of Zeeman splittings and of any circularly polarized light
fields~\cite{PhysRevLett.107.145301}.  In 2D, the insulating state formed by filling a $\mathbb{Z}_2$ topological band exhibits a quantized Hall effect for the spin-current. A similar quantized spin Hall response also arises in a related setting without spin-orbit coupling, in which spin-up and spin-down fill Chern bands with equal and opposite Chern number: such energy bands have been realized (for bosons) in cold atoms~\cite{Aidelsburger2013} using laser-assisted tunneling to generate a Harper-Hofstadter model with equal and opposite fluxes for the two spin states~\cite{Kennedy2013}.

Cold atomic gases provide a natural setting with which to explore certain forms of
topological band that are more difficult to implement in solid state materials.
The sublattice (``chiral'') symmetry
is very fragile in solid state systems, as it is typically broken by
any disorder potential. However, it can arise readily for optical lattice potentials 
in cold atom gases for which disorder can be
negligible~\cite{PhysRevB.85.195116,PhysRevLett.113.033002}.
Similarly, there are predicted to be topological insulators
beyond those classified in the ten-fold way,  arising from 
lattice symmetries which rely on a high spatial regularity that can arise in cold gases.
These include topological invariants stabilized by crystalline lattice symmetries~\cite{PhysRevLett.106.106802},
and the Hopf
insulator~\cite{PhysRevLett.101.186805,PhysRevB.88.201105}, which also relies on a form of translational 
symmetry for stability~\cite{PhysRevB.95.161116}.
Note also that cold gases allow topology in dimensions higher than $d=3$ to be explored, through the use of synthetic dimensions provided by internal degrees of freedom or by
viewing a phase degree of freedom as an additional
quasimomentum. Recent experimental work~\cite{Lohse4D} has used pumping to
demonstrate the topological response of an effective 4D
quantum Hall system~\cite{Zhang4dqhe} based on theoretical proposal of~\textcite{Price2015}.

An area of growing interest in solid state settings concerns so-called
topological metals, or semi-metals~\cite{RevModPhys.88.035005}. (There exist analogous
topologically stable forms of gapless superconductors.) As for
topological insulators, the classification of topological metals
depends also on the dimensionality and the existence (or absence) of
symmetries. However, since metals involve bands that are only
partially filled by fermions,
they cannot be characterized by the topological invariants used for
insulators, which involve integrals over the filled energy bands.
Instead, topological metals can be characterized by topological
invariants defined in terms of integrals over the Fermi surface which
separates filled from empty states~\cite{volovik}.
One example of a topological metal is provided by the 2D honeycomb
lattice with nearest-neighbor hopping. This realizes the
band structure of graphene in which there are two Dirac points in BZ, each of which leads to a Fermi surface when the Fermi energy lies close to the Dirac
point. Each of these Fermi surfaces is a topological metal,
characterized by the Berry phase of Bloch states around the Fermi
surface, which is $\pi$. This value is a topological invariant,
\ie it is robust to continuous changes of the underlying parameters,
provided the system retains both time-reversal symmetry and inversion
symmetry. (These symmetries ensure that the Berry curvature of the
bands vanishes, except for singular delta-function contributions at
the Dirac nodes which lead to the $\pi$ Berry phases.)  The Dirac
node is topologically stable, and can only be created or annihilated
if it merges within another Dirac point.
A related example that arises in 3D is the ``Weyl
point''. This can be viewed as a point
source of Berry curvature in reciprocal space. The integral of the
flux of Berry {\it curvature} through any closed 2D surface in reciprocal space is $2\pi {\cal C}$, where
${\cal C}$ is required to be an integer. If this surface encloses a Weyl point, 
$|{\cal C}|=1$. The Weyl point is topologically stable to deformations
of the Hamiltonian, without any symmetry requirements, unless it
annihilates with a second Weyl point of opposite charge.
Optical lattice with Weyl points can be constructed using the
laser-induced tunneling methods described in Sec.~\ref{sec:hopping}
~\cite{Dubcek:2015}.

\subsection{Far-from-equilibrium dynamics}

Cold atomic gases readily allow the study of coherent quantum dynamics in far-from-equilibrium settings. Starting with a cold gas at thermal equilibrium, a sudden change of the Hamiltonian -- a so-called
 ``quantum quench" -- typically leaves the system in a far-from-equilibrium state. 
Here we discuss some of the consequences of quantum quenches between Hamiltonians for which the ground states have different {\it topological} character. We focus on cases of non-interacting particles in topological energy bands.

\begin{figure}
\includegraphics[width=0.99\columnwidth]{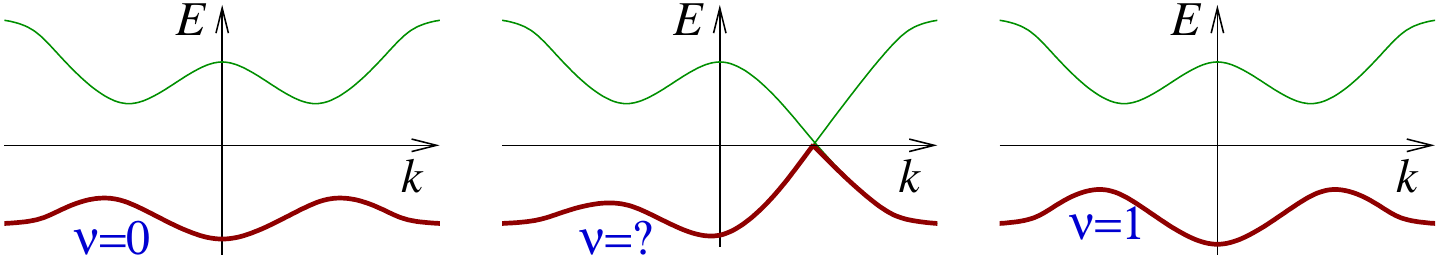}
\caption{Schematic illustration of the change of band Hamiltonian that causes the topological index of the lowest band $\nu$ to change, via the closing of a band gap.}
\label{fig:bandclose}
\end{figure}

Consider first an optical lattice  potential which is varied slowly in time, such that the  topological invariant of the lowest energy band  differs between initial and final Hamiltonians.
 If the relevant topological invariant is {\it symmetry-protected}, it is
  possible to have a smooth evolution between these two cases provided that the Hamiltonian breaks the symmetry at intervening times. An example of this is the pumping sequence of the RM model, which breaks the
  chiral symmetry of the SSH model and therefore allows a smooth evolution between the topologically distinct phases of the SSH model.
 When there is no symmetry protection (\eg  Chern bands in
  2D), then
  the change in band topology requires the band gap to close at some
  intermediate time, as illustrated in 
Fig.~\ref{fig:bandclose}.
  For a BEC formed close to the band minimum, such a change in band  topology need not induce any phase transition: 
  the Bloch wavefunctions of those states which the bosons occupy can evolve smoothly,
allowing  adiabatic evolution of the BEC, albeit into a very different local wavefunction. This allows, for example, 
the
 adiabatic formation of a dense vortex
  lattice~\cite{PhysRevA.88.033603}. However, for non-interacting fermions that fill the lowest energy band, the ground states of the initial and final Hamiltonians have different topological characters,
  so these two states must be separated by a phase transition.
  
The far-from-equilibrium dynamics following a quantum quench between Hamiltonians whose ground states have different topologies has been explored for
non-interacting fermions which fill a band.
One striking result is that the
topological invariant of the many-body state is often preserved
 under unitary time evolution. This has been shown for topological
 superfluids~\cite{PhysRevB.88.104511,PhysRevLett.113.076403,PhysRevE.93.062117},
 and for Chern bands~\cite{dAlessio2015,PhysRevLett.115.236403}. (Different behavior can arise for symmetry-protected topological invariants~\cite{mcginley2018}.)
In interpreting this result it is crucial to distinguish between the topology of the {\it Hamiltonian} and the topology of the {\it many-body state}. 
The former is defined as the topological invariant constructed for the
 lowest energy band of the Hamiltonian, with Bloch wavefunctions $|u^{(0)}_\vv{q}\rangle$, while the latter is defined as the topological invariant constructed from the wavefunctions which the fermions
 occupy $|u^{{\rm s}}_\vv{q}\rangle$. These two sets of wavefunctions need only coincide if the many-body wavefunction is the ground state of the Hamiltonian. Out of equilibrium, for example following a quantum quench, the Bloch wavefunction occupied by the particle at momentum $\vv{q}$ need not be an eigenstate of the Hamiltonian, so  it becomes time-dependent $|u^{\rm s}_\vv{q}(t)\rangle$. We consider situations in which the periodicity of the lattice is preserved, such that wavevector $\vv{q}$ remains a good quantum number. That the topological invariant of the {\it state} is preserved  is guaranteed provided the unitary evolution of the Bloch states $|u^{\rm s}_\vv{q}(t)\rangle  = \E^{-\I \hat{H}_\vv{q} t/\hbar} |u^{\rm s}_\vv{q}(0)\rangle$ is smooth in momentum space  $\vv{q}$,
 which is true for short-range hopping. (For topological invariants that are symmetry-protected, it is also required that the symmetry is not broken either explicitly or dynamically~\cite{mcginley2018}.) That said, at long times, the  Bloch wavefunctions of the many-body state $|u^{\rm s}_\vv{q}(t)\rangle$ will become rapidly varying as a function of $\vv{q}$.  When the variation in $\vv{q}$ is so fast that this cannot be viewed as smooth on the scale of $(2\pi/L)$, with $L$ the typical sample dimension, then the bulk topological invariant becomes ill-defined.  Thus for any finite system, there is an upper timescale after the quench for which it is meaningful to expect the 
 Chern number to be preserved. Simple estimates lead to the conclusion that  
this time is of order $L/v$, with $v$ a characteristic group velocity of the final Hamiltonian~\cite{PhysRevB.94.155104}.
Such systems can still be characterized by the Bott index~\cite{loringhastings}, which provides a real-space formulation of the Chern index applicable also to finite-sized systems. This has been used to find protocols for preparing non-equilibrium (Floquet) systems by which the Bott index undergoes transitions between topological values~\cite{dalessio,Ye2017}.

Although the topological invariant of the wavefunction is unchanged following the quench to a new Hamiltonian with different topology, this does not mean that there are no observable consequences of the new Hamiltonian.  Local physical observables can be strongly influenced by the new Hamiltonian, despite the fact that the (non-local) topological invariant of the state is unchanged. Indeed, theory shows that under a quantum quench of the Haldane model, the edge current quickly adapts to become close to that of the ground state of the final Hamiltonian~\cite{PhysRevLett.115.236403}. Furthermore, by following the dynamics of the Bloch wavefunctions $|u^{\rm s}_\vv{q}(t)\rangle$ detailed information on  the final Hamiltonian can be recovered~\cite{theorylinking,2017arXiv171005289G}. This has been demonstrated in experiments by the Hamburg group~\cite{hamburg_phasetransition,hamburglinking}, using the band-mapping techniques described in Sec.~\ref{subsubsec:localmeasureberry} to reconstruct the dynamical evolution of the occupied wavefunctions $|u^{\rm s}_\vv{q}(t)\rangle$. The experiments use a two-band model, for which the  Hamiltonian at $\vv{q}$ may be written 
\begin{equation}
\hat{H}_\vv{q} = h_0(\vv{q})\hat{1} -\vv{h}(\vv{q}) \cdot \hat{\vv{\sigma}}\,.
\end{equation}
The wavefunction of the fermion with $\vv{q}$ can be represented by a three-component unit vector $\vv{e}(\vv{q})$. If $\vv{e}(\vv{q})$ is aligned with $\vv{h}(\vv{q})$, then this is an energy eigenstate of the Hamiltonian. In the quench experiment, the system is prepared with $\vv{e}^{\rm initial}(\vv{q}) = (0,0,-1)$ for all wavevectors [the ground state of an initial Hamiltonian with $\vv{h}^{\rm initial}(\vv{q}) = (0,0,1)$] and then the Hamiltonian is changed to its final value  $\vv{h}(\vv{q})$. For general $\vv{q}$ this is not aligned with $\vv{e}^{\rm initial}(\vv{q})$ so the wavefunction evolves in time, precessing around the local $\vv{h}(\vv{q})$. This leads to the appearance of non-zero components of the vector $(e_x,e_y)$ which indicate interband coherences.  The evolution can lead to the creation of vortex-antivortex pairs in these components $(e_x,e_y)$. The appearance of each vortex-antivortex pair was shown to be associated with a cusp in the Loschmidt echo, hence giving rise to the characteristic feature of a  ``dynamical phase transition"~\cite{heylreview}. Such singular features can arise even for quenches within a single topological phase, so are not connected to the topological phase transition itself.
Instead, the change in the topology of the Hamiltonian can be found by tracing the time evolution of  $\vv{e}(\vv{q},t)$ and constructing the linking number of the trajectories (in $\vv{q}$ and $t$) of any two values of $\vv{e}$ [\eg $\vv{e}=(1,0,0)$ and $(0,0,1)$]~\cite{theorylinking}. This linking number is the Hopf index of the 
map $\vv{e}(\vv{q},t)$. This procedure has been successfully carried out in experiments~\cite{hamburglinking}.
A related approach has recently been used to demonstrate the topological character of the effective Hamiltonian in a spin-orbit coupled BEC \cite{sun2018uncover}.

The above considerations rely on the assumption that the fermions are non-interacting. It will be of interest to explore the extent to which these, or similar, approaches can be applied in the presence of interparticle interactions. Being gapped phases of matter, associated with filled bands, the ground states of the Hamiltonian are expected to be robust to weak interactions. However, even weak interactions can lead to the generation of entanglement between single-particle states at different wavevectors under  far-from-equilibrium dynamics, which may be viewed as a form of decoherence.

Moving beyond the phases of non-interacting fermions in topological bands, it will be of interest to explore similar quench dynamics in topological phases that arise only because of strong interparticle interactions. This is an area where results remain limited. Recent work on the Haldane phase of a spin-1 chain (a symmetry protected topological phase of this interacting quantum spin system) has shown that the ``string order" that characterizes the topological phase is lost following the quench~\cite{PhysRevB.94.024302}, suggesting a difference from the non-interacting fermion cases described above. One important difference concerns the role of symmetry-protection under dynamical evolution~\cite{mcginley2018}.

\subsection{Invariants in Floquet-Bloch systems}

The topological invariants we have focused on are those of static
single-particle Hamiltonians, in which the spatial periodicity leads
to the existence of a Bloch Hamiltonian that depends on a
quasimomentum $\vv{q}$ within a BZ.  If the Hamiltonian is time-varying, but
periodic with period $T\equiv 2\pi/\omega$, then energy is replaced by a Floquet
quasi-energy that is defined up to the addition of integer multiples
of $\hbar\omega$.  The combination of both temporal and spatial periodicities causes Floquet-Bloch systems to have topological invariants that are  distinct from
those of static Hamiltonians.

\begin{figure}[t]
\begin{center}
\includegraphics[width=8cm]{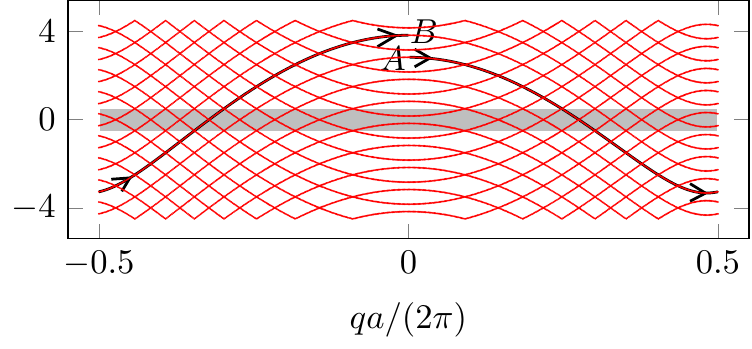}
\includegraphics[width=8cm]{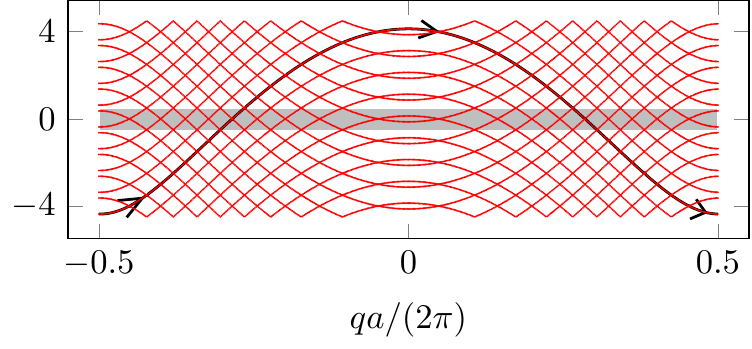}
\end{center}
\caption{Floquet energy spectrum in units of $\hbar \omega$ for the driven Rice-Mele model in the topological (top) and non-topological (bottom) cases. The parameters $(J'/J,\Delta/J)$ are varied along a circle of radius 0.5 with angular frequency $\omega=0.23\,J/\hbar$. The circle is centered on the point $(2,0)$ in the non-topological case and on the point $(1,0)$, \ie the vortex in Fig.~\ref{fig:zakphaserm1}, in the topological one. In the topological case a particle starting in state $A$ at momentum $q=0$ arrives in $B$ after an adiabatic motion across the full BZ, leading to the net current (\ref{eq:current_Floquet}) for a filled band. The grey area highlights a given ``temporal Brillouin zone", of energy width $\hbar \omega$. }  
\label{fig:floquetwind}
\end{figure}

(1) The periodicity of the Floquet spectrum allows the Floquet bands to
wind in quasi-energy, by an integer multiple of $h/T$ as $\vv{q}$ runs
over the BZ. This winding in quasi-energy gives rise to topological invariants of Floquet-Bloch bands~\cite{kitagawa} 
that are absent in static settings.
Figure~\ref{fig:floquetwind} shows an example of a Floquet-Bloch band in
a 1D system that winds in quasi-energy once across the BZ. 

To understand the physical significance of such situations, it is instructive to consider the band of
Fig.~\ref{fig:floquetwind} (top) to be filled with non-interacting
fermions. [How one might prepare such a state is itself
an interesting question~\cite{lindnerbergrudner,Dauphin2017}. 
Indeed, non-adiabatic effects associated with the switch-on of a pump can lead to deviation from quantization~\cite{PhysRevLett.120.106601}.]  The net current carried by the filled band is
\begin{equation}
I \equiv \frac{L}{2\pi} \int_{-\pi/a}^{\pi/a}
\frac{1}{\hbar}\frac{\mbox{d} \epsilon_q}{\mbox{d}q} \, \mbox{d}q  =  \frac{L}{T}\,,
\end{equation}
where $a$ is the lattice constant and $L$ is the total length of the system. The number of particles in the filled band is $N = L/a$, so the current per particle is
\begin{equation}
\frac{I}{N} = \frac{a}{T}\,.
\label{eq:current_Floquet}
\end{equation}
Thus, the current is equivalent to a displacement of each particle by a lattice constant $a$ in each period of the cycle $T$. For a band that winds in quasi-energy $s$ times, the current is quantized at $sa$ in each cycle $T$.
The quantized current from a filled Floquet-Bloch band that winds $s$ times in quasi-energy precisely
matches the current expected from an adiabatic pump, as described in Sec.~\ref{subsec:pumping}, operated
cyclically with period $T$ and with Chern number $s$.  Indeed, settings in which adiabatic
Thouless pumping occurs give rise to Floquet-Bloch spectra in which a band
winds in quasi-energy. (The adiabaticity condition corresponds to the
relevant Floquet-Bloch band {\it crossing} with all
Floquet-Bloch bands that wind in the opposite sense.)  However, the
general structure of the Floquet-Bloch states and the associated topological invariants~\cite{kitagawa} are not restricted to
such adiabatic settings.

\begin{figure}
\includegraphics[width=0.9\columnwidth]{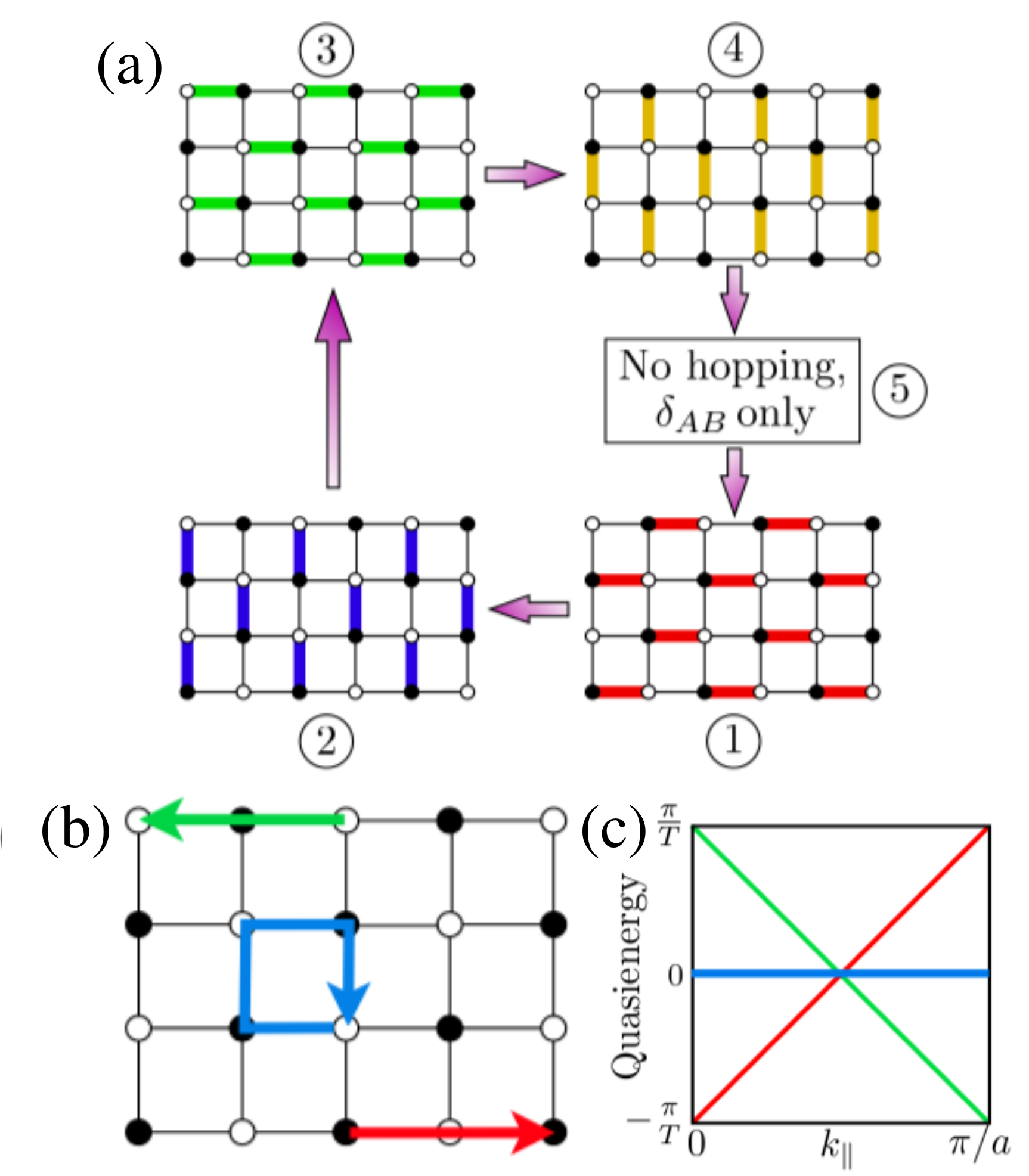}
\caption{Simple model that shows anomalous edge states. (a) A cycle consists of four steps in which specific sets of bonds are active and one step in which there is only a sublattice energy offset $\delta_{AB}$. (b) Trajectories of particles initially in the bulk (blue) or on the edges (red, green). (c) Floquet spectrum showing the dispersionless bulk band (blue) and the anomalous edge states (red, green).
Courtesy of~\protect\textcite{PhysRevX.3.031005}.}
\label{fig:floquetblochsquaremodel}
\end{figure}

\begin{figure}
\includegraphics[width=0.9\columnwidth]{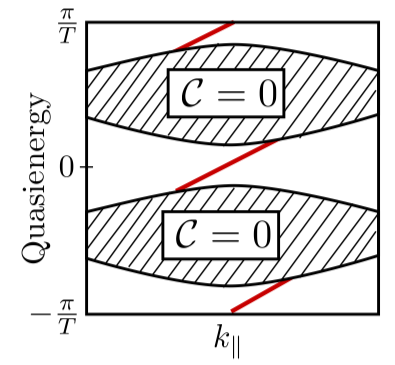}
\caption{For Floquet-Bloch systems, the topological invariant of the
  bulk band determines the change in the number of edge modes as
  quasi-energy passes through the band. The periodicity in quasi-energy
  allows the existence of edge states even in settings where the bulk
  bands are all topologically trivial. Courtesy of~\protect\textcite{PhysRevX.3.031005}.}
\label{fig:floquetblochtopology}
\end{figure}

(2) New features also arise when one considers the edge states on systems with a boundary.
One finds that finite-size systems can have protected edge
states, at quasi-energies between the bulk bands, even if the topological invariants
  constructed from the Floquet Hamiltonian $\hat{H}^F_\vv{q}$ are
  trivial for all of the bulk energy bands~\cite{kitagawa}. 
  A simple example of a 2D model which exhibits such ``anomalous edge states'' 
is  illustrated in Fig.~\ref{fig:floquetblochsquaremodel}.  One period, $T$,
  is broken into four sub-periods $T/4$, during which the tunneling amplitude
  $J$ is turned on only for a subset of the bonds. This tunneling is
  chosen to satisfy $(J/\hbar)(T/4) = \pi$ such that a particle that
  starts on one side of the active bond is transferred to the other
  side of the bond during the time $T/4$.  For particles in the bulk, the net action after all four
  parts of the cycle is to return the particle to its starting point (up to some overall phase). Thus the Floquet operator for time evolution over one cycle [See Eqn.~(\ref{eq:floquetoperator})] is
  proportional to the identity and
  has a bulk Floquet-Bloch band structure with constant quasi-energy
  and whose wavefunctions have trivial topological character. However,
  a particle that starts on the edge of the system is transported along
  the edge over one period; this motion appears as dispersive (chiral)
  edge states in the Floquet spectrum of the finite-size system, Fig.~\ref{fig:floquetblochsquaremodel}c.
The understanding of such anomalous edge states is
that the
topological invariant of a band constructed from the Floquet operator determines the {\it change} in the
number of edge states as the quasi-energy passes through the band~\cite{PhysRevX.3.031005}. An
edge state can pass through a set of topologically trivial bands and
satisfy periodicity in quasi-energy and quasi-momentum (see
Fig.~\ref{fig:floquetblochtopology}).
To compute the number of anomalous edge states requires one to know not just the stroboscopic
evolution defined by the Floquet operator, but also
the full time evolution operator at intermediate times, from which an additional invariant
can be constructed~\cite{PhysRevX.3.031005}. Anomalous edge states have not yet been seen in atomic systems, but have been observed in experiments on light propagation in photonic structures~\cite{Mukherjee2017,Maczewsky2017}.

(3) The concept of particle-hole symmetry needs be generalized for
Floquet-Bloch systems.  For static Hamiltonians, particle-hole
symmetry ($E\to -E$) stabilizes (edge) modes at energies $E=0$ (SSH
model or Majorana mode in the Kitaev model). The periodicity of the
Floquet quasi-energy $\epsilon$ under $\epsilon \to \epsilon+\hbar\omega$ means
that symmetry $\epsilon\to\epsilon$ arises for $\epsilon=0$ or
$\epsilon=\hbar\omega/2$. Indeed, topologically stable edge modes at
quasi-energies $\epsilon=\hbar\omega/2$ have been theoretically demonstrated in periodically
driven lattices models with sublattice (chiral) symmetry analogous to
the SSH model~\cite{PhysRevB.90.125143}, and in Floquet
superfluid systems where they appear as Majorana modes~\cite{PhysRevLett.106.220402}.

Recent works have constructed a general theory of the topological classification of time-varying single-particle Hamiltonians~\cite{nathanrudner,PhysRevB.93.115429,royharper2016}.
Extensions of these ideas to interacting, many-body phases is an active area of theoretical research.

\subsection{Open systems}

So far we focused our studies on the topological properties of the ground state of an isolated system. The possibility to introduce a coupling between this system and an environment opens new possibilities and raises new questions that we now discuss. We restrict our presentation to studies in direct relation with atomic gas implementations. 

Consider first the case where the environment is at a non-zero temperature $T$. If the coupling is sufficiently weak, the energy levels of the system remain relevant and its steady-state is now a statistical mixture of these levels.  Since bands with various topologies that were empty at $T=0$ now acquire a finite population, the Chern number calculated via a thermal average will not be an integer anymore. One could naively conclude that the system loses all its topological properties when $T$ becomes non-negligible with respect to the gap protecting the bands that are populated at zero temperature. 

However, one may look for more subtle topological invariants that can be associated with the density matrix describing the system at nonzero $T$. A possible direction consists in generalizing the notion of geometric phase, using the concept of parallel transport for density matrices introduced by \textcite{uhlmann1986parallel}. This line was investigated by \textcite{viyuela2014two}  and \textcite{huang2014topological}, who could derive in this way a classification of topological phases at nonzero $T$. It was subsequently  revisited with a somewhat different perspective by \textcite{budich2015topology}, who pointed out possible ambiguities of the previous approaches for a model 2D system. Another direction consists of establishing an equivalence between classes of density matrices using local unitary operations~\cite{chen2010local}, and deducing from this equivalence the desired classification of topological phases. It was developed for the particular case of mixed Gaussian states of free fermions by~\textcite{diehl2011topology} and \textcite{Bardyn:2013}, and recently generalized by \textcite{grusdt2017topological}  (see also \textcite{van2014classification} for the specific case of 1D systems). Independently of the tool that is used to define the topological class of density matrices, it is worth emphasizing that on the one hand these topological features are in principle observable, but on the other hand their relation with usual physical quantities associated with the response  of the system is still the subject of ongoing research~\cite{budich2015dissipative}.

Working with open systems also offers the possibility of creating novel topological states that emerge from the dissipative coupling itself. Here we take the reservoir at $T=0$ for simplicity. In the conceptually simplest version of the scheme,  the coupling is engineered so that the system ends up after some relaxation time in a pure state, often called a {dark state}. ``Topology by dissipation" is achieved when this dark state possesses nontrivial topological properties. For a coupling compatible with the Born-Markov approximation (see \eg \textcite{Gardiner:2004book} and \textcite{daley2014quantum}), the master equation describing the evolution of the density operator $\rho$ of the system can be written in the Lindblad form~\cite{lindblad1976generators} $\dot \rho={\cal L}(\rho)$ where the linear operator ${\cal L}$ -- the Liouvillian -- acts in operator space: 
\begin{equation}
{\cal L}(\rho)={}i[\rho,H]+\frac{1}{2}\sum_j \left( 2L_j\rho L_j^\dagger -L_j^\dagger L_j\rho - \rho L_j^\dagger L_j \right).
\label{eq:Liouville}
\end{equation}
Here $H$ is the Hamiltonian of the system in the absence of coupling to the reservoir, and the $L_j$'s are the Lindblad operators describing the various coupling channels to the bath. If one neglects for simplicity the Hamiltonian evolution of the system ($H=0$), one finds that a pure state $|\psi\rangle$ satisfying $L_j|\psi\rangle=0$ for all $j$ is  an eigenstate of the Liouvillian with eigenvalue 0 and thus a dark state.  The topology associated with this state can be readily inferred from the fact that it is the ground state of the parent Hamiltonian $H'=\sum_j L_j^\dagger L_j$. Conditions for the existence and uniqueness of such dark states were discussed in the context of fermionic setups by \textcite{bardyn2012majorana} and \textcite{Bardyn:2013}. 

The steady-state of the master equation will be protected against small perturbations if it is an isolated point  in the spectrum of the Liouvillian. One defines in this case the ``damping gap" as the smallest rate at which deviations from the steady-state are washed out. In this dissipative context, the damping gap plays a role that is formally equivalent to the energy gap in the Hamiltonian context. 

The concepts of topology by dissipation and damping gap can be generalized to the case where the steady-state of the master equation associated to (\ref{eq:Liouville}) is a mixed state. Its analysis is relatively simple for free fermion systems on a lattice, assuming that the Lindblad operators $L_j$ are linear functions of the on-site creation and annihilation operators. This corresponds for example to the case where dissipation occurs via an exchange of particles with a reservoir that is described by a classical mean-field.  In this case, the system density operator is Gaussian and the already mentioned tools developed by \textcite{Bardyn:2013} can be used to characterize its topological properties. Interestingly for such systems, a change of the topological properties of the system can occur either when the damping gap closes, or when the ``purity gap" closes. This notion of purity gap closure, first introduced by \textcite{diehl2011topology} and later generalized by \textcite{budich2015topology}, corresponds to a situation where there exist bulk modes associated with a subspace (\eg a given $\vv{q}$ in momentum space) in which the state is completely mixed.  

Examples of topologically nontrivial pure or mixed states and their corresponding edge modes were presented  by \textcite{diehl2011topology} and \textcite{bardyn2012majorana} in a 1D and a 2D setup, respectively. 
\textcite{budich2015dissipative} pointed out a remarkable property of steady-states involving a mixed state: they may possess a non-trivial topological character even when all Lindblad operators $L_j$ are local in space. This cannot occur when the steady-state is pure, an impossibility that is reminiscent from the  fact that in the Hamiltonian framework,  Wannier functions of a  topologically non-trivial filled band (non-zero Chern number) must decay slowly, \ie, algebraically, in space~\cite{thoulesswannier}.

The concept of topological pumping can also be extended to the case of open systems, as shown by \textcite{linzner2016reservoir} and \textcite{hu2017exceptional}.  Using a proper engineering of the Liouvillian (\ref{eq:Liouville}) of a fermionic 1D chain, \textcite{linzner2016reservoir} could generalize the result of Sec.~\ref{subsubsec:Exp_transport_pumping} for the Rice--Mele model, and prove the quantization of the variation of the many-body polarization  after a closed loop in parameter space.  As for the other topological features described in this section, this quantization holds even when the steady-state of the master equation is a mixed state. 

Finally we note that dissipation also offers a new route for revealing an existing topological order. \textcite{rudner2009topological} considered the motion of a single particle on a 1D bipartite (AB) lattice similar to the one at the basis of the SSH model. Here, dissipation corresponds to a non-zero decay rate of the particle when it resides on the sites of the sublattice A. The particle is assumed to start on the site ${\rm B}_0$, \ie the ${\rm B}$ sublattice site in the lattice cell $m=0$.  Here, one is interested in the average displacement $\langle \Delta m\rangle =\sum_m mP_m$, where $P_m$ is the probability that the particle decays from the A site of the $m$th lattice cell. Quite remarkably this average displacement is quantized and can only take the values $0$ and $1$. This binary result corresponds to the two topological classes that we identified above for the SSH model within the Hamiltonian framework. This result holds for any values of the decay rate and on-site energies, but it stops being valid if one introduces a dissipative component in the hopping process between the two sublattices. It was confirmed experimentally with a setup in the photonic context by \textcite{zeuner2015observation}. These authors used a lattice of evanescently coupled optical wave guides, in which losses/dissipation were engineered by bending the waveguides. The scheme of \textcite{rudner2009topological} was recently generalized theoretically to multipartite lattices by \textcite{rakovszky2017detecting}, who used a weak measurement of the particle position as the source of dissipation.

\section{Summary}

We have summarized the methods that have been used  to engineer topological bands for cold atomic gases, and the main observables that have allowed  characterizations of their geometrical and topological properties.
Most experimental studies so far have been at the single-particle level.
Theory suggests many interesting possibilities for
novel many-body phases in regimes where  interparticle interactions become strong. Such systems are in regimes where theoretical understanding is very limited, so experimental investigation will be particularly valuable.
 Accessing this regime for large gases will require careful management of heating -- from lattice modulation methods or from Raman-coupling of internal states -- as well as the development of robust detection schemes to uncover the underlying order.
There is already progress in this direction. It seems likely to provide a rich vein to explore, with much scope for experimental discoveries and surprises.

\acknowledgments{We are grateful to the innumerable colleagues from whom we have learned so much. We also thank J. Beugnon, B. Gadway, F. Gerbier, N. Goldman, Z. Hadzibabic, S. Nascimbene, and T. Porto for comments on an earlier drafts of this paper. NRC was supported by EPSRC Grant Nos. EP/K030094/1 and   EP/P009565/1 and by the Simons Foundation. JD was supported by the ERC Synergy grant UQUAM.
IBS was partially supported by the ARO's atomtronics MURI, the AFOSR's Quantum Matter MURI, NIST, and the NSF through the PFC at the JQI.}

\appendix

\section{Topological bands in one dimension}

\label{app:1dmodels}

\subsection{Edges states in the SSH model}
\label{subsec:SSH}

The Su-Schrieffer-Heeger (SSH) model was introduced to describe the electronic structure of polyacetylene~\cite{Su:1979}. This molecule has  alternating single- and double-bonds along the carbon chain,  which are represented in a tight-binding model with one orbital per carbon atom (site) by alternating tunnel couplings, $J$ and $J'$, between the sites, see Fig.~\ref{fig:sshmodel1}. There are two sites in the unit cell, which we label A and B and that are assigned the same energy in the SSH model (this constraint is relaxed in the Rice--Mele model). The single-particle Hamiltonian is given in 
(\ref{eq:SSH_Hamiltonian}) and we emphasized in the main text the existence of two distinct topological classes, corresponding to $J<J'$ (winding number $N=0$) and $J>J'$ ($N=1$). 

This classification may appear rather formal for an infinite chain, since the labeling in terms of A and B sites is arbitrary and one can exchange  the roles of $J$ and $J'$ without changing the physical system. However its physical relevance appears very clearly if one consider a finite or semi-infinite chain. Then the two classes correspond to different possibilities for the edge state(s) of the chain, as an illustration of the general bulk-edge correspondence. In the following we first describe the case of a semi-infinite chain for which analytical calculations are straightforward, and then discuss the case of a finite chain.

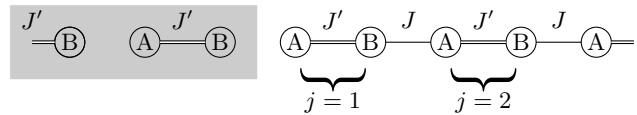
\begin{figure}[t]
\begin{center}
\begin{tikzpicture}
\draw [white,fill=gray!40]  (-0.8,-0.5) rectangle (2.5,0.5);
\draw (0,0) circle (0.2) ;
\draw (0,0) circle (0.2) ;
\draw (1,0) circle (0.2) ;
\draw (2,0) circle (0.2) ;
\draw (3,0) circle (0.2) ;
\draw (4,0) circle (0.2) ;
\draw (5,0) circle (0.2) ;
\draw (6,0) circle (0.2) ;
\draw (7,0) circle (0.2) ;
\draw (0,0) node{B};
\draw (1,0) node{A};
\draw (2,0) node{B};
\draw (3,0) node{A};
\draw (4,0) node{B};
\draw (5,0) node{A};
\draw (6,0) node{B};
\draw (7,0) node{A};
\draw [double] (-0.2,0) -- (-0.5,0) ;
\draw  [double] (1.2,0) -- (1.8,0) ;
\draw  [double] (3.2,0) -- (3.8,0) ;
\draw                (4.2,0) -- (4.8,0) ;
\draw [double]  (5.2,0) -- (5.8,0) ;
\draw                (6.2,0) -- (6.8,0) ;
\draw  [double] (7.2,0) -- (7.5,0) ;
\draw (-0.5,0.3) node{$J'$};
\draw (1.5,0.3) node{$J'$};
\draw (3.5,0.3) node{$J'$};
\draw (4.5,0.3) node{$J$};
\draw (5.5,0.3) node{$J'$};
\draw (6.5,0.3) node{$J$};
\draw (3.5,-0.5) node[rotate=90] {\Huge \{};
\draw (5.5,-0.5) node[rotate=90] {\Huge \{};
\draw (3.5,-0.8) node{$j=1$};
\draw (5.5,-0.8) node{$j=2$};
\end{tikzpicture}
\end{center}
\caption{A semi-infinite SSH model. The left region (grey area) has $J=0$, hence a winding number $N=0$. The right region may have $N=0$ or $N=1$, depending on the ratio $J/J'$. In the latter case a zero-energy edge state resides close to the boundary between the two domains. }
\label{fig:ssh_semi_infinite}
\end{figure}

\paragraph{Semi-infinite chain.}
We consider a semi-infinite lattice, with unit cells labeled by $j=1,2,3,4\ldots$. Such a boundary could be formed  by imposing a very large potential on the unit cells at $j\leq 0$ on an infinite lattice. However, to retain the chiral symmetry it is convenient to view this
as a boundary to a region, at $j\leq 0$, where the intercell tunneling $J=0$: this
causes the $j=1$ unit cell to be disconnected from any sites with
$j\leq 0$ (see Fig \ref{fig:ssh_semi_infinite}). The region on the left, $j\leq 0$, has  $J=0$ so its winding number is $N=0$.

For the semi-infinite SSH chain (with $j\geq 1$) the eigenvalue condition is
\begin{eqnarray}
E\psi^{\rm A}_j & = & \left\{
 \begin{array}{ll} -J' \psi^{\rm B}_{j} - J \psi^{\rm B}_{j-1} & j> 1 \\ 
 -J' \psi^{\rm B}_{j}  & j= 1\end{array}
\right.
\label{eq:sshedge1}
\\
E\psi^{\rm B}_j & = & -J \psi^{\rm A}_{j+1} - J' \psi^{\rm B}_{j} \qquad j\geq  1\,.
\label{eq:sshedge2}
\end{eqnarray}
Assuming that there is only one edge state in this problem (which can be checked analytically), the chiral symmetry entails that it has a zero energy. 
Setting $E=0$ in (\ref{eq:sshedge1},\ref{eq:sshedge2})
one readily finds the solution
\begin{equation}
\left(\begin{array}{l} \psi^{\rm A}_j\\\psi^{\rm B}_j \end{array}
\right) \propto
\left(\begin{array}{c} (-J'/J)^j \\0  \end{array}\right).
\end{equation}
This is a localized (normalizable) state provided $J'<J$, \ie provided the insulator to the right of the boundary, $j\geq 1$, is in the topological phase with $N=1$. Recalling that the boundary is to a region with $J=0$, and hence $N=0$, we establish the existence of an $E=0$ edge state if and only if the insulators on either side have differing topological index $N$. The length $\xi$ over which this edge state is localized is $\xi \sim a \ln(J/J')$. Note that this edge state only occupies one type of site (here the A sites), in agreement with the general requirement of the chiral symmetry for a state $|\psi\rangle$ at zero energy: $\hat{U}|\Psi\rangle = (\hat{P}_A-\hat{P}_B)|\Psi\rangle \propto |\Psi\rangle$.
This solution is the discrete version of the edge mode, Eq.~\ref{eq:jrzero1}, of the continuum model derived in Sec.~\ref{sec:sshedge}.

\begin{figure}[t]
\begin{center}
\includegraphics{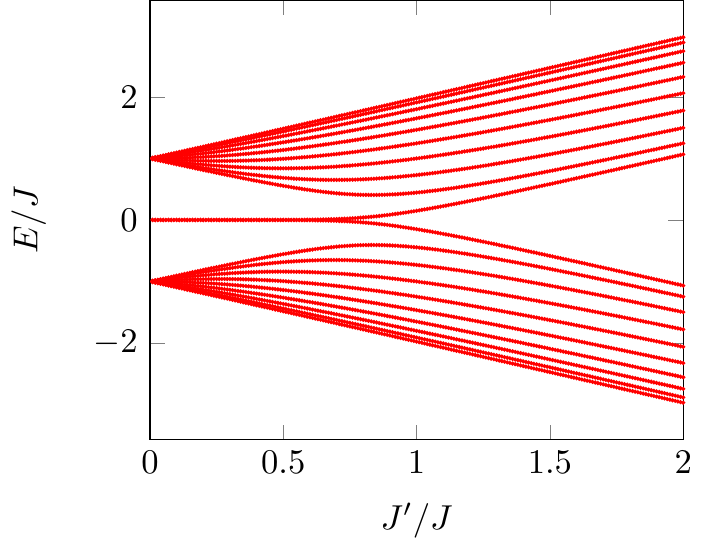}
\end{center}
\caption{Energy spectrum of a SSH chain of $M=10$ dimers. A pair of edge states close to zero energy exists when $J'/J$ is notably below 1. When $J'/J$ increases above 1, these edge states are gradually transformed into bulk states.  Figure adapted from \textcite{Delplace:2011}.}
\label{fig:Delplace}
\end{figure}

\paragraph{Finite chain.}

One can also consider a chain  with $M$ unit cells (Figure~\ref{fig:Delplace}). For $J' < J$ and a length $Ma$ much larger than the extension $\xi$ of the edge state found for a semi-infinite chain, we expect now two edge states of energy $\sim 0$, each localized at one end of the chain. When $J'/J$ increases and approaches unity, this pair of edge states gradually hybridize and they finally merge into the quasi-continua corresponding to the two bands for $J'/J$ above unity. 

In practice, generating a boundary that preserves chiral symmetry could be very delicate to achieve. Thus edge states of cold atom implementations of the SSH model are likely to be shifted away from zero energy. In contrast, the topological protection of the Majorana modes of the Kitaev chain that we will describe below (\S~\ref{subsec:Kitaev_model}) does not require any such local fine-tuning. There the protecting symmetry is the exact particle-hole symmetry of the Bogoliubov--de Gennes Hamiltonian (\ref{eq:bdgphsymmetry}.

\subsection{Gauge invariance and Zak phase}

\label{subsubsec:gaugeinvariancezak}

Strictly speaking the Zak phase is not a gauge invariant quantity. Here we illustrate this point using again the SSH model. In a local gauge transform, one can redefine the positions of the sites A$_j$ and B$_j$ to be $ja+\xi_{\rm A}$ and $jb+\xi_{\rm B}$, leading to the modified form of the Bloch wave form:
\begin{equation}
|\psi_q\rangle \deq \sum_j \E^{\I (j a+\xi_{\rm A})q} \, u_q^{\rm A}\, |{\rm A}_j\rangle \;+\; \E^{\I (j a+\xi_{\rm B})q} \, u_q^{\rm B}\, |{\rm B}_j\rangle .
\end{equation}
The vector $\vv{h}(q)$ defining the Hamiltonian in reciprocal space (\ref{eq:sshham1}) is then changed into
\begin{equation}
h_x(q)+\I h_y(q)=\left( J'+J\E^{\I qa}\right) \E^{\I q(\xi_{\rm A}-\xi_{\rm B})},
\end{equation}
so that the Hamiltonian $\hat H_q$ is not $q-$periodic anymore. The  phase $q(\xi_{\rm A}-\xi_{\rm B})$ is added to $\phi_q$ in the definition (\ref{eq:1dwavefunction1}) of the eigenstates of $\hat H_q$. This in turns adds the quantity $\pi (\xi_{\rm B}-\xi_{\rm A})/a$ to the Zak phase [see \eg \textcite{Atala2013}].

To decrease the ambiguity in the definition of the Zak phase, one may insist on keeping the $q$-periodicity of  $\hat H(q)$, and thus $\vv{h}(q)$, over the BZ. This restricts the local gauge transform described above to  $\xi_{\rm B}-\xi_{\rm A}=N'a$, where $N'$ is an integer. The Zak phase then recovers the structure  $\pi\,\times\,$integer, but the corresponding winding number is increased by $N'$. In practice this has no impact because physical consequences only involve differences of winding numbers or phases between regions of possibly different topology, which are gauge-invariant.

\subsection{Time-reversal symmetry of the SSH model}

Additional symmetries can be exploited to simplify the calculation (or measurement) of the topological invariant. The SSH model has  time-reversal symmetry, such that $\hat{H}_q = \hat{H}^*_{-q}$ and therefore
 $\phi_q = - \phi_{-q}$. 
This allows the winding number (\ref{eq:windingnumber1}) to be written as an integral over half of the BZ
\begin{eqnarray}
N & = & \frac{1}{\pi} \int_{0}^{\pi/a} 
\frac{\partial \phi_q}{\partial q} dq = \frac{1}{\pi} \left(\phi_{\pi/a} - \phi_0\right).
\label{eq:halfBZwinding}
\end{eqnarray}
Thus, information on the winding number $N$ can be obtained by measuring $\phi_{q}$  at just two points, $q=0,\pi/a$. 
Since  the angle $\phi_{q}$ is only defined modulo  $2\pi$,
this method can only determine if $N$ is even or odd.
 Nevertheless, this  partial information can be useful, notably for $\mathbb{Z}_2$ topological invariants~\cite{fu2007topological}.

\subsection{Adiabatic pumping for the Rice--Mele model}

\label{appsubsubsec:pumping}

The Rice--Mele model generalizes the SSH model to the case where the energies of sites A and B may differ by a quantity $2\Delta$. As explained in Sec.~\ref{subsec:pumping}, the vortex in
$\phi_{\rm Zak}$ around the gap-closing point ($\Delta=0$ and
$J'/J=1$) is a topological invariant which can be viewed as a Chern
number in the periodic 2D space formed from crystal momentum
$-\pi/a< q \leq \pi/a$ and time $0< t \leq T$.
This Chern number counts the number of particles that 
are pumped through the system under this adiabatic cycle~\cite{Thouless:1983}. 

To establish the link between the winding of $\phi_{\rm Zak}$ and the transported mass, we start from the fact that the lattice remains at any time periodic, so that the Bloch theorem ensures that the quasi-momentum $q$ is conserved during the evolution. Let us assume a slow evolution and a non-closing gap, so that transitions to upper bands are negligible. The evolution operator $\hat U(T)$ for a closed cycle of duration $T$ in parameter space ($J'/J,\Delta/J$) is diagonal in the Bloch state basis $|\psi_q\rangle$: $|\psi_q\rangle \to \hat U(T) |\psi_q\rangle=\E^{\I \gamma(q)} |\psi_q\rangle $. Using the canonical commutation relation, the evolution of the position operator $\hat x$ follows: $\hat U^\dagger(T)\,\hat x\, \hat U(T)=\hat x-\partial_q\gamma(\hat q)$,
from which we can deduce the global displacement $\Delta x$ of the atomic cloud by averaging over all states in the band. Now there are two contributions to $\gamma(q)$: (i) The dynamical phase related to the energy $E(q)$, which has a vanishing contribution to $\Delta x$ because the average of the group velocity $\partial_q E(q)$ over the band is zero; (ii) Zak phase $\I \int_0^T
\langle u_q|\partial_t u_q\rangle\;\D t$ which leads after average over the band to
\begin{equation}
\Delta x= \frac{a}{2\pi} \int_0^T \partial_t \phi_{\rm Zak}(t)\ \D t.
\end{equation}
Because $\phi_{\rm Zak}(T)$  equals $\phi_{\rm Zak}(0)$ plus the winding number corresponding to the vortex of Fig. \ref{fig:zakphaserm1}, this proves the quantization of the displacement in units of the lattice period $a$.

\subsection{The Kitaev model for topological superconductors}
\label{subsec:Kitaev_model}

\begin{figure}[t]
\begin{center}
\includegraphics{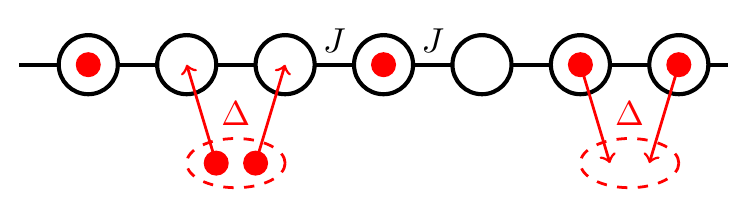}
\end{center}
\caption{Kitaev model: A 1D chain of identical sites with nearest-neighbor hopping and a coherent coupling to a superfluid reservoir that injects and removes pairs of fermions on neighboring sites.}
\label{fig:Kitaev_chain}
\end{figure}

In this model one considers{} spinless fermions moving on a 1D chain of identical sites with the hopping coefficient $J$ between nearest neighbors (figure \ref{fig:Kitaev_chain}). We assume that the chain is coherently coupled to a superfluid reservoir that can inject and remove pairs of fermions on neighboring sites. This coupling is characterized by the real parameter $\Delta$, which stands for the $p$-wave superconducting gap induced by the reservoir. It also sets the chemical potential $\mu$ which controls the total population of the chain. We are therefore interested in the eigenstates and the corresponding energies of
\begin{eqnarray}
\nonumber
\hat{H}-\mu\hat{N}  & = & \sum_{j}
\left\{- J\left(\hat{c}^\dag_j \hat{c}_{j+1} +
\hat{c}^\dag_{j+1} \hat{c}_{j}  \right) - \mu\, \hat{c}^\dag_j \hat{c}_{j}\right.
\\
& &\qquad  \left. +\ \Delta 
\left(\hat{c}_j \hat{c}_{j+1} +
\hat{c}^\dag_{j+1} \hat{c}^\dag_{j}  \right)\right\}.
\label{eq:kitaevrealspaceapp}
\end{eqnarray} 
For an infinite chain (or a finite chain with periodic boundary conditions), translation invariance ensures that $\hat{H}-\mu\hat{N}$ takes a simple form in momentum space. It can be  written in the standard Bogoliubov--de Gennes form 
  \begin{equation}
\hat{H}-\mu\hat{N}=\frac{1}{2}\sum_q 
\begin{pmatrix}\hat{\tilde c}_q^\dagger, \hat{\tilde c}_{-q}\end{pmatrix} {\mathcal H}^{\rm BdG}_q \begin{pmatrix} \hat{\tilde c}_q \\ \hat{\tilde c}_{-q}^\dagger \end{pmatrix}
\end{equation}
where the operator $\hat{\tilde c}_q^\dagger \propto \sum_j \E^{\I qja}\;\hat c_j^\dagger$ creates a particle  with quasimomentum $q$ and each Hamiltonian ${\mathcal H}^{\rm BdG}_q$ is a $2\times 2$ matrix:
\begin{equation}
{\mathcal H}^{\rm BdG}_q= -\vv{h}(q)\cdot  {\vv{\sigma}}\quad 
\mbox{with}\ \left\{  
\begin{array}{l}
h_y(q)=-2\Delta\,\sin(qa)\\
h_z(q)=2J\cos(qa)+\mu
\end{array}
\right. 
\label{eq:Kitaev0}
\end{equation}
and $h_x=0$. Although the present physical problem is different from the SSH and Rice--Mele models, we recover in (\ref{eq:Kitaev0}) a Hamiltonian in reciprocal space which has a similar structure. In particular the search for distinct topological phases can be performed by analyzing the trajectory of the vector $\vv{h}(q)$ when $q$ travels across the BZ.

\begin{figure}[t]
\begin{center}
\includegraphics[width=0.9\columnwidth]{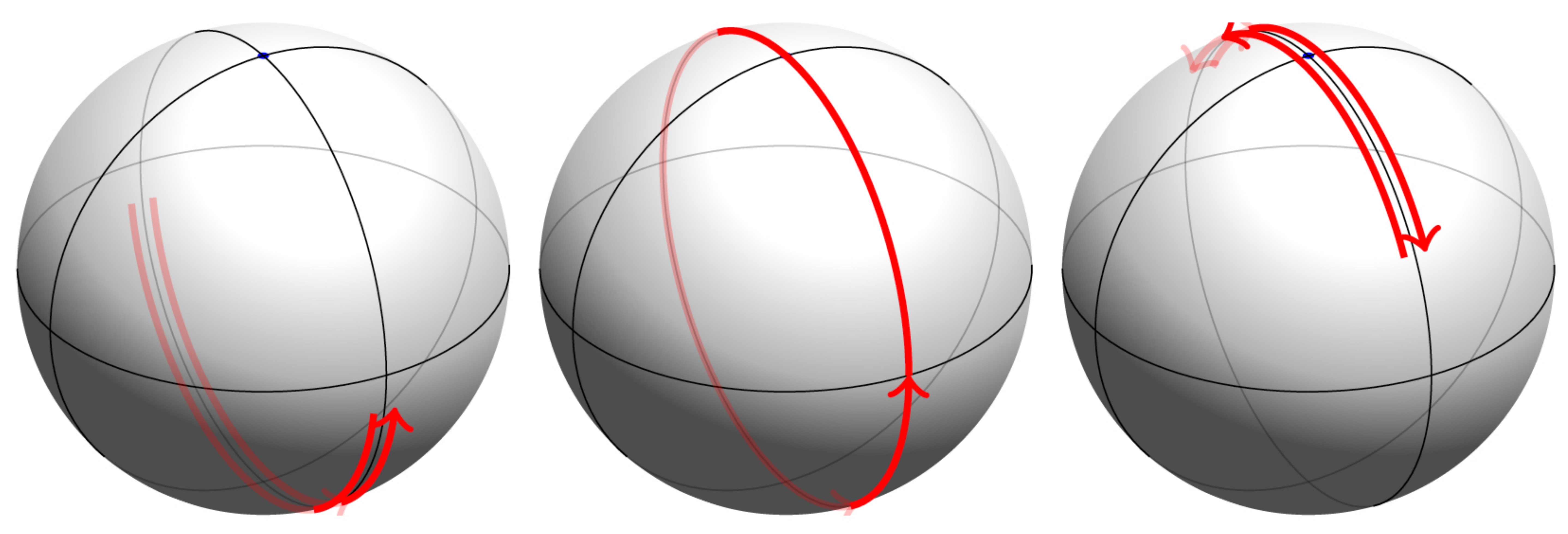}
\end{center}
\caption{Distinct topological phases of the Kitaev model, evidenced by the trajectory of $\vv{h}(q)/|\vv{h}(q)|$ on the unit sphere, as $q$ scans the Brillouin zone. Left ($\mu<-2J$) and right ($\mu>2J$): topologically trivial phase, with a zero winding around the origin. Middle ($|\mu|<2J$): topological phase.}
\label{fig:Kitaev_spheres}
\end{figure}

More precisely we see from (\ref{eq:Kitaev0}) that the vector $\vv{h}$ always lies in a plane (here $yz$), which makes the discussion formally similar to the SSH model. The quasiparticle excitation spectrum  is set by $|\vv{h}(q)|$: this is gapped for all $q$ in the BZ  
provided $|\mu| \neq 2J$. However there are two topologically distinct phases (figure \ref{fig:Kitaev_spheres}). For $|\mu| < 2J$ the 2-component vector $\vv{h}(q)$ encircles the origin, winding by $2\pi$ as $q$ runs over the BZ: this is the {\it topological superconducting} phase, for which there exist localized (Majorana) modes on the boundaries of a finite sample. For  $|\mu| > 2J$ the vector $\vv{h}(q)$ does not encircle the origin so  has vanishing winding number: this is the non-topological superconducting phase. These two topological phases cannot be smoothly connected without causing $|\vv{h}(q)|=0$ at some point in the BZ, \ie a closing of the quasiparticle gap.

\begin{figure}[t]
\begin{center}
\includegraphics[width=0.8\columnwidth]{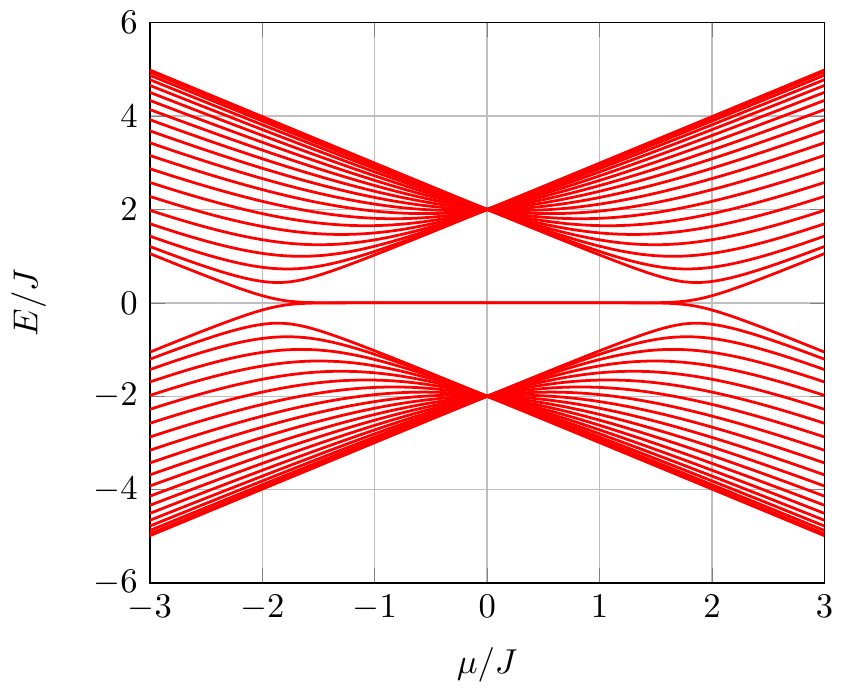}
\end{center}
\caption{Quasi-particle spectrum of $\hat H-\mu \hat N$ [Eq. (\ref{eq:kitaevrealspaceapp})] for an open chain of 20 sites and $\Delta=J$. The pair of solutions $(E,-E)$ with $E\approx 0$ that appears in the topological region $|\mu|<2J$ represents the Majorana zero energy modes at the two ends of the chain.}
\label{fig:Kitaev_spectrum}
\end{figure}

The existence of edge states for an open chain of $M$ sites can be simply revealed by taking the particular case $\Delta=J$ and $\mu=0$~\cite{Kitaev:2000} (the energy spectrum as function of $\mu$ is shown in Fig. \ref{fig:Kitaev_spectrum}). In this case, one can perform a canonical transformation using the new fermionic operators:
\begin{equation}
\hat C_j=\frac{\I}{2}\left( \hat c_j^\dagger-\hat c_j+\hat c_{j+1}^\dagger +\hat c_{j+1} \right)
\end{equation}
for $j=1,\ldots,M-1$ and
\begin{equation}
\hat C_M=\frac{\I}{2}\left( \hat c_M^\dagger-\hat c_M+\hat c_{1}^\dagger +\hat c_{1}\right)
\end{equation}
so that the $\{\hat C_j,\hat C_j^\dagger\}$ satisfy canonical fermionic commutation rules. The Hamiltonian (\ref{eq:kitaevrealspaceapp}) can then be written 
\begin{equation}
\hat H - \mu \hat N =2J \sum_{j=1}^{M-1}  \left( \hat C_j^\dagger \hat C_j -\frac{1}{2}\right).
\label{eq:Kitaev_simple}
\end{equation}
A ground state of $\hat H - \mu \hat N$ is obtained by solving 
\begin{equation}
\hat C_j |\psi_0\rangle=0, \qquad j=1,\ldots,M-1.
\label{eq:eq_GS}
\end{equation}
It is separated from the excited states by the gap $2J$, as expected from the above analysis for the infinite chain: For $J=\Delta$ and $\mu=0$,  $|\vv{h}(q)|=2J$ is indeed independent of $q$. The existence of edge states in this case originates from the fact that the non-local fermion mode $(\hat C_M,\hat C_M^\dagger)$, acting on both ends of the chain, does not contribute to the Hamiltonian (\ref{eq:Kitaev_simple}). It can therefore be filled or emptied at no energy cost, leading to two independent ground states. More precisely, if $|\psi_0\rangle$ is the ground state satisfying $\hat C_M |\psi_0\rangle=0$, then $|\psi_1\rangle=\hat C_M^\dagger |\psi_0\rangle$ is also a ground state. The states $|\psi_0\rangle$ and $|\psi_1\rangle$ correspond to different global parities of the total fermion number.  The situation is thus different from a usual superconductor, where the ground state is non-degenerate and represents a condensate of Cooper pairs, hence a superposition of states with an even number of fermions.

As discussed in the main text, the gapless edge modes of the topological superconductor are better viewed in terms of the Majorana operators 
\begin{eqnarray}
\hat{\gamma}_1 & \equiv & \hat C_M+\hat C_M^\dagger =\I(\hat{c}^\dag_M-\hat{c}_M) \\
\hat{\gamma}_2 & \equiv & \I\left( \hat C_M-\hat C_M^\dagger\right)= -(\hat{c}^\dag_1+\hat{c}_1) 
\end{eqnarray}
in view of the fact that these Majorana operators are spatially localized on each end of the chain.

\section{Floquet systems and the Magnus expansion}

\label{app:floquet}

We mention here several key results for Floquet systems of use in the main text. We refer the reader to \textcite{eckardtreview} for a comprehensive recent account.

Consider a Floquet system, defined by a time-varying Hamiltonian
$\hat{H}(t)$ that is periodic $\hat{H}(t+T) = \hat{H}(t)$ with period
$T=2\pi/\omega$. The time evolution  over an
integer number of periods, from $t_0$ to $t_0+NT$, is described by the unitary operator
\begin{eqnarray}
\hat{U}(t_0, t_0+ NT) & \equiv & {\mathcal T} \exp{\left[{-\frac{\I}{\hbar}\int_{t_0}^{t_0+NT}\hat{H}(t')\D t'}\right]}
\end{eqnarray}
where ${\mathcal T}$ denotes time ordering. This can be written
\begin{eqnarray}
\hat{U}(t_0, t_0+ NT) & = & \left[\hat{U}(t_0,t_0+ T)\right]^N \equiv \E^{-\frac{\I}{\hbar}\hat{H}^F_{t_0} T}
\label{eq:floquetoperator}
\end{eqnarray}
which defines the effective Floquet Hamiltonian
$\hat{H}^F_{t_0}$ in terms of the (logarithm of the) time evolution operator
over a single period $T$. Clearly
$\hat{H}^F_{t_0}$ is undefined up to the addition of integer multiples of
$2\pi \hbar/T =
\hbar\omega$, so its eigenvalues -- defining the Floquet spectrum -- have a periodicity in energy of $\hbar\omega$.  The Floquet Hamiltonian describes the ``stroboscopic" evolution of the system, \ie  at the selected times $t=t_0,t_0+T,\ldots
t_0+NT$. This is relevant for understanding the dynamics over timescales long compared to the drive period $T$.

In addition to this stroboscopic evolution, the system undergoes a ``micromotion" on a timescale of the period $T$.
This causes the Floquet Hamiltonian
$\hat{H}^F_{t_0}$ to depend on the time
during the cycle, via a $t_0$-dependent unitary transformation (the Floquet spectrum is therefore  invariant).  A convenient way to account for this
micromotion
is to write the general time-evolution operator, from time $t_{\rm i}$ to $t_{\rm f}$, 
as~\cite{goldmandalibard} 
\begin{equation}
\hat{U}(t_{\rm i},t_{\rm f}) = \E^{-\I\hat{K}(t_{\rm f})} \E^{-\frac{\I}{\hbar}\hat{H}^{\rm eff}(t_{\rm f}-t_{\rm i})}
 \E^{\I \hat{K}(t_{\rm i})}\,.
\end{equation}
This can be achieved with a time-independent effective Hamiltonian $\hat{H}^{\rm eff}$ 
and a
``kick'' operator that is periodic in time $\hat{K}(t) = \hat{K}(t+T)$ and that 
has vanishing average over one period~\cite{goldmandalibard}. The kick operator takes account of the micromotion (related to how $t_{\rm i}$ and $t_{\rm f}$ 
fall in the period $T$) and the time-independent effective Hamiltonian $\hat{H}^{\rm eff}$ controls the long-time behavior.

In general, the determination of the effective Hamiltonian $\hat{H}_{\rm eff}$ (or the closely related $\hat{H}^F_{t_0}$)  is a difficult task. However, for a drive frequency, $\omega$, that is large compared to other frequency scales, the effective Hamiltonian can often be approximated by the Magnus expansion in powers of $1/\omega$~\cite{eckardtreview}. Writing the Hamiltonian in terms of its harmonics
\begin{equation}
\hat{H}(t)
 = \sum_{m = -\infty}^\infty \hat{H}_m \E^{\I m \omega t}
 = \hat{H}_0 + \sum_{m = 1}^\infty \hat{H}_m \E^{\I m \omega t}+\hat{H}_{-m} \E^{-\I m \omega t}
\end{equation}
where $\hat{H}_m^\dag = \hat{H}_{-m}$, the Magnus expansion leads to~\cite{goldmandalibard}
\begin{widetext}
\begin{eqnarray}
\hat{H}_{\rm eff} 
& = & \hat{H}_0 + \frac{1}{\hbar\omega} \sum_{m = 1}^\infty \frac{1}{m}[\hat{H}_m,\hat{H}_{-m}] +
  \frac{1}{2(\hbar\omega)^2} \sum_{m = 1}^\infty \frac{1}{m^2}\left[[\hat{H}_m,\hat{H}_0],\hat{H}_{m}\right]
+\left[[\hat{H}_{-m},\hat{H}_0],\hat{H}_{m}\right] 
  + \ldots \,.
\label{eq:magnus}
\end{eqnarray}
\end{widetext}

The Magnus expansion underpins several results used in the main text. 

(i) The leading term, $\hat{H}_0$, is the time-averaged Hamiltonian, as
used in our discussion of inertial forces in Sec.~\ref{sec:force} to
construct non-zero Peierls phase factors for tunneling.

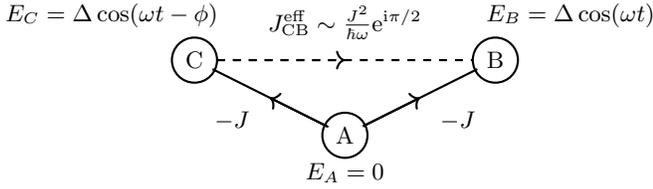
\begin{figure}

\begin{center}
\begin{tikzpicture}
\draw [thick] (4,0) circle (0.3) ;

\draw [thick] (6,1) circle (0.3) ;

\draw [thick] (2,1) circle (0.3) ;

\draw (4,0) node{A};
\draw (6,1) node{B};
\draw (2,1) node{C};

\draw [thick] (3.78,0.11) -- (2.24,0.88) ;
\draw  [thick,->] (3.76,0.12) -- (3,0.5) ;
\draw  [thick,->] (4.24,0.12) -- (5,0.5) ;

\draw  [thick] (4.24,0.12) -- (5.76,0.88) ;

\draw  [thick, dashed] (2.3,1) -- (5.7,1) ;
\draw  [thick, dashed,->] (2.3,1) -- (4,1) ;

\draw (4,1.5) node{$J_{\rm CB}^{\rm eff} \sim \frac{J^2}{\hbar\omega}{\rm e}^{{\rm i} \pi/2}$};

\draw (2.5,0.2) node{$-J$};
\draw (5.5,0.2) node{$-J$};
\draw (0.9,1.6) node{$E_C=\Delta \cos(\omega t-\phi)$};
\draw (7.0,1.6) node{$E_B=\Delta \cos(\omega t)$};
\draw (4.0,-0.5) node{$E_A=0$};
\end{tikzpicture}
\end{center}

\caption{Hopping on three sites of the honeycomb lattice, with time-varying on site energies. The unitary transformation $(\ref{eq:unitary_magnus})$ maps this problem to 
time-varying tunneling matrix elements between nearest-neighbors, A$\leftrightarrow$B and A$\leftrightarrow$C.
Then the first order correction in $1/\omega$ from the Magnus expansion leads to a tunneling term of order $J^2/(\hbar\omega)$ between next-nearest-neighbors, B$\leftrightarrow$C, 
with nonzero Peierls phase factor. [See Eq.~(\ref{eq:nnn}).]}

\label{fig:haldane3site}

\end{figure}

(ii) The first order term is important in generating the next-nearest-neighbor hopping from circular shaking of the honeycomb lattice, as required to simulate the Haldane model. Consider the three-site system shown in Fig.~\ref{fig:haldane3site} with on-site energies $E_\alpha$, $\alpha=A,B,C$,  a constant tunnel matrix element $-J$ along the links $AB$ and $AC$, and no `bare' tunneling between $B$ and $C$:
\begin{equation}
\hat H =-J\left( |B\rangle \langle A|+ |C\rangle \langle A|+\mbox{h.c.}\right)+\sum_{\alpha}E_\alpha \hat P_\alpha
\label{eq:modulated_energies}
\end{equation}
with the projectors $\hat P_\alpha =|\alpha\rangle \langle \alpha|$.  Shaking at a frequency $\omega$ leads to sinusoidally varying energy offsets that we model as $E_B(t)=\Delta \cos(\omega t)$ and $E_C(t)=\Delta \cos(\omega t-\phi)$, whereas $E_A(t)$ is set to zero by convention. The phase offset $\phi$ arises from the circular shaking and the angle between the two nearest-neighbor bonds (typically $\phi =2\pi/3$). Using the unitary transformation generalizing (\ref{eq:force:gauge}) 
\begin{equation}
\hat U(t)=\sum_\alpha \E^{\I\int_0^t E_\alpha(t')\;\D t'/\hbar}\;\hat P_\alpha,
\label{eq:unitary_magnus}
\end{equation}
we convert these energy modulations into time-varying phase factors on the hopping:
\begin{eqnarray}
\frac{\hat{\tilde H}(t)}{-J} & = & 
\E^{\I\frac{\Delta}{\hbar\omega} \sin \omega t}|B\rangle \langle A| 
\nonumber
\\
& & +
  \E^{\I\frac{\Delta}{\hbar\omega} \sin (\omega t - \phi)}|C\rangle \langle A| + \mbox{h.c.}\,.
\end{eqnarray}
Expanding in terms of the harmonics leads to
\begin{eqnarray}
\frac{\hat{H}_m}{-J} & = & {\mathcal J}_m(\Delta/\hbar\omega) |B\rangle \langle A| + {\mathcal J}_{-m}(\Delta/\hbar\omega) |A\rangle \langle B|
\nonumber
\\
 & &  \!\!\!\!\!\!\!\!\!\! \!\!\!\!\!\!\!\!\!\! \!\!\!\!\!\!\!\!\!\!
  +  \E^{-\I  m \phi} \left[ {\mathcal J}_m(\Delta/\hbar\omega) |C\rangle \langle A| 
 +{\mathcal J}_{-m} (\Delta/\hbar\omega) |A\rangle \langle C|\right]
\end{eqnarray}
where ${\mathcal J}_m$ are Bessel functions. Computing the first order correction to the effective Hamiltonian (\ref{eq:magnus}), one finds
\begin{eqnarray}
\hat{H}_{\rm eff}^{(1)} & \equiv & \frac{1}{\hbar\omega} \sum_{m = 1}^\infty \frac{1}{m}[\hat{H}_m,\hat{H}_{-m}]\\
 & = &  - J^{\rm eff}\left[\E^{\I\pi/2} |B\rangle \langle C|  + \E^{-\I\pi/2} |B\rangle \langle C|\right]   \\
J^{\rm eff} & = & \frac{2 J^2}{\hbar\omega}  \sum_{m = 1}^\infty \frac{\sin(m\phi)\left[{\mathcal J}_m(\Delta/\hbar\omega) \right]^2}{m} \,.
\label{eq:nnn}
\end{eqnarray}
This describes a next-nearest-neighbor tunneling term, between B and C sites, which inherits a Peierls phase factor of $\pi/2$. This phase arises
from the phase offset of the circular shaking.
A particle that encircles the plaquette, A$\to$B$\to$C$\to$A, picks up a gauge-invariant phase $\psi_{\rm AB} = -\pi/2$, placing the system in the regime where the Haldane model has topological bands. It is interesting to note that for this $1/\omega$ expansion, the leading term in $J_{\rm eff}$ corresponds to the $m=1$ contribution in (\ref{eq:nnn}), hence scales as $1/\omega^3$. Had we used directly the Magnus expansion (\ref{eq:magnus}) for the initial Hamiltonian (\ref{eq:modulated_energies}), this would have required us to go up to the 3rd order of the expansion, which would have been quite involved. Fortunately the unitary transformation (\ref{eq:unitary_magnus}) involves an integral of the on-site energies over time, which provides a gain of a factor $1/\omega$. The Magnus expansion at order 1 applied to $\hat {\tilde H}$ is then sufficient to obtain the relevant effective Hamiltonian.

(iii) The second order term in Eq.~(\ref{eq:magnus}), of order $1/\omega^2$,  is responsible for the non-local interactions discussed in Sec.~\ref{subsec:nonlocal}.
These arise from the contribution of the Hubbard interaction $U$ to $\hat{H}_0$, and from the oscillating tunneling matrix elements in $\hat{H}_{m\neq 0}$ in the double commutator,
leading the term of order $J^2U/(\hbar\omega)^2$ of Eq.~(\ref{eq:nonlocalsite}). We refer the reader to \textcite{eckardt:2015} for a comprehensive analysis of such terms.

\section{Light matter interaction}\label{app:alkali}
Here we describe the general form of light-matter must take for two-photon interactions, and describe the specific structure of this interaction for alkali atoms.  To briefly summarize what follows, the light matter interaction can be divided into contributions from irreducible rank-0, -1, and -2 spherical tensor operators.  In alkali atoms the rank-0 and -1 contributions dominate, but in heaver atoms (or sufficiently close to atomic resonance) the rank-2 contribution can be large.  In addition, a similar description of single photon transitions (as would be relevant to systems coupling ground and meta-stable excited states) gives only the rank-0 and rank-1 contributions.

Now consider a system of ultracold alkali atoms in their electronic ground state manifold illuminated by one or several laser fields which non-resonantly couple the ground states with the lowest electronic excited states.  In the presence of an external magnetic field only, the light-matter Hamiltonian for the atomic ground state manifold is  
\begin{align}
\hat H_0 =& A_{\rm hf} \hat{\vv{I}}\!\cdot\!\hat{\vv{J}} + \frac{\mu_B}{\hbar}\vv{B}\!\cdot\!\left(g_J\hat{\vv{J}} + g_I\hat{\vv{I}}\right)\,,
\end{align}
where $A_{\rm hf}$ is the magnetic dipole hyperfine coefficient; and $\mu_B$ is the Bohr magneton.  The Zeeman term includes separate contributions from $\hat{\vv{J}}=\hat{\vv{L}}+\hat{\vv{S}}$ (the sum of the orbital $\hat{\vv{L}}$ and electronic spin $\hat{\vv{S}}$ angular momentum) and the nuclear angular momentum $\hat{\vv{I}}$, along with their respective Land\'e $g$-factors.  We next consider the additional contributions to the atomic Hamiltonian resulting with off-resonant interaction with laser fields. 

As was observed in Refs.~\cite{Deutsch1998,Dudarev2004,Sebby-Strabley2006}, conventional spin independent (scalar, $U_s$) optical potentials acquire additional spin-dependent terms near atomic resonance: the rank-1 (vector, $U_v$) and rank-2 tensor light shifts~\cite{Deutsch1998}.  For the alkali atoms, adiabatic elimination of the excited states labeled by $j =1/2$ (D1) and $j=3/2$ (D2) yields an effective atom-light coupling Hamiltonian for the ground state atoms (with $j=1/2$):
\begin{align}
\hat H_L&= \hat H_0\!+\!\hat H_1\!+\!\hat H_2 = u_s(\vv{E}^*\!\cdot\!\vv{E}) + \frac{iu_v(\vv{E}^*\!\times\!\vv{E})}{\hbar} \cdot\vv{J} + \hat H_2.
\end{align}
The rank-2 term $\hat H_2$ is negligible for the parameters of interest and henceforth neglected. Here $\vv{E}$ is the optical electric field; $u_v=-2u_s\Delta_{\rm FS}/3(\omega-\omega_0)$ determines the vector light shift;  $\Delta_{\rm FS} = \omega_{3/2}-\omega_{1/2}$ is the fine-structure splitting; $\hbar\omega_{1/2}$ and $\hbar\omega_{3/2}$ are the D1 and D2 transition energies; and $\omega_0=(2\omega_{1/2}+\omega_{3/2})/3$ is a suitable average.  $u_s$ sets the scale of the light shift and proportional to the atoms ac polarizability.  
 
The contributions from the scalar and vector light shifts featured in  $H_L$ can be independently specified with informed choices of laser frequency $\omega$ and intensity.  Evidently, the vector light shift is a contribution to the total Hamiltonian acting like an effective magnetic field
\begin{align}
\vv{B}_{\rm eff} &= \frac{iu_v(\vv{E}^*\!\times\!\vv{E})}{\mu_B g_J}
\end{align}
that acts on $\hat{\vv{J}}$ and not the nuclear spin $\hat{\vv{I}}$.  Instead of using the full Breit-Rabi equation~\cite{Breit1931} for the Zeeman energies, we assume that the Zeeman shifts are small in comparison with the hyperfine splitting -- the linear, or anomalous, Zeeman regime -- in which case, the effective Hamiltonian for a single manifold of total angular momentum $\hat{\vv{F}}=\hat{\vv{J}} + \hat{\vv{I}}$ states is
\begin{align}
H_0 + H_L =&\ u_s(\vv{E}^*\!\cdot\!\vv{E}) + \frac{\mu_B g_F}{\hbar}\left(\vv{B} + \vv{B}_{\rm eff}\right)\!\cdot\!\hat{\vv{F}}\,.
\end{align}
Notice that $\vv{B}_{\rm eff}$ acts as a true magnetic field and adds vectorially with $\vv{B}$, and since $|g_I/g_J|\approx .0005$ in the alkali atoms, we safely neglected a contribution $-\mu_B g_I \vv{B}_{\rm eff}\cdot\hat{\vv{I}}/ \hbar$ to the atomic Hamiltonian.   
 We also introduced the hyperfine Land\'e $g$-factor $g_F$.  In $^{87}{\rm Rb}$'s lowest energy manifold with $f=1$, for which $j=1/2$ and $i=3/2$, we get $g_F = -g_J/4\approx-1/2$.  In the following, we always consider a single angular momentum manifold labeled by $f$, and select its energy at zero field as the zero of energy.

\paragraph*{Bichromatic light field.}
Consider an ensemble of ultracold atoms subjected to a magnetic field $\vv{B} = B_0 {\vv{e}_z}$. The atoms are illuminated by several lasers with frequencies $\omega$ and $\omega+\delta\omega$, where $\delta\omega\approx |g_F \mu_B B_0/\hbar|$ differs by a small detuning $\delta=g_F \mu_B B_0/\hbar-\delta\omega$ from the linear Zeeman shift between $m_F$ states (where $|\delta|\ll\delta\omega$).  In this case, the complex electric field $\vv{E} = \vv{E}_{\omega_-}\exp(-\I\omega t) + \vv{E}_{\omega_+}\exp\left[-\I(\omega+\delta\omega) t\right]$ contributes to the combined magnetic field, giving
\begin{align}
\vv{B} + \vv{B}_{\rm eff} =& B_0 {\vv{e}_z} + \frac{\I u_v}{\mu_{\rm B} g_J}\bigg[\left(\vv{E}_{\omega_-}^*\!\times\!\vv{E}_{\omega_-}\right) + \left(\vv{E}^*_{\omega_+}\!\times\!\vv{E}_{\omega_+}\right) \nonumber \\
& + \left(\vv{E}_{\omega_-}^*\!\times\!\vv{E}_{\omega_+}\right)e^{-\I\delta\omega t} 
+ \left(\vv{E}^*_{\omega_+}\!\times\!\vv{E}_{\omega_-}\right)e^{\I\delta\omega t}
\bigg]\,.
\end{align} 
The first two terms of $ \vv{B}_{\rm eff}$ add to the static bias field $B_0 {\vv{e}_z}$, and the remaining two time-dependent terms describe transitions between different $m_F$ levels.  Provided $B_0\gg\left|\vv{B}_{\rm eff}\right|$ and $\delta\omega$ are large compared to the kinetic energy scales, the Hamiltonian can be simplified by time-averaging to zero the time-dependent terms in the scalar light shift and making the rotating wave approximation (RWA) to eliminate the time-dependence of the coupling fields.  The resulting contribution to the Hamiltonian
\begin{align}
\hat H_{\rm RWA} &= U(\vv{r}) \hat 1 + {\boldsymbol \kappa}(\vv{r})\!\cdot\!\hat{\vv{F}},\label{eq:RWA_Hamiltonian}
\end{align}
where we identify the scalar potential
\begin{align}
U(\vv{r}) &= u_s \left(\vv{E}_{\omega_-}^*\!\cdot\!\vv{E}_{\omega_-} + \vv{E}^*_{\omega_+}\!\cdot\!\vv{E}_{\omega_+}\right)
\end{align}
and the RWA effective magnetic field.  This expression is valid for $g_F>0$ (for $g_F<0$ the sign of the ${\vv{e}_x}$ and $\I{\vv{e}_y}$ terms would both be  positive, owing to selecting the opposite complex terms in the RWA).
\begin{align}
{\boldsymbol \kappa} =& \left[\delta + i \frac{u_v}{\hbar}\left(\vv{E}_{\omega_-}^*\!\times\!\vv{E}_{\omega_-}+\vv{E}^*_{\omega_+}\!\times\!\vv{E}_{\omega_+}\right)\!\cdot\!{\vv{e}_z}\right]{\vv{e}_z}\nonumber\\
&- \frac{u_v}{\hbar}{\rm Im}\left[\left(\vv{E}_{\omega_-}^*\!\times\!\vv{E}_{\omega_+}\right)\!\cdot\!\left({\vv{e}_x}-i{\vv{e}_y}\right)\right]{\vv{e}_x} \\
&-\frac{u_v}{\hbar}{\rm Re}\left[\left(\vv{E}_{\omega_-}^*\!\times\!\vv{E}_{\omega_+}\right)\!\cdot\!\left({\vv{e}_x}-i{\vv{e}_y}\right)\right]{\vv{e}_y}\nonumber.
\end{align}
Although this effective coupling is directly derived from the initial vector light shifts, $\vv{\kappa}$ is composed of both static and resonant couplings in a way that goes beyond the restrictive $\vv{B}_{\rm eff} \propto i\vv{E}^*\!\times\!\vv{E}$ form.  This enables topological state-dependent lattices.

\section{Berry curvature and unit cell geometry}
\label{app:embedding}

As noted in the main text, it can be convenient to perform a unitary transformation of the Bloch Hamiltonian $\hat{H}_{\vv{q}}' = \hat{\cal U}_\vv{q}\hat{H}_\vv{q}\hat{\cal U}^\dag_\vv{q}$, with $\hat{\cal U}_\vv{q}$ a wavevector-dependent unitary operator. Such transformations can be used to render the Hamiltonian periodic in the BZ, but also can relate different choices of unit cell without change of periodicity. They must leave all physical observables unchanged. The energy spectrum is  invariant under this unitary transformation, $E'_{\vv{q}} =E_{\vv{q}}$, while the Bloch states transform as 
$|u'_\vv{q}\rangle = \hat{\cal U}_\vv{q}
|u_\vv{q}\rangle$. (In this appendix we drop the band index for clarity.)  The new Berry connection in reciprocal space is
\begin{eqnarray}
\vv{A}'   & = &  \I\langle u'_\vv{q}| \vv{\nabla}_\vv{q} |u'_\vv{q}\rangle     =  \vv{A} + \I \langle u_\vv{q}| \hat{{\cal U}}^\dag_\vv{q} \left[ \vv{\nabla}_\vv{q} \hat{{\cal U}}_\vv{q}\right] |u_\vv{q}\rangle \,.
 \end{eqnarray}
It is interesting to note that, in general, this transformation causes the {\it Berry  curvature} to change.  How can one reconcile this with the fact that  the Berry curvature has measurable physical consequences, \eg within semiclassical dynamics (Sec.~\ref{subsubsec:anomalousvelocity})?
As we shall see, the answer lies in noting that these unitary transformations can lead to changes of the positions of the orbitals within the unit cell, \ie changing the internal geometry of the unit cell.

We shall illustrate this for a two-band model, with a unitary transformation 
\begin{equation}
\hat{\cal U}_\vv{q} = \exp\left(\frac{1}{2}\I\vv{q}\cdot\vv{\rho}\,\hat{\sigma}_z\right)\,.
\label{eq:unitarysigmaz}
\end{equation}
This is the transformation used in our discussion of the Haldane model (\ref{eq:haldanenn}). It leads to the change
\begin{eqnarray}
\vv{A}'(\vv{q})    &   = &   \vv{A}(\vv{q}) -  \frac{1}{2}\vv{\rho}  \langle \hat{\sigma}_z\rangle \end{eqnarray}
of the Berry connection, where we have defined $\langle \hat{\sigma}_z\rangle\equiv \langle u_\vv{q}| \hat{\sigma}_z |u_\vv{q}\rangle$. Hence, the Berry curvature becomes
\begin{equation}
\vv{\Omega}'(\vv{q}) \equiv \vv{\nabla}_\vv{q} \times \vv{A}'   =   \vv{\Omega}(\vv{q}) -  \frac{1}{2} \left[ \vv{\nabla}_\vv{q} \langle \hat{\sigma}_z\rangle
\right] \times \vv{\rho} \,.
\label{eq:omegaprime}
 \end{equation}
In general, the Berry curvature changes, $\vv{\Omega}'(\vv{q})\neq \vv{\Omega}(\vv{q})$. However, since the difference is a total derivative, its integral over the BZ vanishes,  so there is no change in the Chern number of the band.

For a consistent application of the unitary transformation, one must consider how all relevant physical quantities transform. Semiclassical dynamics describes the velocity 
 of a wavepacket centered on $\vv{q}$ and $\vv{r}$ in response to a uniform force. 
A  uniform force in the original basis arises from a potential
\begin{equation}
\hat{V} = -\vv{F} \cdot\hat{\vv{r}}\,.
\label{eq:v}
\end{equation}
Under the unitary  transformation this becomes
\begin{equation}
 \hat{\cal U}_\vv{q}\hat{V}  \hat{\cal U}_\vv{q}^\dag= -\vv{F} \cdot \hat{\vv{r}}- \frac{1}{2}\vv{F} \cdot\vv{\rho}\,\hat{\sigma}_z\,.
\label{eq:vprime}
\end{equation}
We have used the fact that, with $\vv{q}$ replaced by the momentum operator, the unitary operator (\ref{eq:unitarysigmaz})  effects translations in real space, albeit in a spin-dependent manner,  $\hat{\vv{r}} \to \hat{\vv{r}}+\frac{1}{2}\vv{\rho}\hat{\sigma}_z$. 
Indeed, this transformation encodes the important physical effect of the unitary transformation: it amounts to spatial displacement of the orbitals within the unit cell, here separating $\sigma_z = \pm 1$  by  $\vv{\rho}$.
In the new basis the force $\bm{F}$ has two effects. First, it provides a uniform potential gradient which leads to 
$\hbar \dot{\vv{q}} = \vv{F}$ as usual. Second, it applies a state-dependent potential that shifts the energy of the wavepacket by $- \frac{1}{2}\vv{F} \cdot\vv{\rho}\langle\hat{\sigma}_z\rangle$ with the expectation value taken for the wavepacket's momentum $\vv{q}$. This second contribution can be incorporated as an $\vv{F}$-dependent change in the dispersion relation
\begin{equation}
E'_\vv{q} = E_\vv{q} - \frac{1}{2}\vv{F} \cdot\vv{\rho}\langle\hat{\sigma}_z\rangle\,.
\label{eq:eprime}
\end{equation}
The displacement of orbitals within the unit cell also  changes  the velocity operator. In the new basis the velocity  is
\begin{eqnarray}
\hat{\vv{v}}'
 & \equiv & \frac{1}{\hbar}\vv{\nabla}_\vv{q}  \hat{H}'_\vv{q} =  \frac{1}{\hbar} \vv{\nabla}_\vv{q} \left[
 \hat{\cal U}_\vv{q}\hat{H}_\vv{q}\hat{\cal U}^\dag_\vv{q}\right]
\\
 & = &  \frac{1}{\hbar} \hat{\cal U}_\vv{q} \left[\vv{\nabla}_\vv{q}  \hat{H}_\vv{q}\right] \hat{\cal U}_\vv{q}^\dag - \frac{\I}{2\hbar} \vv{\rho}  \left[\hat{H}'_\vv{q},\hat{\sigma}_z\right] 
\\
  & = & \hat{\cal U}_\vv{q}\hat{\vv{v}} \hat{\cal U}_\vv{q}^\dag - \frac{1}{2}\vv{\rho}\,\hat{\dot{\sigma}}_z \,.
\end{eqnarray}
There is a correction to the velocity $\hat{\vv{v}}\equiv  \frac{1}{\hbar}\vv{\nabla}_\vv{q}  \hat{H}_\vv{q} $ beyond the mere unitary transformation, $\hat{\cal U}_\vv{q}\hat{\vv{v}} \hat{\cal U}_\vv{q}^\dag$. In the last line this correction is written 
in terms of the rate of change of $\hat{\sigma}_z$
(in the Heisenberg picture) to indicate its physical origin:  the motion of a particle between orbitals within the unit cell corresponds to motion through space with nonzero velocity. 
In the semiclassical dynamics of the wavepacket, this gives the correction
\begin{equation}
\vv{v}' =\vv {v} -  \frac{1}{2}\vv{\rho}\,\langle {\hat{\dot{\sigma}}}_z \rangle = 
\vv{v} -  \frac{1}{2}\vv{\rho}\,\dot{\vv{q}}\cdot \vv{\nabla}_\vv{q} 
\langle \hat{{\sigma}}_z \rangle \,.
\label{eq:velprime}
\end{equation}
Starting from the standard semiclassical dynamics  (\ref{eq:anomalous}) in the original (unprimed) basis,
and using the transformations (\ref{eq:omegaprime}),  (\ref{eq:eprime}), and  (\ref{eq:velprime}), one recovers the 
correct semiclassical dynamics in the new (primed) basis
\begin{eqnarray}
{\vv{v}}'&=&\frac{1}{\hbar}\vv{\nabla}_{\vv{q}}E'_\vv{q} + {\vv{\Omega'}}\times \dot{\vv{q}} \,.
\end{eqnarray}
Thus, unitary transformations of this form 
lead to changes in the positions of the orbitals within the unit cell: these changes in cell geometry modify the effect of the forces applied, and  the velocity on the scale of the unit cell. These modifications compensate the change  in the Berry curvature  to  recover the correct semiclassical dynamics.



\begin{thebibliography}{259}%
\makeatletter
\providecommand \@ifxundefined [1]{%
 \@ifx{#1\undefined}
}%
\providecommand \@ifnum [1]{%
 \ifnum #1\expandafter \@firstoftwo
 \else \expandafter \@secondoftwo
 \fi
}%
\providecommand \@ifx [1]{%
 \ifx #1\expandafter \@firstoftwo
 \else \expandafter \@secondoftwo
 \fi
}%
\providecommand \natexlab [1]{#1}%
\providecommand \enquote  [1]{``#1''}%
\providecommand \bibnamefont  [1]{#1}%
\providecommand \bibfnamefont [1]{#1}%
\providecommand \citenamefont [1]{#1}%
\providecommand \href@noop [0]{\@secondoftwo}%
\providecommand \href [0]{\begingroup \@sanitize@url \@href}%
\providecommand \@href[1]{\@@startlink{#1}\@@href}%
\providecommand \@@href[1]{\endgroup#1\@@endlink}%
\providecommand \@sanitize@url [0]{\catcode `\\12\catcode `\$12\catcode
  `\&12\catcode `\#12\catcode `\^12\catcode `\_12\catcode `\%12\relax}%
\providecommand \@@startlink[1]{}%
\providecommand \@@endlink[0]{}%
\providecommand \url  [0]{\begingroup\@sanitize@url \@url }%
\providecommand \@url [1]{\endgroup\@href {#1}{\urlprefix }}%
\providecommand \urlprefix  [0]{URL }%
\providecommand \Eprint [0]{\href }%
\providecommand \doibase [0]{http://dx.doi.org/}%
\providecommand \selectlanguage [0]{\@gobble}%
\providecommand \bibinfo  [0]{\@secondoftwo}%
\providecommand \bibfield  [0]{\@secondoftwo}%
\providecommand \translation [1]{[#1]}%
\providecommand \BibitemOpen [0]{}%
\providecommand \bibitemStop [0]{}%
\providecommand \bibitemNoStop [0]{.\EOS\space}%
\providecommand \EOS [0]{\spacefactor3000\relax}%
\providecommand \BibitemShut  [1]{\csname bibitem#1\endcsname}%
\let\auto@bib@innerbib\@empty
\bibitem [{\citenamefont {Abanin}\ \emph {et~al.}(2016)\citenamefont {Abanin},
  \citenamefont {Roeck},\ and\ \citenamefont {Huveneers}}]{abanin2016}%
  \BibitemOpen
  \bibfield  {author} {\bibinfo {author} {\bibnamefont {Abanin}, \bibfnamefont
  {Dmitry~A}}, \bibinfo {author} {\bibfnamefont {Wojciech~De}\ \bibnamefont
  {Roeck}}, \ and\ \bibinfo {author} {\bibfnamefont {Francois}\ \bibnamefont
  {Huveneers}}} (\bibinfo {year} {2016}),\ \bibfield  {title} {\enquote
  {\bibinfo {title} {Theory of many-body localization in periodically driven
  systems},}\ }\href {\doibase 10.1016/j.aop.2016.03.010} {\bibfield  {journal}
  {\bibinfo  {journal} {Annals of Physics}\ }\textbf {\bibinfo {volume}
  {372}},\ \bibinfo {pages} {1 -- 11}}\BibitemShut {NoStop}%
\bibitem [{\citenamefont {Aidelsburger}\ \emph {et~al.}(2013)\citenamefont
  {Aidelsburger}, \citenamefont {Atala}, \citenamefont {Lohse}, \citenamefont
  {Barreiro}, \citenamefont {Paredes},\ and\ \citenamefont
  {Bloch}}]{Aidelsburger2013}%
  \BibitemOpen
  \bibfield  {author} {\bibinfo {author} {\bibnamefont {Aidelsburger},
  \bibfnamefont {M}}, \bibinfo {author} {\bibfnamefont {M.}~\bibnamefont
  {Atala}}, \bibinfo {author} {\bibfnamefont {M.}~\bibnamefont {Lohse}},
  \bibinfo {author} {\bibfnamefont {J.~T.}\ \bibnamefont {Barreiro}}, \bibinfo
  {author} {\bibfnamefont {B.}~\bibnamefont {Paredes}}, \ and\ \bibinfo
  {author} {\bibfnamefont {I.}~\bibnamefont {Bloch}}} (\bibinfo {year}
  {2013}),\ \bibfield  {title} {\enquote {\bibinfo {title} {Realization of the
  {H}ofstadter {H}amiltonian with ultracold atoms in optical lattices},}\
  }\href {\doibase 10.1103/PhysRevLett.111.185301} {\bibfield  {journal}
  {\bibinfo  {journal} {Phys. Rev. Lett.}\ }\textbf {\bibinfo {volume} {111}},\
  \bibinfo {pages} {185301}}\BibitemShut {NoStop}%
\bibitem [{\citenamefont {Aidelsburger}\ \emph {et~al.}(2015)\citenamefont
  {Aidelsburger}, \citenamefont {Lohse}, \citenamefont {Schweizer},
  \citenamefont {Atala}, \citenamefont {Barreiro}, \citenamefont
  {Nascimb{\`e}ne}, \citenamefont {Cooper}, \citenamefont {Bloch},\ and\
  \citenamefont {Goldman}}]{munichchern}%
  \BibitemOpen
  \bibfield  {author} {\bibinfo {author} {\bibnamefont {Aidelsburger},
  \bibfnamefont {M}}, \bibinfo {author} {\bibfnamefont {M.}~\bibnamefont
  {Lohse}}, \bibinfo {author} {\bibfnamefont {C.}~\bibnamefont {Schweizer}},
  \bibinfo {author} {\bibfnamefont {M.}~\bibnamefont {Atala}}, \bibinfo
  {author} {\bibfnamefont {J.~T.}\ \bibnamefont {Barreiro}}, \bibinfo {author}
  {\bibfnamefont {S.}~\bibnamefont {Nascimb{\`e}ne}}, \bibinfo {author}
  {\bibfnamefont {N.~R.}\ \bibnamefont {Cooper}}, \bibinfo {author}
  {\bibfnamefont {I.}~\bibnamefont {Bloch}}, \ and\ \bibinfo {author}
  {\bibfnamefont {N.}~\bibnamefont {Goldman}}} (\bibinfo {year} {2015}),\
  \bibfield  {title} {\enquote {\bibinfo {title} {Measuring the {C}hern number
  of {H}ofstadter bands with ultracold bosonic atoms},}\ }\href {\doibase
  10.1038/nphys3171} {\bibfield  {journal} {\bibinfo  {journal} {Nature
  Physics}\ }\textbf {\bibinfo {volume} {111}},\ \bibinfo {pages}
  {162--166}}\BibitemShut {NoStop}%
\bibitem [{\citenamefont {{Aidelsburger}}\ \emph {et~al.}(2017)\citenamefont
  {{Aidelsburger}}, \citenamefont {{Nascimbene}},\ and\ \citenamefont
  {{Goldman}}}]{Aidelsburger:2017}%
  \BibitemOpen
  \bibfield  {author} {\bibinfo {author} {\bibnamefont {{Aidelsburger}},
  \bibfnamefont {M}}, \bibinfo {author} {\bibfnamefont {S.}~\bibnamefont
  {{Nascimbene}}}, \ and\ \bibinfo {author} {\bibfnamefont {N.}~\bibnamefont
  {{Goldman}}}} (\bibinfo {year} {2017}),\ \bibfield  {title} {\enquote
  {\bibinfo {title} {{Artificial gauge fields in materials and engineered
  systems}},}\ }\href@noop {} {\bibfield  {journal} {\bibinfo  {journal} {ArXiv
  e-prints}\ }}\Eprint {http://arxiv.org/abs/1710.00851} {arXiv:1710.00851}
  \BibitemShut {NoStop}%
\bibitem [{\citenamefont {Alba}\ \emph {et~al.}(2011)\citenamefont {Alba},
  \citenamefont {Fernandez-Gonzalvo}, \citenamefont {Mur-Petit}, \citenamefont
  {Pachos},\ and\ \citenamefont {Garcia-Ripoll}}]{Alba:2011}%
  \BibitemOpen
  \bibfield  {author} {\bibinfo {author} {\bibnamefont {Alba}, \bibfnamefont
  {E}}, \bibinfo {author} {\bibfnamefont {X.}~\bibnamefont
  {Fernandez-Gonzalvo}}, \bibinfo {author} {\bibfnamefont {J.}~\bibnamefont
  {Mur-Petit}}, \bibinfo {author} {\bibfnamefont {J.~K.}\ \bibnamefont
  {Pachos}}, \ and\ \bibinfo {author} {\bibfnamefont {J.~J.}\ \bibnamefont
  {Garcia-Ripoll}}} (\bibinfo {year} {2011}),\ \bibfield  {title} {\enquote
  {\bibinfo {title} {Seeing topological order in time-of-flight
  measurements},}\ }\href {\doibase 10.1103/PhysRevLett.107.235301} {\bibfield
  {journal} {\bibinfo  {journal} {Phys. Rev. Lett.}\ }\textbf {\bibinfo
  {volume} {107}},\ \bibinfo {pages} {235301}}\BibitemShut {NoStop}%
\bibitem [{\citenamefont {Alexandradinata}\ \emph {et~al.}(2014)\citenamefont
  {Alexandradinata}, \citenamefont {Dai},\ and\ \citenamefont
  {Bernevig}}]{Alexandradinata:2014}%
  \BibitemOpen
  \bibfield  {author} {\bibinfo {author} {\bibnamefont {Alexandradinata},
  \bibfnamefont {A}}, \bibinfo {author} {\bibfnamefont {Xi}~\bibnamefont
  {Dai}}, \ and\ \bibinfo {author} {\bibfnamefont {B.~Andrei}\ \bibnamefont
  {Bernevig}}} (\bibinfo {year} {2014}),\ \bibfield  {title} {\enquote
  {\bibinfo {title} {Wilson-loop characterization of inversion-symmetric
  topological insulators},}\ }\href {\doibase 10.1103/PhysRevB.89.155114}
  {\bibfield  {journal} {\bibinfo  {journal} {Phys. Rev. B}\ }\textbf {\bibinfo
  {volume} {89}},\ \bibinfo {pages} {155114}}\BibitemShut {NoStop}%
\bibitem [{\citenamefont {Altland}\ and\ \citenamefont
  {Zirnbauer}(1997)}]{Altland1997}%
  \BibitemOpen
  \bibfield  {author} {\bibinfo {author} {\bibnamefont {Altland}, \bibfnamefont
  {Alexander}}, \ and\ \bibinfo {author} {\bibfnamefont {Martin~R}\
  \bibnamefont {Zirnbauer}}} (\bibinfo {year} {1997}),\ \bibfield  {title}
  {\enquote {\bibinfo {title} {Nonstandard symmetry classes in mesoscopic
  normal-superconducting hybrid structures},}\ }\href {\doibase
  10.1103/PhysRevB.55.1142} {\bibfield  {journal} {\bibinfo  {journal} {Phys.
  Rev. B}\ }\textbf {\bibinfo {volume} {55}}~(\bibinfo {number} {2}),\ \bibinfo
  {pages} {1142}}\BibitemShut {NoStop}%
\bibitem [{\citenamefont {Anderson}\ \emph {et~al.}(2013)\citenamefont
  {Anderson}, \citenamefont {Spielman},\ and\ \citenamefont
  {Juzeli{\=u}nas}}]{Anderson2013a}%
  \BibitemOpen
  \bibfield  {author} {\bibinfo {author} {\bibnamefont {Anderson},
  \bibfnamefont {Brandon~M}}, \bibinfo {author} {\bibfnamefont {I.~B.}\
  \bibnamefont {Spielman}}, \ and\ \bibinfo {author} {\bibfnamefont
  {Gediminas}\ \bibnamefont {Juzeli{\=u}nas}}} (\bibinfo {year} {2013}),\
  \bibfield  {title} {\enquote {\bibinfo {title} {{Magnetically Generated
  Spin-Orbit Coupling for Ultracold Atoms}},}\ }\href {\doibase
  10.1103/PhysRevLett.111.125301} {\bibfield  {journal} {\bibinfo  {journal}
  {Phys. Rev. Lett.}\ }\textbf {\bibinfo {volume} {111}}~(\bibinfo {number}
  {12}),\ \bibinfo {pages} {125301}}\BibitemShut {NoStop}%
\bibitem [{\citenamefont {Arun}\ \emph {et~al.}(2016)\citenamefont {Arun},
  \citenamefont {Sohal}, \citenamefont {Hickey},\ and\ \citenamefont
  {Paramekanti}}]{PhysRevB.93.115110}%
  \BibitemOpen
  \bibfield  {author} {\bibinfo {author} {\bibnamefont {Arun}, \bibfnamefont
  {V~S}}, \bibinfo {author} {\bibfnamefont {R.}~\bibnamefont {Sohal}}, \bibinfo
  {author} {\bibfnamefont {C.}~\bibnamefont {Hickey}}, \ and\ \bibinfo {author}
  {\bibfnamefont {A.}~\bibnamefont {Paramekanti}}} (\bibinfo {year} {2016}),\
  \bibfield  {title} {\enquote {\bibinfo {title} {Mean field study of the
  topological {H}aldane-{H}ubbard model of spin-$\frac{1}{2}$ fermions},}\
  }\href {\doibase 10.1103/PhysRevB.93.115110} {\bibfield  {journal} {\bibinfo
  {journal} {Phys. Rev. B}\ }\textbf {\bibinfo {volume} {93}},\ \bibinfo
  {pages} {115110}}\BibitemShut {NoStop}%
\bibitem [{\citenamefont {Asb\'oth}\ \emph {et~al.}(2014)\citenamefont
  {Asb\'oth}, \citenamefont {Tarasinski},\ and\ \citenamefont
  {Delplace}}]{PhysRevB.90.125143}%
  \BibitemOpen
  \bibfield  {author} {\bibinfo {author} {\bibnamefont {Asb\'oth},
  \bibfnamefont {J~K}}, \bibinfo {author} {\bibfnamefont {B.}~\bibnamefont
  {Tarasinski}}, \ and\ \bibinfo {author} {\bibfnamefont {P.}~\bibnamefont
  {Delplace}}} (\bibinfo {year} {2014}),\ \bibfield  {title} {\enquote
  {\bibinfo {title} {Chiral symmetry and bulk-boundary correspondence in
  periodically driven one-dimensional systems},}\ }\href {\doibase
  10.1103/PhysRevB.90.125143} {\bibfield  {journal} {\bibinfo  {journal} {Phys.
  Rev. B}\ }\textbf {\bibinfo {volume} {90}},\ \bibinfo {pages}
  {125143}}\BibitemShut {NoStop}%
\bibitem [{\citenamefont {Ashcroft}\ and\ \citenamefont
  {Mermin}(1976)}]{Ashcroft:1976}%
  \BibitemOpen
  \bibfield  {author} {\bibinfo {author} {\bibnamefont {Ashcroft},
  \bibfnamefont {N~W}}, \ and\ \bibinfo {author} {\bibfnamefont {N.~D.}\
  \bibnamefont {Mermin}}} (\bibinfo {year} {1976}),\ \href@noop {} {\emph
  {\bibinfo {title} {Solid State Physics}}}\ (\bibinfo  {publisher} {Holt,
  Rinehardt and Winston},\ \bibinfo {address} {New York})\BibitemShut {NoStop}%
\bibitem [{\citenamefont {Ashoori}\ \emph {et~al.}(1992)\citenamefont
  {Ashoori}, \citenamefont {Stormer}, \citenamefont {Pfeiffer}, \citenamefont
  {Baldwin},\ and\ \citenamefont {West}}]{Ashoori1992}%
  \BibitemOpen
  \bibfield  {author} {\bibinfo {author} {\bibnamefont {Ashoori}, \bibfnamefont
  {R~C}}, \bibinfo {author} {\bibfnamefont {H.~L.}\ \bibnamefont {Stormer}},
  \bibinfo {author} {\bibfnamefont {L.~N.}\ \bibnamefont {Pfeiffer}}, \bibinfo
  {author} {\bibfnamefont {K.~W.}\ \bibnamefont {Baldwin}}, \ and\ \bibinfo
  {author} {\bibfnamefont {K.}~\bibnamefont {West}}} (\bibinfo {year} {1992}),\
  \bibfield  {title} {\enquote {\bibinfo {title} {{Edge magnetoplasmons in the
  time domain}},}\ }\href {\doibase 10.1103/PhysRevB.45.3894} {\bibfield
  {journal} {\bibinfo  {journal} {Phys. Rev. B}\ }\textbf {\bibinfo {volume}
  {45}}~(\bibinfo {number} {7}),\ \bibinfo {pages} {3894--3897}}\BibitemShut
  {NoStop}%
\bibitem [{\citenamefont {Atala}\ \emph {et~al.}(2013)\citenamefont {Atala},
  \citenamefont {Aidelsburger}, \citenamefont {Barreiro}, \citenamefont
  {Abanin}, \citenamefont {Kitagawa}, \citenamefont {Demler},\ and\
  \citenamefont {Bloch}}]{Atala2013}%
  \BibitemOpen
  \bibfield  {author} {\bibinfo {author} {\bibnamefont {Atala}, \bibfnamefont
  {Marcos}}, \bibinfo {author} {\bibfnamefont {Monika}\ \bibnamefont
  {Aidelsburger}}, \bibinfo {author} {\bibfnamefont {Julio~T.}\ \bibnamefont
  {Barreiro}}, \bibinfo {author} {\bibfnamefont {Dmitry}\ \bibnamefont
  {Abanin}}, \bibinfo {author} {\bibfnamefont {Takuya}\ \bibnamefont
  {Kitagawa}}, \bibinfo {author} {\bibfnamefont {Eugene}\ \bibnamefont
  {Demler}}, \ and\ \bibinfo {author} {\bibfnamefont {Immanuel}\ \bibnamefont
  {Bloch}}} (\bibinfo {year} {2013}),\ \bibfield  {title} {\enquote {\bibinfo
  {title} {Direct measurement of the {Z}ak phase in topological {B}loch
  bands},}\ }\href {http://dx.doi.org/10.1038/nphys2790} {\bibfield  {journal}
  {\bibinfo  {journal} {Nat Phys}\ }\textbf {\bibinfo {volume} {9}}~(\bibinfo
  {number} {12}),\ \bibinfo {pages} {795--800}}\BibitemShut {NoStop}%
\bibitem [{\citenamefont {Atala}\ \emph {et~al.}(2014)\citenamefont {Atala},
  \citenamefont {Aidelsburger}, \citenamefont {Lohse}, \citenamefont
  {Barreiro}, \citenamefont {Paredes},\ and\ \citenamefont
  {Bloch}}]{Atala2014}%
  \BibitemOpen
  \bibfield  {author} {\bibinfo {author} {\bibnamefont {Atala}, \bibfnamefont
  {Marcos}}, \bibinfo {author} {\bibfnamefont {Monika}\ \bibnamefont
  {Aidelsburger}}, \bibinfo {author} {\bibfnamefont {Michael}\ \bibnamefont
  {Lohse}}, \bibinfo {author} {\bibfnamefont {Julio~T.}\ \bibnamefont
  {Barreiro}}, \bibinfo {author} {\bibfnamefont {Belen}\ \bibnamefont
  {Paredes}}, \ and\ \bibinfo {author} {\bibfnamefont {Immanuel}\ \bibnamefont
  {Bloch}}} (\bibinfo {year} {2014}),\ \bibfield  {title} {\enquote {\bibinfo
  {title} {Observation of chiral currents with ultracold atoms in bosonic
  ladders},}\ }\href {http://dx.doi.org/10.1038/nphys2998} {\bibfield
  {journal} {\bibinfo  {journal} {Nat Phys}\ }\textbf {\bibinfo {volume}
  {10}}~(\bibinfo {number} {8}),\ \bibinfo {pages} {588--593}}\BibitemShut
  {NoStop}%
\bibitem [{\citenamefont {Bakr}\ \emph {et~al.}(2009)\citenamefont {Bakr},
  \citenamefont {Gillen}, \citenamefont {Peng}, \citenamefont {F{\"o}lling},\
  and\ \citenamefont {Greiner}}]{Bakr2009}%
  \BibitemOpen
  \bibfield  {author} {\bibinfo {author} {\bibnamefont {Bakr}, \bibfnamefont
  {Waseem~S}}, \bibinfo {author} {\bibfnamefont {Jonathon~I.}\ \bibnamefont
  {Gillen}}, \bibinfo {author} {\bibfnamefont {Amy}\ \bibnamefont {Peng}},
  \bibinfo {author} {\bibfnamefont {Simon}\ \bibnamefont {F{\"o}lling}}, \ and\
  \bibinfo {author} {\bibfnamefont {Markus}\ \bibnamefont {Greiner}}} (\bibinfo
  {year} {2009}),\ \bibfield  {title} {\enquote {\bibinfo {title} {A quantum
  gas microscope for detecting single atoms in a hubbard-regime optical
  lattice},}\ }\href {\doibase 10.1038/nature08482} {\bibfield  {journal}
  {\bibinfo  {journal} {Nature}\ }\textbf {\bibinfo {volume} {462}},\ \bibinfo
  {pages} {74 EP --}}\BibitemShut {NoStop}%
\bibitem [{\citenamefont {Baranov}\ \emph {et~al.}(2012)\citenamefont
  {Baranov}, \citenamefont {Dalmonte}, \citenamefont {Pupillo},\ and\
  \citenamefont {Zoller}}]{baranov2012condensed}%
  \BibitemOpen
  \bibfield  {author} {\bibinfo {author} {\bibnamefont {Baranov}, \bibfnamefont
  {Mikhail~A}}, \bibinfo {author} {\bibfnamefont {Marcello}\ \bibnamefont
  {Dalmonte}}, \bibinfo {author} {\bibfnamefont {Guido}\ \bibnamefont
  {Pupillo}}, \ and\ \bibinfo {author} {\bibfnamefont {Peter}\ \bibnamefont
  {Zoller}}} (\bibinfo {year} {2012}),\ \bibfield  {title} {\enquote {\bibinfo
  {title} {Condensed matter theory of dipolar quantum gases},}\ }\href
  {\doibase 10.1021/cr2003568} {\bibfield  {journal} {\bibinfo  {journal}
  {Chemical Reviews}\ }\textbf {\bibinfo {volume} {112}}~(\bibinfo {number}
  {9}),\ \bibinfo {pages} {5012--5061}}\BibitemShut {NoStop}%
\bibitem [{\citenamefont {Bardyn}\ \emph {et~al.}(2013)\citenamefont {Bardyn},
  \citenamefont {Baranov}, \citenamefont {Kraus}, \citenamefont {Rico},
  \citenamefont {{\.I}mamo{\u g}lu}, \citenamefont {Zoller},\ and\
  \citenamefont {Diehl}}]{Bardyn:2013}%
  \BibitemOpen
  \bibfield  {author} {\bibinfo {author} {\bibnamefont {Bardyn}, \bibfnamefont
  {C-E}}, \bibinfo {author} {\bibfnamefont {M~A}\ \bibnamefont {Baranov}},
  \bibinfo {author} {\bibfnamefont {C~V}\ \bibnamefont {Kraus}}, \bibinfo
  {author} {\bibfnamefont {E}~\bibnamefont {Rico}}, \bibinfo {author}
  {\bibfnamefont {A}~\bibnamefont {{\.I}mamo{\u g}lu}}, \bibinfo {author}
  {\bibfnamefont {P}~\bibnamefont {Zoller}}, \ and\ \bibinfo {author}
  {\bibfnamefont {S}~\bibnamefont {Diehl}}} (\bibinfo {year} {2013}),\
  \bibfield  {title} {\enquote {\bibinfo {title} {Topology by dissipation},}\
  }\href {http://stacks.iop.org/1367-2630/15/i=8/a=085001} {\bibfield
  {journal} {\bibinfo  {journal} {New Journal of Physics}\ }\textbf {\bibinfo
  {volume} {15}}~(\bibinfo {number} {8}),\ \bibinfo {pages}
  {085001}}\BibitemShut {NoStop}%
\bibitem [{\citenamefont {Bardyn}\ \emph {et~al.}(2012)\citenamefont {Bardyn},
  \citenamefont {Baranov}, \citenamefont {Rico}, \citenamefont
  {{\.I}mamo{\u{g}}lu}, \citenamefont {Zoller},\ and\ \citenamefont
  {Diehl}}]{bardyn2012majorana}%
  \BibitemOpen
  \bibfield  {author} {\bibinfo {author} {\bibnamefont {Bardyn}, \bibfnamefont
  {C-E}}, \bibinfo {author} {\bibfnamefont {MA}~\bibnamefont {Baranov}},
  \bibinfo {author} {\bibfnamefont {E}~\bibnamefont {Rico}}, \bibinfo {author}
  {\bibfnamefont {A}~\bibnamefont {{\.I}mamo{\u{g}}lu}}, \bibinfo {author}
  {\bibfnamefont {P}~\bibnamefont {Zoller}}, \ and\ \bibinfo {author}
  {\bibfnamefont {S}~\bibnamefont {Diehl}}} (\bibinfo {year} {2012}),\
  \bibfield  {title} {\enquote {\bibinfo {title} {Majorana modes in
  driven-dissipative atomic superfluids with a zero {C}hern number},}\ }\href
  {\doibase 10.1103/PhysRevLett.109.130402} {\bibfield  {journal} {\bibinfo
  {journal} {Phys. Rev. Lett.}\ }\textbf {\bibinfo {volume} {109}}~(\bibinfo
  {number} {13}),\ \bibinfo {pages} {130402}}\BibitemShut {NoStop}%
\bibitem [{\citenamefont {Baur}\ and\ \citenamefont
  {Cooper}(2013)}]{PhysRevA.88.033603}%
  \BibitemOpen
  \bibfield  {author} {\bibinfo {author} {\bibnamefont {Baur}, \bibfnamefont
  {Stefan~K}}, \ and\ \bibinfo {author} {\bibfnamefont {Nigel~R.}\ \bibnamefont
  {Cooper}}} (\bibinfo {year} {2013}),\ \bibfield  {title} {\enquote {\bibinfo
  {title} {Adiabatic preparation of vortex lattices},}\ }\href {\doibase
  10.1103/PhysRevA.88.033603} {\bibfield  {journal} {\bibinfo  {journal} {Phys.
  Rev. A}\ }\textbf {\bibinfo {volume} {88}},\ \bibinfo {pages}
  {033603}}\BibitemShut {NoStop}%
\bibitem [{\citenamefont {Behrmann}\ \emph {et~al.}(2016)\citenamefont
  {Behrmann}, \citenamefont {Liu},\ and\ \citenamefont
  {Bergholtz}}]{PhysRevLett.116.216802}%
  \BibitemOpen
  \bibfield  {author} {\bibinfo {author} {\bibnamefont {Behrmann},
  \bibfnamefont {J\"org}}, \bibinfo {author} {\bibfnamefont {Zhao}\
  \bibnamefont {Liu}}, \ and\ \bibinfo {author} {\bibfnamefont {Emil~J.}\
  \bibnamefont {Bergholtz}}} (\bibinfo {year} {2016}),\ \bibfield  {title}
  {\enquote {\bibinfo {title} {Model fractional {C}hern insulators},}\ }\href
  {\doibase 10.1103/PhysRevLett.116.216802} {\bibfield  {journal} {\bibinfo
  {journal} {Phys. Rev. Lett.}\ }\textbf {\bibinfo {volume} {116}},\ \bibinfo
  {pages} {216802}}\BibitemShut {NoStop}%
\bibitem [{\citenamefont {Bena}\ and\ \citenamefont
  {Montambaux}(2009)}]{Bena2009}%
  \BibitemOpen
  \bibfield  {author} {\bibinfo {author} {\bibnamefont {Bena}, \bibfnamefont
  {Cristina}}, \ and\ \bibinfo {author} {\bibfnamefont {Gilles}\ \bibnamefont
  {Montambaux}}} (\bibinfo {year} {2009}),\ \bibfield  {title} {\enquote
  {\bibinfo {title} {Remarks on the tight-binding model of graphene},}\ }\href
  {http://stacks.iop.org/1367-2630/11/i=9/a=095003} {\bibfield  {journal}
  {\bibinfo  {journal} {New Journal of Physics}\ }\textbf {\bibinfo {volume}
  {11}}~(\bibinfo {number} {9}),\ \bibinfo {pages} {095003}}\BibitemShut
  {NoStop}%
\bibitem [{\citenamefont {{B\'eri}}\ and\ \citenamefont
  {Cooper}(2011)}]{PhysRevLett.107.145301}%
  \BibitemOpen
  \bibfield  {author} {\bibinfo {author} {\bibnamefont {{B\'eri}},
  \bibfnamefont {B}}, \ and\ \bibinfo {author} {\bibfnamefont {N.~R.}\
  \bibnamefont {Cooper}}} (\bibinfo {year} {2011}),\ \bibfield  {title}
  {\enquote {\bibinfo {title} {{Z}$_2$ topological insulators in ultracold
  atomic gases},}\ }\href {\doibase 10.1103/PhysRevLett.107.145301} {\bibfield
  {journal} {\bibinfo  {journal} {Phys. Rev. Lett.}\ }\textbf {\bibinfo
  {volume} {107}},\ \bibinfo {pages} {145301}}\BibitemShut {NoStop}%
\bibitem [{\citenamefont {Bermudez}\ and\ \citenamefont
  {Porras}(2015)}]{bermudezporras}%
  \BibitemOpen
  \bibfield  {author} {\bibinfo {author} {\bibnamefont {Bermudez},
  \bibfnamefont {Alejandro}}, \ and\ \bibinfo {author} {\bibfnamefont {Diego}\
  \bibnamefont {Porras}}} (\bibinfo {year} {2015}),\ \bibfield  {title}
  {\enquote {\bibinfo {title} {Interaction-dependent photon-assisted tunneling
  in optical lattices: a quantum simulator of strongly-correlated electrons and
  dynamical gauge fields},}\ }\href
  {http://stacks.iop.org/1367-2630/17/i=10/a=103021} {\bibfield  {journal}
  {\bibinfo  {journal} {New Journal of Physics}\ }\textbf {\bibinfo {volume}
  {17}}~(\bibinfo {number} {10}),\ \bibinfo {pages} {103021}}\BibitemShut
  {NoStop}%
\bibitem [{\citenamefont {Bernevig}\ and\ \citenamefont
  {Hughes}(2013)}]{bernevigbook}%
  \BibitemOpen
  \bibfield  {author} {\bibinfo {author} {\bibnamefont {Bernevig},
  \bibfnamefont {B~Andrei}}, \ and\ \bibinfo {author} {\bibfnamefont
  {Taylor~L.}\ \bibnamefont {Hughes}}} (\bibinfo {year} {2013}),\ \href@noop {}
  {\emph {\bibinfo {title} {Topological insulators and topological
  superconductors}}}\ (\bibinfo  {publisher} {Princeton university
  press})\BibitemShut {NoStop}%
\bibitem [{\citenamefont {Berry}(1984)}]{Berry}%
  \BibitemOpen
  \bibfield  {author} {\bibinfo {author} {\bibnamefont {Berry}, \bibfnamefont
  {M~V}}} (\bibinfo {year} {1984}),\ \bibfield  {title} {\enquote {\bibinfo
  {title} {Quantal phase factors accompanying adiabatic changes},}\ }\href
  {\doibase 10.1098/rspa.1984.0023} {\bibfield  {journal} {\bibinfo  {journal}
  {Proceedings of the Royal Society of London. A. Mathematical and Physical
  Sciences}\ }\textbf {\bibinfo {volume} {392}}~(\bibinfo {number} {1802}),\
  \bibinfo {pages} {45--57}}\BibitemShut {NoStop}%
\bibitem [{\citenamefont {Bianco}\ and\ \citenamefont
  {Resta}(2011)}]{Bianco2011}%
  \BibitemOpen
  \bibfield  {author} {\bibinfo {author} {\bibnamefont {Bianco}, \bibfnamefont
  {Raffaello}}, \ and\ \bibinfo {author} {\bibfnamefont {Raffaele}\
  \bibnamefont {Resta}}} (\bibinfo {year} {2011}),\ \bibfield  {title}
  {\enquote {\bibinfo {title} {Mapping topological order in coordinate
  space},}\ }\href {\doibase 10.1103/PhysRevB.84.241106} {\bibfield  {journal}
  {\bibinfo  {journal} {Phys. Rev. B}\ }\textbf {\bibinfo {volume} {84}},\
  \bibinfo {pages} {241106}}\BibitemShut {NoStop}%
\bibitem [{\citenamefont {Bilitewski}\ and\ \citenamefont
  {Cooper}(2015{\natexlab{a}})}]{bilitewskiharper2015}%
  \BibitemOpen
  \bibfield  {author} {\bibinfo {author} {\bibnamefont {Bilitewski},
  \bibfnamefont {Thomas}}, \ and\ \bibinfo {author} {\bibfnamefont {Nigel~R.}\
  \bibnamefont {Cooper}}} (\bibinfo {year} {2015}{\natexlab{a}}),\ \bibfield
  {title} {\enquote {\bibinfo {title} {Population dynamics in a {F}loquet
  realization of the {H}arper-{H}ofstadter {H}amiltonian},}\ }\href {\doibase
  10.1103/PhysRevA.91.063611} {\bibfield  {journal} {\bibinfo  {journal} {Phys.
  Rev. A}\ }\textbf {\bibinfo {volume} {91}},\ \bibinfo {pages}
  {063611}}\BibitemShut {NoStop}%
\bibitem [{\citenamefont {Bilitewski}\ and\ \citenamefont
  {Cooper}(2015{\natexlab{b}})}]{bilitewski2015}%
  \BibitemOpen
  \bibfield  {author} {\bibinfo {author} {\bibnamefont {Bilitewski},
  \bibfnamefont {Thomas}}, \ and\ \bibinfo {author} {\bibfnamefont {Nigel~R.}\
  \bibnamefont {Cooper}}} (\bibinfo {year} {2015}{\natexlab{b}}),\ \bibfield
  {title} {\enquote {\bibinfo {title} {Scattering theory for {F}loquet-{B}loch
  states},}\ }\href {\doibase 10.1103/PhysRevA.91.033601} {\bibfield  {journal}
  {\bibinfo  {journal} {Phys. Rev. A}\ }\textbf {\bibinfo {volume} {91}},\
  \bibinfo {pages} {033601}}\BibitemShut {NoStop}%
\bibitem [{\citenamefont {Bilitewski}\ and\ \citenamefont
  {Cooper}(2016)}]{bilitewskicoopersd}%
  \BibitemOpen
  \bibfield  {author} {\bibinfo {author} {\bibnamefont {Bilitewski},
  \bibfnamefont {Thomas}}, \ and\ \bibinfo {author} {\bibfnamefont {Nigel~R.}\
  \bibnamefont {Cooper}}} (\bibinfo {year} {2016}),\ \bibfield  {title}
  {\enquote {\bibinfo {title} {Synthetic dimensions in the strong-coupling
  limit: Supersolids and pair superfluids},}\ }\href {\doibase
  10.1103/PhysRevA.94.023630} {\bibfield  {journal} {\bibinfo  {journal} {Phys.
  Rev. A}\ }\textbf {\bibinfo {volume} {94}},\ \bibinfo {pages}
  {023630}}\BibitemShut {NoStop}%
\bibitem [{\citenamefont {Boada}\ \emph {et~al.}(2012)\citenamefont {Boada},
  \citenamefont {Celi}, \citenamefont {Latorre},\ and\ \citenamefont
  {Lewenstein}}]{Boada2012}%
  \BibitemOpen
  \bibfield  {author} {\bibinfo {author} {\bibnamefont {Boada}, \bibfnamefont
  {O}}, \bibinfo {author} {\bibfnamefont {A}~\bibnamefont {Celi}}, \bibinfo
  {author} {\bibfnamefont {JI}~\bibnamefont {Latorre}}, \ and\ \bibinfo
  {author} {\bibfnamefont {M}~\bibnamefont {Lewenstein}}} (\bibinfo {year}
  {2012}),\ \bibfield  {title} {\enquote {\bibinfo {title} {Quantum simulation
  of an extra dimension},}\ }\href {\doibase 10.1103/PhysRevLett.108.133001}
  {\bibfield  {journal} {\bibinfo  {journal} {Phys. Rev. Lett.}\ }\textbf
  {\bibinfo {volume} {108}}~(\bibinfo {number} {13}),\ \bibinfo {pages}
  {133001}}\BibitemShut {NoStop}%
\bibitem [{\citenamefont {Breit}\ and\ \citenamefont {Rabi}(1931)}]{Breit1931}%
  \BibitemOpen
  \bibfield  {author} {\bibinfo {author} {\bibnamefont {Breit}, \bibfnamefont
  {G}}, \ and\ \bibinfo {author} {\bibfnamefont {I.~I.}\ \bibnamefont {Rabi}}}
  (\bibinfo {year} {1931}),\ \bibfield  {title} {\enquote {\bibinfo {title}
  {{Measurement of Nuclear Spin}},}\ }\href {\doibase
  10.1103/PhysRev.38.2082.2} {\bibfield  {journal} {\bibinfo  {journal} {Phys.
  Rev.}\ }\textbf {\bibinfo {volume} {38}}~(\bibinfo {number} {11}),\ \bibinfo
  {pages} {2082--2083}}\BibitemShut {NoStop}%
\bibitem [{\citenamefont {Bromley}\ \emph {et~al.}(2018)\citenamefont
  {Bromley}, \citenamefont {Kolkowitz}, \citenamefont {Bothwell}, \citenamefont
  {Kedar}, \citenamefont {Safavi-Naini}, \citenamefont {Wall}, \citenamefont
  {Salomon}, \citenamefont {Rey},\ and\ \citenamefont {Ye}}]{Bromley2018}%
  \BibitemOpen
  \bibfield  {author} {\bibinfo {author} {\bibnamefont {Bromley}, \bibfnamefont
  {S~L}}, \bibinfo {author} {\bibfnamefont {S.}~\bibnamefont {Kolkowitz}},
  \bibinfo {author} {\bibfnamefont {T.}~\bibnamefont {Bothwell}}, \bibinfo
  {author} {\bibfnamefont {D.}~\bibnamefont {Kedar}}, \bibinfo {author}
  {\bibfnamefont {A.}~\bibnamefont {Safavi-Naini}}, \bibinfo {author}
  {\bibfnamefont {M.~L.}\ \bibnamefont {Wall}}, \bibinfo {author}
  {\bibfnamefont {C.}~\bibnamefont {Salomon}}, \bibinfo {author} {\bibfnamefont
  {A.~M.}\ \bibnamefont {Rey}}, \ and\ \bibinfo {author} {\bibfnamefont
  {J.}~\bibnamefont {Ye}}} (\bibinfo {year} {2018}),\ \bibfield  {title}
  {\enquote {\bibinfo {title} {Dynamics of interacting fermions under
  spin-orbit coupling in an optical lattice clock},}\ }\href {\doibase
  10.1038/s41567-017-0029-0} {\bibfield  {journal} {\bibinfo  {journal} {Nature
  Physics}\ }\textbf {\bibinfo {volume} {14}}~(\bibinfo {number} {4}),\
  \bibinfo {pages} {399--404}}\BibitemShut {NoStop}%
\bibitem [{\citenamefont {Budich}\ and\ \citenamefont
  {Diehl}(2015)}]{budich2015topology}%
  \BibitemOpen
  \bibfield  {author} {\bibinfo {author} {\bibnamefont {Budich}, \bibfnamefont
  {Jan~Carl}}, \ and\ \bibinfo {author} {\bibfnamefont {Sebastian}\
  \bibnamefont {Diehl}}} (\bibinfo {year} {2015}),\ \bibfield  {title}
  {\enquote {\bibinfo {title} {Topology of density matrices},}\ }\href
  {\doibase 10.1103/PhysRevB.91.165140} {\bibfield  {journal} {\bibinfo
  {journal} {Phys. Rev. B}\ }\textbf {\bibinfo {volume} {91}}~(\bibinfo
  {number} {16}),\ \bibinfo {pages} {165140}}\BibitemShut {NoStop}%
\bibitem [{\citenamefont {Budich}\ \emph {et~al.}(2015)\citenamefont {Budich},
  \citenamefont {Zoller},\ and\ \citenamefont {Diehl}}]{budich2015dissipative}%
  \BibitemOpen
  \bibfield  {author} {\bibinfo {author} {\bibnamefont {Budich}, \bibfnamefont
  {Jan~Carl}}, \bibinfo {author} {\bibfnamefont {Peter}\ \bibnamefont
  {Zoller}}, \ and\ \bibinfo {author} {\bibfnamefont {Sebastian}\ \bibnamefont
  {Diehl}}} (\bibinfo {year} {2015}),\ \bibfield  {title} {\enquote {\bibinfo
  {title} {Dissipative preparation of {C}hern insulators},}\ }\href {\doibase
  10.1103/PhysRevA.91.042117} {\bibfield  {journal} {\bibinfo  {journal} {Phys.
  Rev. A}\ }\textbf {\bibinfo {volume} {91}}~(\bibinfo {number} {4}),\ \bibinfo
  {pages} {042117}}\BibitemShut {NoStop}%
\bibitem [{\citenamefont {Bukov}\ \emph {et~al.}(2015)\citenamefont {Bukov},
  \citenamefont {Gopalakrishnan}, \citenamefont {Knap},\ and\ \citenamefont
  {Demler}}]{PhysRevLett.115.205301}%
  \BibitemOpen
  \bibfield  {author} {\bibinfo {author} {\bibnamefont {Bukov}, \bibfnamefont
  {Marin}}, \bibinfo {author} {\bibfnamefont {Sarang}\ \bibnamefont
  {Gopalakrishnan}}, \bibinfo {author} {\bibfnamefont {Michael}\ \bibnamefont
  {Knap}}, \ and\ \bibinfo {author} {\bibfnamefont {Eugene}\ \bibnamefont
  {Demler}}} (\bibinfo {year} {2015}),\ \bibfield  {title} {\enquote {\bibinfo
  {title} {Prethermal {F}loquet steady states and instabilities in the
  periodically driven, weakly interacting {B}ose-{H}ubbard model},}\ }\href
  {\doibase 10.1103/PhysRevLett.115.205301} {\bibfield  {journal} {\bibinfo
  {journal} {Phys. Rev. Lett.}\ }\textbf {\bibinfo {volume} {115}},\ \bibinfo
  {pages} {205301}}\BibitemShut {NoStop}%
\bibitem [{\citenamefont {Caio}\ \emph {et~al.}(2015)\citenamefont {Caio},
  \citenamefont {Cooper},\ and\ \citenamefont
  {Bhaseen}}]{PhysRevLett.115.236403}%
  \BibitemOpen
  \bibfield  {author} {\bibinfo {author} {\bibnamefont {Caio}, \bibfnamefont
  {M~D}}, \bibinfo {author} {\bibfnamefont {N.~R.}\ \bibnamefont {Cooper}}, \
  and\ \bibinfo {author} {\bibfnamefont {M.~J.}\ \bibnamefont {Bhaseen}}}
  (\bibinfo {year} {2015}),\ \bibfield  {title} {\enquote {\bibinfo {title}
  {Quantum quenches in {C}hern insulators},}\ }\href {\doibase
  10.1103/PhysRevLett.115.236403} {\bibfield  {journal} {\bibinfo  {journal}
  {Phys. Rev. Lett.}\ }\textbf {\bibinfo {volume} {115}},\ \bibinfo {pages}
  {236403}}\BibitemShut {NoStop}%
\bibitem [{\citenamefont {Caio}\ \emph {et~al.}(2016)\citenamefont {Caio},
  \citenamefont {Cooper},\ and\ \citenamefont {Bhaseen}}]{PhysRevB.94.155104}%
  \BibitemOpen
  \bibfield  {author} {\bibinfo {author} {\bibnamefont {Caio}, \bibfnamefont
  {M~D}}, \bibinfo {author} {\bibfnamefont {N.~R.}\ \bibnamefont {Cooper}}, \
  and\ \bibinfo {author} {\bibfnamefont {M.~J.}\ \bibnamefont {Bhaseen}}}
  (\bibinfo {year} {2016}),\ \bibfield  {title} {\enquote {\bibinfo {title}
  {Hall response and edge current dynamics in {C}hern insulators out of
  equilibrium},}\ }\href {\doibase 10.1103/PhysRevB.94.155104} {\bibfield
  {journal} {\bibinfo  {journal} {Phys. Rev. B}\ }\textbf {\bibinfo {volume}
  {94}},\ \bibinfo {pages} {155104}}\BibitemShut {NoStop}%
\bibitem [{\citenamefont {Calvanese~Strinati}\ \emph
  {et~al.}(2017)\citenamefont {Calvanese~Strinati}, \citenamefont {Cornfeld},
  \citenamefont {Rossini}, \citenamefont {Barbarino}, \citenamefont {Dalmonte},
  \citenamefont {Fazio}, \citenamefont {Sela},\ and\ \citenamefont
  {Mazza}}]{2016arXiv161206682C}%
  \BibitemOpen
  \bibfield  {author} {\bibinfo {author} {\bibnamefont {Calvanese~Strinati},
  \bibfnamefont {Marcello}}, \bibinfo {author} {\bibfnamefont {Eyal}\
  \bibnamefont {Cornfeld}}, \bibinfo {author} {\bibfnamefont {Davide}\
  \bibnamefont {Rossini}}, \bibinfo {author} {\bibfnamefont {Simone}\
  \bibnamefont {Barbarino}}, \bibinfo {author} {\bibfnamefont {Marcello}\
  \bibnamefont {Dalmonte}}, \bibinfo {author} {\bibfnamefont {Rosario}\
  \bibnamefont {Fazio}}, \bibinfo {author} {\bibfnamefont {Eran}\ \bibnamefont
  {Sela}}, \ and\ \bibinfo {author} {\bibfnamefont {Leonardo}\ \bibnamefont
  {Mazza}}} (\bibinfo {year} {2017}),\ \bibfield  {title} {\enquote {\bibinfo
  {title} {Laughlin-like states in bosonic and fermionic atomic synthetic
  ladders},}\ }\href {\doibase 10.1103/PhysRevX.7.021033} {\bibfield  {journal}
  {\bibinfo  {journal} {Phys. Rev. X}\ }\textbf {\bibinfo {volume} {7}},\
  \bibinfo {pages} {021033}}\BibitemShut {NoStop}%
\bibitem [{\citenamefont {Calvanese~Strinati}\ \emph
  {et~al.}(2016)\citenamefont {Calvanese~Strinati}, \citenamefont {Mazza},
  \citenamefont {Endres}, \citenamefont {Rossini},\ and\ \citenamefont
  {Fazio}}]{PhysRevB.94.024302}%
  \BibitemOpen
  \bibfield  {author} {\bibinfo {author} {\bibnamefont {Calvanese~Strinati},
  \bibfnamefont {Marcello}}, \bibinfo {author} {\bibfnamefont {Leonardo}\
  \bibnamefont {Mazza}}, \bibinfo {author} {\bibfnamefont {Manuel}\
  \bibnamefont {Endres}}, \bibinfo {author} {\bibfnamefont {Davide}\
  \bibnamefont {Rossini}}, \ and\ \bibinfo {author} {\bibfnamefont {Rosario}\
  \bibnamefont {Fazio}}} (\bibinfo {year} {2016}),\ \bibfield  {title}
  {\enquote {\bibinfo {title} {Destruction of string order after a quantum
  quench},}\ }\href {\doibase 10.1103/PhysRevB.94.024302} {\bibfield  {journal}
  {\bibinfo  {journal} {Phys. Rev. B}\ }\textbf {\bibinfo {volume} {94}},\
  \bibinfo {pages} {024302}}\BibitemShut {NoStop}%
\bibitem [{\citenamefont {Canovi}\ \emph {et~al.}(2016)\citenamefont {Canovi},
  \citenamefont {Kollar},\ and\ \citenamefont {Eckstein}}]{PhysRevE.93.012130}%
  \BibitemOpen
  \bibfield  {author} {\bibinfo {author} {\bibnamefont {Canovi}, \bibfnamefont
  {Elena}}, \bibinfo {author} {\bibfnamefont {Marcus}\ \bibnamefont {Kollar}},
  \ and\ \bibinfo {author} {\bibfnamefont {Martin}\ \bibnamefont {Eckstein}}}
  (\bibinfo {year} {2016}),\ \bibfield  {title} {\enquote {\bibinfo {title}
  {Stroboscopic prethermalization in weakly interacting periodically driven
  systems},}\ }\href {\doibase 10.1103/PhysRevE.93.012130} {\bibfield
  {journal} {\bibinfo  {journal} {Phys. Rev. E}\ }\textbf {\bibinfo {volume}
  {93}},\ \bibinfo {pages} {012130}}\BibitemShut {NoStop}%
\bibitem [{\citenamefont {Cazalilla}\ \emph {et~al.}(2005)\citenamefont
  {Cazalilla}, \citenamefont {Barber\'{a}n},\ and\ \citenamefont
  {Cooper}}]{cazalilla:121303}%
  \BibitemOpen
  \bibfield  {author} {\bibinfo {author} {\bibnamefont {Cazalilla},
  \bibfnamefont {M~A}}, \bibinfo {author} {\bibfnamefont {N.}~\bibnamefont
  {Barber\'{a}n}}, \ and\ \bibinfo {author} {\bibfnamefont {N.~R.}\
  \bibnamefont {Cooper}}} (\bibinfo {year} {2005}),\ \bibfield  {title}
  {\enquote {\bibinfo {title} {Edge excitations and topological order in a
  rotating {B}ose gas},}\ }\href {\doibase 10.1103/PhysRevB.71.121303}
  {\bibfield  {journal} {\bibinfo  {journal} {Phys. Rev. B}\ }\textbf {\bibinfo
  {volume} {71}}~(\bibinfo {number} {12}),\ \bibinfo {eid}
  {121303}}\BibitemShut {NoStop}%
\bibitem [{\citenamefont {Celi}\ \emph {et~al.}(2014)\citenamefont {Celi},
  \citenamefont {Massignan}, \citenamefont {Ruseckas}, \citenamefont {Goldman},
  \citenamefont {Spielman}, \citenamefont {Juzeli{\=u}nas},\ and\ \citenamefont
  {Lewenstein}}]{Celi:2014}%
  \BibitemOpen
  \bibfield  {author} {\bibinfo {author} {\bibnamefont {Celi}, \bibfnamefont
  {A}}, \bibinfo {author} {\bibfnamefont {P.}~\bibnamefont {Massignan}},
  \bibinfo {author} {\bibfnamefont {J.}~\bibnamefont {Ruseckas}}, \bibinfo
  {author} {\bibfnamefont {N.}~\bibnamefont {Goldman}}, \bibinfo {author}
  {\bibfnamefont {I.~B.}\ \bibnamefont {Spielman}}, \bibinfo {author}
  {\bibfnamefont {G.}~\bibnamefont {Juzeli{\=u}nas}}, \ and\ \bibinfo {author}
  {\bibfnamefont {M.}~\bibnamefont {Lewenstein}}} (\bibinfo {year} {2014}),\
  \bibfield  {title} {\enquote {\bibinfo {title} {Synthetic gauge fields in
  synthetic dimensions},}\ }\href {\doibase 10.1103/PhysRevLett.112.043001}
  {\bibfield  {journal} {\bibinfo  {journal} {Phys. Rev. Lett.}\ }\textbf
  {\bibinfo {volume} {112}},\ \bibinfo {pages} {043001}}\BibitemShut {NoStop}%
\bibitem [{\citenamefont {Chandran}\ and\ \citenamefont
  {Sondhi}(2016)}]{PhysRevB.93.174305}%
  \BibitemOpen
  \bibfield  {author} {\bibinfo {author} {\bibnamefont {Chandran},
  \bibfnamefont {Anushya}}, \ and\ \bibinfo {author} {\bibfnamefont {S.~L.}\
  \bibnamefont {Sondhi}}} (\bibinfo {year} {2016}),\ \bibfield  {title}
  {\enquote {\bibinfo {title} {Interaction-stabilized steady states in the
  driven {O}({$N$}) model},}\ }\href {\doibase 10.1103/PhysRevB.93.174305}
  {\bibfield  {journal} {\bibinfo  {journal} {Phys. Rev. B}\ }\textbf {\bibinfo
  {volume} {93}},\ \bibinfo {pages} {174305}}\BibitemShut {NoStop}%
\bibitem [{\citenamefont {Chen}\ \emph {et~al.}(2018)\citenamefont {Chen},
  \citenamefont {Lin}, \citenamefont {Chen}, \citenamefont {Chiu},
  \citenamefont {Wang}, \citenamefont {Chen}, \citenamefont {Huang},
  \citenamefont {Yip}, \citenamefont {Kawaguchi},\ and\ \citenamefont
  {Lin}}]{Chen2018}%
  \BibitemOpen
  \bibfield  {author} {\bibinfo {author} {\bibnamefont {Chen}, \bibfnamefont
  {H-R}}, \bibinfo {author} {\bibfnamefont {K.-Y.}\ \bibnamefont {Lin}},
  \bibinfo {author} {\bibfnamefont {P.-K.}\ \bibnamefont {Chen}}, \bibinfo
  {author} {\bibfnamefont {N.-C.}\ \bibnamefont {Chiu}}, \bibinfo {author}
  {\bibfnamefont {J.-B.}\ \bibnamefont {Wang}}, \bibinfo {author}
  {\bibfnamefont {C.-A.}\ \bibnamefont {Chen}}, \bibinfo {author}
  {\bibfnamefont {P.-P.}\ \bibnamefont {Huang}}, \bibinfo {author}
  {\bibfnamefont {S.-K.}\ \bibnamefont {Yip}}, \bibinfo {author} {\bibfnamefont
  {Yuki}\ \bibnamefont {Kawaguchi}}, \ and\ \bibinfo {author} {\bibfnamefont
  {Y.-J.}\ \bibnamefont {Lin}}} (\bibinfo {year} {2018}),\ \bibfield  {title}
  {\enquote {\bibinfo {title} {Spin--orbital-angular-momentum coupled
  {B}ose-{E}instein condensates},}\ }\href {\doibase
  10.1103/PhysRevLett.121.113204} {\bibfield  {journal} {\bibinfo  {journal}
  {Phys. Rev. Lett.}\ }\textbf {\bibinfo {volume} {121}},\ \bibinfo {pages}
  {113204}}\BibitemShut {NoStop}%
\bibitem [{\citenamefont {Chen}\ \emph {et~al.}(2010)\citenamefont {Chen},
  \citenamefont {Gu},\ and\ \citenamefont {Wen}}]{chen2010local}%
  \BibitemOpen
  \bibfield  {author} {\bibinfo {author} {\bibnamefont {Chen}, \bibfnamefont
  {Xie}}, \bibinfo {author} {\bibfnamefont {Zheng-Cheng}\ \bibnamefont {Gu}}, \
  and\ \bibinfo {author} {\bibfnamefont {Xiao-Gang}\ \bibnamefont {Wen}}}
  (\bibinfo {year} {2010}),\ \bibfield  {title} {\enquote {\bibinfo {title}
  {Local unitary transformation, long-range quantum entanglement, wave function
  renormalization, and topological order},}\ }\href {\doibase
  10.1103/PhysRevB.82.155138} {\bibfield  {journal} {\bibinfo  {journal} {Phys.
  Rev. B}\ }\textbf {\bibinfo {volume} {82}}~(\bibinfo {number} {15}),\
  \bibinfo {pages} {155138}}\BibitemShut {NoStop}%
\bibitem [{\citenamefont {Chiu}\ \emph {et~al.}(2016)\citenamefont {Chiu},
  \citenamefont {Teo}, \citenamefont {Schnyder},\ and\ \citenamefont
  {Ryu}}]{RevModPhys.88.035005}%
  \BibitemOpen
  \bibfield  {author} {\bibinfo {author} {\bibnamefont {Chiu}, \bibfnamefont
  {Ching-Kai}}, \bibinfo {author} {\bibfnamefont {Jeffrey C.~Y.}\ \bibnamefont
  {Teo}}, \bibinfo {author} {\bibfnamefont {Andreas~P.}\ \bibnamefont
  {Schnyder}}, \ and\ \bibinfo {author} {\bibfnamefont {Shinsei}\ \bibnamefont
  {Ryu}}} (\bibinfo {year} {2016}),\ \bibfield  {title} {\enquote {\bibinfo
  {title} {Classification of topological quantum matter with symmetries},}\
  }\href {\doibase 10.1103/RevModPhys.88.035005} {\bibfield  {journal}
  {\bibinfo  {journal} {Rev. Mod. Phys.}\ }\textbf {\bibinfo {volume} {88}},\
  \bibinfo {pages} {035005}}\BibitemShut {NoStop}%
\bibitem [{\citenamefont {Choudhury}\ and\ \citenamefont
  {Mueller}(2014)}]{PhysRevA.90.013621}%
  \BibitemOpen
  \bibfield  {author} {\bibinfo {author} {\bibnamefont {Choudhury},
  \bibfnamefont {Sayan}}, \ and\ \bibinfo {author} {\bibfnamefont {Erich~J.}\
  \bibnamefont {Mueller}}} (\bibinfo {year} {2014}),\ \bibfield  {title}
  {\enquote {\bibinfo {title} {Stability of a {F}loquet {B}ose-{E}instein
  condensate in a one-dimensional optical lattice},}\ }\href {\doibase
  10.1103/PhysRevA.90.013621} {\bibfield  {journal} {\bibinfo  {journal} {Phys.
  Rev. A}\ }\textbf {\bibinfo {volume} {90}},\ \bibinfo {pages}
  {013621}}\BibitemShut {NoStop}%
\bibitem [{\citenamefont {Choudhury}\ and\ \citenamefont
  {Mueller}(2015)}]{PhysRevA.91.023624}%
  \BibitemOpen
  \bibfield  {author} {\bibinfo {author} {\bibnamefont {Choudhury},
  \bibfnamefont {Sayan}}, \ and\ \bibinfo {author} {\bibfnamefont {Erich~J.}\
  \bibnamefont {Mueller}}} (\bibinfo {year} {2015}),\ \bibfield  {title}
  {\enquote {\bibinfo {title} {Transverse collisional instabilities of a
  {B}ose-{E}instein condensate in a driven one-dimensional lattice},}\ }\href
  {\doibase 10.1103/PhysRevA.91.023624} {\bibfield  {journal} {\bibinfo
  {journal} {Phys. Rev. A}\ }\textbf {\bibinfo {volume} {91}},\ \bibinfo
  {pages} {023624}}\BibitemShut {NoStop}%
\bibitem [{\citenamefont {Cooper}(2008)}]{cooperadvancesreview}%
  \BibitemOpen
  \bibfield  {author} {\bibinfo {author} {\bibnamefont {Cooper}, \bibfnamefont
  {N~R}}} (\bibinfo {year} {2008}),\ \bibfield  {title} {\enquote {\bibinfo
  {title} {Rapidly rotating atomic gases},}\ }\href {\doibase
  10.1080/00018730802564122} {\bibfield  {journal} {\bibinfo  {journal}
  {Advances in Physics}\ }\textbf {\bibinfo {volume} {57}}~(\bibinfo {number}
  {6}),\ \bibinfo {pages} {539--616}}\BibitemShut {NoStop}%
\bibitem [{\citenamefont {Cooper}(2011)}]{Cooper2011}%
  \BibitemOpen
  \bibfield  {author} {\bibinfo {author} {\bibnamefont {Cooper}, \bibfnamefont
  {N~R}}} (\bibinfo {year} {2011}),\ \bibfield  {title} {\enquote {\bibinfo
  {title} {Optical flux lattices for ultracold atomic gases},}\ }\href
  {\doibase 10.1103/PhysRevLett.106.175301} {\bibfield  {journal} {\bibinfo
  {journal} {Phys. Rev. Lett.}\ }\textbf {\bibinfo {volume} {106}}~(\bibinfo
  {number} {17}),\ \bibinfo {pages} {175301}}\BibitemShut {NoStop}%
\bibitem [{\citenamefont {Cooper}\ and\ \citenamefont
  {Dalibard}(2011)}]{Cooper2011a}%
  \BibitemOpen
  \bibfield  {author} {\bibinfo {author} {\bibnamefont {Cooper}, \bibfnamefont
  {N~R}}, \ and\ \bibinfo {author} {\bibfnamefont {J.}~\bibnamefont
  {Dalibard}}} (\bibinfo {year} {2011}),\ \bibfield  {title} {\enquote
  {\bibinfo {title} {Optical flux lattices for two-photon dressed states},}\
  }\href {http://stacks.iop.org/0295-5075/95/i=6/a=66004} {\bibfield  {journal}
  {\bibinfo  {journal} {Europhysics Letters}\ }\textbf {\bibinfo {volume}
  {95}}~(\bibinfo {number} {6}),\ \bibinfo {pages} {66004}}\BibitemShut
  {NoStop}%
\bibitem [{\citenamefont {Cooper}\ and\ \citenamefont
  {Moessner}(2012)}]{coopermoessner}%
  \BibitemOpen
  \bibfield  {author} {\bibinfo {author} {\bibnamefont {Cooper}, \bibfnamefont
  {N~R}}, \ and\ \bibinfo {author} {\bibfnamefont {R.}~\bibnamefont
  {Moessner}}} (\bibinfo {year} {2012}),\ \bibfield  {title} {\enquote
  {\bibinfo {title} {Designing topological bands in reciprocal space},}\ }\href
  {\doibase 10.1103/PhysRevLett.109.215302} {\bibfield  {journal} {\bibinfo
  {journal} {Phys. Rev. Lett.}\ }\textbf {\bibinfo {volume} {109}},\ \bibinfo
  {pages} {215302}}\BibitemShut {NoStop}%
\bibitem [{\citenamefont {Cooper}\ and\ \citenamefont
  {Dalibard}(2013)}]{cooperdalibard2013}%
  \BibitemOpen
  \bibfield  {author} {\bibinfo {author} {\bibnamefont {Cooper}, \bibfnamefont
  {Nigel~R}}, \ and\ \bibinfo {author} {\bibfnamefont {Jean}\ \bibnamefont
  {Dalibard}}} (\bibinfo {year} {2013}),\ \bibfield  {title} {\enquote
  {\bibinfo {title} {Reaching fractional quantum {H}all states with optical
  flux lattices},}\ }\href {\doibase 10.1103/PhysRevLett.110.185301} {\bibfield
   {journal} {\bibinfo  {journal} {Phys. Rev. Lett.}\ }\textbf {\bibinfo
  {volume} {110}},\ \bibinfo {pages} {185301}}\BibitemShut {NoStop}%
\bibitem [{\citenamefont {Cooper}\ and\ \citenamefont
  {Simon}(2015)}]{coopersimonexclusion}%
  \BibitemOpen
  \bibfield  {author} {\bibinfo {author} {\bibnamefont {Cooper}, \bibfnamefont
  {Nigel~R}}, \ and\ \bibinfo {author} {\bibfnamefont {Steven~H.}\ \bibnamefont
  {Simon}}} (\bibinfo {year} {2015}),\ \bibfield  {title} {\enquote {\bibinfo
  {title} {Signatures of fractional exclusion statistics in the spectroscopy of
  quantum {H}all droplets},}\ }\href {\doibase 10.1103/PhysRevLett.114.106802}
  {\bibfield  {journal} {\bibinfo  {journal} {Phys. Rev. Lett.}\ }\textbf
  {\bibinfo {volume} {114}},\ \bibinfo {pages} {106802}}\BibitemShut {NoStop}%
\bibitem [{\citenamefont {Cornfeld}\ and\ \citenamefont
  {Sela}(2015)}]{PhysRevB.92.115446}%
  \BibitemOpen
  \bibfield  {author} {\bibinfo {author} {\bibnamefont {Cornfeld},
  \bibfnamefont {Eyal}}, \ and\ \bibinfo {author} {\bibfnamefont {Eran}\
  \bibnamefont {Sela}}} (\bibinfo {year} {2015}),\ \bibfield  {title} {\enquote
  {\bibinfo {title} {Chiral currents in one-dimensional fractional quantum
  {H}all states},}\ }\href {\doibase 10.1103/PhysRevB.92.115446} {\bibfield
  {journal} {\bibinfo  {journal} {Phys. Rev. B}\ }\textbf {\bibinfo {volume}
  {92}},\ \bibinfo {pages} {115446}}\BibitemShut {NoStop}%
\bibitem [{\citenamefont {Creutz}(1999)}]{Creutz:1999}%
  \BibitemOpen
  \bibfield  {author} {\bibinfo {author} {\bibnamefont {Creutz}, \bibfnamefont
  {Michael}}} (\bibinfo {year} {1999}),\ \bibfield  {title} {\enquote {\bibinfo
  {title} {End states, ladder compounds, and domain-wall fermions},}\ }\href
  {\doibase 10.1103/PhysRevLett.83.2636} {\bibfield  {journal} {\bibinfo
  {journal} {Phys. Rev. Lett.}\ }\textbf {\bibinfo {volume} {83}},\ \bibinfo
  {pages} {2636--2639}}\BibitemShut {NoStop}%
\bibitem [{\citenamefont {D'Alessio}\ and\ \citenamefont
  {Polkovnikov}(2013)}]{dalessio}%
  \BibitemOpen
  \bibfield  {author} {\bibinfo {author} {\bibnamefont {D'Alessio},
  \bibfnamefont {Luca}}, \ and\ \bibinfo {author} {\bibfnamefont {Anatoli}\
  \bibnamefont {Polkovnikov}}} (\bibinfo {year} {2013}),\ \bibfield  {title}
  {\enquote {\bibinfo {title} {Many-body energy localization transition in
  periodically driven systems},}\ }\href {\doibase
  https://doi.org/10.1016/j.aop.2013.02.011} {\bibfield  {journal} {\bibinfo
  {journal} {Annals of Physics}\ }\textbf {\bibinfo {volume} {333}},\ \bibinfo
  {pages} {19 -- 33}}\BibitemShut {NoStop}%
\bibitem [{\citenamefont {D'{A}lessio}\ and\ \citenamefont
  {Rigol}(2015)}]{dAlessio2015}%
  \BibitemOpen
  \bibfield  {author} {\bibinfo {author} {\bibnamefont {D'{A}lessio},
  \bibfnamefont {Luca}}, \ and\ \bibinfo {author} {\bibfnamefont {Marcos}\
  \bibnamefont {Rigol}}} (\bibinfo {year} {2015}),\ \bibfield  {title}
  {\enquote {\bibinfo {title} {Dynamical preparation of {F}loquet {C}hern
  insulators},}\ }\href {http://dx.doi.org/10.1038/ncomms9336} {\bibfield
  {journal} {\bibinfo  {journal} {Nature Communications}\ }\textbf {\bibinfo
  {volume} {6}},\ \bibinfo {pages} {8336}}\BibitemShut {NoStop}%
\bibitem [{\citenamefont {Daley}(2014)}]{daley2014quantum}%
  \BibitemOpen
  \bibfield  {author} {\bibinfo {author} {\bibnamefont {Daley}, \bibfnamefont
  {Andrew~J}}} (\bibinfo {year} {2014}),\ \bibfield  {title} {\enquote
  {\bibinfo {title} {Quantum trajectories and open many-body quantum
  systems},}\ }\href {\doibase 10.1080/00018732.2014.933502} {\bibfield
  {journal} {\bibinfo  {journal} {Advances in Physics}\ }\textbf {\bibinfo
  {volume} {63}}~(\bibinfo {number} {2}),\ \bibinfo {pages}
  {77--149}}\BibitemShut {NoStop}%
\bibitem [{\citenamefont {Dalfovo}\ \emph {et~al.}(1999)\citenamefont
  {Dalfovo}, \citenamefont {Giorgini}, \citenamefont {Pitaevskii},\ and\
  \citenamefont {Stringari}}]{Dalfovo:1999}%
  \BibitemOpen
  \bibfield  {author} {\bibinfo {author} {\bibnamefont {Dalfovo}, \bibfnamefont
  {Franco}}, \bibinfo {author} {\bibfnamefont {Stefano}\ \bibnamefont
  {Giorgini}}, \bibinfo {author} {\bibfnamefont {Lev~P.}\ \bibnamefont
  {Pitaevskii}}, \ and\ \bibinfo {author} {\bibfnamefont {Sandro}\ \bibnamefont
  {Stringari}}} (\bibinfo {year} {1999}),\ \bibfield  {title} {\enquote
  {\bibinfo {title} {Theory of {B}ose-{E}instein condensation in trapped
  gases},}\ }\href {\doibase 10.1103/RevModPhys.71.463} {\bibfield  {journal}
  {\bibinfo  {journal} {Rev. Mod. Phys.}\ }\textbf {\bibinfo {volume} {71}},\
  \bibinfo {pages} {463--512}}\BibitemShut {NoStop}%
\bibitem [{\citenamefont {Dalibard}(2016)}]{Dalibard:2016}%
  \BibitemOpen
  \bibfield  {author} {\bibinfo {author} {\bibnamefont {Dalibard},
  \bibfnamefont {J}}} (\bibinfo {year} {2016}),\ \bibfield  {title} {\enquote
  {\bibinfo {title} {Introduction to the physics of artificial gauge fields},}\
  }in\ \href@noop {} {\emph {\bibinfo {booktitle} {Quantum Matter at Ultralow
  Temperatures}}},\ \bibinfo {editor} {edited by\ \bibinfo {editor}
  {\bibfnamefont {M.}~\bibnamefont {Inguscio}}, \bibinfo {editor}
  {\bibfnamefont {W.}~\bibnamefont {Ketterle}}, \ and\ \bibinfo {editor}
  {\bibfnamefont {S.}~\bibnamefont {Stringari}}}\ (\bibinfo  {publisher} {IOS
  Press})\ pp.\ \bibinfo {pages} {1--62}\BibitemShut {NoStop}%
\bibitem [{\citenamefont {Dareau}\ \emph {et~al.}(2017)\citenamefont {Dareau},
  \citenamefont {Levy}, \citenamefont {Aguilera}, \citenamefont {Bouganne},
  \citenamefont {Akkermans}, \citenamefont {Gerbier},\ and\ \citenamefont
  {Beugnon}}]{Dareau2017}%
  \BibitemOpen
  \bibfield  {author} {\bibinfo {author} {\bibnamefont {Dareau}, \bibfnamefont
  {A}}, \bibinfo {author} {\bibfnamefont {E.}~\bibnamefont {Levy}}, \bibinfo
  {author} {\bibfnamefont {M.~Bosch}\ \bibnamefont {Aguilera}}, \bibinfo
  {author} {\bibfnamefont {R.}~\bibnamefont {Bouganne}}, \bibinfo {author}
  {\bibfnamefont {E.}~\bibnamefont {Akkermans}}, \bibinfo {author}
  {\bibfnamefont {F.}~\bibnamefont {Gerbier}}, \ and\ \bibinfo {author}
  {\bibfnamefont {J.}~\bibnamefont {Beugnon}}} (\bibinfo {year} {2017}),\
  \bibfield  {title} {\enquote {\bibinfo {title} {Revealing the topology of
  quasicrystals with a diffraction experiment},}\ }\href {\doibase
  10.1103/PhysRevLett.119.215304} {\bibfield  {journal} {\bibinfo  {journal}
  {Phys. Rev. Lett.}\ }\textbf {\bibinfo {volume} {119}},\ \bibinfo {pages}
  {215304}}\BibitemShut {NoStop}%
\bibitem [{\citenamefont {Dauphin}\ and\ \citenamefont
  {Goldman}(2013)}]{Dauphin2013}%
  \BibitemOpen
  \bibfield  {author} {\bibinfo {author} {\bibnamefont {Dauphin}, \bibfnamefont
  {Alexandre}}, \ and\ \bibinfo {author} {\bibfnamefont {Nathan}\ \bibnamefont
  {Goldman}}} (\bibinfo {year} {2013}),\ \bibfield  {title} {\enquote {\bibinfo
  {title} {Extracting the {C}hern number from the dynamics of a {F}ermi gas:
  Implementing a quantum {H}all bar for cold atoms},}\ }\href {\doibase
  10.1103/PhysRevLett.111.135302} {\bibfield  {journal} {\bibinfo  {journal}
  {Phys. Rev. Lett.}\ }\textbf {\bibinfo {volume} {111}},\ \bibinfo {pages}
  {135302}}\BibitemShut {NoStop}%
\bibitem [{\citenamefont {Dauphin}\ \emph {et~al.}(2017)\citenamefont
  {Dauphin}, \citenamefont {Tran}, \citenamefont {Lewenstein},\ and\
  \citenamefont {Goldman}}]{Dauphin2017}%
  \BibitemOpen
  \bibfield  {author} {\bibinfo {author} {\bibnamefont {Dauphin}, \bibfnamefont
  {Alexandre}}, \bibinfo {author} {\bibfnamefont {Duc-Thanh}\ \bibnamefont
  {Tran}}, \bibinfo {author} {\bibfnamefont {Maciej}\ \bibnamefont
  {Lewenstein}}, \ and\ \bibinfo {author} {\bibfnamefont {Nathan}\ \bibnamefont
  {Goldman}}} (\bibinfo {year} {2017}),\ \bibfield  {title} {\enquote {\bibinfo
  {title} {Loading ultracold gases in topological {F}loquet bands: the fate of
  current and center-of-mass responses},}\ }\href
  {http://stacks.iop.org/2053-1583/4/i=2/a=024010} {\bibfield  {journal}
  {\bibinfo  {journal} {2D Materials}\ }\textbf {\bibinfo {volume}
  {4}}~(\bibinfo {number} {2}),\ \bibinfo {pages} {024010}}\BibitemShut
  {NoStop}%
\bibitem [{\citenamefont {Delplace}\ \emph {et~al.}(2011)\citenamefont
  {Delplace}, \citenamefont {Ullmo},\ and\ \citenamefont
  {Montambaux}}]{Delplace:2011}%
  \BibitemOpen
  \bibfield  {author} {\bibinfo {author} {\bibnamefont {Delplace},
  \bibfnamefont {P}}, \bibinfo {author} {\bibfnamefont {D.}~\bibnamefont
  {Ullmo}}, \ and\ \bibinfo {author} {\bibfnamefont {G.}~\bibnamefont
  {Montambaux}}} (\bibinfo {year} {2011}),\ \bibfield  {title} {\enquote
  {\bibinfo {title} {Zak phase and the existence of edge states in graphene},}\
  }\href {\doibase 10.1103/PhysRevB.84.195452} {\bibfield  {journal} {\bibinfo
  {journal} {Phys. Rev. B}\ }\textbf {\bibinfo {volume} {84}},\ \bibinfo
  {pages} {195452}}\BibitemShut {NoStop}%
\bibitem [{\citenamefont {Deng}\ \emph {et~al.}(2013)\citenamefont {Deng},
  \citenamefont {Wang}, \citenamefont {Shen},\ and\ \citenamefont
  {Duan}}]{PhysRevB.88.201105}%
  \BibitemOpen
  \bibfield  {author} {\bibinfo {author} {\bibnamefont {Deng}, \bibfnamefont
  {D-L}}, \bibinfo {author} {\bibfnamefont {S.-T.}\ \bibnamefont {Wang}},
  \bibinfo {author} {\bibfnamefont {C.}~\bibnamefont {Shen}}, \ and\ \bibinfo
  {author} {\bibfnamefont {L.-M.}\ \bibnamefont {Duan}}} (\bibinfo {year}
  {2013}),\ \bibfield  {title} {\enquote {\bibinfo {title} {Hopf insulators and
  their topologically protected surface states},}\ }\href {\doibase
  10.1103/PhysRevB.88.201105} {\bibfield  {journal} {\bibinfo  {journal} {Phys.
  Rev. B}\ }\textbf {\bibinfo {volume} {88}},\ \bibinfo {pages}
  {201105}}\BibitemShut {NoStop}%
\bibitem [{\citenamefont {Deutsch}\ and\ \citenamefont
  {Jessen}(1998)}]{Deutsch1998}%
  \BibitemOpen
  \bibfield  {author} {\bibinfo {author} {\bibnamefont {Deutsch}, \bibfnamefont
  {Ivan~H}}, \ and\ \bibinfo {author} {\bibfnamefont {Poul~S.}\ \bibnamefont
  {Jessen}}} (\bibinfo {year} {1998}),\ \bibfield  {title} {\enquote {\bibinfo
  {title} {{Quantum-state control in optical lattices}},}\ }\href {\doibase
  10.1103/PhysRevA.57.1972} {\bibfield  {journal} {\bibinfo  {journal} {Phys.
  Rev. A}\ }\textbf {\bibinfo {volume} {57}}~(\bibinfo {number} {3}),\ \bibinfo
  {pages} {1972--1986}}\BibitemShut {NoStop}%
\bibitem [{\citenamefont {Dhar}\ \emph {et~al.}(2013)\citenamefont {Dhar},
  \citenamefont {Mishra}, \citenamefont {Maji}, \citenamefont {Pai},
  \citenamefont {Mukerjee},\ and\ \citenamefont
  {Paramekanti}}]{PhysRevB.87.174501}%
  \BibitemOpen
  \bibfield  {author} {\bibinfo {author} {\bibnamefont {Dhar}, \bibfnamefont
  {Arya}}, \bibinfo {author} {\bibfnamefont {Tapan}\ \bibnamefont {Mishra}},
  \bibinfo {author} {\bibfnamefont {Maheswar}\ \bibnamefont {Maji}}, \bibinfo
  {author} {\bibfnamefont {R.~V.}\ \bibnamefont {Pai}}, \bibinfo {author}
  {\bibfnamefont {Subroto}\ \bibnamefont {Mukerjee}}, \ and\ \bibinfo {author}
  {\bibfnamefont {Arun}\ \bibnamefont {Paramekanti}}} (\bibinfo {year}
  {2013}),\ \bibfield  {title} {\enquote {\bibinfo {title} {Chiral {M}ott
  insulator with staggered loop currents in the fully frustrated
  {B}ose-{H}ubbard model},}\ }\href {\doibase 10.1103/PhysRevB.87.174501}
  {\bibfield  {journal} {\bibinfo  {journal} {Phys. Rev. B}\ }\textbf {\bibinfo
  {volume} {87}},\ \bibinfo {pages} {174501}}\BibitemShut {NoStop}%
\bibitem [{\citenamefont {Diehl}\ \emph {et~al.}(2011)\citenamefont {Diehl},
  \citenamefont {Rico}, \citenamefont {Baranov},\ and\ \citenamefont
  {Zoller}}]{diehl2011topology}%
  \BibitemOpen
  \bibfield  {author} {\bibinfo {author} {\bibnamefont {Diehl}, \bibfnamefont
  {Sebastian}}, \bibinfo {author} {\bibfnamefont {Enrique}\ \bibnamefont
  {Rico}}, \bibinfo {author} {\bibfnamefont {Mikhail~A}\ \bibnamefont
  {Baranov}}, \ and\ \bibinfo {author} {\bibfnamefont {Peter}\ \bibnamefont
  {Zoller}}} (\bibinfo {year} {2011}),\ \bibfield  {title} {\enquote {\bibinfo
  {title} {Topology by dissipation in atomic quantum wires},}\ }\href {\doibase
  10.1038/nphys2106} {\bibfield  {journal} {\bibinfo  {journal} {Nature
  Physics}\ }\textbf {\bibinfo {volume} {7}}~(\bibinfo {number} {12}),\
  \bibinfo {pages} {971--977}}\BibitemShut {NoStop}%
\bibitem [{\citenamefont {Dub{\v{c}}ek}\ \emph {et~al.}(2015)\citenamefont
  {Dub{\v{c}}ek}, \citenamefont {Kennedy}, \citenamefont {Lu}, \citenamefont
  {Ketterle}, \citenamefont {Solja{\v{c}}i{\'{c}}},\ and\ \citenamefont
  {Buljan}}]{Dubcek:2015}%
  \BibitemOpen
  \bibfield  {author} {\bibinfo {author} {\bibnamefont {Dub{\v{c}}ek},
  \bibfnamefont {Tena}}, \bibinfo {author} {\bibfnamefont {Colin~J.}\
  \bibnamefont {Kennedy}}, \bibinfo {author} {\bibfnamefont {Ling}\
  \bibnamefont {Lu}}, \bibinfo {author} {\bibfnamefont {Wolfgang}\ \bibnamefont
  {Ketterle}}, \bibinfo {author} {\bibfnamefont {Marin}\ \bibnamefont
  {Solja{\v{c}}i{\'{c}}}}, \ and\ \bibinfo {author} {\bibfnamefont {Hrvoje}\
  \bibnamefont {Buljan}}} (\bibinfo {year} {2015}),\ \bibfield  {title}
  {\enquote {\bibinfo {title} {Weyl points in three-dimensional optical
  lattices: Synthetic magnetic monopoles in momentum space},}\ }\href {\doibase
  10.1103/PhysRevLett.114.225301} {\bibfield  {journal} {\bibinfo  {journal}
  {Phys. Rev. Lett.}\ }\textbf {\bibinfo {volume} {114}},\ \bibinfo {pages}
  {225301}}\BibitemShut {NoStop}%
\bibitem [{\citenamefont {Duca}\ \emph {et~al.}(2015)\citenamefont {Duca},
  \citenamefont {Li}, \citenamefont {Reitter}, \citenamefont {Bloch},
  \citenamefont {Schleier-Smith},\ and\ \citenamefont {Schneider}}]{Duca:2015}%
  \BibitemOpen
  \bibfield  {author} {\bibinfo {author} {\bibnamefont {Duca}, \bibfnamefont
  {L}}, \bibinfo {author} {\bibfnamefont {T.}~\bibnamefont {Li}}, \bibinfo
  {author} {\bibfnamefont {M.}~\bibnamefont {Reitter}}, \bibinfo {author}
  {\bibfnamefont {I.}~\bibnamefont {Bloch}}, \bibinfo {author} {\bibfnamefont
  {M.}~\bibnamefont {Schleier-Smith}}, \ and\ \bibinfo {author} {\bibfnamefont
  {U.}~\bibnamefont {Schneider}}} (\bibinfo {year} {2015}),\ \bibfield  {title}
  {\enquote {\bibinfo {title} {An {A}haronov--{B}ohm interferometer for
  determining {B}loch band topology},}\ }\href {\doibase
  10.1126/science.1259052} {\bibfield  {journal} {\bibinfo  {journal}
  {Science}\ }\textbf {\bibinfo {volume} {347}},\ \bibinfo {pages}
  {288}}\BibitemShut {NoStop}%
\bibitem [{\citenamefont {Dudarev}\ \emph {et~al.}(2004)\citenamefont
  {Dudarev}, \citenamefont {Diener}, \citenamefont {Carusotto},\ and\
  \citenamefont {Niu}}]{Dudarev2004}%
  \BibitemOpen
  \bibfield  {author} {\bibinfo {author} {\bibnamefont {Dudarev}, \bibfnamefont
  {A~M}}, \bibinfo {author} {\bibfnamefont {R~B}\ \bibnamefont {Diener}},
  \bibinfo {author} {\bibfnamefont {I}~\bibnamefont {Carusotto}}, \ and\
  \bibinfo {author} {\bibfnamefont {Q}~\bibnamefont {Niu}}} (\bibinfo {year}
  {2004}),\ \bibfield  {title} {\enquote {\bibinfo {title} {{Spin-Orbit
  Coupling and Berry Phase with Ultracold Atoms in 2D Optical Lattices}},}\
  }\href {\doibase 10.1103/PhysRevLett.92.153005} {\bibfield  {journal}
  {\bibinfo  {journal} {Phys. Rev. Lett.}\ }\textbf {\bibinfo {volume}
  {92}}~(\bibinfo {number} {15}),\ \bibinfo {pages} {153005}}\BibitemShut
  {NoStop}%
\bibitem [{\citenamefont {Eckardt}(2017)}]{eckardtreview}%
  \BibitemOpen
  \bibfield  {author} {\bibinfo {author} {\bibnamefont {Eckardt}, \bibfnamefont
  {Andr\'e}}} (\bibinfo {year} {2017}),\ \bibfield  {title} {\enquote {\bibinfo
  {title} {Atomic quantum gases in periodically driven optical lattices},}\
  }\href {\doibase 10.1103/RevModPhys.89.011004} {\bibfield  {journal}
  {\bibinfo  {journal} {Rev. Mod. Phys.}\ }\textbf {\bibinfo {volume} {89}},\
  \bibinfo {pages} {011004}}\BibitemShut {NoStop}%
\bibitem [{\citenamefont {Eckardt}\ and\ \citenamefont
  {Anisimovas}(2015)}]{eckardt:2015}%
  \BibitemOpen
  \bibfield  {author} {\bibinfo {author} {\bibnamefont {Eckardt}, \bibfnamefont
  {Andr{\' e}}}, \ and\ \bibinfo {author} {\bibfnamefont {Egidijus}\
  \bibnamefont {Anisimovas}}} (\bibinfo {year} {2015}),\ \bibfield  {title}
  {\enquote {\bibinfo {title} {High-frequency approximation for periodically
  driven quantum systems from a {F}loquet-space perspective},}\ }\href
  {http://stacks.iop.org/1367-2630/17/i=9/a=093039} {\bibfield  {journal}
  {\bibinfo  {journal} {New Journal of Physics}\ }\textbf {\bibinfo {volume}
  {17}}~(\bibinfo {number} {9}),\ \bibinfo {pages} {093039}}\BibitemShut
  {NoStop}%
\bibitem [{\citenamefont {Edmonds}\ \emph {et~al.}(2013)\citenamefont
  {Edmonds}, \citenamefont {Valiente}, \citenamefont {Juzeli{\=u}nas},
  \citenamefont {Santos},\ and\ \citenamefont {{\"O}hberg}}]{edmonds2013}%
  \BibitemOpen
  \bibfield  {author} {\bibinfo {author} {\bibnamefont {Edmonds}, \bibfnamefont
  {M~J}}, \bibinfo {author} {\bibfnamefont {M.}~\bibnamefont {Valiente}},
  \bibinfo {author} {\bibfnamefont {G.}~\bibnamefont {Juzeli{\=u}nas}},
  \bibinfo {author} {\bibfnamefont {L.}~\bibnamefont {Santos}}, \ and\ \bibinfo
  {author} {\bibfnamefont {P.}~\bibnamefont {{\"O}hberg}}} (\bibinfo {year}
  {2013}),\ \bibfield  {title} {\enquote {\bibinfo {title} {Simulating an
  interacting gauge theory with ultracold {B}ose gases},}\ }\href {\doibase
  10.1103/PhysRevLett.110.085301} {\bibfield  {journal} {\bibinfo  {journal}
  {Phys. Rev. Lett.}\ }\textbf {\bibinfo {volume} {110}},\ \bibinfo {pages}
  {085301}}\BibitemShut {NoStop}%
\bibitem [{\citenamefont {Essin}\ and\ \citenamefont
  {Gurarie}(2012)}]{PhysRevB.85.195116}%
  \BibitemOpen
  \bibfield  {author} {\bibinfo {author} {\bibnamefont {Essin}, \bibfnamefont
  {Andrew~M}}, \ and\ \bibinfo {author} {\bibfnamefont {Victor}\ \bibnamefont
  {Gurarie}}} (\bibinfo {year} {2012}),\ \bibfield  {title} {\enquote {\bibinfo
  {title} {Antiferromagnetic topological insulators in cold atomic gases},}\
  }\href {\doibase 10.1103/PhysRevB.85.195116} {\bibfield  {journal} {\bibinfo
  {journal} {Phys. Rev. B}\ }\textbf {\bibinfo {volume} {85}},\ \bibinfo
  {pages} {195116}}\BibitemShut {NoStop}%
\bibitem [{\citenamefont {Fidkowski}\ \emph {et~al.}(2011)\citenamefont
  {Fidkowski}, \citenamefont {Lutchyn}, \citenamefont {Nayak},\ and\
  \citenamefont {Fisher}}]{Fidkowski2011}%
  \BibitemOpen
  \bibfield  {author} {\bibinfo {author} {\bibnamefont {Fidkowski},
  \bibfnamefont {Lukasz}}, \bibinfo {author} {\bibfnamefont {Roman~M.}\
  \bibnamefont {Lutchyn}}, \bibinfo {author} {\bibfnamefont {Chetan}\
  \bibnamefont {Nayak}}, \ and\ \bibinfo {author} {\bibfnamefont {Matthew
  P.~A.}\ \bibnamefont {Fisher}}} (\bibinfo {year} {2011}),\ \bibfield  {title}
  {\enquote {\bibinfo {title} {Majorana zero modes in one-dimensional quantum
  wires without long-ranged superconducting order},}\ }\href {\doibase
  10.1103/PhysRevB.84.195436} {\bibfield  {journal} {\bibinfo  {journal} {Phys.
  Rev. B}\ }\textbf {\bibinfo {volume} {84}},\ \bibinfo {pages}
  {195436}}\BibitemShut {NoStop}%
\bibitem [{\citenamefont {Fl{\"a}schner}\ \emph {et~al.}(2016)\citenamefont
  {Fl{\"a}schner}, \citenamefont {Rem}, \citenamefont {Tarnowski},
  \citenamefont {Vogel}, \citenamefont {L{\"u}hmann}, \citenamefont
  {Sengstock},\ and\ \citenamefont {Weitenberg}}]{Flaschner:2016}%
  \BibitemOpen
  \bibfield  {author} {\bibinfo {author} {\bibnamefont {Fl{\"a}schner},
  \bibfnamefont {N}}, \bibinfo {author} {\bibfnamefont {BS}~\bibnamefont
  {Rem}}, \bibinfo {author} {\bibfnamefont {M}~\bibnamefont {Tarnowski}},
  \bibinfo {author} {\bibfnamefont {D}~\bibnamefont {Vogel}}, \bibinfo {author}
  {\bibfnamefont {D-S}\ \bibnamefont {L{\"u}hmann}}, \bibinfo {author}
  {\bibfnamefont {K}~\bibnamefont {Sengstock}}, \ and\ \bibinfo {author}
  {\bibfnamefont {C}~\bibnamefont {Weitenberg}}} (\bibinfo {year} {2016}),\
  \bibfield  {title} {\enquote {\bibinfo {title} {Experimental reconstruction
  of the {B}erry curvature in a {F}loquet {B}loch band},}\ }\href {\doibase
  10.1126/science.aad4568} {\bibfield  {journal} {\bibinfo  {journal}
  {Science}\ }\textbf {\bibinfo {volume} {352}}~(\bibinfo {number} {6289}),\
  \bibinfo {pages} {1091--1094}}\BibitemShut {NoStop}%
\bibitem [{\citenamefont {Fl{\"a}schner}\ \emph {et~al.}(2018)\citenamefont
  {Fl{\"a}schner}, \citenamefont {Vogel}, \citenamefont {Tarnowski},
  \citenamefont {Rem}, \citenamefont {L{\"u}hmann}, \citenamefont {Heyl},
  \citenamefont {Budich}, \citenamefont {Mathey}, \citenamefont {Sengstock},\
  and\ \citenamefont {Weitenberg}}]{hamburg_phasetransition}%
  \BibitemOpen
  \bibfield  {author} {\bibinfo {author} {\bibnamefont {Fl{\"a}schner},
  \bibfnamefont {N}}, \bibinfo {author} {\bibfnamefont {D}~\bibnamefont
  {Vogel}}, \bibinfo {author} {\bibfnamefont {M}~\bibnamefont {Tarnowski}},
  \bibinfo {author} {\bibfnamefont {BS}~\bibnamefont {Rem}}, \bibinfo {author}
  {\bibfnamefont {D-S}\ \bibnamefont {L{\"u}hmann}}, \bibinfo {author}
  {\bibfnamefont {M}~\bibnamefont {Heyl}}, \bibinfo {author} {\bibfnamefont
  {JC}~\bibnamefont {Budich}}, \bibinfo {author} {\bibfnamefont
  {L}~\bibnamefont {Mathey}}, \bibinfo {author} {\bibfnamefont {K}~\bibnamefont
  {Sengstock}}, \ and\ \bibinfo {author} {\bibfnamefont {C}~\bibnamefont
  {Weitenberg}}} (\bibinfo {year} {2018}),\ \bibfield  {title} {\enquote
  {\bibinfo {title} {Observation of dynamical vortices after quenches in a
  system with topology},}\ }\href {\doibase 10.1038/s41567-017-0013-8}
  {\bibfield  {journal} {\bibinfo  {journal} {Nature Physics}\ }\textbf
  {\bibinfo {volume} {14}}~(\bibinfo {number} {3}),\ \bibinfo {pages}
  {265}}\BibitemShut {NoStop}%
\bibitem [{\citenamefont {Foster}\ \emph {et~al.}(2013)\citenamefont {Foster},
  \citenamefont {Dzero}, \citenamefont {Gurarie},\ and\ \citenamefont
  {Yuzbashyan}}]{PhysRevB.88.104511}%
  \BibitemOpen
  \bibfield  {author} {\bibinfo {author} {\bibnamefont {Foster}, \bibfnamefont
  {Matthew~S}}, \bibinfo {author} {\bibfnamefont {Maxim}\ \bibnamefont
  {Dzero}}, \bibinfo {author} {\bibfnamefont {Victor}\ \bibnamefont {Gurarie}},
  \ and\ \bibinfo {author} {\bibfnamefont {Emil~A.}\ \bibnamefont
  {Yuzbashyan}}} (\bibinfo {year} {2013}),\ \bibfield  {title} {\enquote
  {\bibinfo {title} {Quantum quench in a $p+ip$ superfluid: Winding numbers and
  topological states far from equilibrium},}\ }\href {\doibase
  10.1103/PhysRevB.88.104511} {\bibfield  {journal} {\bibinfo  {journal} {Phys.
  Rev. B}\ }\textbf {\bibinfo {volume} {88}},\ \bibinfo {pages}
  {104511}}\BibitemShut {NoStop}%
\bibitem [{\citenamefont {Foster}\ \emph {et~al.}(2014)\citenamefont {Foster},
  \citenamefont {Gurarie}, \citenamefont {Dzero},\ and\ \citenamefont
  {Yuzbashyan}}]{PhysRevLett.113.076403}%
  \BibitemOpen
  \bibfield  {author} {\bibinfo {author} {\bibnamefont {Foster}, \bibfnamefont
  {Matthew~S}}, \bibinfo {author} {\bibfnamefont {Victor}\ \bibnamefont
  {Gurarie}}, \bibinfo {author} {\bibfnamefont {Maxim}\ \bibnamefont {Dzero}},
  \ and\ \bibinfo {author} {\bibfnamefont {Emil~A.}\ \bibnamefont
  {Yuzbashyan}}} (\bibinfo {year} {2014}),\ \bibfield  {title} {\enquote
  {\bibinfo {title} {Quench-induced {F}loquet topological $p$-wave
  superfluids},}\ }\href {\doibase 10.1103/PhysRevLett.113.076403} {\bibfield
  {journal} {\bibinfo  {journal} {Phys. Rev. Lett.}\ }\textbf {\bibinfo
  {volume} {113}},\ \bibinfo {pages} {076403}}\BibitemShut {NoStop}%
\bibitem [{\citenamefont {Fruchart}(2016)}]{PhysRevB.93.115429}%
  \BibitemOpen
  \bibfield  {author} {\bibinfo {author} {\bibnamefont {Fruchart},
  \bibfnamefont {Michel}}} (\bibinfo {year} {2016}),\ \bibfield  {title}
  {\enquote {\bibinfo {title} {Complex classes of periodically driven
  topological lattice systems},}\ }\href {\doibase 10.1103/PhysRevB.93.115429}
  {\bibfield  {journal} {\bibinfo  {journal} {Phys. Rev. B}\ }\textbf {\bibinfo
  {volume} {93}},\ \bibinfo {pages} {115429}}\BibitemShut {NoStop}%
\bibitem [{\citenamefont {Fu}(2011)}]{PhysRevLett.106.106802}%
  \BibitemOpen
  \bibfield  {author} {\bibinfo {author} {\bibnamefont {Fu}, \bibfnamefont
  {Liang}}} (\bibinfo {year} {2011}),\ \bibfield  {title} {\enquote {\bibinfo
  {title} {Topological crystalline insulators},}\ }\href {\doibase
  10.1103/PhysRevLett.106.106802} {\bibfield  {journal} {\bibinfo  {journal}
  {Phys. Rev. Lett.}\ }\textbf {\bibinfo {volume} {106}},\ \bibinfo {pages}
  {106802}}\BibitemShut {NoStop}%
\bibitem [{\citenamefont {Fu}\ and\ \citenamefont
  {Kane}(2007)}]{fu2007topological}%
  \BibitemOpen
  \bibfield  {author} {\bibinfo {author} {\bibnamefont {Fu}, \bibfnamefont
  {Liang}}, \ and\ \bibinfo {author} {\bibfnamefont {Charles~L}\ \bibnamefont
  {Kane}}} (\bibinfo {year} {2007}),\ \bibfield  {title} {\enquote {\bibinfo
  {title} {Topological insulators with inversion symmetry},}\ }\href {\doibase
  10.1103/PhysRevB.76.045302} {\bibfield  {journal} {\bibinfo  {journal} {Phys.
  Rev. B}\ }\textbf {\bibinfo {volume} {76}}~(\bibinfo {number} {4}),\ \bibinfo
  {pages} {045302}}\BibitemShut {NoStop}%
\bibitem [{\citenamefont {Galitski}\ and\ \citenamefont
  {Spielman}(2013)}]{Galitski:2013}%
  \BibitemOpen
  \bibfield  {author} {\bibinfo {author} {\bibnamefont {Galitski},
  \bibfnamefont {Victor}}, \ and\ \bibinfo {author} {\bibfnamefont {Ian~B.}\
  \bibnamefont {Spielman}}} (\bibinfo {year} {2013}),\ \bibfield  {title}
  {\enquote {\bibinfo {title} {Spin-orbit coupling in quantum gases},}\ }\href
  {\doibase 10.1038/nature11841} {\bibfield  {journal} {\bibinfo  {journal}
  {Nature}\ }\textbf {\bibinfo {volume} {494}},\ \bibinfo {pages}
  {49}}\BibitemShut {NoStop}%
\bibitem [{\citenamefont {Gardiner}\ and\ \citenamefont
  {Zoller}(2004)}]{Gardiner:2004book}%
  \BibitemOpen
  \bibfield  {author} {\bibinfo {author} {\bibnamefont {Gardiner},
  \bibfnamefont {Crispin}}, \ and\ \bibinfo {author} {\bibfnamefont {Peter}\
  \bibnamefont {Zoller}}} (\bibinfo {year} {2004}),\ \href@noop {} {\emph
  {\bibinfo {title} {Quantum noise}}}\ (\bibinfo  {publisher} {Springer Science
  \& Business Media})\BibitemShut {NoStop}%
\bibitem [{\citenamefont {Ge}\ and\ \citenamefont {Rigol}(2017)}]{Ye2017}%
  \BibitemOpen
  \bibfield  {author} {\bibinfo {author} {\bibnamefont {Ge}, \bibfnamefont
  {Yang}}, \ and\ \bibinfo {author} {\bibfnamefont {Marcos}\ \bibnamefont
  {Rigol}}} (\bibinfo {year} {2017}),\ \bibfield  {title} {\enquote {\bibinfo
  {title} {Topological phase transitions in finite-size periodically driven
  translationally invariant systems},}\ }\href {\doibase
  10.1103/PhysRevA.96.023610} {\bibfield  {journal} {\bibinfo  {journal} {Phys.
  Rev. A}\ }\textbf {\bibinfo {volume} {96}},\ \bibinfo {pages}
  {023610}}\BibitemShut {NoStop}%
\bibitem [{\citenamefont {Goldman}\ and\ \citenamefont
  {Dalibard}(2014)}]{goldmandalibard}%
  \BibitemOpen
  \bibfield  {author} {\bibinfo {author} {\bibnamefont {Goldman}, \bibfnamefont
  {N}}, \ and\ \bibinfo {author} {\bibfnamefont {J.}~\bibnamefont {Dalibard}}}
  (\bibinfo {year} {2014}),\ \bibfield  {title} {\enquote {\bibinfo {title}
  {Periodically driven quantum systems: Effective {H}amiltonians and engineered
  gauge fields},}\ }\href {\doibase 10.1103/PhysRevX.4.031027} {\bibfield
  {journal} {\bibinfo  {journal} {Phys. Rev. X}\ }\textbf {\bibinfo {volume}
  {4}},\ \bibinfo {pages} {031027}}\BibitemShut {NoStop}%
\bibitem [{\citenamefont {Goldman}\ \emph {et~al.}(2016)\citenamefont
  {Goldman}, \citenamefont {Jotzu}, \citenamefont {Messer}, \citenamefont
  {G\"org}, \citenamefont {Desbuquois},\ and\ \citenamefont
  {Esslinger}}]{Goldman:2016}%
  \BibitemOpen
  \bibfield  {author} {\bibinfo {author} {\bibnamefont {Goldman}, \bibfnamefont
  {N}}, \bibinfo {author} {\bibfnamefont {G.}~\bibnamefont {Jotzu}}, \bibinfo
  {author} {\bibfnamefont {M.}~\bibnamefont {Messer}}, \bibinfo {author}
  {\bibfnamefont {F.}~\bibnamefont {G\"org}}, \bibinfo {author} {\bibfnamefont
  {R.}~\bibnamefont {Desbuquois}}, \ and\ \bibinfo {author} {\bibfnamefont
  {T.}~\bibnamefont {Esslinger}}} (\bibinfo {year} {2016}),\ \bibfield  {title}
  {\enquote {\bibinfo {title} {Creating topological interfaces and detecting
  chiral edge modes in a two-dimensional optical lattice},}\ }\href {\doibase
  10.1103/PhysRevA.94.043611} {\bibfield  {journal} {\bibinfo  {journal} {Phys.
  Rev. A}\ }\textbf {\bibinfo {volume} {94}},\ \bibinfo {pages}
  {043611}}\BibitemShut {NoStop}%
\bibitem [{\citenamefont {Goldman}\ \emph {et~al.}(2010)\citenamefont
  {Goldman}, \citenamefont {Satija}, \citenamefont {Nikolic}, \citenamefont
  {Bermudez}, \citenamefont {Martin-Delgado}, \citenamefont {Lewenstein},\ and\
  \citenamefont {Spielman}}]{PhysRevLett.105.255302}%
  \BibitemOpen
  \bibfield  {author} {\bibinfo {author} {\bibnamefont {Goldman}, \bibfnamefont
  {N}}, \bibinfo {author} {\bibfnamefont {I.}~\bibnamefont {Satija}}, \bibinfo
  {author} {\bibfnamefont {P.}~\bibnamefont {Nikolic}}, \bibinfo {author}
  {\bibfnamefont {A.}~\bibnamefont {Bermudez}}, \bibinfo {author}
  {\bibfnamefont {M.~A.}\ \bibnamefont {Martin-Delgado}}, \bibinfo {author}
  {\bibfnamefont {M.}~\bibnamefont {Lewenstein}}, \ and\ \bibinfo {author}
  {\bibfnamefont {I.~B.}\ \bibnamefont {Spielman}}} (\bibinfo {year} {2010}),\
  \bibfield  {title} {\enquote {\bibinfo {title} {Realistic time-reversal
  invariant topological insulators with neutral atoms},}\ }\href {\doibase
  10.1103/PhysRevLett.105.255302} {\bibfield  {journal} {\bibinfo  {journal}
  {Phys. Rev. Lett.}\ }\textbf {\bibinfo {volume} {105}},\ \bibinfo {pages}
  {255302}}\BibitemShut {NoStop}%
\bibitem [{\citenamefont {Goldman}\ \emph {et~al.}(2013)\citenamefont
  {Goldman}, \citenamefont {Dalibard}, \citenamefont {Dauphin}, \citenamefont
  {Gerbier}, \citenamefont {Lewenstein}, \citenamefont {Zoller},\ and\
  \citenamefont {Spielman}}]{Goldman:2013}%
  \BibitemOpen
  \bibfield  {author} {\bibinfo {author} {\bibnamefont {Goldman}, \bibfnamefont
  {Nathan}}, \bibinfo {author} {\bibfnamefont {Jean}\ \bibnamefont {Dalibard}},
  \bibinfo {author} {\bibfnamefont {Alexandre}\ \bibnamefont {Dauphin}},
  \bibinfo {author} {\bibfnamefont {Fabrice}\ \bibnamefont {Gerbier}}, \bibinfo
  {author} {\bibfnamefont {Maciej}\ \bibnamefont {Lewenstein}}, \bibinfo
  {author} {\bibfnamefont {Peter}\ \bibnamefont {Zoller}}, \ and\ \bibinfo
  {author} {\bibfnamefont {Ian~B.}\ \bibnamefont {Spielman}}} (\bibinfo {year}
  {2013}),\ \bibfield  {title} {\enquote {\bibinfo {title} {Direct imaging of
  topological edge states in cold-atom systems},}\ }\href {\doibase
  10.1073/pnas.1300170110} {\bibfield  {journal} {\bibinfo  {journal}
  {Proceedings of the National Academy of Sciences}\ }\textbf {\bibinfo
  {volume} {110}}~(\bibinfo {number} {17}),\ \bibinfo {pages}
  {6736--6741}}\BibitemShut {NoStop}%
\bibitem [{\citenamefont {{Gong}}\ and\ \citenamefont
  {{Ueda}}(2017)}]{2017arXiv171005289G}%
  \BibitemOpen
  \bibfield  {author} {\bibinfo {author} {\bibnamefont {{Gong}}, \bibfnamefont
  {Z}}, \ and\ \bibinfo {author} {\bibfnamefont {M.}~\bibnamefont {{Ueda}}}}
  (\bibinfo {year} {2017}),\ \bibfield  {title} {\enquote {\bibinfo {title}
  {{Entanglement-Spectrum Crossing and Momentum-Time {S}kyrmions in Quench
  Dynamics}},}\ }\href@noop {} {\bibfield  {journal} {\bibinfo  {journal}
  {ArXiv e-prints}\ }}\Eprint {http://arxiv.org/abs/1710.05289}
  {arXiv:1710.05289} \BibitemShut {NoStop}%
\bibitem [{\citenamefont {G{\"o}rg}\ \emph {et~al.}(2018)\citenamefont
  {G{\"o}rg}, \citenamefont {Messer}, \citenamefont {Sandholzer}, \citenamefont
  {Jotzu}, \citenamefont {Desbuquois},\ and\ \citenamefont
  {Esslinger}}]{goerg2017}%
  \BibitemOpen
  \bibfield  {author} {\bibinfo {author} {\bibnamefont {G{\"o}rg},
  \bibfnamefont {Frederik}}, \bibinfo {author} {\bibfnamefont {Michael}\
  \bibnamefont {Messer}}, \bibinfo {author} {\bibfnamefont {Kilian}\
  \bibnamefont {Sandholzer}}, \bibinfo {author} {\bibfnamefont {Gregor}\
  \bibnamefont {Jotzu}}, \bibinfo {author} {\bibfnamefont {R{\'e}mi}\
  \bibnamefont {Desbuquois}}, \ and\ \bibinfo {author} {\bibfnamefont {Tilman}\
  \bibnamefont {Esslinger}}} (\bibinfo {year} {2018}),\ \bibfield  {title}
  {\enquote {\bibinfo {title} {Enhancement and sign change of magnetic
  correlations in a driven quantum many-body system},}\ }\href
  {http://dx.doi.org/10.1038/nature25135} {\bibfield  {journal} {\bibinfo
  {journal} {Nature}\ }\textbf {\bibinfo {volume} {553}},\ \bibinfo {pages}
  {481}}\BibitemShut {NoStop}%
\bibitem [{\citenamefont {Greiner}\ \emph {et~al.}(2001)\citenamefont
  {Greiner}, \citenamefont {Bloch}, \citenamefont {Mandel}, \citenamefont
  {H\"ansch},\ and\ \citenamefont {Esslinger}}]{Greiner:2001}%
  \BibitemOpen
  \bibfield  {author} {\bibinfo {author} {\bibnamefont {Greiner}, \bibfnamefont
  {Markus}}, \bibinfo {author} {\bibfnamefont {Immanuel}\ \bibnamefont
  {Bloch}}, \bibinfo {author} {\bibfnamefont {Olaf}\ \bibnamefont {Mandel}},
  \bibinfo {author} {\bibfnamefont {Theodor~W.}\ \bibnamefont {H\"ansch}}, \
  and\ \bibinfo {author} {\bibfnamefont {Tilman}\ \bibnamefont {Esslinger}}}
  (\bibinfo {year} {2001}),\ \bibfield  {title} {\enquote {\bibinfo {title}
  {Exploring phase coherence in a 2{D} lattice of {B}ose-{E}instein
  condensates},}\ }\href {\doibase 10.1103/PhysRevLett.87.160405} {\bibfield
  {journal} {\bibinfo  {journal} {Phys. Rev. Lett.}\ }\textbf {\bibinfo
  {volume} {87}},\ \bibinfo {pages} {160405}}\BibitemShut {NoStop}%
\bibitem [{\citenamefont {Greschner}\ \emph {et~al.}(2015)\citenamefont
  {Greschner}, \citenamefont {Piraud}, \citenamefont {Heidrich-Meisner},
  \citenamefont {McCulloch}, \citenamefont {Schollw\"ock},\ and\ \citenamefont
  {Vekua}}]{PhysRevLett.115.190402}%
  \BibitemOpen
  \bibfield  {author} {\bibinfo {author} {\bibnamefont {Greschner},
  \bibfnamefont {S}}, \bibinfo {author} {\bibfnamefont {M.}~\bibnamefont
  {Piraud}}, \bibinfo {author} {\bibfnamefont {F.}~\bibnamefont
  {Heidrich-Meisner}}, \bibinfo {author} {\bibfnamefont {I.~P.}\ \bibnamefont
  {McCulloch}}, \bibinfo {author} {\bibfnamefont {U.}~\bibnamefont
  {Schollw\"ock}}, \ and\ \bibinfo {author} {\bibfnamefont {T.}~\bibnamefont
  {Vekua}}} (\bibinfo {year} {2015}),\ \bibfield  {title} {\enquote {\bibinfo
  {title} {Spontaneous increase of magnetic flux and chiral-current reversal in
  bosonic ladders: Swimming against the tide},}\ }\href {\doibase
  10.1103/PhysRevLett.115.190402} {\bibfield  {journal} {\bibinfo  {journal}
  {Phys. Rev. Lett.}\ }\textbf {\bibinfo {volume} {115}},\ \bibinfo {pages}
  {190402}}\BibitemShut {NoStop}%
\bibitem [{\citenamefont {Grosfeld}\ \emph {et~al.}(2007)\citenamefont
  {Grosfeld}, \citenamefont {Cooper}, \citenamefont {Stern},\ and\
  \citenamefont {Ilan}}]{grosfeld:104516}%
  \BibitemOpen
  \bibfield  {author} {\bibinfo {author} {\bibnamefont {Grosfeld},
  \bibfnamefont {Eytan}}, \bibinfo {author} {\bibfnamefont {Nigel~R.}\
  \bibnamefont {Cooper}}, \bibinfo {author} {\bibfnamefont {Ady}\ \bibnamefont
  {Stern}}, \ and\ \bibinfo {author} {\bibfnamefont {Roni}\ \bibnamefont
  {Ilan}}} (\bibinfo {year} {2007}),\ \bibfield  {title} {\enquote {\bibinfo
  {title} {Predicted signatures of $p$-wave superfluid phases and {M}ajorana
  zero modes of fermionic atoms in rf absorption},}\ }\href {\doibase
  10.1103/PhysRevB.76.104516} {\bibfield  {journal} {\bibinfo  {journal} {Phys.
  Rev. B}\ }\textbf {\bibinfo {volume} {76}}~(\bibinfo {number} {10}),\
  \bibinfo {eid} {104516}}\BibitemShut {NoStop}%
\bibitem [{\citenamefont {Grusdt}(2017)}]{grusdt2017topological}%
  \BibitemOpen
  \bibfield  {author} {\bibinfo {author} {\bibnamefont {Grusdt}, \bibfnamefont
  {F}}} (\bibinfo {year} {2017}),\ \bibfield  {title} {\enquote {\bibinfo
  {title} {Topological order of mixed states in correlated quantum many-body
  systems},}\ }\href {\doibase 10.1103/PhysRevB.95.075106} {\bibfield
  {journal} {\bibinfo  {journal} {Phys. Rev. B}\ }\textbf {\bibinfo {volume}
  {95}}~(\bibinfo {number} {7}),\ \bibinfo {pages} {075106}}\BibitemShut
  {NoStop}%
\bibitem [{\citenamefont {Gurarie}\ and\ \citenamefont
  {Radzihovsky}(2007)}]{gurarie2007resonantly}%
  \BibitemOpen
  \bibfield  {author} {\bibinfo {author} {\bibnamefont {Gurarie}, \bibfnamefont
  {V}}, \ and\ \bibinfo {author} {\bibfnamefont {L}~\bibnamefont
  {Radzihovsky}}} (\bibinfo {year} {2007}),\ \bibfield  {title} {\enquote
  {\bibinfo {title} {Resonantly paired fermionic superfluids},}\ }\href
  {\doibase 10.1016/j.aop.2006.10.009} {\bibfield  {journal} {\bibinfo
  {journal} {Annals of Physics}\ }\textbf {\bibinfo {volume} {322}}~(\bibinfo
  {number} {1}),\ \bibinfo {pages} {2--119}}\BibitemShut {NoStop}%
\bibitem [{\citenamefont {Hafezi}\ \emph {et~al.}(2007)\citenamefont {Hafezi},
  \citenamefont {S{\o}rensen}, \citenamefont {Demler},\ and\ \citenamefont
  {Lukin}}]{hafezi-2007}%
  \BibitemOpen
  \bibfield  {author} {\bibinfo {author} {\bibnamefont {Hafezi}, \bibfnamefont
  {M}}, \bibinfo {author} {\bibfnamefont {A.~S.}\ \bibnamefont {S{\o}rensen}},
  \bibinfo {author} {\bibfnamefont {E.}~\bibnamefont {Demler}}, \ and\ \bibinfo
  {author} {\bibfnamefont {M.~D.}\ \bibnamefont {Lukin}}} (\bibinfo {year}
  {2007}),\ \bibfield  {title} {\enquote {\bibinfo {title} {Fractional quantum
  {H}all effect in optical lattices},}\ }\href {\doibase
  10.1103/PhysRevA.76.023613} {\bibfield  {journal} {\bibinfo  {journal} {Phys.
  Rev. A}\ }\textbf {\bibinfo {volume} {76}}~(\bibinfo {number} {2}),\ \bibinfo
  {eid} {023613}}\BibitemShut {NoStop}%
\bibitem [{\citenamefont {Haldane}(1988{\natexlab{a}})}]{haldanehoneycomb}%
  \BibitemOpen
  \bibfield  {author} {\bibinfo {author} {\bibnamefont {Haldane}, \bibfnamefont
  {F~D~M}}} (\bibinfo {year} {1988}{\natexlab{a}}),\ \bibfield  {title}
  {\enquote {\bibinfo {title} {Model for a quantum {H}all effect without landau
  levels: Condensed-matter realization of the `parity anomaly'},}\ }\href
  {\doibase 10.1103/PhysRevLett.61.2015} {\bibfield  {journal} {\bibinfo
  {journal} {Phys. Rev. Lett.}\ }\textbf {\bibinfo {volume} {61}}~(\bibinfo
  {number} {18}),\ \bibinfo {pages} {2015--2018}}\BibitemShut {NoStop}%
\bibitem [{\citenamefont {Haldane}(1988{\natexlab{b}})}]{Haldane:1988}%
  \BibitemOpen
  \bibfield  {author} {\bibinfo {author} {\bibnamefont {Haldane}, \bibfnamefont
  {F~D~M}}} (\bibinfo {year} {1988}{\natexlab{b}}),\ \bibfield  {title}
  {\enquote {\bibinfo {title} {Model for a quantum {H}all effect without
  {L}andau levels: Condensed-matter realization of the ``parity anomaly"},}\
  }\href {\doibase 10.1103/PhysRevLett.61.2015} {\bibfield  {journal} {\bibinfo
   {journal} {Phys. Rev. Lett.}\ }\textbf {\bibinfo {volume} {61}}~(\bibinfo
  {number} {18}),\ \bibinfo {pages} {2015--2018}}\BibitemShut {NoStop}%
\bibitem [{\citenamefont {Haldane}(1991)}]{haldaneexclusion}%
  \BibitemOpen
  \bibfield  {author} {\bibinfo {author} {\bibnamefont {Haldane}, \bibfnamefont
  {F~D~M}}} (\bibinfo {year} {1991}),\ \bibfield  {title} {\enquote {\bibinfo
  {title} {`'{F}ractional statistics' in arbitrary dimensions: A generalization
  of the {P}auli principle},}\ }\href {\doibase 10.1103/PhysRevLett.67.937}
  {\bibfield  {journal} {\bibinfo  {journal} {Phys. Rev. Lett.}\ }\textbf
  {\bibinfo {volume} {67}},\ \bibinfo {pages} {937--940}}\BibitemShut {NoStop}%
\bibitem [{\citenamefont {Halperin}(1982)}]{halperingauge}%
  \BibitemOpen
  \bibfield  {author} {\bibinfo {author} {\bibnamefont {Halperin},
  \bibfnamefont {B~I}}} (\bibinfo {year} {1982}),\ \bibfield  {title} {\enquote
  {\bibinfo {title} {Quantized {H}all conductance, current-carrying edge
  states, and the existence of extended states in a two-dimensional disordered
  potential},}\ }\href {\doibase 10.1103/PhysRevB.25.2185} {\bibfield
  {journal} {\bibinfo  {journal} {Phys. Rev. B}\ }\textbf {\bibinfo {volume}
  {25}}~(\bibinfo {number} {4}),\ \bibinfo {pages} {2185--2190}}\BibitemShut
  {NoStop}%
\bibitem [{\citenamefont {Haroche}\ \emph {et~al.}(1970)\citenamefont
  {Haroche}, \citenamefont {Cohen-Tannoudji}, \citenamefont {Audoin},\ and\
  \citenamefont {Schermann}}]{Haroche1970}%
  \BibitemOpen
  \bibfield  {author} {\bibinfo {author} {\bibnamefont {Haroche}, \bibfnamefont
  {S}}, \bibinfo {author} {\bibfnamefont {C.}~\bibnamefont {Cohen-Tannoudji}},
  \bibinfo {author} {\bibfnamefont {C.}~\bibnamefont {Audoin}}, \ and\ \bibinfo
  {author} {\bibfnamefont {J.~P.}\ \bibnamefont {Schermann}}} (\bibinfo {year}
  {1970}),\ \bibfield  {title} {\enquote {\bibinfo {title} {Modified {Z}eeman
  hyperfine spectra observed in {H}$^{1}$ and {R}b$^{87}$ ground states
  interacting with a nonresonant rf field},}\ }\href {\doibase
  10.1103/PhysRevLett.24.861} {\bibfield  {journal} {\bibinfo  {journal} {Phys.
  Rev. Lett.}\ }\textbf {\bibinfo {volume} {24}},\ \bibinfo {pages}
  {861--864}}\BibitemShut {NoStop}%
\bibitem [{\citenamefont {Harper}(1955)}]{Harper1955}%
  \BibitemOpen
  \bibfield  {author} {\bibinfo {author} {\bibnamefont {Harper}, \bibfnamefont
  {P~G}}} (\bibinfo {year} {1955}),\ \bibfield  {title} {\enquote {\bibinfo
  {title} {The general motion of conduction electrons in a uniform magnetic
  field, with application to the diamagnetism of metals},}\ }\href
  {http://stacks.iop.org/0370-1298/68/i=10/a=305} {\bibfield  {journal}
  {\bibinfo  {journal} {Proceedings of the Physical Society. Section A}\
  }\textbf {\bibinfo {volume} {68}}~(\bibinfo {number} {10}),\ \bibinfo {pages}
  {879}}\BibitemShut {NoStop}%
\bibitem [{\citenamefont {Hasan}\ and\ \citenamefont {Kane}(2010)}]{hasankane}%
  \BibitemOpen
  \bibfield  {author} {\bibinfo {author} {\bibnamefont {Hasan}, \bibfnamefont
  {M~Z}}, \ and\ \bibinfo {author} {\bibfnamefont {C.~L.}\ \bibnamefont
  {Kane}}} (\bibinfo {year} {2010}),\ \bibfield  {title} {\enquote {\bibinfo
  {title} {Colloquium: Topological insulators},}\ }\href {\doibase
  10.1103/RevModPhys.82.3045} {\bibfield  {journal} {\bibinfo  {journal} {Rev.
  Mod. Phys.}\ }\textbf {\bibinfo {volume} {82}}~(\bibinfo {number} {4}),\
  \bibinfo {pages} {3045--3067}}\BibitemShut {NoStop}%
\bibitem [{\citenamefont {Hatsugai}(1993)}]{Hatsugai:1993}%
  \BibitemOpen
  \bibfield  {author} {\bibinfo {author} {\bibnamefont {Hatsugai},
  \bibfnamefont {Yasuhiro}}} (\bibinfo {year} {1993}),\ \bibfield  {title}
  {\enquote {\bibinfo {title} {Chern number and edge states in the integer
  quantum {H}all effect},}\ }\href {\doibase 10.1103/PhysRevLett.71.3697}
  {\bibfield  {journal} {\bibinfo  {journal} {Phys. Rev. Lett.}\ }\textbf
  {\bibinfo {volume} {71}},\ \bibinfo {pages} {3697--3700}}\BibitemShut
  {NoStop}%
\bibitem [{\citenamefont {Hauke}\ \emph {et~al.}(2014)\citenamefont {Hauke},
  \citenamefont {Lewenstein},\ and\ \citenamefont {Eckardt}}]{Hauke:2014}%
  \BibitemOpen
  \bibfield  {author} {\bibinfo {author} {\bibnamefont {Hauke}, \bibfnamefont
  {Philipp}}, \bibinfo {author} {\bibfnamefont {Maciej}\ \bibnamefont
  {Lewenstein}}, \ and\ \bibinfo {author} {\bibfnamefont {Andr\'e}\
  \bibnamefont {Eckardt}}} (\bibinfo {year} {2014}),\ \bibfield  {title}
  {\enquote {\bibinfo {title} {Tomography of band insulators from quench
  dynamics},}\ }\href {\doibase 10.1103/PhysRevLett.113.045303} {\bibfield
  {journal} {\bibinfo  {journal} {Phys. Rev. Lett.}\ }\textbf {\bibinfo
  {volume} {113}},\ \bibinfo {pages} {045303}}\BibitemShut {NoStop}%
\bibitem [{\citenamefont {He}\ \emph {et~al.}(2017)\citenamefont {He},
  \citenamefont {Grusdt}, \citenamefont {Kaufman}, \citenamefont {Greiner},\
  and\ \citenamefont {Vishwanath}}]{2017arXiv170300430H}%
  \BibitemOpen
  \bibfield  {author} {\bibinfo {author} {\bibnamefont {He}, \bibfnamefont
  {Yin-Chen}}, \bibinfo {author} {\bibfnamefont {Fabian}\ \bibnamefont
  {Grusdt}}, \bibinfo {author} {\bibfnamefont {Adam}\ \bibnamefont {Kaufman}},
  \bibinfo {author} {\bibfnamefont {Markus}\ \bibnamefont {Greiner}}, \ and\
  \bibinfo {author} {\bibfnamefont {Ashvin}\ \bibnamefont {Vishwanath}}}
  (\bibinfo {year} {2017}),\ \bibfield  {title} {\enquote {\bibinfo {title}
  {Realizing and adiabatically preparing bosonic integer and fractional quantum
  {H}all states in optical lattices},}\ }\href {\doibase
  10.1103/PhysRevB.96.201103} {\bibfield  {journal} {\bibinfo  {journal} {Phys.
  Rev. B}\ }\textbf {\bibinfo {volume} {96}},\ \bibinfo {pages}
  {201103}}\BibitemShut {NoStop}%
\bibitem [{\citenamefont {Heyl}(2018)}]{heylreview}%
  \BibitemOpen
  \bibfield  {author} {\bibinfo {author} {\bibnamefont {Heyl}, \bibfnamefont
  {Markus}}} (\bibinfo {year} {2018}),\ \bibfield  {title} {\enquote {\bibinfo
  {title} {Dynamical quantum phase transitions: a review},}\ }\href
  {http://stacks.iop.org/0034-4885/81/i=5/a=054001} {\bibfield  {journal}
  {\bibinfo  {journal} {Reports on Progress in Physics}\ }\textbf {\bibinfo
  {volume} {81}}~(\bibinfo {number} {5}),\ \bibinfo {pages}
  {054001}}\BibitemShut {NoStop}%
\bibitem [{\citenamefont {Hickey}\ \emph {et~al.}(2016)\citenamefont {Hickey},
  \citenamefont {Cincio}, \citenamefont {Papi{\'{c}}},\ and\ \citenamefont
  {Paramekanti}}]{PhysRevLett.116.137202}%
  \BibitemOpen
  \bibfield  {author} {\bibinfo {author} {\bibnamefont {Hickey}, \bibfnamefont
  {Ciar{\'a}n}}, \bibinfo {author} {\bibfnamefont {Lukasz}\ \bibnamefont
  {Cincio}}, \bibinfo {author} {\bibfnamefont {Zlatko}\ \bibnamefont
  {Papi{\'{c}}}}, \ and\ \bibinfo {author} {\bibfnamefont {Arun}\ \bibnamefont
  {Paramekanti}}} (\bibinfo {year} {2016}),\ \bibfield  {title} {\enquote
  {\bibinfo {title} {Haldane-{H}ubbard {M}ott insulator: From tetrahedral spin
  crystal to chiral spin liquid},}\ }\href {\doibase
  10.1103/PhysRevLett.116.137202} {\bibfield  {journal} {\bibinfo  {journal}
  {Phys. Rev. Lett.}\ }\textbf {\bibinfo {volume} {116}},\ \bibinfo {pages}
  {137202}}\BibitemShut {NoStop}%
\bibitem [{\citenamefont {Hofstadter}(1976)}]{Hofstadter1976}%
  \BibitemOpen
  \bibfield  {author} {\bibinfo {author} {\bibnamefont {Hofstadter},
  \bibfnamefont {Douglas~R}}} (\bibinfo {year} {1976}),\ \bibfield  {title}
  {\enquote {\bibinfo {title} {{Energy levels and wave functions of Bloch
  electrons in rational and irrational magnetic fields}},}\ }\href {\doibase
  10.1103/PhysRevB.14.2239} {\bibfield  {journal} {\bibinfo  {journal} {Phys.
  Rev. B}\ }\textbf {\bibinfo {volume} {14}}~(\bibinfo {number} {6}),\ \bibinfo
  {pages} {2239--2249}}\BibitemShut {NoStop}%
\bibitem [{\citenamefont {van Houten}\ \emph {et~al.}(1989)\citenamefont {van
  Houten}, \citenamefont {Beenakker}, \citenamefont {Williamson}, \citenamefont
  {Broekaart}, \citenamefont {van Loosdrecht}, \citenamefont {van Wees},
  \citenamefont {Mooij}, \citenamefont {Foxon},\ and\ \citenamefont
  {Harris}}]{Houten1989}%
  \BibitemOpen
  \bibfield  {author} {\bibinfo {author} {\bibnamefont {van Houten},
  \bibfnamefont {H}}, \bibinfo {author} {\bibfnamefont {C~W~J}\ \bibnamefont
  {Beenakker}}, \bibinfo {author} {\bibfnamefont {J~G}\ \bibnamefont
  {Williamson}}, \bibinfo {author} {\bibfnamefont {M~E~I}\ \bibnamefont
  {Broekaart}}, \bibinfo {author} {\bibfnamefont {P~H~M}\ \bibnamefont {van
  Loosdrecht}}, \bibinfo {author} {\bibfnamefont {B~J}\ \bibnamefont {van
  Wees}}, \bibinfo {author} {\bibfnamefont {J~E}\ \bibnamefont {Mooij}},
  \bibinfo {author} {\bibfnamefont {C~T}\ \bibnamefont {Foxon}}, \ and\
  \bibinfo {author} {\bibfnamefont {J~J}\ \bibnamefont {Harris}}} (\bibinfo
  {year} {1989}),\ \bibfield  {title} {\enquote {\bibinfo {title} {{Coherent
  electron focusing with quantum point contacts in a two-dimensional electron
  gas}},}\ }\href {\doibase 10.1103/PhysRevB.39.8556} {\bibfield  {journal}
  {\bibinfo  {journal} {Phys. Rev. B}\ }\textbf {\bibinfo {volume}
  {39}}~(\bibinfo {number} {12}),\ \bibinfo {pages} {8556--8575}}\BibitemShut
  {NoStop}%
\bibitem [{\citenamefont {Hu}\ \emph {et~al.}(2017)\citenamefont {Hu},
  \citenamefont {Wang}, \citenamefont {Shum},\ and\ \citenamefont
  {Chong}}]{hu2017exceptional}%
  \BibitemOpen
  \bibfield  {author} {\bibinfo {author} {\bibnamefont {Hu}, \bibfnamefont
  {Wenchao}}, \bibinfo {author} {\bibfnamefont {Hailong}\ \bibnamefont {Wang}},
  \bibinfo {author} {\bibfnamefont {Perry~Ping}\ \bibnamefont {Shum}}, \ and\
  \bibinfo {author} {\bibfnamefont {YD}~\bibnamefont {Chong}}} (\bibinfo {year}
  {2017}),\ \bibfield  {title} {\enquote {\bibinfo {title} {Exceptional points
  in a non-{H}ermitian topological pump},}\ }\href {\doibase
  10.1103/PhysRevB.95.184306} {\bibfield  {journal} {\bibinfo  {journal} {Phys.
  Rev. B}\ }\textbf {\bibinfo {volume} {95}}~(\bibinfo {number} {18}),\
  \bibinfo {pages} {184306}}\BibitemShut {NoStop}%
\bibitem [{\citenamefont {Huang}\ \emph {et~al.}(2016)\citenamefont {Huang},
  \citenamefont {Meng}, \citenamefont {Wang}, \citenamefont {Peng},
  \citenamefont {Zhang}, \citenamefont {Chen}, \citenamefont {Li},
  \citenamefont {Zhou},\ and\ \citenamefont {Zhang}}]{Huang:2016}%
  \BibitemOpen
  \bibfield  {author} {\bibinfo {author} {\bibnamefont {Huang}, \bibfnamefont
  {Lianghui}}, \bibinfo {author} {\bibfnamefont {Zengming}\ \bibnamefont
  {Meng}}, \bibinfo {author} {\bibfnamefont {Pengjun}\ \bibnamefont {Wang}},
  \bibinfo {author} {\bibfnamefont {Peng}\ \bibnamefont {Peng}}, \bibinfo
  {author} {\bibfnamefont {Shao-Liang}\ \bibnamefont {Zhang}}, \bibinfo
  {author} {\bibfnamefont {Liangchao}\ \bibnamefont {Chen}}, \bibinfo {author}
  {\bibfnamefont {Donghao}\ \bibnamefont {Li}}, \bibinfo {author}
  {\bibfnamefont {Qi}~\bibnamefont {Zhou}}, \ and\ \bibinfo {author}
  {\bibfnamefont {Jing}\ \bibnamefont {Zhang}}} (\bibinfo {year} {2016}),\
  \bibfield  {title} {\enquote {\bibinfo {title} {Experimental realization of
  two-dimensional synthetic spin-orbit coupling in ultracold {F}ermi gases},}\
  }\href {\doibase 10.1038/nphys3672} {\bibfield  {journal} {\bibinfo
  {journal} {Nature Physics}\ }\textbf {\bibinfo {volume} {12}}~(\bibinfo
  {number} {6}),\ \bibinfo {pages} {540--544}}\BibitemShut {NoStop}%
\bibitem [{\citenamefont {Huang}\ and\ \citenamefont
  {Arovas}(2014)}]{huang2014topological}%
  \BibitemOpen
  \bibfield  {author} {\bibinfo {author} {\bibnamefont {Huang}, \bibfnamefont
  {Zhoushen}}, \ and\ \bibinfo {author} {\bibfnamefont {Daniel~P}\ \bibnamefont
  {Arovas}}} (\bibinfo {year} {2014}),\ \bibfield  {title} {\enquote {\bibinfo
  {title} {Topological indices for open and thermal systems via {U}hlmann's
  phase},}\ }\href {\doibase 10.1103/PhysRevLett.113.076407} {\bibfield
  {journal} {\bibinfo  {journal} {Phys. Rev. Lett.}\ }\textbf {\bibinfo
  {volume} {113}}~(\bibinfo {number} {7}),\ \bibinfo {pages}
  {076407}}\BibitemShut {NoStop}%
\bibitem [{\citenamefont {Huber}\ and\ \citenamefont
  {Altman}(2010)}]{PhysRevB.82.184502}%
  \BibitemOpen
  \bibfield  {author} {\bibinfo {author} {\bibnamefont {Huber}, \bibfnamefont
  {Sebastian~D}}, \ and\ \bibinfo {author} {\bibfnamefont {Ehud}\ \bibnamefont
  {Altman}}} (\bibinfo {year} {2010}),\ \bibfield  {title} {\enquote {\bibinfo
  {title} {Bose condensation in flat bands},}\ }\href {\doibase
  10.1103/PhysRevB.82.184502} {\bibfield  {journal} {\bibinfo  {journal} {Phys.
  Rev. B}\ }\textbf {\bibinfo {volume} {82}},\ \bibinfo {pages}
  {184502}}\BibitemShut {NoStop}%
\bibitem [{\citenamefont {H{\"u}gel}\ and\ \citenamefont
  {Paredes}(2014)}]{Hugel2014}%
  \BibitemOpen
  \bibfield  {author} {\bibinfo {author} {\bibnamefont {H{\"u}gel},
  \bibfnamefont {Dario}}, \ and\ \bibinfo {author} {\bibfnamefont {Bel{\'e}n}\
  \bibnamefont {Paredes}}} (\bibinfo {year} {2014}),\ \bibfield  {title}
  {\enquote {\bibinfo {title} {{Chiral ladders and the edges of quantum Hall
  insulators}},}\ }\href {\doibase 10.1103/PhysRevA.89.023619} {\bibfield
  {journal} {\bibinfo  {journal} {Phys. Rev. A}\ }\textbf {\bibinfo {volume}
  {89}}~(\bibinfo {number} {2}),\ \bibinfo {pages} {023619}}\BibitemShut
  {NoStop}%
\bibitem [{\citenamefont {Iemini}\ \emph {et~al.}(2017)\citenamefont {Iemini},
  \citenamefont {Mazza}, \citenamefont {Fallani}, \citenamefont {Zoller},
  \citenamefont {Fazio},\ and\ \citenamefont {Dalmonte}}]{Iemini:2017}%
  \BibitemOpen
  \bibfield  {author} {\bibinfo {author} {\bibnamefont {Iemini}, \bibfnamefont
  {F}}, \bibinfo {author} {\bibfnamefont {L.}~\bibnamefont {Mazza}}, \bibinfo
  {author} {\bibfnamefont {L.}~\bibnamefont {Fallani}}, \bibinfo {author}
  {\bibfnamefont {P.}~\bibnamefont {Zoller}}, \bibinfo {author} {\bibfnamefont
  {R.}~\bibnamefont {Fazio}}, \ and\ \bibinfo {author} {\bibfnamefont
  {M.}~\bibnamefont {Dalmonte}}} (\bibinfo {year} {2017}),\ \bibfield  {title}
  {\enquote {\bibinfo {title} {Majorana quasiparticles protected by
  ${\mathbb{z}}_{2}$ angular momentum conservation},}\ }\href {\doibase
  10.1103/PhysRevLett.118.200404} {\bibfield  {journal} {\bibinfo  {journal}
  {Phys. Rev. Lett.}\ }\textbf {\bibinfo {volume} {118}},\ \bibinfo {pages}
  {200404}}\BibitemShut {NoStop}%
\bibitem [{\citenamefont {Iemini}\ \emph {et~al.}(2015)\citenamefont {Iemini},
  \citenamefont {Mazza}, \citenamefont {Rossini}, \citenamefont {Fazio},\ and\
  \citenamefont {Diehl}}]{Iemini:2015}%
  \BibitemOpen
  \bibfield  {author} {\bibinfo {author} {\bibnamefont {Iemini}, \bibfnamefont
  {Fernando}}, \bibinfo {author} {\bibfnamefont {Leonardo}\ \bibnamefont
  {Mazza}}, \bibinfo {author} {\bibfnamefont {Davide}\ \bibnamefont {Rossini}},
  \bibinfo {author} {\bibfnamefont {Rosario}\ \bibnamefont {Fazio}}, \ and\
  \bibinfo {author} {\bibfnamefont {Sebastian}\ \bibnamefont {Diehl}}}
  (\bibinfo {year} {2015}),\ \bibfield  {title} {\enquote {\bibinfo {title}
  {Localized {M}ajorana-like modes in a number-conserving setting: An exactly
  solvable model},}\ }\href {\doibase 10.1103/PhysRevLett.115.156402}
  {\bibfield  {journal} {\bibinfo  {journal} {Phys. Rev. Lett.}\ }\textbf
  {\bibinfo {volume} {115}},\ \bibinfo {pages} {156402}}\BibitemShut {NoStop}%
\bibitem [{\citenamefont {Imri{\v{s}}ka}\ \emph {et~al.}(2016)\citenamefont
  {Imri{\v{s}}ka}, \citenamefont {Wang},\ and\ \citenamefont
  {Troyer}}]{PhysRevB.94.035109}%
  \BibitemOpen
  \bibfield  {author} {\bibinfo {author} {\bibnamefont {Imri{\v{s}}ka},
  \bibfnamefont {Jakub}}, \bibinfo {author} {\bibfnamefont {Lei}\ \bibnamefont
  {Wang}}, \ and\ \bibinfo {author} {\bibfnamefont {Matthias}\ \bibnamefont
  {Troyer}}} (\bibinfo {year} {2016}),\ \bibfield  {title} {\enquote {\bibinfo
  {title} {First-order topological phase transition of the {H}aldane-{H}ubbard
  model},}\ }\href {\doibase 10.1103/PhysRevB.94.035109} {\bibfield  {journal}
  {\bibinfo  {journal} {Phys. Rev. B}\ }\textbf {\bibinfo {volume} {94}},\
  \bibinfo {pages} {035109}}\BibitemShut {NoStop}%
\bibitem [{\citenamefont {Jackiw}\ and\ \citenamefont
  {Rebbi}(1976)}]{Jackiw:1976}%
  \BibitemOpen
  \bibfield  {author} {\bibinfo {author} {\bibnamefont {Jackiw}, \bibfnamefont
  {R}}, \ and\ \bibinfo {author} {\bibfnamefont {C.}~\bibnamefont {Rebbi}}}
  (\bibinfo {year} {1976}),\ \bibfield  {title} {\enquote {\bibinfo {title}
  {Solitons with fermion number \textonehalf{}},}\ }\href {\doibase
  10.1103/PhysRevD.13.3398} {\bibfield  {journal} {\bibinfo  {journal} {Phys.
  Rev. D}\ }\textbf {\bibinfo {volume} {13}},\ \bibinfo {pages}
  {3398--3409}}\BibitemShut {NoStop}%
\bibitem [{\citenamefont {Jaksch}\ and\ \citenamefont
  {Zoller}(2003)}]{Jaksch2003}%
  \BibitemOpen
  \bibfield  {author} {\bibinfo {author} {\bibnamefont {Jaksch}, \bibfnamefont
  {D}}, \ and\ \bibinfo {author} {\bibfnamefont {P.}~\bibnamefont {Zoller}}}
  (\bibinfo {year} {2003}),\ \bibfield  {title} {\enquote {\bibinfo {title}
  {{Creation of effective magnetic fields in optical lattices: the {H}ofstadter
  butterfly for cold neutral atoms}},}\ }\href
  {http://stacks.iop.org/1367-2630/5/56} {\bibfield  {journal} {\bibinfo
  {journal} {New Journal of Physics}\ }\textbf {\bibinfo {volume} {5}},\
  \bibinfo {pages} {56}}\BibitemShut {NoStop}%
\bibitem [{\citenamefont {Jiang}\ \emph {et~al.}(2011)\citenamefont {Jiang},
  \citenamefont {Kitagawa}, \citenamefont {Alicea}, \citenamefont {Akhmerov},
  \citenamefont {Pekker}, \citenamefont {Refael}, \citenamefont {Cirac},
  \citenamefont {Demler}, \citenamefont {Lukin},\ and\ \citenamefont
  {Zoller}}]{PhysRevLett.106.220402}%
  \BibitemOpen
  \bibfield  {author} {\bibinfo {author} {\bibnamefont {Jiang}, \bibfnamefont
  {Liang}}, \bibinfo {author} {\bibfnamefont {Takuya}\ \bibnamefont
  {Kitagawa}}, \bibinfo {author} {\bibfnamefont {Jason}\ \bibnamefont
  {Alicea}}, \bibinfo {author} {\bibfnamefont {A.~R.}\ \bibnamefont
  {Akhmerov}}, \bibinfo {author} {\bibfnamefont {David}\ \bibnamefont
  {Pekker}}, \bibinfo {author} {\bibfnamefont {Gil}\ \bibnamefont {Refael}},
  \bibinfo {author} {\bibfnamefont {J.~Ignacio}\ \bibnamefont {Cirac}},
  \bibinfo {author} {\bibfnamefont {Eugene}\ \bibnamefont {Demler}}, \bibinfo
  {author} {\bibfnamefont {Mikhail~D.}\ \bibnamefont {Lukin}}, \ and\ \bibinfo
  {author} {\bibfnamefont {Peter}\ \bibnamefont {Zoller}}} (\bibinfo {year}
  {2011}),\ \bibfield  {title} {\enquote {\bibinfo {title} {Majorana fermions
  in equilibrium and in driven cold-atom quantum wires},}\ }\href {\doibase
  10.1103/PhysRevLett.106.220402} {\bibfield  {journal} {\bibinfo  {journal}
  {Phys. Rev. Lett.}\ }\textbf {\bibinfo {volume} {106}},\ \bibinfo {pages}
  {220402}}\BibitemShut {NoStop}%
\bibitem [{\citenamefont {Jim{\'e}nez-Garc{\'\i}a}\ and\ \citenamefont
  {Spielman}(2013)}]{Jimenez-Garcia2013}%
  \BibitemOpen
  \bibfield  {author} {\bibinfo {author} {\bibnamefont
  {Jim{\'e}nez-Garc{\'\i}a}, \bibfnamefont {K}}, \ and\ \bibinfo {author}
  {\bibfnamefont {I.~B.}\ \bibnamefont {Spielman}}} (\bibinfo {year} {2013}),\
  \enquote {\bibinfo {title} {Annual review of cold atoms and molecules: Volume
  2},}\ \ (\bibinfo  {publisher} {World Scientific})\ pp.\ \bibinfo {pages}
  {145--191}\BibitemShut {NoStop}%
\bibitem [{\citenamefont {Jotzu}\ \emph {et~al.}(2014)\citenamefont {Jotzu},
  \citenamefont {Messer}, \citenamefont {Desbuquois}, \citenamefont {Lebrat},
  \citenamefont {Uehlinger}, \citenamefont {Greif},\ and\ \citenamefont
  {Esslinger}}]{Jotzu:2014}%
  \BibitemOpen
  \bibfield  {author} {\bibinfo {author} {\bibnamefont {Jotzu}, \bibfnamefont
  {Gregor}}, \bibinfo {author} {\bibfnamefont {Michael}\ \bibnamefont
  {Messer}}, \bibinfo {author} {\bibfnamefont {R{\'e}mi}\ \bibnamefont
  {Desbuquois}}, \bibinfo {author} {\bibfnamefont {Martin}\ \bibnamefont
  {Lebrat}}, \bibinfo {author} {\bibfnamefont {Thomas}\ \bibnamefont
  {Uehlinger}}, \bibinfo {author} {\bibfnamefont {Daniel}\ \bibnamefont
  {Greif}}, \ and\ \bibinfo {author} {\bibfnamefont {Tilman}\ \bibnamefont
  {Esslinger}}} (\bibinfo {year} {2014}),\ \bibfield  {title} {\enquote
  {\bibinfo {title} {{Experimental realization of the topological Haldane model
  with ultracold fermions}},}\ }\href {\doibase 10.1038/nature13915} {\bibfield
   {journal} {\bibinfo  {journal} {Nature}\ }\textbf {\bibinfo {volume}
  {515}}~(\bibinfo {number} {7526}),\ \bibinfo {pages} {237--240}}\BibitemShut
  {NoStop}%
\bibitem [{\citenamefont {J{\" u}nemann}\ \emph {et~al.}(2017)\citenamefont
  {J{\" u}nemann}, \citenamefont {Piga}, \citenamefont {Ran}, \citenamefont
  {Lewenstein}, \citenamefont {Rizzi},\ and\ \citenamefont
  {Bermudez}}]{PhysRevX.7.031057}%
  \BibitemOpen
  \bibfield  {author} {\bibinfo {author} {\bibnamefont {J{\" u}nemann},
  \bibfnamefont {J}}, \bibinfo {author} {\bibfnamefont {A.}~\bibnamefont
  {Piga}}, \bibinfo {author} {\bibfnamefont {S.-J.}\ \bibnamefont {Ran}},
  \bibinfo {author} {\bibfnamefont {M.}~\bibnamefont {Lewenstein}}, \bibinfo
  {author} {\bibfnamefont {M.}~\bibnamefont {Rizzi}}, \ and\ \bibinfo {author}
  {\bibfnamefont {A.}~\bibnamefont {Bermudez}}} (\bibinfo {year} {2017}),\
  \bibfield  {title} {\enquote {\bibinfo {title} {Exploring interacting
  topological insulators with ultracold atoms: The synthetic {C}reutz-{H}ubbard
  model},}\ }\href {\doibase 10.1103/PhysRevX.7.031057} {\bibfield  {journal}
  {\bibinfo  {journal} {Phys. Rev. X}\ }\textbf {\bibinfo {volume} {7}},\
  \bibinfo {pages} {031057}}\BibitemShut {NoStop}%
\bibitem [{\citenamefont {Juzeli{\=u}nas}\ and\ \citenamefont
  {Spielman}(2012)}]{Juzeliunas2012}%
  \BibitemOpen
  \bibfield  {author} {\bibinfo {author} {\bibnamefont {Juzeli{\=u}nas},
  \bibfnamefont {G}}, \ and\ \bibinfo {author} {\bibfnamefont {I.~B.}\
  \bibnamefont {Spielman}}} (\bibinfo {year} {2012}),\ \bibfield  {title}
  {\enquote {\bibinfo {title} {{Flux lattices reformulated}},}\ }\href
  {\doibase 10.1088/1367-2630/14/12/123022} {\bibfield  {journal} {\bibinfo
  {journal} {New Journal of Physics}\ }\textbf {\bibinfo {volume}
  {14}}~(\bibinfo {number} {12}),\ \bibinfo {pages} {123022}}\BibitemShut
  {NoStop}%
\bibitem [{\citenamefont {Kalmeyer}\ and\ \citenamefont
  {Laughlin}(1987)}]{PhysRevLett.59.2095}%
  \BibitemOpen
  \bibfield  {author} {\bibinfo {author} {\bibnamefont {Kalmeyer},
  \bibfnamefont {V}}, \ and\ \bibinfo {author} {\bibfnamefont {R.~B.}\
  \bibnamefont {Laughlin}}} (\bibinfo {year} {1987}),\ \bibfield  {title}
  {\enquote {\bibinfo {title} {Equivalence of the resonating-valence-bond and
  fractional quantum {H}all states},}\ }\href {\doibase
  10.1103/PhysRevLett.59.2095} {\bibfield  {journal} {\bibinfo  {journal}
  {Phys. Rev. Lett.}\ }\textbf {\bibinfo {volume} {59}},\ \bibinfo {pages}
  {2095--2098}}\BibitemShut {NoStop}%
\bibitem [{\citenamefont {Kapit}\ and\ \citenamefont
  {Mueller}(2010)}]{PhysRevLett.105.215303}%
  \BibitemOpen
  \bibfield  {author} {\bibinfo {author} {\bibnamefont {Kapit}, \bibfnamefont
  {Eliot}}, \ and\ \bibinfo {author} {\bibfnamefont {Erich}\ \bibnamefont
  {Mueller}}} (\bibinfo {year} {2010}),\ \bibfield  {title} {\enquote {\bibinfo
  {title} {Exact parent {H}amiltonian for the quantum {H}all states in a
  lattice},}\ }\href {\doibase 10.1103/PhysRevLett.105.215303} {\bibfield
  {journal} {\bibinfo  {journal} {Phys. Rev. Lett.}\ }\textbf {\bibinfo
  {volume} {105}},\ \bibinfo {pages} {215303}}\BibitemShut {NoStop}%
\bibitem [{\citenamefont {Kennedy}\ \emph {et~al.}(2015)\citenamefont
  {Kennedy}, \citenamefont {Burton}, \citenamefont {Chung},\ and\ \citenamefont
  {Ketterle}}]{Kennedy2015}%
  \BibitemOpen
  \bibfield  {author} {\bibinfo {author} {\bibnamefont {Kennedy}, \bibfnamefont
  {Colin~J}}, \bibinfo {author} {\bibfnamefont {William~Cody}\ \bibnamefont
  {Burton}}, \bibinfo {author} {\bibfnamefont {Woo~Chang}\ \bibnamefont
  {Chung}}, \ and\ \bibinfo {author} {\bibfnamefont {Wolfgang}\ \bibnamefont
  {Ketterle}}} (\bibinfo {year} {2015}),\ \bibfield  {title} {\enquote
  {\bibinfo {title} {Observation of {B}ose-{E}instein condensation in a strong
  synthetic magnetic field},}\ }\href {http://dx.doi.org/10.1038/nphys3421}
  {\bibfield  {journal} {\bibinfo  {journal} {Nat Phys}\ }\textbf {\bibinfo
  {volume} {11}}~(\bibinfo {number} {10}),\ \bibinfo {pages}
  {859--864}}\BibitemShut {NoStop}%
\bibitem [{\citenamefont {Kennedy}\ \emph {et~al.}(2013)\citenamefont
  {Kennedy}, \citenamefont {Siviloglou}, \citenamefont {Miyake}, \citenamefont
  {Burton},\ and\ \citenamefont {Ketterle}}]{Kennedy2013}%
  \BibitemOpen
  \bibfield  {author} {\bibinfo {author} {\bibnamefont {Kennedy}, \bibfnamefont
  {Colin~J}}, \bibinfo {author} {\bibfnamefont {Georgios~A.}\ \bibnamefont
  {Siviloglou}}, \bibinfo {author} {\bibfnamefont {Hirokazu}\ \bibnamefont
  {Miyake}}, \bibinfo {author} {\bibfnamefont {William~Cody}\ \bibnamefont
  {Burton}}, \ and\ \bibinfo {author} {\bibfnamefont {Wolfgang}\ \bibnamefont
  {Ketterle}}} (\bibinfo {year} {2013}),\ \bibfield  {title} {\enquote
  {\bibinfo {title} {Spin-orbit coupling and quantum spin hall effect for
  neutral atoms without spin flips},}\ }\href {\doibase
  10.1103/PhysRevLett.111.225301} {\bibfield  {journal} {\bibinfo  {journal}
  {Phys. Rev. Lett.}\ }\textbf {\bibinfo {volume} {111}},\ \bibinfo {pages}
  {225301}}\BibitemShut {NoStop}%
\bibitem [{\citenamefont {{Kitaev}}(2001)}]{Kitaev:2000}%
  \BibitemOpen
  \bibfield  {author} {\bibinfo {author} {\bibnamefont {{Kitaev}},
  \bibfnamefont {A~Y}}} (\bibinfo {year} {2001}),\ \bibfield  {title} {\enquote
  {\bibinfo {title} {Unpaired {M}ajorana fermions in quantum wires},}\ }\href
  {http://stacks.iop.org/1063-7869/44/i=10S/a=S29} {\bibfield  {journal}
  {\bibinfo  {journal} {Physics Uspekhi}\ }\textbf {\bibinfo {volume} {44}},\
  \bibinfo {pages} {131}}\BibitemShut {NoStop}%
\bibitem [{\citenamefont {Kitagawa}\ \emph {et~al.}(2010)\citenamefont
  {Kitagawa}, \citenamefont {Berg}, \citenamefont {Rudner},\ and\ \citenamefont
  {Demler}}]{kitagawa}%
  \BibitemOpen
  \bibfield  {author} {\bibinfo {author} {\bibnamefont {Kitagawa},
  \bibfnamefont {Takuya}}, \bibinfo {author} {\bibfnamefont {Erez}\
  \bibnamefont {Berg}}, \bibinfo {author} {\bibfnamefont {Mark}\ \bibnamefont
  {Rudner}}, \ and\ \bibinfo {author} {\bibfnamefont {Eugene}\ \bibnamefont
  {Demler}}} (\bibinfo {year} {2010}),\ \bibfield  {title} {\enquote {\bibinfo
  {title} {Topological characterization of periodically driven quantum
  systems},}\ }\href {\doibase 10.1103/PhysRevB.82.235114} {\bibfield
  {journal} {\bibinfo  {journal} {Phys. Rev. B}\ }\textbf {\bibinfo {volume}
  {82}},\ \bibinfo {pages} {235114}}\BibitemShut {NoStop}%
\bibitem [{\citenamefont {Kitagawa}\ \emph {et~al.}(2011)\citenamefont
  {Kitagawa}, \citenamefont {Oka}, \citenamefont {Brataas}, \citenamefont
  {Fu},\ and\ \citenamefont {Demler}}]{PhysRevB.84.235108}%
  \BibitemOpen
  \bibfield  {author} {\bibinfo {author} {\bibnamefont {Kitagawa},
  \bibfnamefont {Takuya}}, \bibinfo {author} {\bibfnamefont {Takashi}\
  \bibnamefont {Oka}}, \bibinfo {author} {\bibfnamefont {Arne}\ \bibnamefont
  {Brataas}}, \bibinfo {author} {\bibfnamefont {Liang}\ \bibnamefont {Fu}}, \
  and\ \bibinfo {author} {\bibfnamefont {Eugene}\ \bibnamefont {Demler}}}
  (\bibinfo {year} {2011}),\ \bibfield  {title} {\enquote {\bibinfo {title}
  {Transport properties of nonequilibrium systems under the application of
  light: Photoinduced quantum {H}all insulators without {L}andau levels},}\
  }\href {\doibase 10.1103/PhysRevB.84.235108} {\bibfield  {journal} {\bibinfo
  {journal} {Phys. Rev. B}\ }\textbf {\bibinfo {volume} {84}},\ \bibinfo
  {pages} {235108}}\BibitemShut {NoStop}%
\bibitem [{\citenamefont {Kolkowitz}\ \emph {et~al.}(2017)\citenamefont
  {Kolkowitz}, \citenamefont {Bromley}, \citenamefont {Bothwell}, \citenamefont
  {Wall}, \citenamefont {Marti}, \citenamefont {Koller}, \citenamefont {Zhang},
  \citenamefont {Rey},\ and\ \citenamefont {Ye}}]{kolkowitz2017spin}%
  \BibitemOpen
  \bibfield  {author} {\bibinfo {author} {\bibnamefont {Kolkowitz},
  \bibfnamefont {S}}, \bibinfo {author} {\bibfnamefont {SL}~\bibnamefont
  {Bromley}}, \bibinfo {author} {\bibfnamefont {T}~\bibnamefont {Bothwell}},
  \bibinfo {author} {\bibfnamefont {ML}~\bibnamefont {Wall}}, \bibinfo {author}
  {\bibfnamefont {GE}~\bibnamefont {Marti}}, \bibinfo {author} {\bibfnamefont
  {AP}~\bibnamefont {Koller}}, \bibinfo {author} {\bibfnamefont
  {X}~\bibnamefont {Zhang}}, \bibinfo {author} {\bibfnamefont {AM}~\bibnamefont
  {Rey}}, \ and\ \bibinfo {author} {\bibfnamefont {J}~\bibnamefont {Ye}}}
  (\bibinfo {year} {2017}),\ \bibfield  {title} {\enquote {\bibinfo {title}
  {Spin--orbit-coupled fermions in an optical lattice clock},}\ }\href
  {\doibase 10.1038/nature20811} {\bibfield  {journal} {\bibinfo  {journal}
  {Nature}\ }\textbf {\bibinfo {volume} {542}}~(\bibinfo {number} {7639}),\
  \bibinfo {pages} {66}}\BibitemShut {NoStop}%
\bibitem [{\citenamefont {Kraus}\ \emph {et~al.}(2012)\citenamefont {Kraus},
  \citenamefont {Lahini}, \citenamefont {Ringel}, \citenamefont {Verbin},\ and\
  \citenamefont {Zilberberg}}]{Kraus2012}%
  \BibitemOpen
  \bibfield  {author} {\bibinfo {author} {\bibnamefont {Kraus}, \bibfnamefont
  {Yaacov~E}}, \bibinfo {author} {\bibfnamefont {Yoav}\ \bibnamefont {Lahini}},
  \bibinfo {author} {\bibfnamefont {Zohar}\ \bibnamefont {Ringel}}, \bibinfo
  {author} {\bibfnamefont {Mor}\ \bibnamefont {Verbin}}, \ and\ \bibinfo
  {author} {\bibfnamefont {Oded}\ \bibnamefont {Zilberberg}}} (\bibinfo {year}
  {2012}),\ \bibfield  {title} {\enquote {\bibinfo {title} {Topological states
  and adiabatic pumping in quasicrystals},}\ }\href {\doibase
  10.1103/PhysRevLett.109.106402} {\bibfield  {journal} {\bibinfo  {journal}
  {Phys. Rev. Lett.}\ }\textbf {\bibinfo {volume} {109}},\ \bibinfo {pages}
  {106402}}\BibitemShut {NoStop}%
\bibitem [{\citenamefont {Kunz}(1986)}]{Kunz1986}%
  \BibitemOpen
  \bibfield  {author} {\bibinfo {author} {\bibnamefont {Kunz}, \bibfnamefont
  {H}}} (\bibinfo {year} {1986}),\ \bibfield  {title} {\enquote {\bibinfo
  {title} {Quantized currents and topological invariants for electrons in
  incommensurate potentials},}\ }\href {\doibase 10.1103/PhysRevLett.57.1095}
  {\bibfield  {journal} {\bibinfo  {journal} {Phys. Rev. Lett.}\ }\textbf
  {\bibinfo {volume} {57}},\ \bibinfo {pages} {1095--1097}}\BibitemShut
  {NoStop}%
\bibitem [{\citenamefont {Lang}\ and\ \citenamefont
  {B\"uchler}(2015)}]{Lang:2015}%
  \BibitemOpen
  \bibfield  {author} {\bibinfo {author} {\bibnamefont {Lang}, \bibfnamefont
  {Nicolai}}, \ and\ \bibinfo {author} {\bibfnamefont {Hans~Peter}\
  \bibnamefont {B\"uchler}}} (\bibinfo {year} {2015}),\ \bibfield  {title}
  {\enquote {\bibinfo {title} {Topological states in a microscopic model of
  interacting fermions},}\ }\href {\doibase 10.1103/PhysRevB.92.041118}
  {\bibfield  {journal} {\bibinfo  {journal} {Phys. Rev. B}\ }\textbf {\bibinfo
  {volume} {92}},\ \bibinfo {pages} {041118}}\BibitemShut {NoStop}%
\bibitem [{\citenamefont {{L{\"a}uchli}}\ and\ \citenamefont
  {{Moessner}}(2015)}]{2015arXiv150404380L}%
  \BibitemOpen
  \bibfield  {author} {\bibinfo {author} {\bibnamefont {{L{\"a}uchli}},
  \bibfnamefont {A~M}}, \ and\ \bibinfo {author} {\bibfnamefont
  {R.}~\bibnamefont {{Moessner}}}} (\bibinfo {year} {2015}),\ \bibfield
  {title} {\enquote {\bibinfo {title} {{Quantum simulations made easy
  plane}},}\ }\href@noop {} {\bibfield  {journal} {\bibinfo  {journal} {ArXiv
  e-prints}\ }}\Eprint {http://arxiv.org/abs/1504.04380} {arXiv:1504.04380}
  \BibitemShut {NoStop}%
\bibitem [{\citenamefont {Laughlin}(1981)}]{laughlingauge}%
  \BibitemOpen
  \bibfield  {author} {\bibinfo {author} {\bibnamefont {Laughlin},
  \bibfnamefont {R~B}}} (\bibinfo {year} {1981}),\ \bibfield  {title} {\enquote
  {\bibinfo {title} {Quantized {H}all conductivity in two dimensions},}\ }\href
  {\doibase 10.1103/PhysRevB.23.5632} {\bibfield  {journal} {\bibinfo
  {journal} {Phys. Rev. B}\ }\textbf {\bibinfo {volume} {23}}~(\bibinfo
  {number} {10}),\ \bibinfo {pages} {5632--5633}}\BibitemShut {NoStop}%
\bibitem [{\citenamefont {Lazarides}\ \emph {et~al.}(2014)\citenamefont
  {Lazarides}, \citenamefont {Das},\ and\ \citenamefont
  {Moessner}}]{PhysRevE.90.012110}%
  \BibitemOpen
  \bibfield  {author} {\bibinfo {author} {\bibnamefont {Lazarides},
  \bibfnamefont {Achilleas}}, \bibinfo {author} {\bibfnamefont {Arnab}\
  \bibnamefont {Das}}, \ and\ \bibinfo {author} {\bibfnamefont {Roderich}\
  \bibnamefont {Moessner}}} (\bibinfo {year} {2014}),\ \bibfield  {title}
  {\enquote {\bibinfo {title} {Equilibrium states of generic quantum systems
  subject to periodic driving},}\ }\href {\doibase 10.1103/PhysRevE.90.012110}
  {\bibfield  {journal} {\bibinfo  {journal} {Phys. Rev. E}\ }\textbf {\bibinfo
  {volume} {90}},\ \bibinfo {pages} {012110}}\BibitemShut {NoStop}%
\bibitem [{\citenamefont {LeBlanc}\ \emph {et~al.}(2015)\citenamefont
  {LeBlanc}, \citenamefont {Jim{\'e}nez-Garc{\'\i}a}, \citenamefont {Williams},
  \citenamefont {Beeler}, \citenamefont {Phillips},\ and\ \citenamefont
  {Spielman}}]{LeBlanc2015}%
  \BibitemOpen
  \bibfield  {author} {\bibinfo {author} {\bibnamefont {LeBlanc}, \bibfnamefont
  {L~J}}, \bibinfo {author} {\bibfnamefont {K}~\bibnamefont
  {Jim{\'e}nez-Garc{\'\i}a}}, \bibinfo {author} {\bibfnamefont {R~A}\
  \bibnamefont {Williams}}, \bibinfo {author} {\bibfnamefont {M~C}\
  \bibnamefont {Beeler}}, \bibinfo {author} {\bibfnamefont {W.~D.}\
  \bibnamefont {Phillips}}, \ and\ \bibinfo {author} {\bibfnamefont {I.~B.}\
  \bibnamefont {Spielman}}} (\bibinfo {year} {2015}),\ \bibfield  {title}
  {\enquote {\bibinfo {title} {{Gauge matters: observing the vortex-nucleation
  transition in a {B}ose condensate}},}\ }\href {\doibase
  10.1088/1367-2630/17/6/065016} {\bibfield  {journal} {\bibinfo  {journal}
  {New Journal of Physics}\ }\textbf {\bibinfo {volume} {17}}~(\bibinfo
  {number} {6}),\ \bibinfo {pages} {065016}}\BibitemShut {NoStop}%
\bibitem [{\citenamefont {Leder}\ \emph {et~al.}(2016)\citenamefont {Leder},
  \citenamefont {Grossert}, \citenamefont {Sitta}, \citenamefont {Genske},
  \citenamefont {Rosch},\ and\ \citenamefont {Weitz}}]{Leder:2016}%
  \BibitemOpen
  \bibfield  {author} {\bibinfo {author} {\bibnamefont {Leder}, \bibfnamefont
  {Martin}}, \bibinfo {author} {\bibfnamefont {Christopher}\ \bibnamefont
  {Grossert}}, \bibinfo {author} {\bibfnamefont {Lukas}\ \bibnamefont {Sitta}},
  \bibinfo {author} {\bibfnamefont {Maximilian}\ \bibnamefont {Genske}},
  \bibinfo {author} {\bibfnamefont {Achim}\ \bibnamefont {Rosch}}, \ and\
  \bibinfo {author} {\bibfnamefont {Martin}\ \bibnamefont {Weitz}}} (\bibinfo
  {year} {2016}),\ \bibfield  {title} {\enquote {\bibinfo {title} {Real-space
  imaging of a topologically protected edge state with ultracold atoms in an
  amplitude-chirped optical lattice},}\ }\href {\doibase 10.1038/ncomms13112}
  {\bibfield  {journal} {\bibinfo  {journal} {Nature communications}\ }\textbf
  {\bibinfo {volume} {7}},\ \bibinfo {pages} {13112}}\BibitemShut {NoStop}%
\bibitem [{\citenamefont {Lellouch}\ \emph {et~al.}(2017)\citenamefont
  {Lellouch}, \citenamefont {Bukov}, \citenamefont {Demler},\ and\
  \citenamefont {Goldman}}]{PhysRevX.7.021015}%
  \BibitemOpen
  \bibfield  {author} {\bibinfo {author} {\bibnamefont {Lellouch},
  \bibfnamefont {S}}, \bibinfo {author} {\bibfnamefont {M.}~\bibnamefont
  {Bukov}}, \bibinfo {author} {\bibfnamefont {E.}~\bibnamefont {Demler}}, \
  and\ \bibinfo {author} {\bibfnamefont {N.}~\bibnamefont {Goldman}}} (\bibinfo
  {year} {2017}),\ \bibfield  {title} {\enquote {\bibinfo {title} {Parametric
  instability rates in periodically driven band systems},}\ }\href {\doibase
  10.1103/PhysRevX.7.021015} {\bibfield  {journal} {\bibinfo  {journal} {Phys.
  Rev. X}\ }\textbf {\bibinfo {volume} {7}},\ \bibinfo {pages}
  {021015}}\BibitemShut {NoStop}%
\bibitem [{\citenamefont {Li}\ \emph {et~al.}(2016)\citenamefont {Li},
  \citenamefont {Duca}, \citenamefont {Reitter}, \citenamefont {Grusdt},
  \citenamefont {Demler}, \citenamefont {Endres}, \citenamefont
  {Schleier-Smith}, \citenamefont {Bloch},\ and\ \citenamefont
  {Schneider}}]{Li:2016}%
  \BibitemOpen
  \bibfield  {author} {\bibinfo {author} {\bibnamefont {Li}, \bibfnamefont
  {Tracy}}, \bibinfo {author} {\bibfnamefont {Lucia}\ \bibnamefont {Duca}},
  \bibinfo {author} {\bibfnamefont {Martin}\ \bibnamefont {Reitter}}, \bibinfo
  {author} {\bibfnamefont {Fabian}\ \bibnamefont {Grusdt}}, \bibinfo {author}
  {\bibfnamefont {Eugene}\ \bibnamefont {Demler}}, \bibinfo {author}
  {\bibfnamefont {Manuel}\ \bibnamefont {Endres}}, \bibinfo {author}
  {\bibfnamefont {Monika}\ \bibnamefont {Schleier-Smith}}, \bibinfo {author}
  {\bibfnamefont {Immanuel}\ \bibnamefont {Bloch}}, \ and\ \bibinfo {author}
  {\bibfnamefont {Ulrich}\ \bibnamefont {Schneider}}} (\bibinfo {year}
  {2016}),\ \bibfield  {title} {\enquote {\bibinfo {title} {Bloch state
  tomography using {W}ilson lines},}\ }\href {\doibase 10.1126/science.aad5812}
  {\bibfield  {journal} {\bibinfo  {journal} {Science}\ }\textbf {\bibinfo
  {volume} {352}}~(\bibinfo {number} {6289}),\ \bibinfo {pages}
  {1094--1097}}\BibitemShut {NoStop}%
\bibitem [{\citenamefont {Lignier}\ \emph {et~al.}(2007)\citenamefont
  {Lignier}, \citenamefont {Sias}, \citenamefont {Ciampini}, \citenamefont
  {Singh}, \citenamefont {Zenesini}, \citenamefont {Morsch},\ and\
  \citenamefont {Arimondo}}]{Lignier2007}%
  \BibitemOpen
  \bibfield  {author} {\bibinfo {author} {\bibnamefont {Lignier}, \bibfnamefont
  {H}}, \bibinfo {author} {\bibfnamefont {C.}~\bibnamefont {Sias}}, \bibinfo
  {author} {\bibfnamefont {D.}~\bibnamefont {Ciampini}}, \bibinfo {author}
  {\bibfnamefont {Y.}~\bibnamefont {Singh}}, \bibinfo {author} {\bibfnamefont
  {A.}~\bibnamefont {Zenesini}}, \bibinfo {author} {\bibfnamefont
  {O.}~\bibnamefont {Morsch}}, \ and\ \bibinfo {author} {\bibfnamefont
  {E.}~\bibnamefont {Arimondo}}} (\bibinfo {year} {2007}),\ \bibfield  {title}
  {\enquote {\bibinfo {title} {Dynamical control of matter-wave tunneling in
  periodic potentials},}\ }\href {\doibase 10.1103/PhysRevLett.99.220403}
  {\bibfield  {journal} {\bibinfo  {journal} {Phys. Rev. Lett.}\ }\textbf
  {\bibinfo {volume} {99}},\ \bibinfo {pages} {220403}}\BibitemShut {NoStop}%
\bibitem [{\citenamefont {Lim}\ \emph {et~al.}(2014)\citenamefont {Lim},
  \citenamefont {Fuchs},\ and\ \citenamefont {Montambaux}}]{Lim2014}%
  \BibitemOpen
  \bibfield  {author} {\bibinfo {author} {\bibnamefont {Lim}, \bibfnamefont
  {Lih-King}}, \bibinfo {author} {\bibfnamefont {Jean-No\"el}\ \bibnamefont
  {Fuchs}}, \ and\ \bibinfo {author} {\bibfnamefont {Gilles}\ \bibnamefont
  {Montambaux}}} (\bibinfo {year} {2014}),\ \bibfield  {title} {\enquote
  {\bibinfo {title} {Mass and chirality inversion of a {D}irac cone pair in
  {S}t\"uckelberg interferometry},}\ }\href {\doibase
  10.1103/PhysRevLett.112.155302} {\bibfield  {journal} {\bibinfo  {journal}
  {Phys. Rev. Lett.}\ }\textbf {\bibinfo {volume} {112}},\ \bibinfo {pages}
  {155302}}\BibitemShut {NoStop}%
\bibitem [{\citenamefont {Lim}\ \emph {et~al.}(2015)\citenamefont {Lim},
  \citenamefont {Fuchs},\ and\ \citenamefont {Montambaux}}]{Lim2015}%
  \BibitemOpen
  \bibfield  {author} {\bibinfo {author} {\bibnamefont {Lim}, \bibfnamefont
  {Lih-King}}, \bibinfo {author} {\bibfnamefont {Jean-No\"el}\ \bibnamefont
  {Fuchs}}, \ and\ \bibinfo {author} {\bibfnamefont {Gilles}\ \bibnamefont
  {Montambaux}}} (\bibinfo {year} {2015}),\ \bibfield  {title} {\enquote
  {\bibinfo {title} {Geometry of {B}loch states probed by {S}t\"uckelberg
  interferometry},}\ }\href {\doibase 10.1103/PhysRevA.92.063627} {\bibfield
  {journal} {\bibinfo  {journal} {Phys. Rev. A}\ }\textbf {\bibinfo {volume}
  {92}},\ \bibinfo {pages} {063627}}\BibitemShut {NoStop}%
\bibitem [{\citenamefont {Lindblad}(1976)}]{lindblad1976generators}%
  \BibitemOpen
  \bibfield  {author} {\bibinfo {author} {\bibnamefont {Lindblad},
  \bibfnamefont {Goran}}} (\bibinfo {year} {1976}),\ \bibfield  {title}
  {\enquote {\bibinfo {title} {On the generators of quantum dynamical
  semigroups},}\ }\href {\doibase 10.1007/BF01608499} {\bibfield  {journal}
  {\bibinfo  {journal} {Communications in Mathematical Physics}\ }\textbf
  {\bibinfo {volume} {48}}~(\bibinfo {number} {2}),\ \bibinfo {pages}
  {119--130}}\BibitemShut {NoStop}%
\bibitem [{\citenamefont {Lindner}\ \emph {et~al.}(2017)\citenamefont
  {Lindner}, \citenamefont {Berg},\ and\ \citenamefont
  {Rudner}}]{lindnerbergrudner}%
  \BibitemOpen
  \bibfield  {author} {\bibinfo {author} {\bibnamefont {Lindner}, \bibfnamefont
  {Netanel~H}}, \bibinfo {author} {\bibfnamefont {Erez}\ \bibnamefont {Berg}},
  \ and\ \bibinfo {author} {\bibfnamefont {Mark~S.}\ \bibnamefont {Rudner}}}
  (\bibinfo {year} {2017}),\ \bibfield  {title} {\enquote {\bibinfo {title}
  {Universal chiral quasisteady states in periodically driven many-body
  systems},}\ }\href {\doibase 10.1103/PhysRevX.7.011018} {\bibfield  {journal}
  {\bibinfo  {journal} {Phys. Rev. X}\ }\textbf {\bibinfo {volume} {7}},\
  \bibinfo {pages} {011018}}\BibitemShut {NoStop}%
\bibitem [{\citenamefont {Linzner}\ \emph {et~al.}(2016)\citenamefont
  {Linzner}, \citenamefont {Wawer}, \citenamefont {Grusdt},\ and\ \citenamefont
  {Fleischhauer}}]{linzner2016reservoir}%
  \BibitemOpen
  \bibfield  {author} {\bibinfo {author} {\bibnamefont {Linzner}, \bibfnamefont
  {Dominik}}, \bibinfo {author} {\bibfnamefont {Lukas}\ \bibnamefont {Wawer}},
  \bibinfo {author} {\bibfnamefont {Fabian}\ \bibnamefont {Grusdt}}, \ and\
  \bibinfo {author} {\bibfnamefont {Michael}\ \bibnamefont {Fleischhauer}}}
  (\bibinfo {year} {2016}),\ \bibfield  {title} {\enquote {\bibinfo {title}
  {Reservoir-induced {T}houless pumping and symmetry-protected topological
  order in open quantum chains},}\ }\href {\doibase 10.1103/PhysRevB.94.201105}
  {\bibfield  {journal} {\bibinfo  {journal} {Phys. Rev. B}\ }\textbf {\bibinfo
  {volume} {94}}~(\bibinfo {number} {20}),\ \bibinfo {pages}
  {201105}}\BibitemShut {NoStop}%
\bibitem [{\citenamefont {Liu}\ \emph {et~al.}(2017)\citenamefont {Liu},
  \citenamefont {Vafa},\ and\ \citenamefont {Xu}}]{PhysRevB.95.161116}%
  \BibitemOpen
  \bibfield  {author} {\bibinfo {author} {\bibnamefont {Liu}, \bibfnamefont
  {Chunxiao}}, \bibinfo {author} {\bibfnamefont {Farzan}\ \bibnamefont {Vafa}},
  \ and\ \bibinfo {author} {\bibfnamefont {Cenke}\ \bibnamefont {Xu}}}
  (\bibinfo {year} {2017}),\ \bibfield  {title} {\enquote {\bibinfo {title}
  {Symmetry-protected topological {H}opf insulator and its generalizations},}\
  }\href {\doibase 10.1103/PhysRevB.95.161116} {\bibfield  {journal} {\bibinfo
  {journal} {Phys. Rev. B}\ }\textbf {\bibinfo {volume} {95}},\ \bibinfo
  {pages} {161116}}\BibitemShut {NoStop}%
\bibitem [{\citenamefont {Liu}\ \emph {et~al.}(2013)\citenamefont {Liu},
  \citenamefont {Law}, \citenamefont {Ng},\ and\ \citenamefont
  {Lee}}]{Liu:2013}%
  \BibitemOpen
  \bibfield  {author} {\bibinfo {author} {\bibnamefont {Liu}, \bibfnamefont
  {Xiong-Jun}}, \bibinfo {author} {\bibfnamefont {Kam-Tuen}\ \bibnamefont
  {Law}}, \bibinfo {author} {\bibfnamefont {Tai-Kai}\ \bibnamefont {Ng}}, \
  and\ \bibinfo {author} {\bibfnamefont {Patrick~A}\ \bibnamefont {Lee}}}
  (\bibinfo {year} {2013}),\ \bibfield  {title} {\enquote {\bibinfo {title}
  {Detecting topological phases in cold atoms},}\ }\href {\doibase
  10.1103/PhysRevLett.111.120402} {\bibfield  {journal} {\bibinfo  {journal}
  {Phys. Rev. Lett.}\ }\textbf {\bibinfo {volume} {111}}~(\bibinfo {number}
  {12}),\ \bibinfo {pages} {120402}}\BibitemShut {NoStop}%
\bibitem [{\citenamefont {Livi}\ \emph {et~al.}(2016)\citenamefont {Livi},
  \citenamefont {Cappellini}, \citenamefont {Diem}, \citenamefont {Franchi},
  \citenamefont {Clivati}, \citenamefont {Frittelli}, \citenamefont {Levi},
  \citenamefont {Calonico}, \citenamefont {Catani}, \citenamefont {Inguscio}
  \emph {et~al.}}]{livi2016synthetic}%
  \BibitemOpen
  \bibfield  {author} {\bibinfo {author} {\bibnamefont {Livi}, \bibfnamefont
  {LF}}, \bibinfo {author} {\bibfnamefont {G}~\bibnamefont {Cappellini}},
  \bibinfo {author} {\bibfnamefont {M}~\bibnamefont {Diem}}, \bibinfo {author}
  {\bibfnamefont {L}~\bibnamefont {Franchi}}, \bibinfo {author} {\bibfnamefont
  {C}~\bibnamefont {Clivati}}, \bibinfo {author} {\bibfnamefont
  {M}~\bibnamefont {Frittelli}}, \bibinfo {author} {\bibfnamefont
  {F}~\bibnamefont {Levi}}, \bibinfo {author} {\bibfnamefont {D}~\bibnamefont
  {Calonico}}, \bibinfo {author} {\bibfnamefont {J}~\bibnamefont {Catani}},
  \bibinfo {author} {\bibfnamefont {M}~\bibnamefont {Inguscio}},  \emph
  {et~al.}} (\bibinfo {year} {2016}),\ \bibfield  {title} {\enquote {\bibinfo
  {title} {Synthetic dimensions and spin-orbit coupling with an optical clock
  transition},}\ }\href {\doibase 10.1103/PhysRevLett.117.220401} {\bibfield
  {journal} {\bibinfo  {journal} {Phys. Rev. Lett.}\ }\textbf {\bibinfo
  {volume} {117}}~(\bibinfo {number} {22}),\ \bibinfo {pages}
  {220401}}\BibitemShut {NoStop}%
\bibitem [{\citenamefont {Lohse}\ \emph {et~al.}(2016)\citenamefont {Lohse},
  \citenamefont {Schweizer}, \citenamefont {Zilberberg}, \citenamefont
  {Aidelsburger},\ and\ \citenamefont {Bloch}}]{Lohse:2016}%
  \BibitemOpen
  \bibfield  {author} {\bibinfo {author} {\bibnamefont {Lohse}, \bibfnamefont
  {M}}, \bibinfo {author} {\bibfnamefont {C.}~\bibnamefont {Schweizer}},
  \bibinfo {author} {\bibfnamefont {O.}~\bibnamefont {Zilberberg}}, \bibinfo
  {author} {\bibfnamefont {M.}~\bibnamefont {Aidelsburger}}, \ and\ \bibinfo
  {author} {\bibfnamefont {I.}~\bibnamefont {Bloch}}} (\bibinfo {year}
  {2016}),\ \bibfield  {title} {\enquote {\bibinfo {title} {A {T}houless
  quantum pump with ultracold bosonic atoms in an optical superlattice},}\
  }\href {http://dx.doi.org/10.1038/nphys3584} {\bibfield  {journal} {\bibinfo
  {journal} {Nat Phys}\ }\textbf {\bibinfo {volume} {12}}~(\bibinfo {number}
  {4}),\ \bibinfo {pages} {350--354}}\BibitemShut {NoStop}%
\bibitem [{\citenamefont {Lohse}\ \emph {et~al.}(2018)\citenamefont {Lohse},
  \citenamefont {Schweizer}, \citenamefont {Price}, \citenamefont
  {Zilberberg},\ and\ \citenamefont {Bloch}}]{Lohse4D}%
  \BibitemOpen
  \bibfield  {author} {\bibinfo {author} {\bibnamefont {Lohse}, \bibfnamefont
  {Michael}}, \bibinfo {author} {\bibfnamefont {Christian}\ \bibnamefont
  {Schweizer}}, \bibinfo {author} {\bibfnamefont {Hannah~M}\ \bibnamefont
  {Price}}, \bibinfo {author} {\bibfnamefont {Oded}\ \bibnamefont
  {Zilberberg}}, \ and\ \bibinfo {author} {\bibfnamefont {Immanuel}\
  \bibnamefont {Bloch}}} (\bibinfo {year} {2018}),\ \bibfield  {title}
  {\enquote {\bibinfo {title} {Exploring 4{D} quantum {H}all physics with a
  2{D} topological charge pump},}\ }\href
  {http://dx.doi.org/10.1038/nature25000} {\bibfield  {journal} {\bibinfo
  {journal} {Nature}\ }\textbf {\bibinfo {volume} {553}}~(\bibinfo {number}
  {7686}),\ \bibinfo {pages} {55}}\BibitemShut {NoStop}%
\bibitem [{\citenamefont {Loring}\ and\ \citenamefont
  {Hastings}(2010)}]{loringhastings}%
  \BibitemOpen
  \bibfield  {author} {\bibinfo {author} {\bibnamefont {Loring}, \bibfnamefont
  {T~A}}, \ and\ \bibinfo {author} {\bibfnamefont {M.~B.}\ \bibnamefont
  {Hastings}}} (\bibinfo {year} {2010}),\ \bibfield  {title} {\enquote
  {\bibinfo {title} {Disordered topological insulators via {C}$^*$-algebras},}\
  }\href {http://stacks.iop.org/0295-5075/92/i=6/a=67004} {\bibfield  {journal}
  {\bibinfo  {journal} {EPL (Europhysics Letters)}\ }\textbf {\bibinfo {volume}
  {92}}~(\bibinfo {number} {6}),\ \bibinfo {pages} {67004}}\BibitemShut
  {NoStop}%
\bibitem [{\citenamefont {Lu}\ \emph {et~al.}(2016)\citenamefont {Lu},
  \citenamefont {Schemmer}, \citenamefont {Aycock}, \citenamefont {Genkina},
  \citenamefont {Sugawa},\ and\ \citenamefont {Spielman}}]{Lu:2016}%
  \BibitemOpen
  \bibfield  {author} {\bibinfo {author} {\bibnamefont {Lu}, \bibfnamefont
  {H-I}}, \bibinfo {author} {\bibfnamefont {M.}~\bibnamefont {Schemmer}},
  \bibinfo {author} {\bibfnamefont {L.~M.}\ \bibnamefont {Aycock}}, \bibinfo
  {author} {\bibfnamefont {D.}~\bibnamefont {Genkina}}, \bibinfo {author}
  {\bibfnamefont {S.}~\bibnamefont {Sugawa}}, \ and\ \bibinfo {author}
  {\bibfnamefont {I.~B.}\ \bibnamefont {Spielman}}} (\bibinfo {year} {2016}),\
  \bibfield  {title} {\enquote {\bibinfo {title} {Geometrical pumping with a
  {B}ose-{E}instein condensate},}\ }\href {\doibase
  10.1103/PhysRevLett.116.200402} {\bibfield  {journal} {\bibinfo  {journal}
  {Phys. Rev. Lett.}\ }\textbf {\bibinfo {volume} {116}},\ \bibinfo {pages}
  {200402}}\BibitemShut {NoStop}%
\bibitem [{\citenamefont {Luo}\ \emph {et~al.}(2016)\citenamefont {Luo},
  \citenamefont {Wu}, \citenamefont {Chen}, \citenamefont {Guan}, \citenamefont
  {Gao}, \citenamefont {Xu}, \citenamefont {You},\ and\ \citenamefont
  {Wang}}]{Luo2016}%
  \BibitemOpen
  \bibfield  {author} {\bibinfo {author} {\bibnamefont {Luo}, \bibfnamefont
  {Xinyu}}, \bibinfo {author} {\bibfnamefont {Lingna}\ \bibnamefont {Wu}},
  \bibinfo {author} {\bibfnamefont {Jiyao}\ \bibnamefont {Chen}}, \bibinfo
  {author} {\bibfnamefont {Qing}\ \bibnamefont {Guan}}, \bibinfo {author}
  {\bibfnamefont {Kuiyi}\ \bibnamefont {Gao}}, \bibinfo {author} {\bibfnamefont
  {Zhi-Fang}\ \bibnamefont {Xu}}, \bibinfo {author} {\bibfnamefont
  {L.}~\bibnamefont {You}}, \ and\ \bibinfo {author} {\bibfnamefont {Ruquan}\
  \bibnamefont {Wang}}} (\bibinfo {year} {2016}),\ \bibfield  {title} {\enquote
  {\bibinfo {title} {Tunable atomic spin-orbit coupling synthesized with a
  modulating gradient magnetic field},}\ }\href {\doibase 0.1038/srep01937}
  {\bibfield  {journal} {\bibinfo  {journal} {Scientific Reports}\ }\textbf
  {\bibinfo {volume} {6}},\ \bibinfo {pages} {18983}}\BibitemShut {NoStop}%
\bibitem [{\citenamefont {Maczewsky}\ \emph {et~al.}(2017)\citenamefont
  {Maczewsky}, \citenamefont {Zeuner}, \citenamefont {Nolte},\ and\
  \citenamefont {Szameit}}]{Maczewsky2017}%
  \BibitemOpen
  \bibfield  {author} {\bibinfo {author} {\bibnamefont {Maczewsky},
  \bibfnamefont {Lukas~J}}, \bibinfo {author} {\bibfnamefont {Julia~M.}\
  \bibnamefont {Zeuner}}, \bibinfo {author} {\bibfnamefont {Stefan}\
  \bibnamefont {Nolte}}, \ and\ \bibinfo {author} {\bibfnamefont {Alexander}\
  \bibnamefont {Szameit}}} (\bibinfo {year} {2017}),\ \bibfield  {title}
  {\enquote {\bibinfo {title} {Observation of photonic anomalous {F}loquet
  topological insulators},}\ }\href {http://dx.doi.org/10.1038/ncomms13756}
  {\bibfield  {journal} {\bibinfo  {journal} {Nature Communications}\ }\textbf
  {\bibinfo {volume} {8}},\ \bibinfo {pages} {13756}}\BibitemShut {NoStop}%
\bibitem [{\citenamefont {Mancini}\ \emph {et~al.}(2015)\citenamefont
  {Mancini}, \citenamefont {Pagano}, \citenamefont {Cappellini}, \citenamefont
  {Livi}, \citenamefont {Rider}, \citenamefont {Catani}, \citenamefont {Sias},
  \citenamefont {Zoller}, \citenamefont {Inguscio}, \citenamefont {Dalmonte}
  \emph {et~al.}}]{Mancini:2015}%
  \BibitemOpen
  \bibfield  {author} {\bibinfo {author} {\bibnamefont {Mancini}, \bibfnamefont
  {M}}, \bibinfo {author} {\bibfnamefont {G}~\bibnamefont {Pagano}}, \bibinfo
  {author} {\bibfnamefont {G}~\bibnamefont {Cappellini}}, \bibinfo {author}
  {\bibfnamefont {L}~\bibnamefont {Livi}}, \bibinfo {author} {\bibfnamefont
  {M}~\bibnamefont {Rider}}, \bibinfo {author} {\bibfnamefont {J}~\bibnamefont
  {Catani}}, \bibinfo {author} {\bibfnamefont {C}~\bibnamefont {Sias}},
  \bibinfo {author} {\bibfnamefont {P}~\bibnamefont {Zoller}}, \bibinfo
  {author} {\bibfnamefont {M}~\bibnamefont {Inguscio}}, \bibinfo {author}
  {\bibfnamefont {M}~\bibnamefont {Dalmonte}},  \emph {et~al.}} (\bibinfo
  {year} {2015}),\ \bibfield  {title} {\enquote {\bibinfo {title} {Observation
  of chiral edge states with neutral fermions in synthetic {H}all ribbons},}\
  }\href {\doibase 10.1126/science.aaa8736} {\bibfield  {journal} {\bibinfo
  {journal} {Science}\ }\textbf {\bibinfo {volume} {349}}~(\bibinfo {number}
  {6255}),\ \bibinfo {pages} {1510--1513}}\BibitemShut {NoStop}%
\bibitem [{\citenamefont {Mazza}\ \emph {et~al.}(2015)\citenamefont {Mazza},
  \citenamefont {Aidelsburger}, \citenamefont {Tu}, \citenamefont {Goldman},\
  and\ \citenamefont {Burrello}}]{1367-2630-17-10-105001}%
  \BibitemOpen
  \bibfield  {author} {\bibinfo {author} {\bibnamefont {Mazza}, \bibfnamefont
  {Leonardo}}, \bibinfo {author} {\bibfnamefont {Monika}\ \bibnamefont
  {Aidelsburger}}, \bibinfo {author} {\bibfnamefont {Hong-Hao}\ \bibnamefont
  {Tu}}, \bibinfo {author} {\bibfnamefont {Nathan}\ \bibnamefont {Goldman}}, \
  and\ \bibinfo {author} {\bibfnamefont {Michele}\ \bibnamefont {Burrello}}}
  (\bibinfo {year} {2015}),\ \bibfield  {title} {\enquote {\bibinfo {title}
  {Methods for detecting charge fractionalization and winding numbers in an
  interacting fermionic ladder},}\ }\href
  {http://stacks.iop.org/1367-2630/17/i=10/a=105001} {\bibfield  {journal}
  {\bibinfo  {journal} {New Journal of Physics}\ }\textbf {\bibinfo {volume}
  {17}}~(\bibinfo {number} {10}),\ \bibinfo {pages} {105001}}\BibitemShut
  {NoStop}%
\bibitem [{\citenamefont {Mazza}\ \emph {et~al.}(2012)\citenamefont {Mazza},
  \citenamefont {Bermudez}, \citenamefont {Goldman}, \citenamefont {Rizzi},
  \citenamefont {Martin-Delgado},\ and\ \citenamefont
  {Lewenstein}}]{Mazza:2012}%
  \BibitemOpen
  \bibfield  {author} {\bibinfo {author} {\bibnamefont {Mazza}, \bibfnamefont
  {Leonardo}}, \bibinfo {author} {\bibfnamefont {Alejandro}\ \bibnamefont
  {Bermudez}}, \bibinfo {author} {\bibfnamefont {Nathan}\ \bibnamefont
  {Goldman}}, \bibinfo {author} {\bibfnamefont {Matteo}\ \bibnamefont {Rizzi}},
  \bibinfo {author} {\bibfnamefont {Miguel~Angel}\ \bibnamefont
  {Martin-Delgado}}, \ and\ \bibinfo {author} {\bibfnamefont {Maciej}\
  \bibnamefont {Lewenstein}}} (\bibinfo {year} {2012}),\ \bibfield  {title}
  {\enquote {\bibinfo {title} {An optical-lattice-based quantum simulator for
  relativistic field theories and topological insulators},}\ }\href
  {http://stacks.iop.org/1367-2630/14/i=1/a=015007} {\bibfield  {journal}
  {\bibinfo  {journal} {New Journal of Physics}\ }\textbf {\bibinfo {volume}
  {14}}~(\bibinfo {number} {1}),\ \bibinfo {pages} {015007}}\BibitemShut
  {NoStop}%
\bibitem [{\citenamefont {McGinley}\ and\ \citenamefont
  {Cooper}(2018)}]{mcginley2018}%
  \BibitemOpen
  \bibfield  {author} {\bibinfo {author} {\bibnamefont {McGinley},
  \bibfnamefont {Max}}, \ and\ \bibinfo {author} {\bibfnamefont {Nigel~R.}\
  \bibnamefont {Cooper}}} (\bibinfo {year} {2018}),\ \bibfield  {title}
  {\enquote {\bibinfo {title} {Topology of one-dimensional quantum systems out
  of equilibrium},}\ }\href {\doibase 10.1103/PhysRevLett.121.090401}
  {\bibfield  {journal} {\bibinfo  {journal} {Phys. Rev. Lett.}\ }\textbf
  {\bibinfo {volume} {121}},\ \bibinfo {pages} {090401}}\BibitemShut {NoStop}%
\bibitem [{\citenamefont {Meier}\ \emph {et~al.}(2016)\citenamefont {Meier},
  \citenamefont {An},\ and\ \citenamefont {Gadway}}]{Meier2016}%
  \BibitemOpen
  \bibfield  {author} {\bibinfo {author} {\bibnamefont {Meier}, \bibfnamefont
  {Eric~J}}, \bibinfo {author} {\bibfnamefont {Fangzhao~Alex}\ \bibnamefont
  {An}}, \ and\ \bibinfo {author} {\bibfnamefont {Bryce}\ \bibnamefont
  {Gadway}}} (\bibinfo {year} {2016}),\ \bibfield  {title} {\enquote {\bibinfo
  {title} {Observation of the topological soliton state in the
  {S}u-{S}chrieffer-{H}eeger model},}\ }\href
  {http://dx.doi.org/10.1038/ncomms13986} {\bibfield  {journal} {\bibinfo
  {journal} {Nature Communications}\ }\textbf {\bibinfo {volume} {7}},\
  \bibinfo {pages} {13986}}\BibitemShut {NoStop}%
\bibitem [{\citenamefont {Mikitik}\ and\ \citenamefont
  {Sharlai}(1999)}]{Mikitik:1999}%
  \BibitemOpen
  \bibfield  {author} {\bibinfo {author} {\bibnamefont {Mikitik}, \bibfnamefont
  {G~P}}, \ and\ \bibinfo {author} {\bibfnamefont {Yu.~V.}\ \bibnamefont
  {Sharlai}}} (\bibinfo {year} {1999}),\ \bibfield  {title} {\enquote {\bibinfo
  {title} {Manifestation of {B}erry's phase in metal physics},}\ }\href
  {\doibase 10.1103/PhysRevLett.82.2147} {\bibfield  {journal} {\bibinfo
  {journal} {Phys. Rev. Lett.}\ }\textbf {\bibinfo {volume} {82}},\ \bibinfo
  {pages} {2147--2150}}\BibitemShut {NoStop}%
\bibitem [{\citenamefont {Miyake}\ \emph {et~al.}(2013)\citenamefont {Miyake},
  \citenamefont {Siviloglou}, \citenamefont {Kennedy}, \citenamefont {Burton},\
  and\ \citenamefont {Ketterle}}]{Miyake2013}%
  \BibitemOpen
  \bibfield  {author} {\bibinfo {author} {\bibnamefont {Miyake}, \bibfnamefont
  {Hirokazu}}, \bibinfo {author} {\bibfnamefont {Georgios~A.}\ \bibnamefont
  {Siviloglou}}, \bibinfo {author} {\bibfnamefont {Colin~J.}\ \bibnamefont
  {Kennedy}}, \bibinfo {author} {\bibfnamefont {William~Cody}\ \bibnamefont
  {Burton}}, \ and\ \bibinfo {author} {\bibfnamefont {Wolfgang}\ \bibnamefont
  {Ketterle}}} (\bibinfo {year} {2013}),\ \bibfield  {title} {\enquote
  {\bibinfo {title} {Realizing the {H}arper {H}amiltonian with laser-assisted
  tunneling in optical lattices},}\ }\href {\doibase
  10.1103/PhysRevLett.111.185302} {\bibfield  {journal} {\bibinfo  {journal}
  {Phys. Rev. Lett.}\ }\textbf {\bibinfo {volume} {111}},\ \bibinfo {pages}
  {185302}}\BibitemShut {NoStop}%
\bibitem [{\citenamefont {Modugno}\ and\ \citenamefont
  {Pettini}(2017)}]{modugno2017}%
  \BibitemOpen
  \bibfield  {author} {\bibinfo {author} {\bibnamefont {Modugno}, \bibfnamefont
  {Michele}}, \ and\ \bibinfo {author} {\bibfnamefont {Giulio}\ \bibnamefont
  {Pettini}}} (\bibinfo {year} {2017}),\ \bibfield  {title} {\enquote {\bibinfo
  {title} {Correspondence between a shaken honeycomb lattice and the haldane
  model},}\ }\href {\doibase 10.1103/PhysRevA.96.053603} {\bibfield  {journal}
  {\bibinfo  {journal} {Phys. Rev. A}\ }\textbf {\bibinfo {volume} {96}},\
  \bibinfo {pages} {053603}}\BibitemShut {NoStop}%
\bibitem [{\citenamefont {Moessner}\ and\ \citenamefont
  {Sondhi}(2017)}]{moessnersondhi}%
  \BibitemOpen
  \bibfield  {author} {\bibinfo {author} {\bibnamefont {Moessner},
  \bibfnamefont {R}}, \ and\ \bibinfo {author} {\bibfnamefont {S.~L.}\
  \bibnamefont {Sondhi}}} (\bibinfo {year} {2017}),\ \bibfield  {title}
  {\enquote {\bibinfo {title} {Equilibration and order in quantum {F}loquet
  matter},}\ }\href {http://dx.doi.org/10.1038/nphys4106} {\bibfield  {journal}
  {\bibinfo  {journal} {Nature Physics}\ }\textbf {\bibinfo {volume} {13}},\
  \bibinfo {pages} {424}}\BibitemShut {NoStop}%
\bibitem [{\citenamefont {M\"oller}\ and\ \citenamefont
  {Cooper}(2009)}]{mollercooper2009}%
  \BibitemOpen
  \bibfield  {author} {\bibinfo {author} {\bibnamefont {M\"oller},
  \bibfnamefont {G}}, \ and\ \bibinfo {author} {\bibfnamefont {N.~R.}\
  \bibnamefont {Cooper}}} (\bibinfo {year} {2009}),\ \bibfield  {title}
  {\enquote {\bibinfo {title} {Composite fermion theory for bosonic quantum
  {H}all states on lattices},}\ }\href {\doibase
  10.1103/PhysRevLett.103.105303} {\bibfield  {journal} {\bibinfo  {journal}
  {Phys. Rev. Lett.}\ }\textbf {\bibinfo {volume} {103}},\ \bibinfo {pages}
  {105303}}\BibitemShut {NoStop}%
\bibitem [{\citenamefont {M{\" o}ller}\ and\ \citenamefont
  {Cooper}(2010)}]{mollercooper2010}%
  \BibitemOpen
  \bibfield  {author} {\bibinfo {author} {\bibnamefont {M{\" o}ller},
  \bibfnamefont {G}}, \ and\ \bibinfo {author} {\bibfnamefont {N.~R.}\
  \bibnamefont {Cooper}}} (\bibinfo {year} {2010}),\ \bibfield  {title}
  {\enquote {\bibinfo {title} {Condensed ground states of frustrated
  {B}ose-{H}ubbard models},}\ }\href {\doibase 10.1103/PhysRevA.82.063625}
  {\bibfield  {journal} {\bibinfo  {journal} {Phys. Rev. A}\ }\textbf {\bibinfo
  {volume} {82}},\ \bibinfo {pages} {063625}}\BibitemShut {NoStop}%
\bibitem [{\citenamefont {M\"oller}\ and\ \citenamefont
  {Cooper}(2012)}]{mollercooperdice}%
  \BibitemOpen
  \bibfield  {author} {\bibinfo {author} {\bibnamefont {M\"oller},
  \bibfnamefont {G}}, \ and\ \bibinfo {author} {\bibfnamefont {N.~R.}\
  \bibnamefont {Cooper}}} (\bibinfo {year} {2012}),\ \bibfield  {title}
  {\enquote {\bibinfo {title} {Correlated phases of bosons in the flat lowest
  band of the dice lattice},}\ }\href {\doibase 10.1103/PhysRevLett.108.045306}
  {\bibfield  {journal} {\bibinfo  {journal} {Phys. Rev. Lett.}\ }\textbf
  {\bibinfo {volume} {108}},\ \bibinfo {pages} {045306}}\BibitemShut {NoStop}%
\bibitem [{\citenamefont {M\"oller}\ and\ \citenamefont
  {Cooper}(2015)}]{mollercooper2015}%
  \BibitemOpen
  \bibfield  {author} {\bibinfo {author} {\bibnamefont {M\"oller},
  \bibfnamefont {Gunnar}}, \ and\ \bibinfo {author} {\bibfnamefont {Nigel~R.}\
  \bibnamefont {Cooper}}} (\bibinfo {year} {2015}),\ \bibfield  {title}
  {\enquote {\bibinfo {title} {Fractional {C}hern insulators in
  {H}arper-{H}ofstadter bands with higher {C}hern number},}\ }\href {\doibase
  10.1103/PhysRevLett.115.126401} {\bibfield  {journal} {\bibinfo  {journal}
  {Phys. Rev. Lett.}\ }\textbf {\bibinfo {volume} {115}},\ \bibinfo {pages}
  {126401}}\BibitemShut {NoStop}%
\bibitem [{\citenamefont {Moore}\ \emph {et~al.}(2008)\citenamefont {Moore},
  \citenamefont {Ran},\ and\ \citenamefont {Wen}}]{PhysRevLett.101.186805}%
  \BibitemOpen
  \bibfield  {author} {\bibinfo {author} {\bibnamefont {Moore}, \bibfnamefont
  {Joel~E}}, \bibinfo {author} {\bibfnamefont {Ying}\ \bibnamefont {Ran}}, \
  and\ \bibinfo {author} {\bibfnamefont {Xiao-Gang}\ \bibnamefont {Wen}}}
  (\bibinfo {year} {2008}),\ \bibfield  {title} {\enquote {\bibinfo {title}
  {Topological surface states in three-dimensional magnetic insulators},}\
  }\href {\doibase 10.1103/PhysRevLett.101.186805} {\bibfield  {journal}
  {\bibinfo  {journal} {Phys. Rev. Lett.}\ }\textbf {\bibinfo {volume} {101}},\
  \bibinfo {pages} {186805}}\BibitemShut {NoStop}%
\bibitem [{\citenamefont {Mueller}\ \emph {et~al.}(2006)\citenamefont
  {Mueller}, \citenamefont {Ho}, \citenamefont {Ueda},\ and\ \citenamefont
  {Baym}}]{mueller2006}%
  \BibitemOpen
  \bibfield  {author} {\bibinfo {author} {\bibnamefont {Mueller}, \bibfnamefont
  {Erich~J}}, \bibinfo {author} {\bibfnamefont {Tin-Lun}\ \bibnamefont {Ho}},
  \bibinfo {author} {\bibfnamefont {Masahito}\ \bibnamefont {Ueda}}, \ and\
  \bibinfo {author} {\bibfnamefont {Gordon}\ \bibnamefont {Baym}}} (\bibinfo
  {year} {2006}),\ \bibfield  {title} {\enquote {\bibinfo {title}
  {Fragmentation of {B}ose-{E}instein condensates},}\ }\href {\doibase
  10.1103/PhysRevA.74.033612} {\bibfield  {journal} {\bibinfo  {journal} {Phys.
  Rev. A}\ }\textbf {\bibinfo {volume} {74}},\ \bibinfo {pages}
  {033612}}\BibitemShut {NoStop}%
\bibitem [{\citenamefont {Mukherjee}\ \emph {et~al.}(2017)\citenamefont
  {Mukherjee}, \citenamefont {Spracklen}, \citenamefont {Valiente},
  \citenamefont {Andersson}, \citenamefont {{\"O}hberg}, \citenamefont
  {Goldman},\ and\ \citenamefont {Thomson}}]{Mukherjee2017}%
  \BibitemOpen
  \bibfield  {author} {\bibinfo {author} {\bibnamefont {Mukherjee},
  \bibfnamefont {Sebabrata}}, \bibinfo {author} {\bibfnamefont {Alexander}\
  \bibnamefont {Spracklen}}, \bibinfo {author} {\bibfnamefont {Manuel}\
  \bibnamefont {Valiente}}, \bibinfo {author} {\bibfnamefont {Erika}\
  \bibnamefont {Andersson}}, \bibinfo {author} {\bibfnamefont {Patrik}\
  \bibnamefont {{\"O}hberg}}, \bibinfo {author} {\bibfnamefont {Nathan}\
  \bibnamefont {Goldman}}, \ and\ \bibinfo {author} {\bibfnamefont {Robert~R.}\
  \bibnamefont {Thomson}}} (\bibinfo {year} {2017}),\ \bibfield  {title}
  {\enquote {\bibinfo {title} {Experimental observation of anomalous
  topological edge modes in a slowly driven photonic lattice},}\ }\href
  {http://dx.doi.org/10.1038/ncomms13918} {\bibfield  {journal} {\bibinfo
  {journal} {Nature Communications}\ }\textbf {\bibinfo {volume} {8}},\
  \bibinfo {pages} {13918}}\BibitemShut {NoStop}%
\bibitem [{\citenamefont {Nakajima}\ \emph {et~al.}(2016)\citenamefont
  {Nakajima}, \citenamefont {Tomita}, \citenamefont {Taie}, \citenamefont
  {Ichinose}, \citenamefont {Ozawa}, \citenamefont {Wang}, \citenamefont
  {Troyer},\ and\ \citenamefont {Takahashi}}]{Nakajima:2016}%
  \BibitemOpen
  \bibfield  {author} {\bibinfo {author} {\bibnamefont {Nakajima},
  \bibfnamefont {Shuta}}, \bibinfo {author} {\bibfnamefont {Takafumi}\
  \bibnamefont {Tomita}}, \bibinfo {author} {\bibfnamefont {Shintaro}\
  \bibnamefont {Taie}}, \bibinfo {author} {\bibfnamefont {Tomohiro}\
  \bibnamefont {Ichinose}}, \bibinfo {author} {\bibfnamefont {Hideki}\
  \bibnamefont {Ozawa}}, \bibinfo {author} {\bibfnamefont {Lei}\ \bibnamefont
  {Wang}}, \bibinfo {author} {\bibfnamefont {Matthias}\ \bibnamefont {Troyer}},
  \ and\ \bibinfo {author} {\bibfnamefont {Yoshiro}\ \bibnamefont {Takahashi}}}
  (\bibinfo {year} {2016}),\ \bibfield  {title} {\enquote {\bibinfo {title}
  {Topological {T}houless pumping of ultracold fermions},}\ }\href {\doibase
  10.1038/nphys3622} {\bibfield  {journal} {\bibinfo  {journal} {Nature
  Physics}\ }\textbf {\bibinfo {volume} {12}},\ \bibinfo {pages}
  {296}}\BibitemShut {NoStop}%
\bibitem [{\citenamefont {Nascimb{\`e}ne}(2015)}]{nascimbene_toposf}%
  \BibitemOpen
  \bibfield  {author} {\bibinfo {author} {\bibnamefont {Nascimb{\`e}ne},
  \bibfnamefont {Sylvain}}} (\bibinfo {year} {2015}),\ \bibfield  {title}
  {\enquote {\bibinfo {title} {Realizing one-dimensional topological
  superfluids with ultracold atomic gases},}\ }\href {\doibase
  10.1088/0953-4075/46/13/134005} {\bibfield  {journal} {\bibinfo  {journal}
  {J. Phys. B}\ }\textbf {\bibinfo {volume} {78}}~(\bibinfo {number} {2}),\
  \bibinfo {pages} {026001}}\BibitemShut {NoStop}%
\bibitem [{\citenamefont {Nathan}\ and\ \citenamefont
  {Rudner}(2015)}]{nathanrudner}%
  \BibitemOpen
  \bibfield  {author} {\bibinfo {author} {\bibnamefont {Nathan}, \bibfnamefont
  {Frederik}}, \ and\ \bibinfo {author} {\bibfnamefont {Mark~S}\ \bibnamefont
  {Rudner}}} (\bibinfo {year} {2015}),\ \bibfield  {title} {\enquote {\bibinfo
  {title} {Topological singularities and the general classification of
  {F}loquet-{B}loch systems},}\ }\href
  {http://stacks.iop.org/1367-2630/17/i=12/a=125014} {\bibfield  {journal}
  {\bibinfo  {journal} {New Journal of Physics}\ }\textbf {\bibinfo {volume}
  {17}}~(\bibinfo {number} {12}),\ \bibinfo {pages} {125014}}\BibitemShut
  {NoStop}%
\bibitem [{\citenamefont {Nayak}\ \emph {et~al.}(2008)\citenamefont {Nayak},
  \citenamefont {Simon}, \citenamefont {Stern}, \citenamefont {Freedman},\ and\
  \citenamefont {Sarma}}]{nayakrmp}%
  \BibitemOpen
  \bibfield  {author} {\bibinfo {author} {\bibnamefont {Nayak}, \bibfnamefont
  {Chetan}}, \bibinfo {author} {\bibfnamefont {Steven~H.}\ \bibnamefont
  {Simon}}, \bibinfo {author} {\bibfnamefont {Ady}\ \bibnamefont {Stern}},
  \bibinfo {author} {\bibfnamefont {Michael}\ \bibnamefont {Freedman}}, \ and\
  \bibinfo {author} {\bibfnamefont {Sankar~Das}\ \bibnamefont {Sarma}}}
  (\bibinfo {year} {2008}),\ \bibfield  {title} {\enquote {\bibinfo {title}
  {Non-abelian anyons and topological quantum computation},}\ }\href {\doibase
  10.1103/RevModPhys.80.1083} {\bibfield  {journal} {\bibinfo  {journal} {Rev.
  Mod. Phys.}\ }\textbf {\bibinfo {volume} {80}}~(\bibinfo {number} {3}),\
  \bibinfo {eid} {1083}}\BibitemShut {NoStop}%
\bibitem [{\citenamefont {van Nieuwenburg}\ and\ \citenamefont
  {Huber}(2014)}]{van2014classification}%
  \BibitemOpen
  \bibfield  {author} {\bibinfo {author} {\bibnamefont {van Nieuwenburg},
  \bibfnamefont {Evert~PL}}, \ and\ \bibinfo {author} {\bibfnamefont
  {Sebastian~D}\ \bibnamefont {Huber}}} (\bibinfo {year} {2014}),\ \bibfield
  {title} {\enquote {\bibinfo {title} {Classification of mixed-state topology
  in one dimension},}\ }\href {\doibase 10.1103/PhysRevB.90.075141} {\bibfield
  {journal} {\bibinfo  {journal} {Phys. Rev. B}\ }\textbf {\bibinfo {volume}
  {90}}~(\bibinfo {number} {7}),\ \bibinfo {pages} {075141}}\BibitemShut
  {NoStop}%
\bibitem [{\citenamefont {Niu}\ \emph {et~al.}(1985)\citenamefont {Niu},
  \citenamefont {Thouless},\ and\ \citenamefont {Wu}}]{Niu1985}%
  \BibitemOpen
  \bibfield  {author} {\bibinfo {author} {\bibnamefont {Niu}, \bibfnamefont
  {Qian}}, \bibinfo {author} {\bibfnamefont {D.~J.}\ \bibnamefont {Thouless}},
  \ and\ \bibinfo {author} {\bibfnamefont {Yong-Shi}\ \bibnamefont {Wu}}}
  (\bibinfo {year} {1985}),\ \bibfield  {title} {\enquote {\bibinfo {title}
  {Quantized hall conductance as a topological invariant},}\ }\href {\doibase
  10.1103/PhysRevB.31.3372} {\bibfield  {journal} {\bibinfo  {journal} {Phys.
  Rev. B}\ }\textbf {\bibinfo {volume} {31}},\ \bibinfo {pages}
  {3372--3377}}\BibitemShut {NoStop}%
\bibitem [{\citenamefont {{Ozawa}}\ \emph {et~al.}(2018)\citenamefont
  {{Ozawa}}, \citenamefont {{Price}}, \citenamefont {{Amo}}, \citenamefont
  {{Goldman}}, \citenamefont {{Hafezi}}, \citenamefont {{Lu}}, \citenamefont
  {{Rechtsman}}, \citenamefont {{Schuster}}, \citenamefont {{Simon}},
  \citenamefont {{Zilberberg}},\ and\ \citenamefont
  {{Carusotto}}}]{photonicreview}%
  \BibitemOpen
  \bibfield  {author} {\bibinfo {author} {\bibnamefont {{Ozawa}}, \bibfnamefont
  {T}}, \bibinfo {author} {\bibfnamefont {H.~M.}\ \bibnamefont {{Price}}},
  \bibinfo {author} {\bibfnamefont {A.}~\bibnamefont {{Amo}}}, \bibinfo
  {author} {\bibfnamefont {N.}~\bibnamefont {{Goldman}}}, \bibinfo {author}
  {\bibfnamefont {M.}~\bibnamefont {{Hafezi}}}, \bibinfo {author}
  {\bibfnamefont {L.}~\bibnamefont {{Lu}}}, \bibinfo {author} {\bibfnamefont
  {M.}~\bibnamefont {{Rechtsman}}}, \bibinfo {author} {\bibfnamefont
  {D.}~\bibnamefont {{Schuster}}}, \bibinfo {author} {\bibfnamefont
  {J.}~\bibnamefont {{Simon}}}, \bibinfo {author} {\bibfnamefont
  {O.}~\bibnamefont {{Zilberberg}}}, \ and\ \bibinfo {author} {\bibfnamefont
  {I.}~\bibnamefont {{Carusotto}}}} (\bibinfo {year} {2018}),\ \bibfield
  {title} {\enquote {\bibinfo {title} {{Topological Photonics}},}\ }\href@noop
  {} {\bibfield  {journal} {\bibinfo  {journal} {ArXiv e-prints}\ }}\Eprint
  {http://arxiv.org/abs/1802.04173} {arXiv:1802.04173} \BibitemShut {NoStop}%
\bibitem [{\citenamefont {Palmer}\ and\ \citenamefont {Jaksch}(2006)}]{palmer}%
  \BibitemOpen
  \bibfield  {author} {\bibinfo {author} {\bibnamefont {Palmer}, \bibfnamefont
  {R~N}}, \ and\ \bibinfo {author} {\bibfnamefont {D.}~\bibnamefont {Jaksch}}}
  (\bibinfo {year} {2006}),\ \bibfield  {title} {\enquote {\bibinfo {title}
  {High-field fractional quantum {H}all effect in optical lattices},}\ }\href
  {\doibase 10.1103/PhysRevLett.96.180407} {\bibfield  {journal} {\bibinfo
  {journal} {Phys. Rev. Lett.}\ }\textbf {\bibinfo {volume} {96}}~(\bibinfo
  {number} {18}),\ \bibinfo {eid} {180407}}\BibitemShut {NoStop}%
\bibitem [{\citenamefont {Parameswaran}\ \emph {et~al.}(2013)\citenamefont
  {Parameswaran}, \citenamefont {Roy},\ and\ \citenamefont
  {Sondhi}}]{Parameswaran2013816}%
  \BibitemOpen
  \bibfield  {author} {\bibinfo {author} {\bibnamefont {Parameswaran},
  \bibfnamefont {Siddharth~A}}, \bibinfo {author} {\bibfnamefont {Rahul}\
  \bibnamefont {Roy}}, \ and\ \bibinfo {author} {\bibfnamefont {Shivaji~L.}\
  \bibnamefont {Sondhi}}} (\bibinfo {year} {2013}),\ \bibfield  {title}
  {\enquote {\bibinfo {title} {Fractional quantum {H}all physics in topological
  flat bands},}\ }\href {\doibase http://dx.doi.org/10.1016/j.crhy.2013.04.003}
  {\bibfield  {journal} {\bibinfo  {journal} {Comptes Rendus Physique}\
  }\textbf {\bibinfo {volume} {14}}~(\bibinfo {number} {9}),\ \bibinfo {pages}
  {816 -- 839}}\BibitemShut {NoStop}%
\bibitem [{\citenamefont {Pethick}\ and\ \citenamefont
  {Smith}(2002)}]{pethick2002bose}%
  \BibitemOpen
  \bibfield  {author} {\bibinfo {author} {\bibnamefont {Pethick}, \bibfnamefont
  {CJ}}, \ and\ \bibinfo {author} {\bibfnamefont {H.}~\bibnamefont {Smith}}}
  (\bibinfo {year} {2002}),\ \href
  {https://books.google.co.uk/books?id=iBk0G3\_5iIQC} {\emph {\bibinfo {title}
  {Bose-{E}instein Condensation in Dilute Gases}}}\ (\bibinfo  {publisher}
  {Cambridge University Press})\BibitemShut {NoStop}%
\bibitem [{\citenamefont {Petrescu}\ and\ \citenamefont
  {Le~Hur}(2015)}]{PhysRevB.91.054520}%
  \BibitemOpen
  \bibfield  {author} {\bibinfo {author} {\bibnamefont {Petrescu},
  \bibfnamefont {Alexandru}}, \ and\ \bibinfo {author} {\bibfnamefont {Karyn}\
  \bibnamefont {Le~Hur}}} (\bibinfo {year} {2015}),\ \bibfield  {title}
  {\enquote {\bibinfo {title} {Chiral {M}ott insulators, {M}eissner effect, and
  {L}aughlin states in quantum ladders},}\ }\href {\doibase
  10.1103/PhysRevB.91.054520} {\bibfield  {journal} {\bibinfo  {journal} {Phys.
  Rev. B}\ }\textbf {\bibinfo {volume} {91}},\ \bibinfo {pages}
  {054520}}\BibitemShut {NoStop}%
\bibitem [{\citenamefont {Piraud}\ \emph {et~al.}(2015)\citenamefont {Piraud},
  \citenamefont {Heidrich-Meisner}, \citenamefont {McCulloch}, \citenamefont
  {Greschner}, \citenamefont {Vekua},\ and\ \citenamefont
  {Schollw\"ock}}]{PhysRevB.91.140406}%
  \BibitemOpen
  \bibfield  {author} {\bibinfo {author} {\bibnamefont {Piraud}, \bibfnamefont
  {M}}, \bibinfo {author} {\bibfnamefont {F.}~\bibnamefont {Heidrich-Meisner}},
  \bibinfo {author} {\bibfnamefont {I.~P.}\ \bibnamefont {McCulloch}}, \bibinfo
  {author} {\bibfnamefont {S.}~\bibnamefont {Greschner}}, \bibinfo {author}
  {\bibfnamefont {T.}~\bibnamefont {Vekua}}, \ and\ \bibinfo {author}
  {\bibfnamefont {U.}~\bibnamefont {Schollw\"ock}}} (\bibinfo {year} {2015}),\
  \bibfield  {title} {\enquote {\bibinfo {title} {Vortex and meissner phases of
  strongly interacting bosons on a two-leg ladder},}\ }\href {\doibase
  10.1103/PhysRevB.91.140406} {\bibfield  {journal} {\bibinfo  {journal} {Phys.
  Rev. B}\ }\textbf {\bibinfo {volume} {91}},\ \bibinfo {pages}
  {140406}}\BibitemShut {NoStop}%
\bibitem [{\citenamefont {Powell}\ \emph {et~al.}(2011)\citenamefont {Powell},
  \citenamefont {Barnett}, \citenamefont {Sensarma},\ and\ \citenamefont
  {Das~Sarma}}]{PhysRevA.83.013612}%
  \BibitemOpen
  \bibfield  {author} {\bibinfo {author} {\bibnamefont {Powell}, \bibfnamefont
  {Stephen}}, \bibinfo {author} {\bibfnamefont {Ryan}\ \bibnamefont {Barnett}},
  \bibinfo {author} {\bibfnamefont {Rajdeep}\ \bibnamefont {Sensarma}}, \ and\
  \bibinfo {author} {\bibfnamefont {Sankar}\ \bibnamefont {Das~Sarma}}}
  (\bibinfo {year} {2011}),\ \bibfield  {title} {\enquote {\bibinfo {title}
  {Bogoliubov theory of interacting bosons on a lattice in a synthetic magnetic
  field},}\ }\href {\doibase 10.1103/PhysRevA.83.013612} {\bibfield  {journal}
  {\bibinfo  {journal} {Phys. Rev. A}\ }\textbf {\bibinfo {volume} {83}},\
  \bibinfo {pages} {013612}}\BibitemShut {NoStop}%
\bibitem [{\citenamefont {Prange}\ and\ \citenamefont
  {Girvin}(1990)}]{prangeandgirvin}%
  \BibitemOpen
  \bibinfo {editor} {\bibnamefont {Prange}, \bibfnamefont {R~E}}, \ and\
  \bibinfo {editor} {\bibfnamefont {S.~M.}\ \bibnamefont {Girvin}},\ Eds.
  (\bibinfo {year} {1990}),\ \href@noop {} {\emph {\bibinfo {title} {The
  Quantum {H}all Effect}}},\ \bibinfo {edition} {2nd}\ ed.\ (\bibinfo
  {publisher} {Springer-Verlag},\ \bibinfo {address} {Berlin})\BibitemShut
  {NoStop}%
\bibitem [{\citenamefont {Price}\ and\ \citenamefont
  {Cooper}(2012)}]{Price:2012}%
  \BibitemOpen
  \bibfield  {author} {\bibinfo {author} {\bibnamefont {Price}, \bibfnamefont
  {H~M}}, \ and\ \bibinfo {author} {\bibfnamefont {N.~R.}\ \bibnamefont
  {Cooper}}} (\bibinfo {year} {2012}),\ \bibfield  {title} {\enquote {\bibinfo
  {title} {Mapping the {B}erry curvature from semiclassical dynamics in optical
  lattices},}\ }\href {\doibase 10.1103/PhysRevA.85.033620} {\bibfield
  {journal} {\bibinfo  {journal} {Phys. Rev. A}\ }\textbf {\bibinfo {volume}
  {85}},\ \bibinfo {pages} {033620}}\BibitemShut {NoStop}%
\bibitem [{\citenamefont {Price}\ \emph {et~al.}(2015)\citenamefont {Price},
  \citenamefont {Zilberberg}, \citenamefont {Ozawa}, \citenamefont
  {Carusotto},\ and\ \citenamefont {Goldman}}]{Price2015}%
  \BibitemOpen
  \bibfield  {author} {\bibinfo {author} {\bibnamefont {Price}, \bibfnamefont
  {H~M}}, \bibinfo {author} {\bibfnamefont {O.}~\bibnamefont {Zilberberg}},
  \bibinfo {author} {\bibfnamefont {T.}~\bibnamefont {Ozawa}}, \bibinfo
  {author} {\bibfnamefont {I.}~\bibnamefont {Carusotto}}, \ and\ \bibinfo
  {author} {\bibfnamefont {N.}~\bibnamefont {Goldman}}} (\bibinfo {year}
  {2015}),\ \bibfield  {title} {\enquote {\bibinfo {title} {Four-dimensional
  quantum {H}all effect with ultracold atoms},}\ }\href {\doibase
  10.1103/PhysRevLett.115.195303} {\bibfield  {journal} {\bibinfo  {journal}
  {Phys. Rev. Lett.}\ }\textbf {\bibinfo {volume} {115}},\ \bibinfo {pages}
  {195303}}\BibitemShut {NoStop}%
\bibitem [{\citenamefont {Price}\ and\ \citenamefont
  {Cooper}(2013)}]{PhysRevLett.111.220407}%
  \BibitemOpen
  \bibfield  {author} {\bibinfo {author} {\bibnamefont {Price}, \bibfnamefont
  {Hannah~M}}, \ and\ \bibinfo {author} {\bibfnamefont {Nigel~R.}\ \bibnamefont
  {Cooper}}} (\bibinfo {year} {2013}),\ \bibfield  {title} {\enquote {\bibinfo
  {title} {Effects of {B}erry curvature on the collective modes of ultracold
  gases},}\ }\href {\doibase 10.1103/PhysRevLett.111.220407} {\bibfield
  {journal} {\bibinfo  {journal} {Phys. Rev. Lett.}\ }\textbf {\bibinfo
  {volume} {111}},\ \bibinfo {pages} {220407}}\BibitemShut {NoStop}%
\bibitem [{\citenamefont {Price}\ \emph {et~al.}(2017)\citenamefont {Price},
  \citenamefont {Ozawa},\ and\ \citenamefont {Goldman}}]{price_ozawa_goldman}%
  \BibitemOpen
  \bibfield  {author} {\bibinfo {author} {\bibnamefont {Price}, \bibfnamefont
  {Hannah~M}}, \bibinfo {author} {\bibfnamefont {Tomoki}\ \bibnamefont
  {Ozawa}}, \ and\ \bibinfo {author} {\bibfnamefont {Nathan}\ \bibnamefont
  {Goldman}}} (\bibinfo {year} {2017}),\ \bibfield  {title} {\enquote {\bibinfo
  {title} {Synthetic dimensions for cold atoms from shaking a harmonic trap},}\
  }\href {\doibase 10.1103/PhysRevA.95.023607} {\bibfield  {journal} {\bibinfo
  {journal} {Phys. Rev. A}\ }\textbf {\bibinfo {volume} {95}},\ \bibinfo
  {pages} {023607}}\BibitemShut {NoStop}%
\bibitem [{\citenamefont {Privitera}\ \emph {et~al.}(2018)\citenamefont
  {Privitera}, \citenamefont {Russomanno}, \citenamefont {Citro},\ and\
  \citenamefont {Santoro}}]{PhysRevLett.120.106601}%
  \BibitemOpen
  \bibfield  {author} {\bibinfo {author} {\bibnamefont {Privitera},
  \bibfnamefont {Lorenzo}}, \bibinfo {author} {\bibfnamefont {Angelo}\
  \bibnamefont {Russomanno}}, \bibinfo {author} {\bibfnamefont {Roberta}\
  \bibnamefont {Citro}}, \ and\ \bibinfo {author} {\bibfnamefont {Giuseppe~E.}\
  \bibnamefont {Santoro}}} (\bibinfo {year} {2018}),\ \bibfield  {title}
  {\enquote {\bibinfo {title} {Nonadiabatic breaking of topological pumping},}\
  }\href {\doibase 10.1103/PhysRevLett.120.106601} {\bibfield  {journal}
  {\bibinfo  {journal} {Phys. Rev. Lett.}\ }\textbf {\bibinfo {volume} {120}},\
  \bibinfo {pages} {106601}}\BibitemShut {NoStop}%
\bibitem [{\citenamefont {Qi}\ and\ \citenamefont {Zhang}(2011)}]{qizhang}%
  \BibitemOpen
  \bibfield  {author} {\bibinfo {author} {\bibnamefont {Qi}, \bibfnamefont
  {Xiao-Liang}}, \ and\ \bibinfo {author} {\bibfnamefont {Shou-Cheng}\
  \bibnamefont {Zhang}}} (\bibinfo {year} {2011}),\ \bibfield  {title}
  {\enquote {\bibinfo {title} {Topological insulators and superconductors},}\
  }\href {\doibase 10.1103/RevModPhys.83.1057} {\bibfield  {journal} {\bibinfo
  {journal} {Rev. Mod. Phys.}\ }\textbf {\bibinfo {volume} {83}},\ \bibinfo
  {pages} {1057--1110}}\BibitemShut {NoStop}%
\bibitem [{\citenamefont {Ra{\v{c}}i{\=u}nas}\ \emph
  {et~al.}(2016)\citenamefont {Ra{\v{c}}i{\=u}nas}, \citenamefont
  {{\v{Z}}labys}, \citenamefont {Eckardt},\ and\ \citenamefont
  {Anisimovas}}]{PhysRevA.93.043618}%
  \BibitemOpen
  \bibfield  {author} {\bibinfo {author} {\bibnamefont {Ra{\v{c}}i{\=u}nas},
  \bibfnamefont {Mantas}}, \bibinfo {author} {\bibfnamefont {Giedrius}\
  \bibnamefont {{\v{Z}}labys}}, \bibinfo {author} {\bibfnamefont {Andr{\'e}}\
  \bibnamefont {Eckardt}}, \ and\ \bibinfo {author} {\bibfnamefont {Egidijus}\
  \bibnamefont {Anisimovas}}} (\bibinfo {year} {2016}),\ \bibfield  {title}
  {\enquote {\bibinfo {title} {Modified interactions in a {F}loquet topological
  system on a square lattice and their impact on a bosonic fractional {C}hern
  insulator state},}\ }\href {\doibase 0.1103/PhysRevA.93.043618} {\bibfield
  {journal} {\bibinfo  {journal} {Phys. Rev. A}\ }\textbf {\bibinfo {volume}
  {93}}~(\bibinfo {number} {4}),\ \bibinfo {pages} {043618}}\BibitemShut
  {NoStop}%
\bibitem [{\citenamefont {Rakovszky}\ \emph {et~al.}(2017)\citenamefont
  {Rakovszky}, \citenamefont {Asb{\'o}th},\ and\ \citenamefont
  {Alberti}}]{rakovszky2017detecting}%
  \BibitemOpen
  \bibfield  {author} {\bibinfo {author} {\bibnamefont {Rakovszky},
  \bibfnamefont {Tibor}}, \bibinfo {author} {\bibfnamefont {J{\'a}nos~K}\
  \bibnamefont {Asb{\'o}th}}, \ and\ \bibinfo {author} {\bibfnamefont {Andrea}\
  \bibnamefont {Alberti}}} (\bibinfo {year} {2017}),\ \bibfield  {title}
  {\enquote {\bibinfo {title} {Detecting topological invariants in chiral
  symmetric insulators via losses},}\ }\href {\doibase
  10.1103/PhysRevB.95.201407} {\bibfield  {journal} {\bibinfo  {journal} {Phys.
  Rev. B}\ }\textbf {\bibinfo {volume} {95}}~(\bibinfo {number} {20}),\
  \bibinfo {pages} {201407}}\BibitemShut {NoStop}%
\bibitem [{\citenamefont {Read}\ and\ \citenamefont {Green}(2000)}]{ReadG00}%
  \BibitemOpen
  \bibfield  {author} {\bibinfo {author} {\bibnamefont {Read}, \bibfnamefont
  {N}}, \ and\ \bibinfo {author} {\bibfnamefont {D.}~\bibnamefont {Green}}}
  (\bibinfo {year} {2000}),\ \bibfield  {title} {\enquote {\bibinfo {title}
  {Paired states of fermions in two dimensions with breaking of parity and
  time-reversal symmetries and the fractional quantum {H}all effect},}\ }\href
  {\doibase 10.1103/PhysRevB.61.10267} {\bibfield  {journal} {\bibinfo
  {journal} {Phys. Rev. B}\ }\textbf {\bibinfo {volume} {61}},\ \bibinfo
  {pages} {10267--10297}}\BibitemShut {NoStop}%
\bibitem [{\citenamefont {Reitter}\ \emph {et~al.}(2017)\citenamefont
  {Reitter}, \citenamefont {N\"ager}, \citenamefont {Wintersperger},
  \citenamefont {Str\"ater}, \citenamefont {Bloch}, \citenamefont {Eckardt},\
  and\ \citenamefont {Schneider}}]{schneiderdata}%
  \BibitemOpen
  \bibfield  {author} {\bibinfo {author} {\bibnamefont {Reitter}, \bibfnamefont
  {Martin}}, \bibinfo {author} {\bibfnamefont {Jakob}\ \bibnamefont {N\"ager}},
  \bibinfo {author} {\bibfnamefont {Karen}\ \bibnamefont {Wintersperger}},
  \bibinfo {author} {\bibfnamefont {Christoph}\ \bibnamefont {Str\"ater}},
  \bibinfo {author} {\bibfnamefont {Immanuel}\ \bibnamefont {Bloch}}, \bibinfo
  {author} {\bibfnamefont {Andr\'e}\ \bibnamefont {Eckardt}}, \ and\ \bibinfo
  {author} {\bibfnamefont {Ulrich}\ \bibnamefont {Schneider}}} (\bibinfo {year}
  {2017}),\ \bibfield  {title} {\enquote {\bibinfo {title} {Interaction
  dependent heating and atom loss in a periodically driven optical lattice},}\
  }\href {\doibase 10.1103/PhysRevLett.119.200402} {\bibfield  {journal}
  {\bibinfo  {journal} {Phys. Rev. Lett.}\ }\textbf {\bibinfo {volume} {119}},\
  \bibinfo {pages} {200402}}\BibitemShut {NoStop}%
\bibitem [{\citenamefont {Rice}\ and\ \citenamefont {Mele}(1982)}]{Rice:1982}%
  \BibitemOpen
  \bibfield  {author} {\bibinfo {author} {\bibnamefont {Rice}, \bibfnamefont
  {MJ}}, \ and\ \bibinfo {author} {\bibfnamefont {EJ}~\bibnamefont {Mele}}}
  (\bibinfo {year} {1982}),\ \bibfield  {title} {\enquote {\bibinfo {title}
  {Elementary excitations of a linearly conjugated diatomic polymer},}\ }\href
  {\doibase 10.1103/PhysRevLett.49.1455} {\bibfield  {journal} {\bibinfo
  {journal} {Phys. Rev. Lett.}\ }\textbf {\bibinfo {volume} {49}}~(\bibinfo
  {number} {19}),\ \bibinfo {pages} {1455}}\BibitemShut {NoStop}%
\bibitem [{\citenamefont {Roy}\ and\ \citenamefont
  {Harper}(2016)}]{royharper2016}%
  \BibitemOpen
  \bibfield  {author} {\bibinfo {author} {\bibnamefont {Roy}, \bibfnamefont
  {Rahul}}, \ and\ \bibinfo {author} {\bibfnamefont {Fenner}\ \bibnamefont
  {Harper}}} (\bibinfo {year} {2016}),\ \bibfield  {title} {\enquote {\bibinfo
  {title} {Abelian {F}loquet symmetry-protected topological phases in one
  dimension},}\ }\href {\doibase 10.1103/PhysRevB.94.125105} {\bibfield
  {journal} {\bibinfo  {journal} {Phys. Rev. B}\ }\textbf {\bibinfo {volume}
  {94}},\ \bibinfo {pages} {125105}}\BibitemShut {NoStop}%
\bibitem [{\citenamefont {Rudner}\ \emph {et~al.}(2013)\citenamefont {Rudner},
  \citenamefont {Lindner}, \citenamefont {Berg},\ and\ \citenamefont
  {Levin}}]{PhysRevX.3.031005}%
  \BibitemOpen
  \bibfield  {author} {\bibinfo {author} {\bibnamefont {Rudner}, \bibfnamefont
  {Mark~S}}, \bibinfo {author} {\bibfnamefont {Netanel~H.}\ \bibnamefont
  {Lindner}}, \bibinfo {author} {\bibfnamefont {Erez}\ \bibnamefont {Berg}}, \
  and\ \bibinfo {author} {\bibfnamefont {Michael}\ \bibnamefont {Levin}}}
  (\bibinfo {year} {2013}),\ \bibfield  {title} {\enquote {\bibinfo {title}
  {Anomalous edge states and the bulk-edge correspondence for periodically
  driven two-dimensional systems},}\ }\href {\doibase
  10.1103/PhysRevX.3.031005} {\bibfield  {journal} {\bibinfo  {journal} {Phys.
  Rev. X}\ }\textbf {\bibinfo {volume} {3}},\ \bibinfo {pages}
  {031005}}\BibitemShut {NoStop}%
\bibitem [{\citenamefont {Rudner}\ and\ \citenamefont
  {Levitov}(2009)}]{rudner2009topological}%
  \BibitemOpen
  \bibfield  {author} {\bibinfo {author} {\bibnamefont {Rudner}, \bibfnamefont
  {MS}}, \ and\ \bibinfo {author} {\bibfnamefont {LS}~\bibnamefont {Levitov}}}
  (\bibinfo {year} {2009}),\ \bibfield  {title} {\enquote {\bibinfo {title}
  {Topological transition in a non-{H}ermitian quantum walk},}\ }\href
  {\doibase 10.1103/PhysRevLett.102.065703} {\bibfield  {journal} {\bibinfo
  {journal} {Phys. Rev. Lett.}\ }\textbf {\bibinfo {volume} {102}}~(\bibinfo
  {number} {6}),\ \bibinfo {pages} {065703}}\BibitemShut {NoStop}%
\bibitem [{\citenamefont {Ruhman}\ and\ \citenamefont
  {Altman}(2017)}]{Ruhman:2017}%
  \BibitemOpen
  \bibfield  {author} {\bibinfo {author} {\bibnamefont {Ruhman}, \bibfnamefont
  {Jonathan}}, \ and\ \bibinfo {author} {\bibfnamefont {Ehud}\ \bibnamefont
  {Altman}}} (\bibinfo {year} {2017}),\ \bibfield  {title} {\enquote {\bibinfo
  {title} {Topological degeneracy and pairing in a one-dimensional gas of
  spinless fermions},}\ }\href {\doibase 10.1103/PhysRevB.96.085133} {\bibfield
   {journal} {\bibinfo  {journal} {Phys. Rev. B}\ }\textbf {\bibinfo {volume}
  {96}},\ \bibinfo {pages} {085133}}\BibitemShut {NoStop}%
\bibitem [{\citenamefont {Ruhman}\ \emph {et~al.}(2015)\citenamefont {Ruhman},
  \citenamefont {Berg},\ and\ \citenamefont {Altman}}]{Ruhman:2015}%
  \BibitemOpen
  \bibfield  {author} {\bibinfo {author} {\bibnamefont {Ruhman}, \bibfnamefont
  {Jonathan}}, \bibinfo {author} {\bibfnamefont {Erez}\ \bibnamefont {Berg}}, \
  and\ \bibinfo {author} {\bibfnamefont {Ehud}\ \bibnamefont {Altman}}}
  (\bibinfo {year} {2015}),\ \bibfield  {title} {\enquote {\bibinfo {title}
  {Topological states in a one-dimensional {F}ermi gas with attractive
  interaction},}\ }\href {\doibase 10.1103/PhysRevLett.114.100401} {\bibfield
  {journal} {\bibinfo  {journal} {Phys. Rev. Lett.}\ }\textbf {\bibinfo
  {volume} {114}},\ \bibinfo {pages} {100401}}\BibitemShut {NoStop}%
\bibitem [{\citenamefont {Sacramento}(2016)}]{PhysRevE.93.062117}%
  \BibitemOpen
  \bibfield  {author} {\bibinfo {author} {\bibnamefont {Sacramento},
  \bibfnamefont {P~D}}} (\bibinfo {year} {2016}),\ \bibfield  {title} {\enquote
  {\bibinfo {title} {Edge mode dynamics of quenched topological wires},}\
  }\href {\doibase 10.1103/PhysRevE.93.062117} {\bibfield  {journal} {\bibinfo
  {journal} {Phys. Rev. E}\ }\textbf {\bibinfo {volume} {93}},\ \bibinfo
  {pages} {062117}}\BibitemShut {NoStop}%
\bibitem [{\citenamefont {Sau}\ \emph {et~al.}(2011)\citenamefont {Sau},
  \citenamefont {Halperin}, \citenamefont {Flensberg},\ and\ \citenamefont
  {Das~Sarma}}]{Sau2011}%
  \BibitemOpen
  \bibfield  {author} {\bibinfo {author} {\bibnamefont {Sau}, \bibfnamefont
  {Jay~D}}, \bibinfo {author} {\bibfnamefont {B.~I.}\ \bibnamefont {Halperin}},
  \bibinfo {author} {\bibfnamefont {K.}~\bibnamefont {Flensberg}}, \ and\
  \bibinfo {author} {\bibfnamefont {S.}~\bibnamefont {Das~Sarma}}} (\bibinfo
  {year} {2011}),\ \bibfield  {title} {\enquote {\bibinfo {title} {Number
  conserving theory for topologically protected degeneracy in one-dimensional
  fermions},}\ }\href {\doibase 10.1103/PhysRevB.84.144509} {\bibfield
  {journal} {\bibinfo  {journal} {Phys. Rev. B}\ }\textbf {\bibinfo {volume}
  {84}},\ \bibinfo {pages} {144509}}\BibitemShut {NoStop}%
\bibitem [{\citenamefont {Sebby-Strabley}\ \emph {et~al.}(2006)\citenamefont
  {Sebby-Strabley}, \citenamefont {Anderlini}, \citenamefont {Jessen},\ and\
  \citenamefont {Porto}}]{Sebby-Strabley2006}%
  \BibitemOpen
  \bibfield  {author} {\bibinfo {author} {\bibnamefont {Sebby-Strabley},
  \bibfnamefont {J}}, \bibinfo {author} {\bibfnamefont {M.}~\bibnamefont
  {Anderlini}}, \bibinfo {author} {\bibfnamefont {P~S}\ \bibnamefont {Jessen}},
  \ and\ \bibinfo {author} {\bibfnamefont {J.~V.}\ \bibnamefont {Porto}}}
  (\bibinfo {year} {2006}),\ \bibfield  {title} {\enquote {\bibinfo {title}
  {{Lattice of double wells for manipulating pairs of cold atoms}},}\ }\href
  {\doibase 10.1103/PhysRevA.73.033605} {\bibfield  {journal} {\bibinfo
  {journal} {Phys. Rev. A}\ }\textbf {\bibinfo {volume} {73}},\ \bibinfo
  {pages} {033605}}\BibitemShut {NoStop}%
\bibitem [{\citenamefont {Senthil}(2015)}]{senthilreview}%
  \BibitemOpen
  \bibfield  {author} {\bibinfo {author} {\bibnamefont {Senthil}, \bibfnamefont
  {T}}} (\bibinfo {year} {2015}),\ \bibfield  {title} {\enquote {\bibinfo
  {title} {Symmetry-protected topological phases of quantum matter},}\ }\href
  {\doibase 10.1146/annurev-conmatphys-031214-014740} {\bibfield  {journal}
  {\bibinfo  {journal} {Annual Review of Condensed Matter Physics}\ }\textbf
  {\bibinfo {volume} {6}}~(\bibinfo {number} {1}),\ \bibinfo {pages}
  {299--324}}\BibitemShut {NoStop}%
\bibitem [{\citenamefont {Sherson}\ \emph {et~al.}(2010)\citenamefont
  {Sherson}, \citenamefont {Weitenberg}, \citenamefont {Endres}, \citenamefont
  {Cheneau}, \citenamefont {Bloch},\ and\ \citenamefont {Kuhr}}]{Sherson2010}%
  \BibitemOpen
  \bibfield  {author} {\bibinfo {author} {\bibnamefont {Sherson}, \bibfnamefont
  {Jacob~F}}, \bibinfo {author} {\bibfnamefont {Christof}\ \bibnamefont
  {Weitenberg}}, \bibinfo {author} {\bibfnamefont {Manuel}\ \bibnamefont
  {Endres}}, \bibinfo {author} {\bibfnamefont {Marc}\ \bibnamefont {Cheneau}},
  \bibinfo {author} {\bibfnamefont {Immanuel}\ \bibnamefont {Bloch}}, \ and\
  \bibinfo {author} {\bibfnamefont {Stefan}\ \bibnamefont {Kuhr}}} (\bibinfo
  {year} {2010}),\ \bibfield  {title} {\enquote {\bibinfo {title}
  {{Single-atom-resolved fluorescence imaging of an atomic Mott insulator}},}\
  }\href {\doibase 10.1038/nature09378} {\bibfield  {journal} {\bibinfo
  {journal} {Nature}\ }\textbf {\bibinfo {volume} {467}}~(\bibinfo {number}
  {7311}),\ \bibinfo {pages} {68--72}}\BibitemShut {NoStop}%
\bibitem [{\citenamefont {S{\o}rensen}\ \emph {et~al.}(2005)\citenamefont
  {S{\o}rensen}, \citenamefont {Demler},\ and\ \citenamefont
  {Lukin}}]{Sorensen:2005}%
  \BibitemOpen
  \bibfield  {author} {\bibinfo {author} {\bibnamefont {S{\o}rensen},
  \bibfnamefont {Anders~S}}, \bibinfo {author} {\bibfnamefont {Eugene}\
  \bibnamefont {Demler}}, \ and\ \bibinfo {author} {\bibfnamefont {Mikhail~D.}\
  \bibnamefont {Lukin}}} (\bibinfo {year} {2005}),\ \bibfield  {title}
  {\enquote {\bibinfo {title} {Fractional quantum {H}all states of atoms in
  optical lattices},}\ }\href {\doibase 10.1103/PhysRevLett.94.086803}
  {\bibfield  {journal} {\bibinfo  {journal} {Phys. Rev. Lett.}\ }\textbf
  {\bibinfo {volume} {94}}~(\bibinfo {number} {8}),\ \bibinfo {eid}
  {086803}}\BibitemShut {NoStop}%
\bibitem [{\citenamefont {Sterdyniak}\ \emph {et~al.}(2015)\citenamefont
  {Sterdyniak}, \citenamefont {Bernevig}, \citenamefont {Cooper},\ and\
  \citenamefont {Regnault}}]{sterdyniak_2015}%
  \BibitemOpen
  \bibfield  {author} {\bibinfo {author} {\bibnamefont {Sterdyniak},
  \bibfnamefont {A}}, \bibinfo {author} {\bibfnamefont {B.~Andrei}\
  \bibnamefont {Bernevig}}, \bibinfo {author} {\bibfnamefont {Nigel~R.}\
  \bibnamefont {Cooper}}, \ and\ \bibinfo {author} {\bibfnamefont
  {N.}~\bibnamefont {Regnault}}} (\bibinfo {year} {2015}),\ \bibfield  {title}
  {\enquote {\bibinfo {title} {Interacting bosons in topological optical flux
  lattices},}\ }\href {\doibase 10.1103/PhysRevB.91.035115} {\bibfield
  {journal} {\bibinfo  {journal} {Phys. Rev. B}\ }\textbf {\bibinfo {volume}
  {91}},\ \bibinfo {pages} {035115}}\BibitemShut {NoStop}%
\bibitem [{\citenamefont {Stern}(2008)}]{adysternreview}%
  \BibitemOpen
  \bibfield  {author} {\bibinfo {author} {\bibnamefont {Stern}, \bibfnamefont
  {Ady}}} (\bibinfo {year} {2008}),\ \bibfield  {title} {\enquote {\bibinfo
  {title} {Anyons and the quantum {H}all effect -- a pedagogical review},}\
  }\href {\doibase https://doi.org/10.1016/j.aop.2007.10.008} {\bibfield
  {journal} {\bibinfo  {journal} {Annals of Physics}\ }\textbf {\bibinfo
  {volume} {323}}~(\bibinfo {number} {1}),\ \bibinfo {pages} {204 --
  249}}\BibitemShut {NoStop}%
\bibitem [{\citenamefont {Stone}\ and\ \citenamefont
  {Goldbart}(2009)}]{stonegoldbart}%
  \BibitemOpen
  \bibfield  {author} {\bibinfo {author} {\bibnamefont {Stone}, \bibfnamefont
  {M}}, \ and\ \bibinfo {author} {\bibfnamefont {P.}~\bibnamefont {Goldbart}}}
  (\bibinfo {year} {2009}),\ \href@noop {} {\emph {\bibinfo {title}
  {Mathematics for Physics: A Guided Tour for Graduate Students}}}\ (\bibinfo
  {publisher} {Cambridge University Press})\BibitemShut {NoStop}%
\bibitem [{\citenamefont {Straley}\ and\ \citenamefont
  {Barnett}(1993)}]{PhysRevB.48.3309}%
  \BibitemOpen
  \bibfield  {author} {\bibinfo {author} {\bibnamefont {Straley}, \bibfnamefont
  {Joseph~P}}, \ and\ \bibinfo {author} {\bibfnamefont {George~Michael}\
  \bibnamefont {Barnett}}} (\bibinfo {year} {1993}),\ \bibfield  {title}
  {\enquote {\bibinfo {title} {Phase diagram for a {J}osephson network in a
  magnetic field},}\ }\href {\doibase 10.1103/PhysRevB.48.3309} {\bibfield
  {journal} {\bibinfo  {journal} {Phys. Rev. B}\ }\textbf {\bibinfo {volume}
  {48}},\ \bibinfo {pages} {3309--3315}}\BibitemShut {NoStop}%
\bibitem [{\citenamefont {{Str{\"a}ter}}\ and\ \citenamefont
  {{Eckardt}}(2016)}]{2016arXiv160400850S}%
  \BibitemOpen
  \bibfield  {author} {\bibinfo {author} {\bibnamefont {{Str{\"a}ter}},
  \bibfnamefont {C}}, \ and\ \bibinfo {author} {\bibfnamefont {A.}~\bibnamefont
  {{Eckardt}}}} (\bibinfo {year} {2016}),\ \bibfield  {title} {\enquote
  {\bibinfo {title} {Interband heating processes in a periodically driven
  optical lattice},}\ }\href {\doibase 10.1515/zna-2016-0129} {\bibfield
  {journal} {\bibinfo  {journal} {Zeitschrift f{\"u}r Naturforschung A}\
  }\textbf {\bibinfo {volume} {71}},\ \bibinfo {pages} {909--920}}\BibitemShut
  {NoStop}%
\bibitem [{\citenamefont {Stringari}(1996)}]{PhysRevLett.77.2360}%
  \BibitemOpen
  \bibfield  {author} {\bibinfo {author} {\bibnamefont {Stringari},
  \bibfnamefont {S}}} (\bibinfo {year} {1996}),\ \bibfield  {title} {\enquote
  {\bibinfo {title} {Collective excitations of a trapped {B}ose-condensed
  gas},}\ }\href {\doibase 10.1103/PhysRevLett.77.2360} {\bibfield  {journal}
  {\bibinfo  {journal} {Phys. Rev. Lett.}\ }\textbf {\bibinfo {volume} {77}},\
  \bibinfo {pages} {2360--2363}}\BibitemShut {NoStop}%
\bibitem [{\citenamefont {Struck}\ \emph {et~al.}(2012)\citenamefont {Struck},
  \citenamefont {{\"O}lschl{\"a}ger}, \citenamefont {Weinberg}, \citenamefont
  {Hauke}, \citenamefont {Simonet}, \citenamefont {Eckardt}, \citenamefont
  {Lewenstein}, \citenamefont {Sengstock},\ and\ \citenamefont
  {Windpassinger}}]{Struck2012}%
  \BibitemOpen
  \bibfield  {author} {\bibinfo {author} {\bibnamefont {Struck}, \bibfnamefont
  {J}}, \bibinfo {author} {\bibfnamefont {C}~\bibnamefont
  {{\"O}lschl{\"a}ger}}, \bibinfo {author} {\bibfnamefont {M}~\bibnamefont
  {Weinberg}}, \bibinfo {author} {\bibfnamefont {P}~\bibnamefont {Hauke}},
  \bibinfo {author} {\bibfnamefont {J}~\bibnamefont {Simonet}}, \bibinfo
  {author} {\bibfnamefont {A}~\bibnamefont {Eckardt}}, \bibinfo {author}
  {\bibfnamefont {M.}~\bibnamefont {Lewenstein}}, \bibinfo {author}
  {\bibfnamefont {K.}~\bibnamefont {Sengstock}}, \ and\ \bibinfo {author}
  {\bibfnamefont {P}~\bibnamefont {Windpassinger}}} (\bibinfo {year} {2012}),\
  \bibfield  {title} {\enquote {\bibinfo {title} {Tunable gauge potential for
  neutral and spinless particles in driven optical lattices},}\ }\href
  {\doibase 10.1103/PhysRevLett.108.225304} {\bibfield  {journal} {\bibinfo
  {journal} {Phys. Rev. Lett.}\ }\textbf {\bibinfo {volume} {108}},\ \bibinfo
  {pages} {225304}}\BibitemShut {NoStop}%
\bibitem [{\citenamefont {Struck}\ \emph {et~al.}(2013)\citenamefont {Struck},
  \citenamefont {Weinberg}, \citenamefont {{\"O}lschl{\"a}ger}, \citenamefont
  {Windpassinger}, \citenamefont {Simonet}, \citenamefont {Sengstock},
  \citenamefont {Hoppner}, \citenamefont {Hauke}, \citenamefont {Eckardt},
  \citenamefont {Lewenstein},\ and\ \citenamefont {Mathey}}]{Struck2013}%
  \BibitemOpen
  \bibfield  {author} {\bibinfo {author} {\bibnamefont {Struck}, \bibfnamefont
  {J}}, \bibinfo {author} {\bibfnamefont {M}~\bibnamefont {Weinberg}}, \bibinfo
  {author} {\bibfnamefont {C}~\bibnamefont {{\"O}lschl{\"a}ger}}, \bibinfo
  {author} {\bibfnamefont {P}~\bibnamefont {Windpassinger}}, \bibinfo {author}
  {\bibfnamefont {J}~\bibnamefont {Simonet}}, \bibinfo {author} {\bibfnamefont
  {K.}~\bibnamefont {Sengstock}}, \bibinfo {author} {\bibfnamefont
  {R}~\bibnamefont {Hoppner}}, \bibinfo {author} {\bibfnamefont
  {P}~\bibnamefont {Hauke}}, \bibinfo {author} {\bibfnamefont {A}~\bibnamefont
  {Eckardt}}, \bibinfo {author} {\bibfnamefont {M.}~\bibnamefont {Lewenstein}},
  \ and\ \bibinfo {author} {\bibfnamefont {L}~\bibnamefont {Mathey}}} (\bibinfo
  {year} {2013}),\ \bibfield  {title} {\enquote {\bibinfo {title} {{Engineering
  Ising-XY spin-models in a triangular lattice using tunable artificial gauge
  fields}},}\ }\href {http://dx.doi.org/10.1038/nphys2750} {\bibfield
  {journal} {\bibinfo  {journal} {Nat Phys}\ }\textbf {\bibinfo {volume}
  {9}}~(\bibinfo {number} {11}),\ \bibinfo {pages} {738--743}}\BibitemShut
  {NoStop}%
\bibitem [{\citenamefont {Stuhl}\ \emph {et~al.}(2015)\citenamefont {Stuhl},
  \citenamefont {Lu}, \citenamefont {Aycock}, \citenamefont {Genkina},\ and\
  \citenamefont {Spielman}}]{Stuhl:2015}%
  \BibitemOpen
  \bibfield  {author} {\bibinfo {author} {\bibnamefont {Stuhl}, \bibfnamefont
  {BK}}, \bibinfo {author} {\bibfnamefont {H-I}\ \bibnamefont {Lu}}, \bibinfo
  {author} {\bibfnamefont {LM}~\bibnamefont {Aycock}}, \bibinfo {author}
  {\bibfnamefont {D}~\bibnamefont {Genkina}}, \ and\ \bibinfo {author}
  {\bibfnamefont {IB}~\bibnamefont {Spielman}}} (\bibinfo {year} {2015}),\
  \bibfield  {title} {\enquote {\bibinfo {title} {Visualizing edge states with
  an atomic {B}ose gas in the quantum {H}all regime},}\ }\href {\doibase
  10.1126/science.aaa8515} {\bibfield  {journal} {\bibinfo  {journal}
  {Science}\ }\textbf {\bibinfo {volume} {349}}~(\bibinfo {number} {6255}),\
  \bibinfo {pages} {1514--1518}}\BibitemShut {NoStop}%
\bibitem [{\citenamefont {Su}\ \emph {et~al.}(1979)\citenamefont {Su},
  \citenamefont {Schrieffer},\ and\ \citenamefont {Heeger}}]{Su:1979}%
  \BibitemOpen
  \bibfield  {author} {\bibinfo {author} {\bibnamefont {Su}, \bibfnamefont
  {W~P}}, \bibinfo {author} {\bibfnamefont {J.~R.}\ \bibnamefont {Schrieffer}},
  \ and\ \bibinfo {author} {\bibfnamefont {A.~J.}\ \bibnamefont {Heeger}}}
  (\bibinfo {year} {1979}),\ \bibfield  {title} {\enquote {\bibinfo {title}
  {Solitons in polyacetylene},}\ }\href {\doibase 10.1103/PhysRevLett.42.1698}
  {\bibfield  {journal} {\bibinfo  {journal} {Phys. Rev. Lett.}\ }\textbf
  {\bibinfo {volume} {42}},\ \bibinfo {pages} {1698--1701}}\BibitemShut
  {NoStop}%
\bibitem [{\citenamefont {Sun}\ \emph {et~al.}(2017)\citenamefont {Sun},
  \citenamefont {Wang}, \citenamefont {Xu}, \citenamefont {Yi}, \citenamefont
  {Zhang}, \citenamefont {Wu}, \citenamefont {Deng}, \citenamefont {Liu},
  \citenamefont {Chen},\ and\ \citenamefont {Pan}}]{Sun2017}%
  \BibitemOpen
  \bibfield  {author} {\bibinfo {author} {\bibnamefont {Sun}, \bibfnamefont
  {Wei}}, \bibinfo {author} {\bibfnamefont {Bao-Zong}\ \bibnamefont {Wang}},
  \bibinfo {author} {\bibfnamefont {Xiao-Tian}\ \bibnamefont {Xu}}, \bibinfo
  {author} {\bibfnamefont {Chang-Rui}\ \bibnamefont {Yi}}, \bibinfo {author}
  {\bibfnamefont {Long}\ \bibnamefont {Zhang}}, \bibinfo {author}
  {\bibfnamefont {Zhan}\ \bibnamefont {Wu}}, \bibinfo {author} {\bibfnamefont
  {Youjin}\ \bibnamefont {Deng}}, \bibinfo {author} {\bibfnamefont {Xiong-Jun}\
  \bibnamefont {Liu}}, \bibinfo {author} {\bibfnamefont {Shuai}\ \bibnamefont
  {Chen}}, \ and\ \bibinfo {author} {\bibfnamefont {Jian-Wei}\ \bibnamefont
  {Pan}}} (\bibinfo {year} {2017}),\ \bibfield  {title} {\enquote {\bibinfo
  {title} {Long-lived 2d spin-orbit coupled topological {B}ose gas},}\
  }\href@noop {} {\bibfield  {journal} {\bibinfo  {journal} {ArXiv eprints}\
  }}\Eprint {http://arxiv.org/abs/1710.00717} {arXiv:1710.00717} \BibitemShut
  {NoStop}%
\bibitem [{\citenamefont {Sun}\ \emph {et~al.}(2018)\citenamefont {Sun},
  \citenamefont {Yi}, \citenamefont {Wang}, \citenamefont {Zhang},
  \citenamefont {Sanders}, \citenamefont {Xu}, \citenamefont {Wang},
  \citenamefont {Schmiedmayer}, \citenamefont {Deng}, \citenamefont {Liu} \emph
  {et~al.}}]{sun2018uncover}%
  \BibitemOpen
  \bibfield  {author} {\bibinfo {author} {\bibnamefont {Sun}, \bibfnamefont
  {Wei}}, \bibinfo {author} {\bibfnamefont {Chang-Rui}\ \bibnamefont {Yi}},
  \bibinfo {author} {\bibfnamefont {Bao-Zong}\ \bibnamefont {Wang}}, \bibinfo
  {author} {\bibfnamefont {Wei-Wei}\ \bibnamefont {Zhang}}, \bibinfo {author}
  {\bibfnamefont {Barry~C}\ \bibnamefont {Sanders}}, \bibinfo {author}
  {\bibfnamefont {Xiao-Tian}\ \bibnamefont {Xu}}, \bibinfo {author}
  {\bibfnamefont {Zong-Yao}\ \bibnamefont {Wang}}, \bibinfo {author}
  {\bibfnamefont {J{\"o}rg}\ \bibnamefont {Schmiedmayer}}, \bibinfo {author}
  {\bibfnamefont {Youjin}\ \bibnamefont {Deng}}, \bibinfo {author}
  {\bibfnamefont {Xiong-Jun}\ \bibnamefont {Liu}},  \emph {et~al.}} (\bibinfo
  {year} {2018}),\ \bibfield  {title} {\enquote {\bibinfo {title} {Uncover
  topology by quantum quench dynamics},}\ }\href@noop {} {\bibfield  {journal}
  {\bibinfo  {journal} {ArXiv e-prints}\ }}\Eprint
  {http://arxiv.org/abs/1804.08226} {arXiv:1804.08226} \BibitemShut {NoStop}%
\bibitem [{\citenamefont {Taddia}\ \emph {et~al.}(2017)\citenamefont {Taddia},
  \citenamefont {Cornfeld}, \citenamefont {Rossini}, \citenamefont {Mazza},
  \citenamefont {Sela},\ and\ \citenamefont {Fazio}}]{Taddia:2017}%
  \BibitemOpen
  \bibfield  {author} {\bibinfo {author} {\bibnamefont {Taddia}, \bibfnamefont
  {Luca}}, \bibinfo {author} {\bibfnamefont {Eyal}\ \bibnamefont {Cornfeld}},
  \bibinfo {author} {\bibfnamefont {Davide}\ \bibnamefont {Rossini}}, \bibinfo
  {author} {\bibfnamefont {Leonardo}\ \bibnamefont {Mazza}}, \bibinfo {author}
  {\bibfnamefont {Eran}\ \bibnamefont {Sela}}, \ and\ \bibinfo {author}
  {\bibfnamefont {Rosario}\ \bibnamefont {Fazio}}} (\bibinfo {year} {2017}),\
  \bibfield  {title} {\enquote {\bibinfo {title} {Topological fractional
  pumping with alkaline-earth-like atoms in synthetic lattices},}\ }\href
  {\doibase 10.1103/PhysRevLett.118.230402} {\bibfield  {journal} {\bibinfo
  {journal} {Phys. Rev. Lett.}\ }\textbf {\bibinfo {volume} {118}},\ \bibinfo
  {pages} {230402}}\BibitemShut {NoStop}%
\bibitem [{\citenamefont {Tai}\ \emph {et~al.}(2017)\citenamefont {Tai},
  \citenamefont {Lukin}, \citenamefont {Rispoli}, \citenamefont {Schittko},
  \citenamefont {Menke}, \citenamefont {Borgnia}, \citenamefont {Preiss},
  \citenamefont {Grusdt}, \citenamefont {Kaufman},\ and\ \citenamefont
  {Greiner}}]{tai2017}%
  \BibitemOpen
  \bibfield  {author} {\bibinfo {author} {\bibnamefont {Tai}, \bibfnamefont
  {M~Eric}}, \bibinfo {author} {\bibfnamefont {Alexander}\ \bibnamefont
  {Lukin}}, \bibinfo {author} {\bibfnamefont {Matthew}\ \bibnamefont
  {Rispoli}}, \bibinfo {author} {\bibfnamefont {Robert}\ \bibnamefont
  {Schittko}}, \bibinfo {author} {\bibfnamefont {Tim}\ \bibnamefont {Menke}},
  \bibinfo {author} {\bibfnamefont {Dan}\ \bibnamefont {Borgnia}}, \bibinfo
  {author} {\bibfnamefont {Philipp~M.}\ \bibnamefont {Preiss}}, \bibinfo
  {author} {\bibfnamefont {Fabian}\ \bibnamefont {Grusdt}}, \bibinfo {author}
  {\bibfnamefont {Adam~M.}\ \bibnamefont {Kaufman}}, \ and\ \bibinfo {author}
  {\bibfnamefont {Markus}\ \bibnamefont {Greiner}}} (\bibinfo {year} {2017}),\
  \bibfield  {title} {\enquote {\bibinfo {title} {Microscopy of the interacting
  {H}arper--{H}ofstadter model in the two-body limit},}\ }\href
  {http://dx.doi.org/10.1038/nature22811} {\bibfield  {journal} {\bibinfo
  {journal} {Nature}\ }\textbf {\bibinfo {volume} {546}},\ \bibinfo {pages}
  {519 -- 523}}\BibitemShut {NoStop}%
\bibitem [{\citenamefont {Tanese}\ \emph {et~al.}(2014)\citenamefont {Tanese},
  \citenamefont {Gurevich}, \citenamefont {Baboux}, \citenamefont {Jacqmin},
  \citenamefont {Lema{\^\i}tre}, \citenamefont {Galopin}, \citenamefont
  {Sagnes}, \citenamefont {Amo}, \citenamefont {Bloch},\ and\ \citenamefont
  {Akkermans}}]{tanese2014fractal}%
  \BibitemOpen
  \bibfield  {author} {\bibinfo {author} {\bibnamefont {Tanese}, \bibfnamefont
  {Dimitrii}}, \bibinfo {author} {\bibfnamefont {Evgeni}\ \bibnamefont
  {Gurevich}}, \bibinfo {author} {\bibfnamefont {Florent}\ \bibnamefont
  {Baboux}}, \bibinfo {author} {\bibfnamefont {Thibaut}\ \bibnamefont
  {Jacqmin}}, \bibinfo {author} {\bibfnamefont {Aristide}\ \bibnamefont
  {Lema{\^\i}tre}}, \bibinfo {author} {\bibfnamefont {Elisabeth}\ \bibnamefont
  {Galopin}}, \bibinfo {author} {\bibfnamefont {Isabelle}\ \bibnamefont
  {Sagnes}}, \bibinfo {author} {\bibfnamefont {Alberto}\ \bibnamefont {Amo}},
  \bibinfo {author} {\bibfnamefont {Jacqueline}\ \bibnamefont {Bloch}}, \ and\
  \bibinfo {author} {\bibfnamefont {Eric}\ \bibnamefont {Akkermans}}} (\bibinfo
  {year} {2014}),\ \bibfield  {title} {\enquote {\bibinfo {title} {Fractal
  energy spectrum of a polariton gas in a {F}ibonacci quasiperiodic
  potential},}\ }\href {\doibase 10.1103/PhysRevLett.112.146404} {\bibfield
  {journal} {\bibinfo  {journal} {Phys. Rev. Lett}\ }\textbf {\bibinfo {volume}
  {112}}~(\bibinfo {number} {14}),\ \bibinfo {pages} {146404}}\BibitemShut
  {NoStop}%
\bibitem [{\citenamefont {{Tarnowski}}\ \emph {et~al.}(2017)\citenamefont
  {{Tarnowski}}, \citenamefont {{Nur {\"U}nal}}, \citenamefont
  {{Fl{\"a}schner}}, \citenamefont {{Rem}}, \citenamefont {{Eckardt}},
  \citenamefont {{Sengstock}},\ and\ \citenamefont
  {{Weitenberg}}}]{hamburglinking}%
  \BibitemOpen
  \bibfield  {author} {\bibinfo {author} {\bibnamefont {{Tarnowski}},
  \bibfnamefont {M}}, \bibinfo {author} {\bibfnamefont {F.}~\bibnamefont {{Nur
  {\"U}nal}}}, \bibinfo {author} {\bibfnamefont {N.}~\bibnamefont
  {{Fl{\"a}schner}}}, \bibinfo {author} {\bibfnamefont {B.~S.}\ \bibnamefont
  {{Rem}}}, \bibinfo {author} {\bibfnamefont {A.}~\bibnamefont {{Eckardt}}},
  \bibinfo {author} {\bibfnamefont {K.}~\bibnamefont {{Sengstock}}}, \ and\
  \bibinfo {author} {\bibfnamefont {C.}~\bibnamefont {{Weitenberg}}}} (\bibinfo
  {year} {2017}),\ \bibfield  {title} {\enquote {\bibinfo {title}
  {{Characterizing topology by dynamics: {C}hern number from linking
  number}},}\ }\href@noop {} {\bibfield  {journal} {\bibinfo  {journal} {ArXiv
  e-prints}\ }}\Eprint {http://arxiv.org/abs/1709.01046} {arXiv:1709.01046}
  \BibitemShut {NoStop}%
\bibitem [{\citenamefont {Tarruell}\ \emph {et~al.}(2012)\citenamefont
  {Tarruell}, \citenamefont {Greif}, \citenamefont {Uehlinger}, \citenamefont
  {Jotzu},\ and\ \citenamefont {Esslinger}}]{Tarruell:2012}%
  \BibitemOpen
  \bibfield  {author} {\bibinfo {author} {\bibnamefont {Tarruell},
  \bibfnamefont {Leticia}}, \bibinfo {author} {\bibfnamefont {Daniel}\
  \bibnamefont {Greif}}, \bibinfo {author} {\bibfnamefont {Thomas}\
  \bibnamefont {Uehlinger}}, \bibinfo {author} {\bibfnamefont {Gregor}\
  \bibnamefont {Jotzu}}, \ and\ \bibinfo {author} {\bibfnamefont {Tilman}\
  \bibnamefont {Esslinger}}} (\bibinfo {year} {2012}),\ \bibfield  {title}
  {\enquote {\bibinfo {title} {Creating, moving and merging {D}irac points with
  a {F}ermi gas in a tunable honeycomb lattice},}\ }\href {\doibase
  10.1038/nature10871} {\bibfield  {journal} {\bibinfo  {journal} {Nature}\
  }\textbf {\bibinfo {volume} {483}},\ \bibinfo {pages} {302}}\BibitemShut
  {NoStop}%
\bibitem [{\citenamefont {Thouless}(1983)}]{Thouless:1983}%
  \BibitemOpen
  \bibfield  {author} {\bibinfo {author} {\bibnamefont {Thouless},
  \bibfnamefont {D~J}}} (\bibinfo {year} {1983}),\ \bibfield  {title} {\enquote
  {\bibinfo {title} {Quantization of particle transport},}\ }\href {\doibase
  10.1103/PhysRevB.27.6083} {\bibfield  {journal} {\bibinfo  {journal} {Phys.
  Rev. B}\ }\textbf {\bibinfo {volume} {27}},\ \bibinfo {pages}
  {6083--6087}}\BibitemShut {NoStop}%
\bibitem [{\citenamefont {Thouless}(1984)}]{thoulesswannier}%
  \BibitemOpen
  \bibfield  {author} {\bibinfo {author} {\bibnamefont {Thouless},
  \bibfnamefont {D~J}}} (\bibinfo {year} {1984}),\ \bibfield  {title} {\enquote
  {\bibinfo {title} {Wannier functions for magnetic sub-bands},}\ }\href
  {\doibase 0.1088/0022-3719/17/12/003} {\bibfield  {journal} {\bibinfo
  {journal} {Journal of Physics C: Solid State Physics}\ }\textbf {\bibinfo
  {volume} {17}}~(\bibinfo {number} {12}),\ \bibinfo {pages}
  {L325}}\BibitemShut {NoStop}%
\bibitem [{\citenamefont {Thouless}\ \emph {et~al.}(1982)\citenamefont
  {Thouless}, \citenamefont {Kohmoto}, \citenamefont {Nightingale},\ and\
  \citenamefont {den Nijs}}]{thoulesschern}%
  \BibitemOpen
  \bibfield  {author} {\bibinfo {author} {\bibnamefont {Thouless},
  \bibfnamefont {D~J}}, \bibinfo {author} {\bibfnamefont {M.}~\bibnamefont
  {Kohmoto}}, \bibinfo {author} {\bibfnamefont {M.~P.}\ \bibnamefont
  {Nightingale}}, \ and\ \bibinfo {author} {\bibfnamefont {M.}~\bibnamefont
  {den Nijs}}} (\bibinfo {year} {1982}),\ \bibfield  {title} {\enquote
  {\bibinfo {title} {Quantized {H}all conductance in a two-dimensional periodic
  potential},}\ }\href {\doibase 10.1103/PhysRevLett.49.405} {\bibfield
  {journal} {\bibinfo  {journal} {Phys. Rev. Lett.}\ }\textbf {\bibinfo
  {volume} {49}}~(\bibinfo {number} {6}),\ \bibinfo {pages}
  {405--408}}\BibitemShut {NoStop}%
\bibitem [{\citenamefont {Tran}\ \emph {et~al.}(2017)\citenamefont {Tran},
  \citenamefont {Dauphin}, \citenamefont {Grushin}, \citenamefont {Zoller},\
  and\ \citenamefont {Goldman}}]{Tran:2017}%
  \BibitemOpen
  \bibfield  {author} {\bibinfo {author} {\bibnamefont {Tran}, \bibfnamefont
  {Duc~Thanh}}, \bibinfo {author} {\bibfnamefont {Alexandre}\ \bibnamefont
  {Dauphin}}, \bibinfo {author} {\bibfnamefont {Adolfo~G}\ \bibnamefont
  {Grushin}}, \bibinfo {author} {\bibfnamefont {Peter}\ \bibnamefont {Zoller}},
  \ and\ \bibinfo {author} {\bibfnamefont {Nathan}\ \bibnamefont {Goldman}}}
  (\bibinfo {year} {2017}),\ \bibfield  {title} {\enquote {\bibinfo {title}
  {Probing topology by `heating': Quantized circular dichroism in ultracold
  atoms},}\ }\href {\doibase 10.1126/sciadv.1701207} {\bibfield  {journal}
  {\bibinfo  {journal} {Science advances}\ }\textbf {\bibinfo {volume}
  {3}}~(\bibinfo {number} {8}),\ \bibinfo {pages} {e1701207}}\BibitemShut
  {NoStop}%
\bibitem [{\citenamefont {Trotzky}\ \emph {et~al.}(2008)\citenamefont
  {Trotzky}, \citenamefont {Cheinet}, \citenamefont {F{\"o}lling},
  \citenamefont {Feld}, \citenamefont {Schnorrberger}, \citenamefont {Rey},
  \citenamefont {Polkovnikov}, \citenamefont {Demler}, \citenamefont {Lukin},\
  and\ \citenamefont {Bloch}}]{Trotzky2008}%
  \BibitemOpen
  \bibfield  {author} {\bibinfo {author} {\bibnamefont {Trotzky}, \bibfnamefont
  {S}}, \bibinfo {author} {\bibfnamefont {P.}~\bibnamefont {Cheinet}}, \bibinfo
  {author} {\bibfnamefont {S.}~\bibnamefont {F{\"o}lling}}, \bibinfo {author}
  {\bibfnamefont {M.}~\bibnamefont {Feld}}, \bibinfo {author} {\bibfnamefont
  {U.}~\bibnamefont {Schnorrberger}}, \bibinfo {author} {\bibfnamefont {A.~M.}\
  \bibnamefont {Rey}}, \bibinfo {author} {\bibfnamefont {A.}~\bibnamefont
  {Polkovnikov}}, \bibinfo {author} {\bibfnamefont {E.~A.}\ \bibnamefont
  {Demler}}, \bibinfo {author} {\bibfnamefont {M.~D.}\ \bibnamefont {Lukin}}, \
  and\ \bibinfo {author} {\bibfnamefont {I.}~\bibnamefont {Bloch}}} (\bibinfo
  {year} {2008}),\ \bibfield  {title} {\enquote {\bibinfo {title}
  {{Time-Resolved Observation and Control of Superexchange Interactions with
  Ultracold Atoms in Optical Lattices}},}\ }\href {\doibase
  0.1126/science.1150841} {\bibfield  {journal} {\bibinfo  {journal} {Science}\
  }\textbf {\bibinfo {volume} {319}}~(\bibinfo {number} {5861}),\ \bibinfo
  {pages} {295--299}}\BibitemShut {NoStop}%
\bibitem [{\citenamefont {Uhlmann}(1986)}]{uhlmann1986parallel}%
  \BibitemOpen
  \bibfield  {author} {\bibinfo {author} {\bibnamefont {Uhlmann}, \bibfnamefont
  {Armin}}} (\bibinfo {year} {1986}),\ \bibfield  {title} {\enquote {\bibinfo
  {title} {Parallel transport and ``quantum holonomy'' along density
  operators},}\ }\href {\doibase 10.1016/0034-4877(86)90055-8} {\bibfield
  {journal} {\bibinfo  {journal} {Reports on Mathematical Physics}\ }\textbf
  {\bibinfo {volume} {24}}~(\bibinfo {number} {2}),\ \bibinfo {pages}
  {229--240}}\BibitemShut {NoStop}%
\bibitem [{\citenamefont {Vanhala}\ \emph {et~al.}(2016)\citenamefont
  {Vanhala}, \citenamefont {Siro}, \citenamefont {Liang}, \citenamefont
  {Troyer}, \citenamefont {Harju},\ and\ \citenamefont
  {T{\"o}rm{\"a}}}]{PhysRevLett.116.225305}%
  \BibitemOpen
  \bibfield  {author} {\bibinfo {author} {\bibnamefont {Vanhala}, \bibfnamefont
  {Tuomas~I}}, \bibinfo {author} {\bibfnamefont {Topi}\ \bibnamefont {Siro}},
  \bibinfo {author} {\bibfnamefont {Long}\ \bibnamefont {Liang}}, \bibinfo
  {author} {\bibfnamefont {Matthias}\ \bibnamefont {Troyer}}, \bibinfo {author}
  {\bibfnamefont {Ari}\ \bibnamefont {Harju}}, \ and\ \bibinfo {author}
  {\bibfnamefont {P{\"a}ivi}\ \bibnamefont {T{\"o}rm{\"a}}}} (\bibinfo {year}
  {2016}),\ \bibfield  {title} {\enquote {\bibinfo {title} {Topological phase
  transitions in the repulsively interacting {H}aldane-{H}ubbard model},}\
  }\href {\doibase 10.1103/PhysRevLett.116.225305} {\bibfield  {journal}
  {\bibinfo  {journal} {Phys. Rev. Lett.}\ }\textbf {\bibinfo {volume} {116}},\
  \bibinfo {pages} {225305}}\BibitemShut {NoStop}%
\bibitem [{\citenamefont {Vasi{\'c}}\ \emph {et~al.}(2015)\citenamefont
  {Vasi{\'c}}, \citenamefont {Petrescu}, \citenamefont {Le~Hur},\ and\
  \citenamefont {Hofstetter}}]{PhysRevB.91.094502}%
  \BibitemOpen
  \bibfield  {author} {\bibinfo {author} {\bibnamefont {Vasi{\'c}},
  \bibfnamefont {Ivana}}, \bibinfo {author} {\bibfnamefont {Alexandru}\
  \bibnamefont {Petrescu}}, \bibinfo {author} {\bibfnamefont {Karyn}\
  \bibnamefont {Le~Hur}}, \ and\ \bibinfo {author} {\bibfnamefont {Walter}\
  \bibnamefont {Hofstetter}}} (\bibinfo {year} {2015}),\ \bibfield  {title}
  {\enquote {\bibinfo {title} {Chiral bosonic phases on the {H}aldane honeycomb
  lattice},}\ }\href {\doibase 10.1103/PhysRevB.91.094502} {\bibfield
  {journal} {\bibinfo  {journal} {Phys. Rev. B}\ }\textbf {\bibinfo {volume}
  {91}}~(\bibinfo {number} {9}),\ \bibinfo {pages} {094502}}\BibitemShut
  {NoStop}%
\bibitem [{\citenamefont {Viyuela}\ \emph {et~al.}(2014)\citenamefont
  {Viyuela}, \citenamefont {Rivas},\ and\ \citenamefont
  {Martin-Delgado}}]{viyuela2014two}%
  \BibitemOpen
  \bibfield  {author} {\bibinfo {author} {\bibnamefont {Viyuela}, \bibfnamefont
  {O}}, \bibinfo {author} {\bibfnamefont {A}~\bibnamefont {Rivas}}, \ and\
  \bibinfo {author} {\bibfnamefont {MA}~\bibnamefont {Martin-Delgado}}}
  (\bibinfo {year} {2014}),\ \bibfield  {title} {\enquote {\bibinfo {title}
  {Two-dimensional density-matrix topological fermionic phases: Topological
  {U}hlmann numbers},}\ }\href {\doibase 10.1103/PhysRevLett.113.076408}
  {\bibfield  {journal} {\bibinfo  {journal} {Phys. Rev. Lett.}\ }\textbf
  {\bibinfo {volume} {113}}~(\bibinfo {number} {7}),\ \bibinfo {pages}
  {076408}}\BibitemShut {NoStop}%
\bibitem [{\citenamefont {Volovik}(2003)}]{volovik}%
  \BibitemOpen
  \bibfield  {author} {\bibinfo {author} {\bibnamefont {Volovik}, \bibfnamefont
  {G~E}}} (\bibinfo {year} {2003}),\ \href@noop {} {\emph {\bibinfo {title}
  {The Universe in a Helium Droplet}}}\ (\bibinfo  {publisher} {Oxford
  University Press})\BibitemShut {NoStop}%
\bibitem [{\citenamefont {Wang}\ \emph {et~al.}(2017)\citenamefont {Wang},
  \citenamefont {Zhang}, \citenamefont {Chen}, \citenamefont {Yu},\ and\
  \citenamefont {Zhai}}]{theorylinking}%
  \BibitemOpen
  \bibfield  {author} {\bibinfo {author} {\bibnamefont {Wang}, \bibfnamefont
  {Ce}}, \bibinfo {author} {\bibfnamefont {Pengfei}\ \bibnamefont {Zhang}},
  \bibinfo {author} {\bibfnamefont {Xin}\ \bibnamefont {Chen}}, \bibinfo
  {author} {\bibfnamefont {Jinlong}\ \bibnamefont {Yu}}, \ and\ \bibinfo
  {author} {\bibfnamefont {Hui}\ \bibnamefont {Zhai}}} (\bibinfo {year}
  {2017}),\ \bibfield  {title} {\enquote {\bibinfo {title} {Scheme to measure
  the topological number of a {C}hern insulator from quench dynamics},}\ }\href
  {\doibase 10.1103/PhysRevLett.118.185701} {\bibfield  {journal} {\bibinfo
  {journal} {Phys. Rev. Lett.}\ }\textbf {\bibinfo {volume} {118}},\ \bibinfo
  {pages} {185701}}\BibitemShut {NoStop}%
\bibitem [{\citenamefont {Wang}\ \emph {et~al.}(2014)\citenamefont {Wang},
  \citenamefont {Deng},\ and\ \citenamefont {Duan}}]{PhysRevLett.113.033002}%
  \BibitemOpen
  \bibfield  {author} {\bibinfo {author} {\bibnamefont {Wang}, \bibfnamefont
  {S-T}}, \bibinfo {author} {\bibfnamefont {D.-L.}\ \bibnamefont {Deng}}, \
  and\ \bibinfo {author} {\bibfnamefont {L.-M.}\ \bibnamefont {Duan}}}
  (\bibinfo {year} {2014}),\ \bibfield  {title} {\enquote {\bibinfo {title}
  {Probe of three-dimensional chiral topological insulators in an optical
  lattice},}\ }\href {\doibase 10.1103/PhysRevLett.113.033002} {\bibfield
  {journal} {\bibinfo  {journal} {Phys. Rev. Lett.}\ }\textbf {\bibinfo
  {volume} {113}},\ \bibinfo {pages} {033002}}\BibitemShut {NoStop}%
\bibitem [{\citenamefont {Weinberg}\ \emph {et~al.}(2015)\citenamefont
  {Weinberg}, \citenamefont {\"Olschl\"ager}, \citenamefont {Str\"ater},
  \citenamefont {Prelle}, \citenamefont {Eckardt}, \citenamefont {Sengstock},\
  and\ \citenamefont {Simonet}}]{PhysRevA.92.043621}%
  \BibitemOpen
  \bibfield  {author} {\bibinfo {author} {\bibnamefont {Weinberg},
  \bibfnamefont {M}}, \bibinfo {author} {\bibfnamefont {C.}~\bibnamefont
  {\"Olschl\"ager}}, \bibinfo {author} {\bibfnamefont {C.}~\bibnamefont
  {Str\"ater}}, \bibinfo {author} {\bibfnamefont {S.}~\bibnamefont {Prelle}},
  \bibinfo {author} {\bibfnamefont {A.}~\bibnamefont {Eckardt}}, \bibinfo
  {author} {\bibfnamefont {K.}~\bibnamefont {Sengstock}}, \ and\ \bibinfo
  {author} {\bibfnamefont {J.}~\bibnamefont {Simonet}}} (\bibinfo {year}
  {2015}),\ \bibfield  {title} {\enquote {\bibinfo {title} {Multiphoton
  interband excitations of quantum gases in driven optical lattices},}\ }\href
  {\doibase 10.1103/PhysRevA.92.043621} {\bibfield  {journal} {\bibinfo
  {journal} {Phys. Rev. A}\ }\textbf {\bibinfo {volume} {92}},\ \bibinfo
  {pages} {043621}}\BibitemShut {NoStop}%
\bibitem [{\citenamefont {Wen}(1995)}]{wenedge}%
  \BibitemOpen
  \bibfield  {author} {\bibinfo {author} {\bibnamefont {Wen}, \bibfnamefont
  {Xiao-Gang}}} (\bibinfo {year} {1995}),\ \bibfield  {title} {\enquote
  {\bibinfo {title} {Topological orders and edge excitations in fractional
  quantum {H}all states},}\ }\href {\doibase 10.1080/00018739500101566}
  {\bibfield  {journal} {\bibinfo  {journal} {Advances in Physics}\ }\textbf
  {\bibinfo {volume} {44}}~(\bibinfo {number} {5}),\ \bibinfo {pages}
  {405--473}}\BibitemShut {NoStop}%
\bibitem [{\citenamefont {Williams}\ \emph {et~al.}(2012)\citenamefont
  {Williams}, \citenamefont {LeBlanc}, \citenamefont {Jimenez-Garcia},
  \citenamefont {Beeler}, \citenamefont {Perry}, \citenamefont {Phillips},\
  and\ \citenamefont {Spielman}}]{Williams20012012}%
  \BibitemOpen
  \bibfield  {author} {\bibinfo {author} {\bibnamefont {Williams},
  \bibfnamefont {R~A}}, \bibinfo {author} {\bibfnamefont {L.~J.}\ \bibnamefont
  {LeBlanc}}, \bibinfo {author} {\bibfnamefont {K.}~\bibnamefont
  {Jimenez-Garcia}}, \bibinfo {author} {\bibfnamefont {M.~C.}\ \bibnamefont
  {Beeler}}, \bibinfo {author} {\bibfnamefont {A.~R.}\ \bibnamefont {Perry}},
  \bibinfo {author} {\bibfnamefont {W.~D.}\ \bibnamefont {Phillips}}, \ and\
  \bibinfo {author} {\bibfnamefont {I.~B.}\ \bibnamefont {Spielman}}} (\bibinfo
  {year} {2012}),\ \bibfield  {title} {\enquote {\bibinfo {title} {Synthetic
  partial waves in ultracold atomic collisions},}\ }\href {\doibase
  10.1126/science.1212652} {\bibfield  {journal} {\bibinfo  {journal}
  {Science}\ }\textbf {\bibinfo {volume} {335}}~(\bibinfo {number} {6066}),\
  \bibinfo {pages} {314--317}}\BibitemShut {NoStop}%
\bibitem [{\citenamefont {Wu}\ \emph {et~al.}(2016)\citenamefont {Wu},
  \citenamefont {Zhang}, \citenamefont {Sun}, \citenamefont {Xu}, \citenamefont
  {Wang}, \citenamefont {Ji}, \citenamefont {Deng}, \citenamefont {Chen},
  \citenamefont {Liu},\ and\ \citenamefont {Pan}}]{Wu:2015}%
  \BibitemOpen
  \bibfield  {author} {\bibinfo {author} {\bibnamefont {Wu}, \bibfnamefont
  {Zhan}}, \bibinfo {author} {\bibfnamefont {Long}\ \bibnamefont {Zhang}},
  \bibinfo {author} {\bibfnamefont {Wei}\ \bibnamefont {Sun}}, \bibinfo
  {author} {\bibfnamefont {Xiao-Tian}\ \bibnamefont {Xu}}, \bibinfo {author}
  {\bibfnamefont {Bao-Zong}\ \bibnamefont {Wang}}, \bibinfo {author}
  {\bibfnamefont {Si-Cong}\ \bibnamefont {Ji}}, \bibinfo {author}
  {\bibfnamefont {Youjin}\ \bibnamefont {Deng}}, \bibinfo {author}
  {\bibfnamefont {Shuai}\ \bibnamefont {Chen}}, \bibinfo {author}
  {\bibfnamefont {Xiong-Jun}\ \bibnamefont {Liu}}, \ and\ \bibinfo {author}
  {\bibfnamefont {Jian-Wei}\ \bibnamefont {Pan}}} (\bibinfo {year} {2016}),\
  \bibfield  {title} {\enquote {\bibinfo {title} {Realization of
  two-dimensional spin-orbit coupling for {B}ose-{E}instein condensates},}\
  }\href {\doibase 10.1126/science.aaf6689} {\bibfield  {journal} {\bibinfo
  {journal} {Science}\ }\textbf {\bibinfo {volume} {354}}~(\bibinfo {number}
  {6308}),\ \bibinfo {pages} {83--88}}\BibitemShut {NoStop}%
\bibitem [{\citenamefont {Wu}\ and\ \citenamefont {Bruun}(2016)}]{Wu2016}%
  \BibitemOpen
  \bibfield  {author} {\bibinfo {author} {\bibnamefont {Wu}, \bibfnamefont
  {Zhigang}}, \ and\ \bibinfo {author} {\bibfnamefont {G.~M.}\ \bibnamefont
  {Bruun}}} (\bibinfo {year} {2016}),\ \bibfield  {title} {\enquote {\bibinfo
  {title} {Topological superfluid in a {F}ermi-{B}ose mixture with a high
  critical temperature},}\ }\href {\doibase 10.1103/PhysRevLett.117.245302}
  {\bibfield  {journal} {\bibinfo  {journal} {Phys. Rev. Lett.}\ }\textbf
  {\bibinfo {volume} {117}},\ \bibinfo {pages} {245302}}\BibitemShut {NoStop}%
\bibitem [{\citenamefont {Xiao}\ \emph {et~al.}(2010)\citenamefont {Xiao},
  \citenamefont {Chang},\ and\ \citenamefont {Niu}}]{Xiao:2010}%
  \BibitemOpen
  \bibfield  {author} {\bibinfo {author} {\bibnamefont {Xiao}, \bibfnamefont
  {Di}}, \bibinfo {author} {\bibfnamefont {Ming-Che}\ \bibnamefont {Chang}}, \
  and\ \bibinfo {author} {\bibfnamefont {Qian}\ \bibnamefont {Niu}}} (\bibinfo
  {year} {2010}),\ \bibfield  {title} {\enquote {\bibinfo {title} {Berry phase
  effects on electronic properties},}\ }\href {\doibase
  10.1103/RevModPhys.82.1959} {\bibfield  {journal} {\bibinfo  {journal} {Rev.
  Mod. Phys.}\ }\textbf {\bibinfo {volume} {82}},\ \bibinfo {pages}
  {1959--2007}}\BibitemShut {NoStop}%
\bibitem [{\citenamefont {Xu}\ \emph {et~al.}(2018)\citenamefont {Xu},
  \citenamefont {Morong}, \citenamefont {Hui}, \citenamefont {Scarola},\ and\
  \citenamefont {DeMarco}}]{2017arXiv171102061X}%
  \BibitemOpen
  \bibfield  {author} {\bibinfo {author} {\bibnamefont {Xu}, \bibfnamefont
  {Wenchao}}, \bibinfo {author} {\bibfnamefont {William}\ \bibnamefont
  {Morong}}, \bibinfo {author} {\bibfnamefont {Hoi-Yin}\ \bibnamefont {Hui}},
  \bibinfo {author} {\bibfnamefont {Vito~W.}\ \bibnamefont {Scarola}}, \ and\
  \bibinfo {author} {\bibfnamefont {Brian}\ \bibnamefont {DeMarco}}} (\bibinfo
  {year} {2018}),\ \bibfield  {title} {\enquote {\bibinfo {title} {Correlated
  spin-flip tunneling in a fermi lattice gas},}\ }\href {\doibase
  10.1103/PhysRevA.98.023623} {\bibfield  {journal} {\bibinfo  {journal} {Phys.
  Rev. A}\ }\textbf {\bibinfo {volume} {98}},\ \bibinfo {pages}
  {023623}}\BibitemShut {NoStop}%
\bibitem [{\citenamefont {Yan}\ \emph {et~al.}(2015)\citenamefont {Yan},
  \citenamefont {Wan},\ and\ \citenamefont {Wang}}]{Yan2015}%
  \BibitemOpen
  \bibfield  {author} {\bibinfo {author} {\bibnamefont {Yan}, \bibfnamefont
  {Zhongbo}}, \bibinfo {author} {\bibfnamefont {Shaolong}\ \bibnamefont {Wan}},
  \ and\ \bibinfo {author} {\bibfnamefont {Zhong}\ \bibnamefont {Wang}}}
  (\bibinfo {year} {2015}),\ \bibfield  {title} {\enquote {\bibinfo {title}
  {Topological superfluid and majorana zero modes in synthetic dimension},}\
  }\href {http://dx.doi.org/10.1038/srep15927} {\bibfield  {journal} {\bibinfo
  {journal} {Scientific Reports}\ }\textbf {\bibinfo {volume} {5}},\ \bibinfo
  {pages} {15927}}\BibitemShut {NoStop}%
\bibitem [{\citenamefont {Zak}(1989)}]{Zak:1989}%
  \BibitemOpen
  \bibfield  {author} {\bibinfo {author} {\bibnamefont {Zak}, \bibfnamefont
  {J}}} (\bibinfo {year} {1989}),\ \bibfield  {title} {\enquote {\bibinfo
  {title} {Berry's phase for energy bands in solids},}\ }\href {\doibase
  10.1103/PhysRevLett.62.2747} {\bibfield  {journal} {\bibinfo  {journal}
  {Phys. Rev. Lett.}\ }\textbf {\bibinfo {volume} {62}},\ \bibinfo {pages}
  {2747--2750}}\BibitemShut {NoStop}%
\bibitem [{\citenamefont {Zaletel}\ \emph {et~al.}(2014)\citenamefont
  {Zaletel}, \citenamefont {Parameswaran}, \citenamefont {R\"uegg},\ and\
  \citenamefont {Altman}}]{PhysRevB.89.155142}%
  \BibitemOpen
  \bibfield  {author} {\bibinfo {author} {\bibnamefont {Zaletel}, \bibfnamefont
  {Michael~P}}, \bibinfo {author} {\bibfnamefont {S.~A.}\ \bibnamefont
  {Parameswaran}}, \bibinfo {author} {\bibfnamefont {Andreas}\ \bibnamefont
  {R\"uegg}}, \ and\ \bibinfo {author} {\bibfnamefont {Ehud}\ \bibnamefont
  {Altman}}} (\bibinfo {year} {2014}),\ \bibfield  {title} {\enquote {\bibinfo
  {title} {Chiral bosonic {M}ott insulator on the frustrated triangular
  lattice},}\ }\href {\doibase 10.1103/PhysRevB.89.155142} {\bibfield
  {journal} {\bibinfo  {journal} {Phys. Rev. B}\ }\textbf {\bibinfo {volume}
  {89}},\ \bibinfo {pages} {155142}}\BibitemShut {NoStop}%
\bibitem [{\citenamefont {Zeng}\ \emph {et~al.}(2015)\citenamefont {Zeng},
  \citenamefont {Wang},\ and\ \citenamefont {Zhai}}]{PhysRevLett.115.095302}%
  \BibitemOpen
  \bibfield  {author} {\bibinfo {author} {\bibnamefont {Zeng}, \bibfnamefont
  {Tian-Sheng}}, \bibinfo {author} {\bibfnamefont {Ce}~\bibnamefont {Wang}}, \
  and\ \bibinfo {author} {\bibfnamefont {Hui}\ \bibnamefont {Zhai}}} (\bibinfo
  {year} {2015}),\ \bibfield  {title} {\enquote {\bibinfo {title} {Charge
  pumping of interacting fermion atoms in the synthetic dimension},}\ }\href
  {\doibase 10.1103/PhysRevLett.115.095302} {\bibfield  {journal} {\bibinfo
  {journal} {Phys. Rev. Lett.}\ }\textbf {\bibinfo {volume} {115}},\ \bibinfo
  {pages} {095302}}\BibitemShut {NoStop}%
\bibitem [{\citenamefont {Zeuner}\ \emph {et~al.}(2015)\citenamefont {Zeuner},
  \citenamefont {Rechtsman}, \citenamefont {Plotnik}, \citenamefont {Lumer},
  \citenamefont {Nolte}, \citenamefont {Rudner}, \citenamefont {Segev},\ and\
  \citenamefont {Szameit}}]{zeuner2015observation}%
  \BibitemOpen
  \bibfield  {author} {\bibinfo {author} {\bibnamefont {Zeuner}, \bibfnamefont
  {Julia~M}}, \bibinfo {author} {\bibfnamefont {Mikael~C}\ \bibnamefont
  {Rechtsman}}, \bibinfo {author} {\bibfnamefont {Yonatan}\ \bibnamefont
  {Plotnik}}, \bibinfo {author} {\bibfnamefont {Yaakov}\ \bibnamefont {Lumer}},
  \bibinfo {author} {\bibfnamefont {Stefan}\ \bibnamefont {Nolte}}, \bibinfo
  {author} {\bibfnamefont {Mark~S}\ \bibnamefont {Rudner}}, \bibinfo {author}
  {\bibfnamefont {Mordechai}\ \bibnamefont {Segev}}, \ and\ \bibinfo {author}
  {\bibfnamefont {Alexander}\ \bibnamefont {Szameit}}} (\bibinfo {year}
  {2015}),\ \bibfield  {title} {\enquote {\bibinfo {title} {Observation of a
  topological transition in the bulk of a non-{H}ermitian system},}\ }\href
  {\doibase 10.1103/PhysRevLett.115.040402} {\bibfield  {journal} {\bibinfo
  {journal} {Phys. Rev. Lett.}\ }\textbf {\bibinfo {volume} {115}}~(\bibinfo
  {number} {4}),\ \bibinfo {pages} {040402}}\BibitemShut {NoStop}%
\bibitem [{\citenamefont {Zhai}(2015)}]{zhaireview}%
  \BibitemOpen
  \bibfield  {author} {\bibinfo {author} {\bibnamefont {Zhai}, \bibfnamefont
  {Hui}}} (\bibinfo {year} {2015}),\ \bibfield  {title} {\enquote {\bibinfo
  {title} {Degenerate quantum gases with spin-orbit coupling: a review},}\
  }\href {\doibase 10.1088/0034-4885/78/2/026001} {\bibfield  {journal}
  {\bibinfo  {journal} {Rep. Prog. Phys.}\ }\textbf {\bibinfo {volume}
  {78}}~(\bibinfo {number} {2}),\ \bibinfo {pages} {026001}}\BibitemShut
  {NoStop}%
\bibitem [{\citenamefont {Zhang}\ \emph {et~al.}(2008)\citenamefont {Zhang},
  \citenamefont {Tewari}, \citenamefont {Lutchyn},\ and\ \citenamefont
  {Das~Sarma}}]{zhang_prl_2008}%
  \BibitemOpen
  \bibfield  {author} {\bibinfo {author} {\bibnamefont {Zhang}, \bibfnamefont
  {Chuanwei}}, \bibinfo {author} {\bibfnamefont {Sumanta}\ \bibnamefont
  {Tewari}}, \bibinfo {author} {\bibfnamefont {Roman~M.}\ \bibnamefont
  {Lutchyn}}, \ and\ \bibinfo {author} {\bibfnamefont {S.}~\bibnamefont
  {Das~Sarma}}} (\bibinfo {year} {2008}),\ \bibfield  {title} {\enquote
  {\bibinfo {title} {${p}_{x}+i{p}_{y}$ superfluid from $s$-wave interactions
  of fermionic cold atoms},}\ }\href {\doibase 10.1103/PhysRevLett.101.160401}
  {\bibfield  {journal} {\bibinfo  {journal} {Phys. Rev. Lett.}\ }\textbf
  {\bibinfo {volume} {101}},\ \bibinfo {pages} {160401}}\BibitemShut {NoStop}%
\bibitem [{\citenamefont {Zhang}\ \emph {et~al.}(2010)\citenamefont {Zhang},
  \citenamefont {Jian}, \citenamefont {Ye},\ and\ \citenamefont
  {Zhai}}]{PhysRevLett.105.155302}%
  \BibitemOpen
  \bibfield  {author} {\bibinfo {author} {\bibnamefont {Zhang}, \bibfnamefont
  {Jian}}, \bibinfo {author} {\bibfnamefont {Chao-Ming}\ \bibnamefont {Jian}},
  \bibinfo {author} {\bibfnamefont {Fei}\ \bibnamefont {Ye}}, \ and\ \bibinfo
  {author} {\bibfnamefont {Hui}\ \bibnamefont {Zhai}}} (\bibinfo {year}
  {2010}),\ \bibfield  {title} {\enquote {\bibinfo {title} {Degeneracy of
  many-body quantum states in an optical lattice under a uniform magnetic
  field},}\ }\href {\doibase 10.1103/PhysRevLett.105.155302} {\bibfield
  {journal} {\bibinfo  {journal} {Phys. Rev. Lett.}\ }\textbf {\bibinfo
  {volume} {105}},\ \bibinfo {pages} {155302}}\BibitemShut {NoStop}%
\bibitem [{\citenamefont {Zhang}\ and\ \citenamefont {Hu}(2001)}]{Zhang4dqhe}%
  \BibitemOpen
  \bibfield  {author} {\bibinfo {author} {\bibnamefont {Zhang}, \bibfnamefont
  {Shou-Cheng}}, \ and\ \bibinfo {author} {\bibfnamefont {Jiangping}\
  \bibnamefont {Hu}}} (\bibinfo {year} {2001}),\ \bibfield  {title} {\enquote
  {\bibinfo {title} {A four-dimensional generalization of the quantum {H}all
  effect},}\ }\href {\doibase 10.1126/science.294.5543.823} {\bibfield
  {journal} {\bibinfo  {journal} {Science}\ }\textbf {\bibinfo {volume}
  {294}}~(\bibinfo {number} {5543}),\ \bibinfo {pages} {823--828}}\BibitemShut
  {NoStop}%
\bibitem [{\citenamefont {Zheng}\ \emph {et~al.}(2015)\citenamefont {Zheng},
  \citenamefont {Shen}, \citenamefont {Wang},\ and\ \citenamefont
  {Zhai}}]{PhysRevB.91.161107}%
  \BibitemOpen
  \bibfield  {author} {\bibinfo {author} {\bibnamefont {Zheng}, \bibfnamefont
  {Wei}}, \bibinfo {author} {\bibfnamefont {Huitao}\ \bibnamefont {Shen}},
  \bibinfo {author} {\bibfnamefont {Zhong}\ \bibnamefont {Wang}}, \ and\
  \bibinfo {author} {\bibfnamefont {Hui}\ \bibnamefont {Zhai}}} (\bibinfo
  {year} {2015}),\ \bibfield  {title} {\enquote {\bibinfo {title}
  {Magnetic-order-driven topological transition in the {H}aldane-{H}ubbard
  model},}\ }\href {\doibase 10.1103/PhysRevB.91.161107} {\bibfield  {journal}
  {\bibinfo  {journal} {Phys. Rev. B}\ }\textbf {\bibinfo {volume} {91}},\
  \bibinfo {pages} {161107}}\BibitemShut {NoStop}%
\end{thebibliography}

%

\end{document}